\documentclass[manuscript]{aastex}
\usepackage{natbib}
\shorttitle{Early infalling satellites in the ErisMod simulations}
\shortauthors{Tomozeiu et al.}

\begin{document}

\title{ The evolution of dwarf galaxy satellites with different dark matter density profiles in the ErisMod simulations. I. The early infalls}

\author{Mihai Tomozeiu, Lucio Mayer}
\affil{Institute for Computational Science, University of Zurich, Winterthurerstrasse 190, CH-8057 Z\"{u}rich, Switzerland}
\affil{Physics Institute, University of Zurich, Winterthurerstrasse 190, CH-8057 Z\"{u}rich, Switzerland}
\email{mihai@physik.uzh.ch}
\and
\author{Thomas Quinn}
\affil{ Astronomy Department, University of Washington, Box 351580, Seattle, WA 98195, USA}

%%%%%%%%%%%%%%%%%%%%%%%%%%%%%%%%%%%%%%%%%%%%%%%
%%%%%%%%%%%%%%%%%%%%%%%%%%%%%%%%%%%%%%%%%%%%%%%
%%%%%%%%%%%%%%%%%%%%%%%%%%%%%%%%%%%%%%%%%%%%%%%

\begin{abstract}

We present the first simulations of tidal stirring of dwarf galaxies in the Local Group carried out in a fully cosmological context. We use the ErisDARK cosmological simulation of a MW-sized galaxy
to identify some of the most massive subhalos ($M_{vir} > 10^8 M_{\odot}$) that fall into the main host before $z=2$.
Subhalos are replaced before infall with high-resolution models of dwarf galaxies comprising a faint stellar disk embedded in a dark matter halo. The set of models contains cuspy halos as well as halos with "cored" profiles (with asymptotic inner slope $\gamma = 0.6$) consistent with  recent results of hydrodynamical simulations of dwarf galaxy formation. The simulations are then run to $z=0$ with as many as 54 million particles and resolution as small as $\sim 4$ pc using the new parallel N-Body code ChaNGa. The stellar components of all satellites are significantly affected by tidal stirring, losing stellar mass and undergoing a morphological transformation towards a pressure supported spheroidal system. However, while some remnants with cuspy halos maintain significant rotational flattening and disk-like features, all the shallow halo models achieve $v/\sigma < 0.5$
and round shapes typical of dSph satellites of the MW and M31. Mass loss
is also enhanced in the latter, and remnants can reach luminosities and velocity dispersions as low as those of Ultra Faint Dwarfs (UFDs)
In two cases a merger transforms the objects quickly into a spheroidal system already before infall.
We argue that cuspy progenitors must be the exception rather than the rule among satellites of the MW
since all the MW and M31 satellites in the luminosity range of our remnants are dSphs, a result matched only in the simulation with "cored" models. This suggests an early and efficient transformation of cusps into cores for Local Group satellites.

\end{abstract}


%%%%%%%%%%%%%%%%%%%%%%%%%%%%%%%%%%%%%%%%%%%%%%%
%%%%%%%%%%%%%%%%%%%%%%%%%%%%%%%%%%%%%%%%%%%%%%%
%%%%%%%%%%%%%%%%%%%%%%%%%%%%%%%%%%%%%%%%%%%%%%% 

\section{Introduction}
Dwarf galaxy satellites of the Milky Way (MW) and M31 are the main focus of "near-field cosmology", providing
a plethora of useful constraints and tests for galaxy formation theories as well as for 
the nature of dark matter in the $\Lambda$CDM paradigm. Yet their origin is still
not completely understood. Nearly all the known satellites, 
located within the expected virial radius of the bright spirals, i.e at $R < 300$ kpc,
are classified as dwarf spheroidal galaxies (dSphs).
They have fairly round stellar components with low or negligible rotation but relatively high
velocity dispersions (\cite{1998ARA&A..36..435M}; \cite{2012AJ....144....4M}). 
At the same time, nearly all of the dwarf galaxies located outside such radius and far from M31 
are gas-rich dwarf irregular galaxies (dIrrs), which have ongoing or
recent star formation, flattened disky shapes, and exhibit substantial rotational support ($v_{rot}/\sigma > 1$).
The different spatial distribution of dwarf galaxy types is known as the morphology-density relation, and reflects
an analogous effect seen in galaxy clusters and rich groups on larger mass scales (\cite{1980ApJ...236..351D}).

Environmental mechanisms have been often invoked 
to transform dIrrs into dSphs (\cite{1998ARA&A..36..435M}; \cite{2012AJ....144....4M}). One of the mechanisms proposed to 
achieve such transformation is tidal stirring, namely 
the cumulative effect of repeated tidal shocks occurring
near the pericenter of the eccentric orbits of satellites (\cite{2010HiA....15..193M}).

The process has been investigated thoroughly with collisionless N-Body simulations adopting high resolution models
of dwarf galaxies interacting with rigid or symmetric models of hosts on orbits with high eccentricities as those
of cosmological subhalos  (\cite{2001ApJ...559..754M}; \cite{2001ApJ...547L.123M} ;\cite{2007MNRAS.378..353K}; \cite{2011ApJ...726...98K}). It has also been 
investigated in combination with ram
pressure mass removal using hydrodynamical simulations of dwarf-main host interaction in which a diffuse 
gaseous halo consistent with observational constraints is added to the dark halo of the main host 
(\cite{2006MNRAS.369.1021M}; \cite{2010HiA....15..193M}). 

These studies established tidal stirring as a major mode of 
morphological evolution of dwarf galaxies, turning them 
from faint rotationally supported disks into even fainter pressure supported spheroidals, thus providing
a natural explanation for the morphology-density relation ubiquitously observed in galaxy groups,
including our own Local Group (\cite{1999AAS...195.0803G}).

Direct tidal heating as well as indirect heating stemming from bar and buckling instabilities 
triggered by tidal shocks have been both shown to play a role in tidal stirring, 
although the latter requires initial disks with mass-to-light ratios at the upper end 
of those observed in dwarf galaxies (\cite{2001ApJ...547L.123M}; \cite{2011ApJ...726...98K}).
These works also determined that ram pressure mass removal aided by tidal shocks 
can rapidly remove the gas content of satellites explaining why dSphs are devoid of gas and why
they can have very high mass-to-light ratios.

Recently the effect of the dark matter density profile of dwarf galaxies on tidal stirring has been studied,
motivated by the new findings of cosmological hydrodynamical simulations of field dwarf 
galaxy formation which have shown that feedback from supernovae and massive stars can turn the 
cuspy halos predicted by $\Lambda$CDM into halos with shallower inner density slopes ($\gamma$ in [0.3, 0.7] range), 
depending on the star formation rate achieved in the galaxy (\cite{2010Natur.463..203G}; \cite{2012MNRAS.421.3464P}; \cite{2014ApJ...792...99S};
\cite{2012MNRAS.422.1231G}; \cite{2013MNRAS.431.1220D}; \cite{2015MNRAS.446.1140T}). There
have also been attempts to show the existence of this effect specifically for dwarf
galaxy satellites, albeit in this case resolution is a potential issue (\citep{2013ApJ...765...22B}).
\cite{2012ApJ...751L..15L} and \cite{2013ApJ...764L..29K} have thus updated the previous tidal stirring
simulations by carrying out N-Body simulations in which disky dwarf galaxies have shallower dark matter
profiles as those arising in the cosmological hydro simulations, and have checked results against
varying resolution. They have found that the transformation by tidal stirring is enhanced significantly when a shallower density profile is adopted ($\gamma =0.2$ or $0.6$). They have
also found that, since mass loss is also enhanced, remnants of  dwarfs with shallow
dark matter profiles can evolve into very faint objects such as Ultra Faint Dwarfs (\cite{2012ApJ...751L..15L}). The enhanced mass loss and transformation follow from the lower binding
energy and more impulsive dynamical response of the system in the case of objects with a shallower halo
profile.

However, these recent simulations, exactly like those of the last decade, are fairly idealized
since they completely lack the cosmological context.  Hence orbits are also idealized, being disconnected
with the epoch of fall inside the host and with the shape of its gravitational potential, which
is just assumed to be spherically symmetric. Instead, cosmological simulations show that the
host is triaxial and grows substantially in mass and radius over the last ten billion years
(\cite{2007ApJ...667..859D}), which implies satellites can be accreted on increasingly wider orbits
over time. Furthermore, in the idealized simulations
subhalos can only interact with the main host, and not between them. Also, for simplicity
these simulations have always explored a fixed or limited mass range for the satellites, rather
than considering the wide range of masses from those of actual subhalos. It is expected that, the initial mass, and 
also the stellar mass relative to that of the halo, ought to have an important effect on the
response of the system to tidal shocks. Furthermore, other studies have shown that mergers
or strong interactions, while rare, can occur among dwarf galaxies just before infall, turning them 
into dwarf spheroidals very quickly (\cite{2011ApJ...740L..24K}), but leaving it completely
open to identify the distinct signature of a remnant of one of such mergers as opposed to one
from tidal stirring. Ideally, all these difficulties should be overcome in a fully cosmological hydro
simulation of a MW sized galaxy, but with current resolutions around 100 pc and $\sim 10^4 M_{\odot}$
in the baryons numerical results cannot be put on firm grounds for objects that should be only a few hundreds parsecs in size.

Therefore, in order to make progress beyond the idealized simulation we devise a numerical strategy
that in the past has only been attempted for simulations of galaxies in clusters (
\cite{1999ASPC..182..491D};
\cite{2005MNRAS.364..607M}).  In this approach, known as the "replacement technique", cosmological subhalos
are replaced with high resolution models of dwarf galaxies within the original
cosmological volume. Models are constructed based on observational
and theoretical constraints, and matched as closely as possible to the total virial mass of the
parent subhalo. After the replacement the simulation is rerun to the present epoch.
In this way galaxy satellites are automatically placed on realistic orbits and
are subject to the full tidal interaction history of the parent subhalos. Due to the computational
facilities and improved speed and parallelism of the codes, the simulations can span the equivalent of more than 10 Gyr of evolution
with this technique, while in previous cluster simulations only a small time span, from $z=0.5$ onwards,
could be covered (\cite{2005MNRAS.364..607M}).
In this paper we present the first replacement simulation carried out for a MW sized halo,
using the ErisDARK cosmological simulation carried out in the concordance $\Lambda$CDM model (\cite{2013ApJ...765...10K}; \cite{2014ApJ...784..161P}). 
The simulations are carried out with state-of-the-art massively
parallel N-Body codes, PKDGRAV2 (\cite{2001PhDT........21S}) and ChaNGa (\cite{changa2008}; \cite{changa2010}; \cite{2015ComAC...2....1M}; http://www-hpcc.astro.washington.edu/tools/changa.html)
on the CRAYXE6 "Monte Rosa" of the Swiss National Supercomputing Centre.

The paper continues with Section 2 where the replacement technique and numerical simulations are presented in details.
Next the analysis methods and results are enumerated in Section 3. Section 4 contains an extensive discussion regarding 
the implications of the results for the origin of dSphs, including their relation to observational data. 
Finally Section 5 includes the conclusion and outlook of the presented work.

\section{Numerical simulations and replacement technique}
\label{sec2}

\subsection{Initial conditions and selection procedure}

The initial conditions of the simulations of this paper are based on the ErisDark run (\cite{2014ApJ...784..161P}), 
which is a dark matter only cosmological gravity run of 40 million particles using the $\Lambda$CDM 
cosmological model ($H_0$ = 73 km $s^{-1} Mpc^{-1}$, $n_s = 0.96$,$ \sigma_8 = 0.76$, $\Omega_m = 0.268$ and $ \Omega_{\lambda} = 0.732$). The cosmological volume contains a 1 Mpc scale zoom-in region with 
three levels of refinement that lead to a minimum mass of the dark matter particle of 1.2 $\cdot$ $10^5 M_{\odot}$ 
and a minimum softening length of 124 pc (\cite{2014ApJ...784..161P}).      
In the zoom-in region an isolated MW-sized halo assembles, with mass consistent with the lower limit for the estimated 
halo mass of the Milky Way, which provides a good fit with current kinematical data on the Milky Way stellar 
halo (\cite{2013ApJ...773L..32R}). The companion hydrodynamical simulations, Eris and ErisLE, have shown to produce realistic 
spiral galaxies by $z=0$, which reproduce nearly all the main properties of the disk and bulge of the Milky Way (\cite{2011ApJ...742...76G}; \cite{2012ASPC..453..289M}; \cite{2013ApJ...772...36G}; \cite{2013ApJ...773...43B}).

Satellite galaxies falling into the main host after the last major merger, which takes place at $z \sim 3$, are considered for replacement 
with hi-resolution dwarf galaxy models including stars.
We analize the entire available set of outputs of the original simulation with the use of the Amiga Halo Finder (AHF) 
(\cite{2004MNRAS.351..410G};\cite{2009ApJS..182..608K}) with a virial overdensity criterion that varies with redshift according to the
cosmology (\cite{1998ApJ...495...80B}). In this paper, in particular, we analyze the earliest infalling population of satellites 
among which there are survivors at $z=0$. In the next paper we will present larger populations of dwarf galaxies which are accreted at later
epochs.

\begin{figure}
\centering
	\includegraphics[width = 0.98\textwidth]{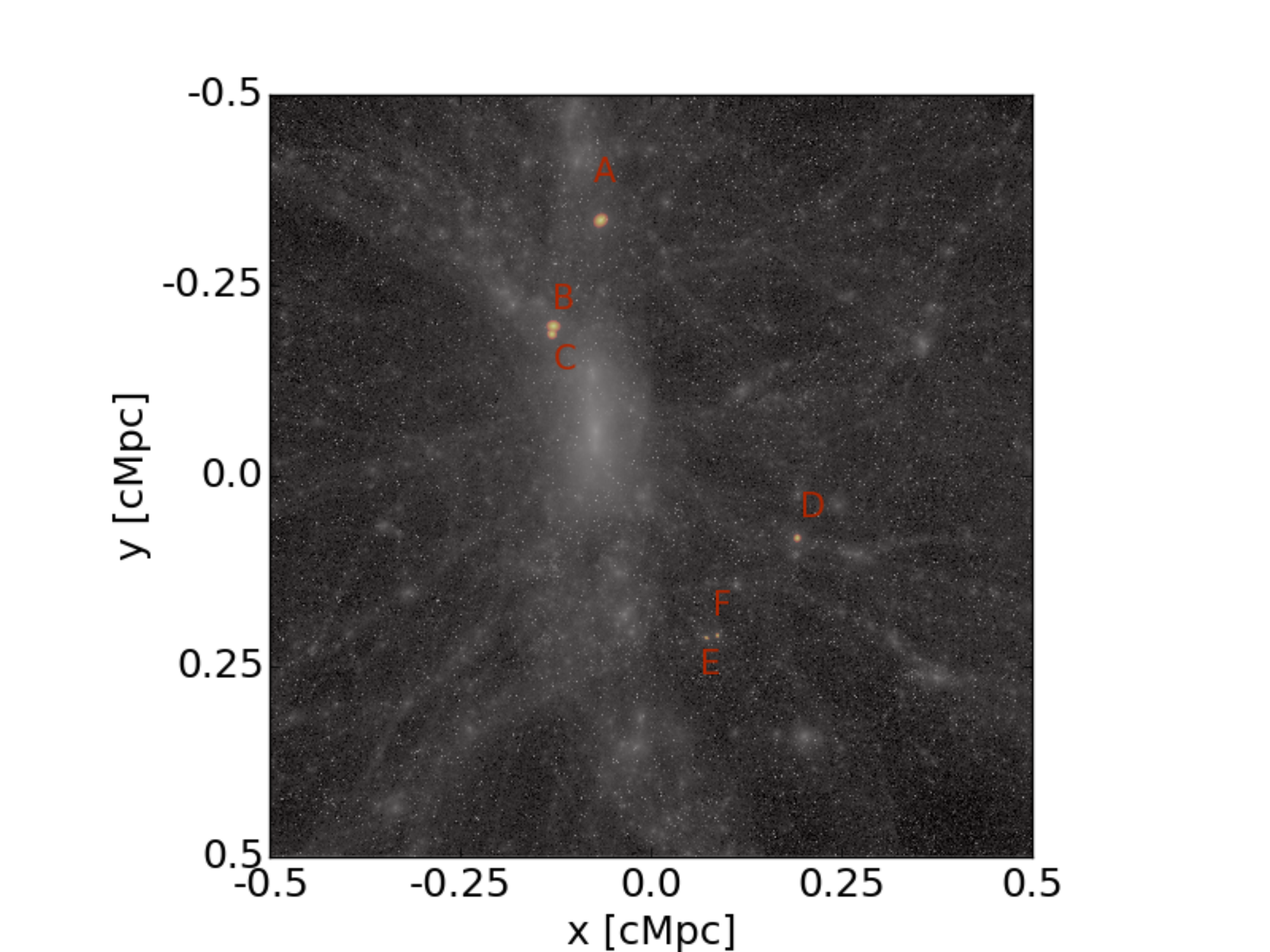}
\caption{\label{fig:figure1}The projected density map of the ErisDark cosmological simulation at redshift 2.87 
for a cubic comoving megaparsec section of the zoom-in region with the same depth is presented. 
The replaced halos are visible due to the presence of the stars marked with red to yellow colors in the image.}
\end{figure}

We thus proceed in the following way. First of all,  among subhaloes 
with bound masses greater than $10^8 M_{sol}$ at z = 0, the satellite with the earliest infall is identified.
The adopted threshold mass ensures that we are considering well resolved objects, with at least $10^3$ particles
(presumably their progenitors are even better resolved at higher redshift, before tidal mass loss begins), to
minimize numerical effects on their internal dynamical evolution (\cite{1996ApJ...457..455M}). 
The object enters the host's virial radius just before $z = 2$, when it is still unrelaxed, being the result of
a merger between two subhalos. We trace the satellite back to  
$z = 2.87$ when the two progenitor merging subhalos are still well separated, namely their virial volumes do not overlap.
Therefore the epoch corresponding to z = $2.87$ is set as the epoch of replacement of the original ErisDark's infalling subhalos with their 
high resolution versions.\\ Next we identify other subhalos that at this epoch are located between 1 and 4 $R_{vir}$ of the host. We
restrict the sample to the most massive ones, having pre-infall masses around $10^9 M_{\odot}$ and above. Among these we reject 60\% 
that cannot be properly identified and replaced because they are either too elongated (axis ratio less than 0.65, 
possibly tidally disturbed by companions) or have a substantial fraction of their mass in subhalos. 
Four objects remain after this rejection procedure, with masses in the range $0.99-4 \times 10^9 M_{\odot}$. 
Note that at $z \sim 2$ this translates into satellites having masses in the range 80-300 times smaller
than the host, which has a virial mass of $3.00 \times 10^{11} M_{\odot}$ and a virial radius of 55.6 kpc at the corresponding epoch (both relatively high 
since ErisDARK has a fast assembly history with no more major mergers after $z=3$,  \cite{2014ApJ...784..161P}). 
We remark that
the lower end of the satellite-to-host mass ratio is still in a regime in which
dynamical friction  is effective in eroding the orbital energy over many Gyrs 
(\cite{1999ApJ...525..720C}), while this is a negligible effect in published idealized simulations that assume 
satellites of similar mass but within a host having a fixed mass $\sim 10^{12} M_{\odot}$ (\cite{2013ApJ...764L..29K}). Finally, we observe that none  of the four objects has a mass above $10^8 M_{sol}$ at z = 0 in the original ErisDark run. 
Therefore they are not part of the largest subhaloes at the final epoch, which can well exceed this mass at low z (Paper II, \cite{2014ApJ...784..161P}).
In total we thus replace 6 subhalos (figure \ref{fig:figure1}), of which 2 are merging during infall and the other four are accreted individually. The halo properties are listed in table \ref{table:table1}.

%%%%%%%%%%%%%%%%%%%%%%%%%%%%%%%%%%%%%%%%%%%%%%%%
\subsection{Model galaxy generation}
 
The replacement models are generated using the GalactICS code (\cite{1995MNRAS.277.1341K}, \cite{2005ApJ...631..838W}, \cite{2008ApJ...679.1239W}), which generates
multi-component equilibrium models of galaxies. Each model galaxy is comprised of two components: the dark matter halo and the stellar disk. 
It is assumed by construction that the initial model has a stellar disk since
this stems from the assumption that disky dwarfs are the progenitors of dSphs, the core idea of tidal stirring.
The code creates halos with negligible angular momentum and a spherically symmetric density profile as described by the relation:

\begin{equation}
\rho_{halo}(r) = \frac{2^{2-\gamma} \sigma_s^2}{4 \pi G r_s^2} \frac{1}{\left(\frac{r}{r_s}\right)^\gamma \left(1+\frac{r}{r_s}\right)^{3-\gamma}} \frac{1}{2} erfc \left(  \frac{r - r_c}{\sqrt{2} r_w}\right).
\end{equation}  

In the previous equation the scale radius and velocity scale are denoted with $r_s$ and $\sigma_s$, and $\gamma$ stands for the cusp coefficient. 
In order to restrain the model to finite dimensions a complementary error function with cut-off radius $r_c$ and width $r_w$ is used.
The velocity distribution is computed by solving the Eddington equation for the prescribed density profile.\\

For each removed halo two replacements are created: one with $\gamma = 1.0$ and another with $\gamma = 0.6$.  
The cut-off radius and width are set to be linear functions of the virial radius of the original object:

\begin{equation}
r_c = r_{vir} - r_w
\end{equation}

\begin{equation}
r_w = \frac {\sqrt{2} r_{vir}}{10}.
\end{equation}

The scale radius is determined indirectly by assuming that the concentration of an average halo of mass $M_{vir}$ at redshift $z$ 
are given by the empirical relations found by \cite{2011ApJ...740..102K}:

\begin{equation}
c(M_{vir}, z) = c_0(z) \left(\frac{M_{vir}}{10^{12} h^{-1}  M_\Sun} \right)^{-0.075} \left( 1 + \left( \frac{M_{vir}}{M_0(z)}\right)^{0.26}\right).
\end{equation}  

With the concentration and virial radius set, the scale length can be determined.

\begin{equation}
r_s = \frac{r_{vir}}{c}
\end{equation}
The two variants corresponding to shallow and steep dark matter density profile have by construct the same scale radius. 
In order to have the same mass enclosed in the same virial volumes with different cusp coefficients the scale velocity is adjust accordingly. 
The values of the characteristic quantities associated to dark matter halos with the virial quantities presented in table \ref{table:table1} 
are listed in table \ref{table:table2}.

%%%%%Stellar part%%%%%
The previously mentioned code can generate a disk component of a galaxy according to the cylindrical spatial distribution:
\begin{equation}
\rho_{stars}\left(r,z\right) = \rho_{0} \exp \left( -\frac{r}{R_d} \right) \left[ sech \left( \frac{z}{z_c} \right) \right] ^2,
\end{equation}
and cylindrical distribution of the radial dispersion:

\begin{equation}
\sigma_r(r) = \sigma_{r0} \left( \exp \left( - \frac{r}{R_d} \right) \right)^{1/2}.
\end{equation}

The reader who is interested in more details on how the code constructs the galactic model is referred to the publication \cite{2008ApJ...679.1239W}.

There are free parameters that have to be set using educated guesses based on available constraints on galactic structure at low mass scales. 
Most importantly, the mass of the stellar disk $M_d$ for each object is $1.2 \times 10^{-3}$ of its virial mass 
Such ratio is chosen to reflect the stellar-to-halo mass ratio, 
$M_{\star}/M_{vir}$, inferred from the abundance matching analysis extrapolated to lower masses (\cite{2013ApJ...770...57B}).     
Past simulations of disk galaxy formation presented in the \cite{2007MNRAS.382.1187K} paper associate to objects with maximum circular velocities in the 24-53 km $s^{-1}$range, 
scale height values in the 0.08-0.18 kpc range. 
Considering the previously mentioned results the values of the scale heights of all the objects is set to the minimum of 0.08 kpc. 
The choice of constant scale height is motivated by the expectation that at the masses of interest the disk thickness
acquires  minimum threshold value.
The value is dictated by the balance between the thermodynamics of disk formation, such as the effect of the cosmic ionizing flux providing
a minimum temperature floor of $\sim 10^4$ K, corresponding to a bulk velocity dispersion $\sim 8-10$ km/s,
and the gravitational potential of the halo. 
For all the stellar disks introduced in the simulation the scale height was set to 0.08 kpc.

With the scale height and stellar mass to total mass ratio assigned, the scale length remains to be determined.  
The square of the scale length is naturally constrained to be proportional to the virial mass of the respective object. 
An observational study provided by \cite{2009A&A...493..871S} finds several  dwarf galaxies with disk scale lengths as low as 0.33 kpc.
Considering the inferred galaxy masses for the latter sample and the values assumed in previous work on tidal stirring
based on various constraints (\cite{2011ApJ...726...98K}), the reference value of $R_d = 0.4 kpc$ was set  
for an object with a virial mass of $10^9 M_{sol}$. Using such reference value as a normalization the following relation 
is used for the calculation of the scale length  as a function of galaxy mass:

\begin{equation}
R_d = 0.4  \left( \frac{M_{vir}}{10^9 M_{sol}}\right)^{0.5} kpc.
\end{equation}
For each pair of objects the parameters of the stellar component are listed in table \ref{table:table2}.

From the original cosmological simulation, quantities such as the virial mass and virial radius are measured (table \ref{table:table1}). 
The virial quantities are used for determining all the parameters required for the generation of the pairs of high resolution galaxies that 
will be used as replacements. Namely the following quantities are determined: ($\sigma_s,r_s, r_c, r_w$) the dark matter halo quantities and 
($M_{stars}, R_d$) the stellar disk quantities. The remaining free parameters are the number of dark matter particles and star particles. 
For all the models created both numbers were set to one million. 
 
\subsection{Stability of the galaxies}
\begin{figure}
\centering
\includegraphics[width = .3\textwidth]{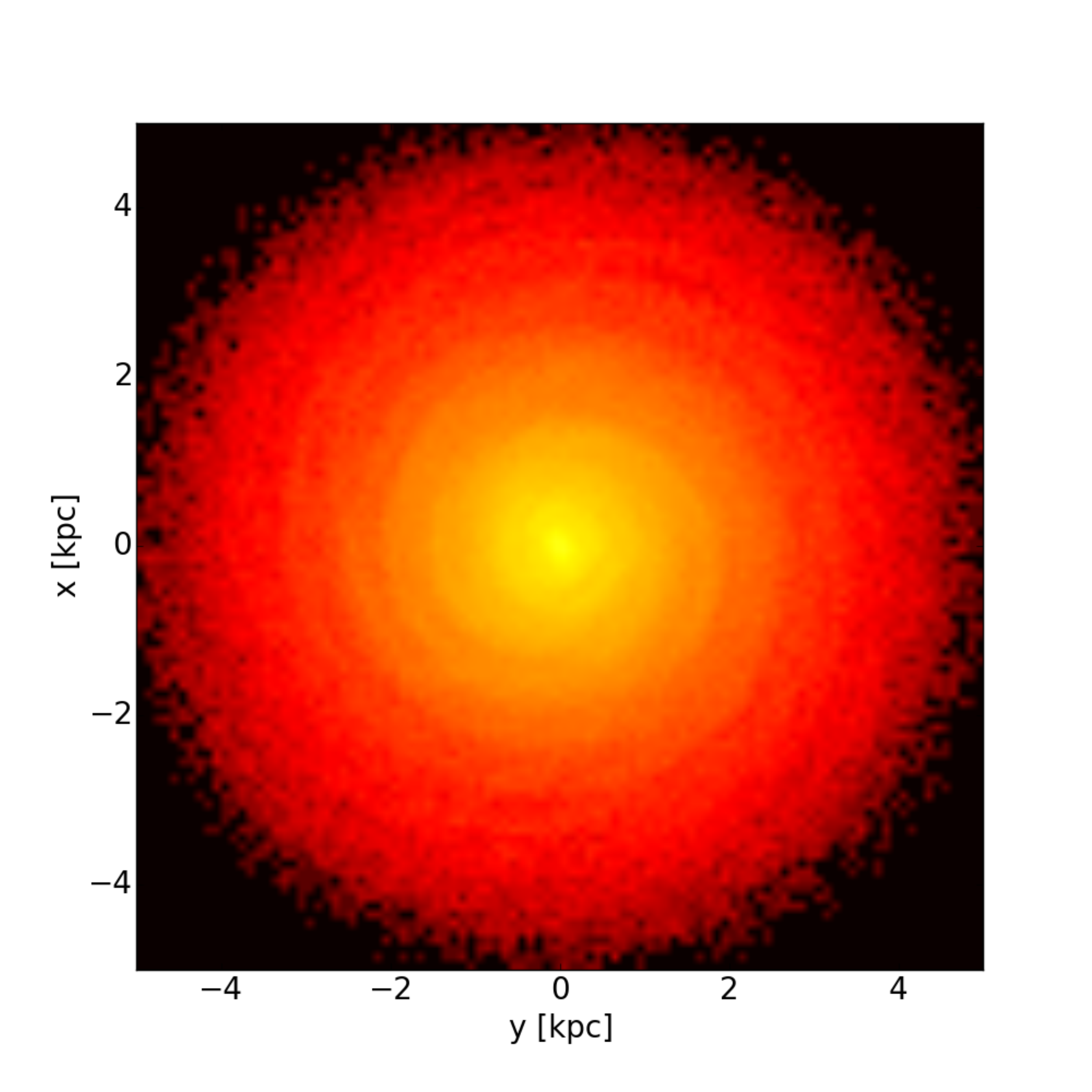}
\includegraphics[width = .3\textwidth]{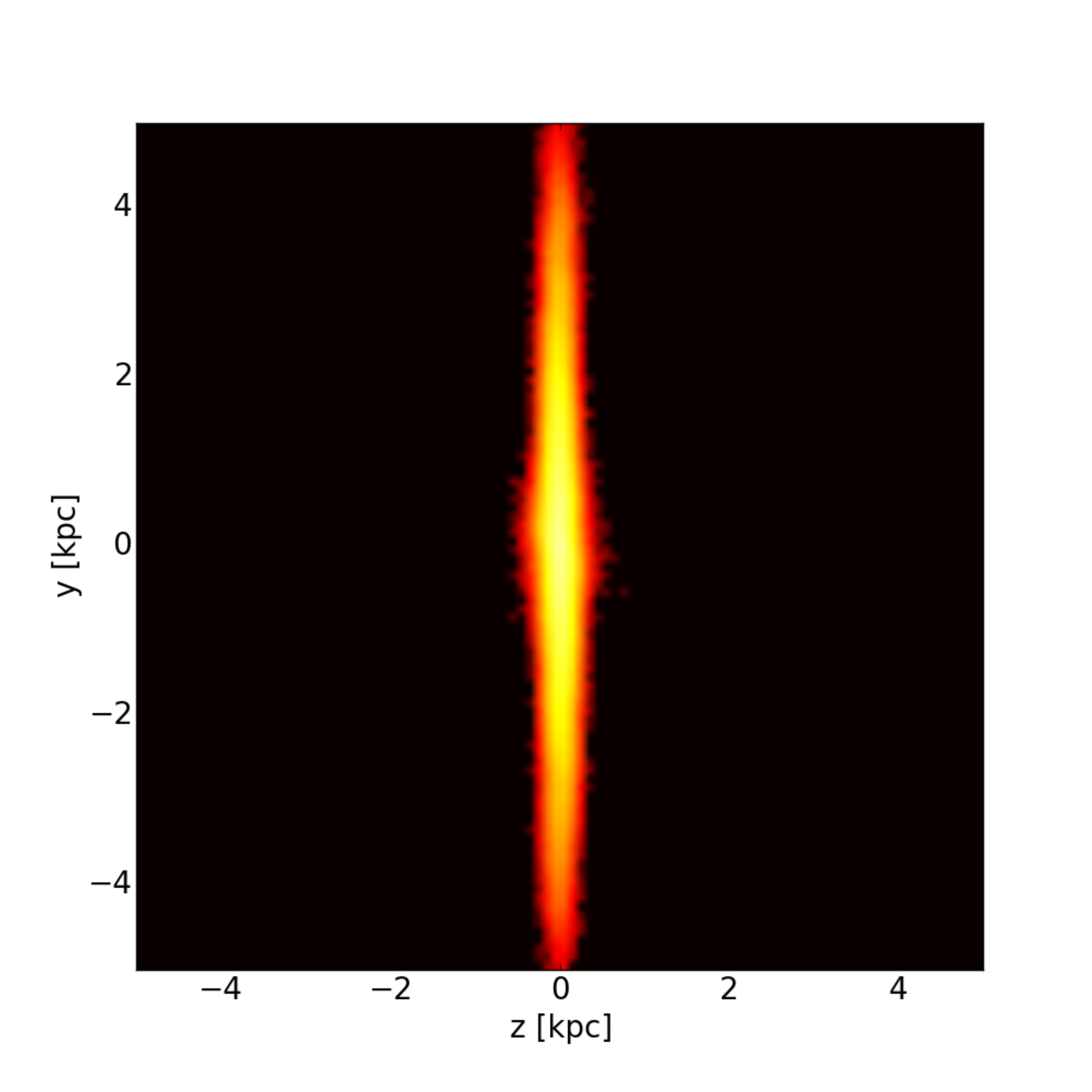}  
\includegraphics[width = .3\textwidth]{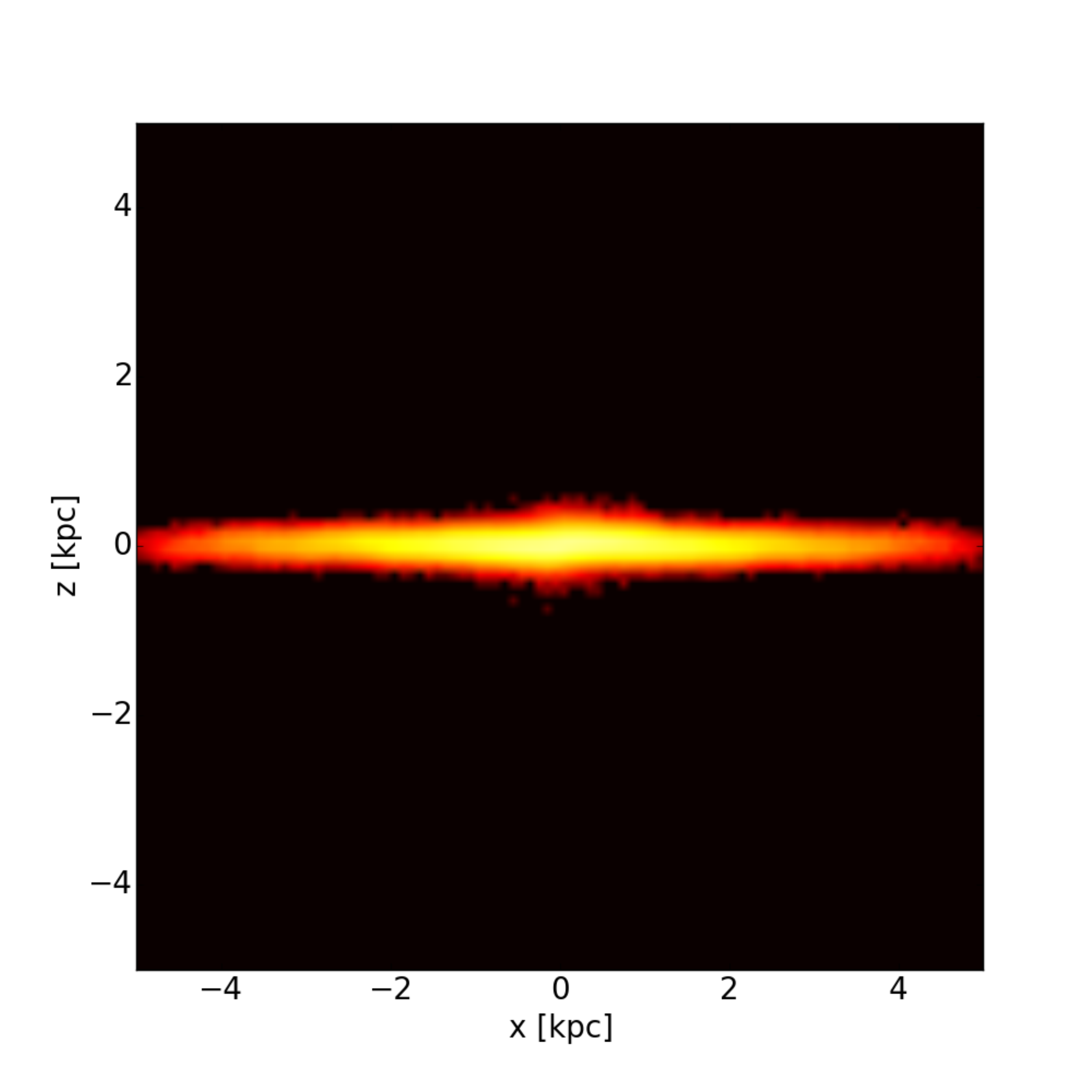}  
\includegraphics[width = .3\textwidth]{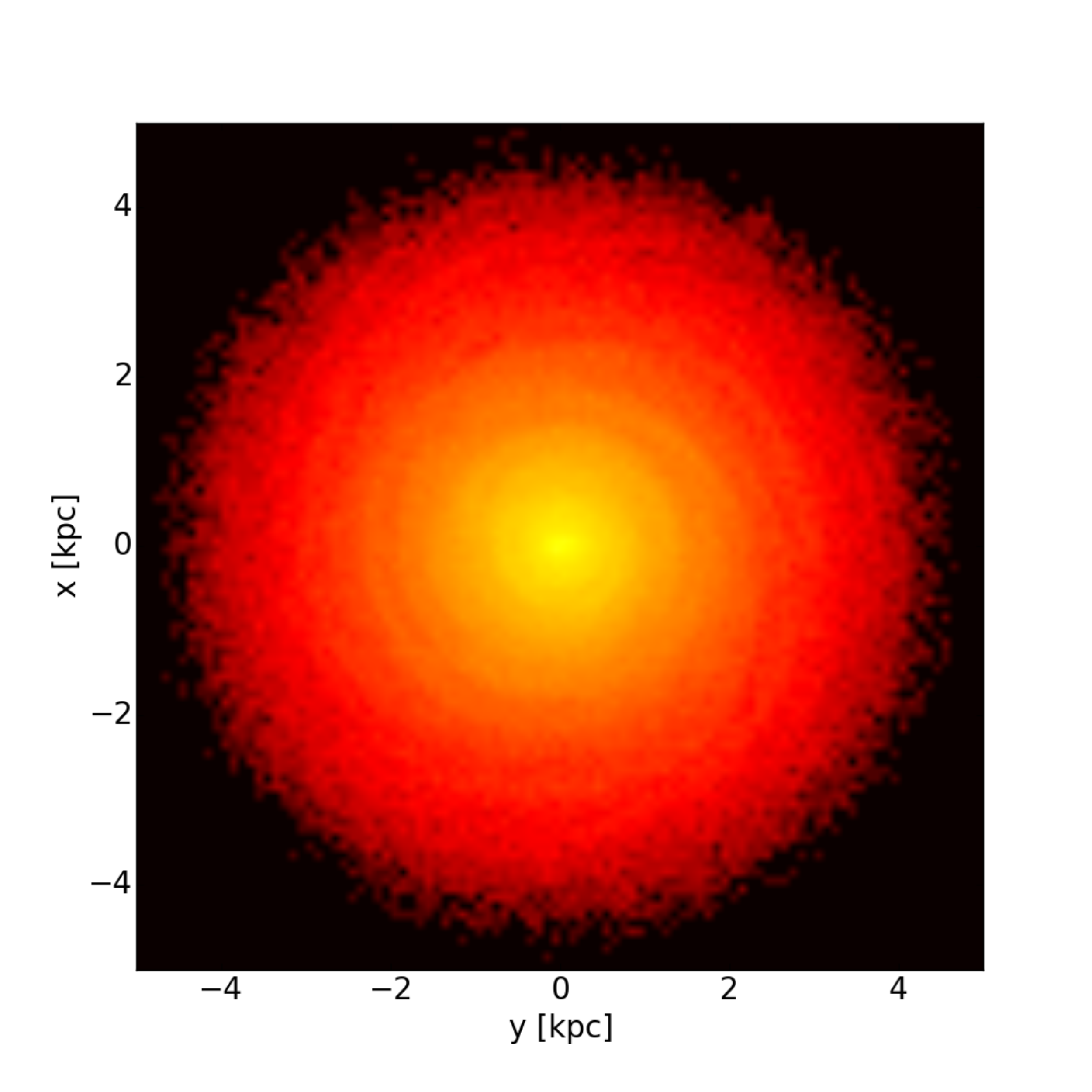}
\includegraphics[width = .3\textwidth]{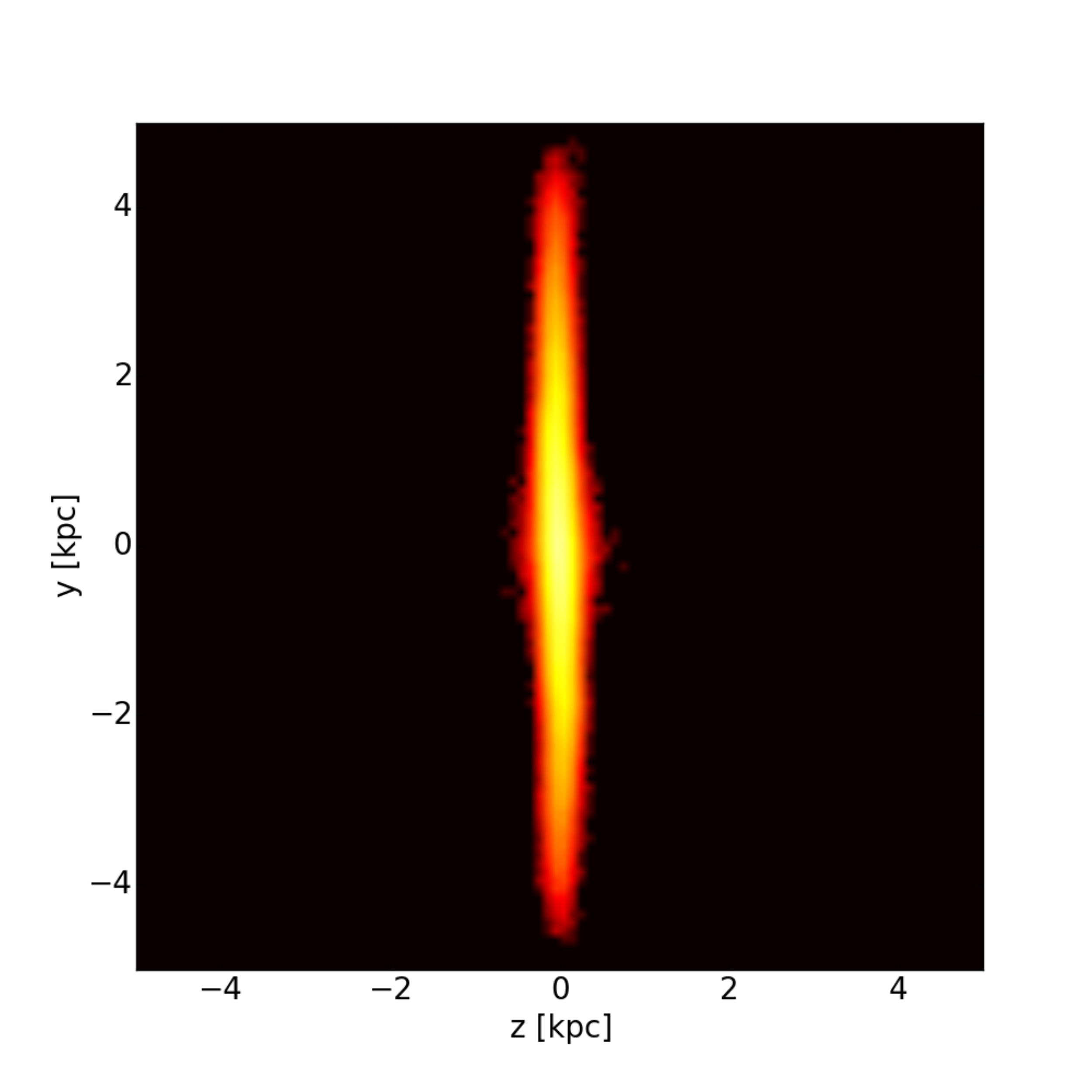}  
\includegraphics[width = .3\textwidth]{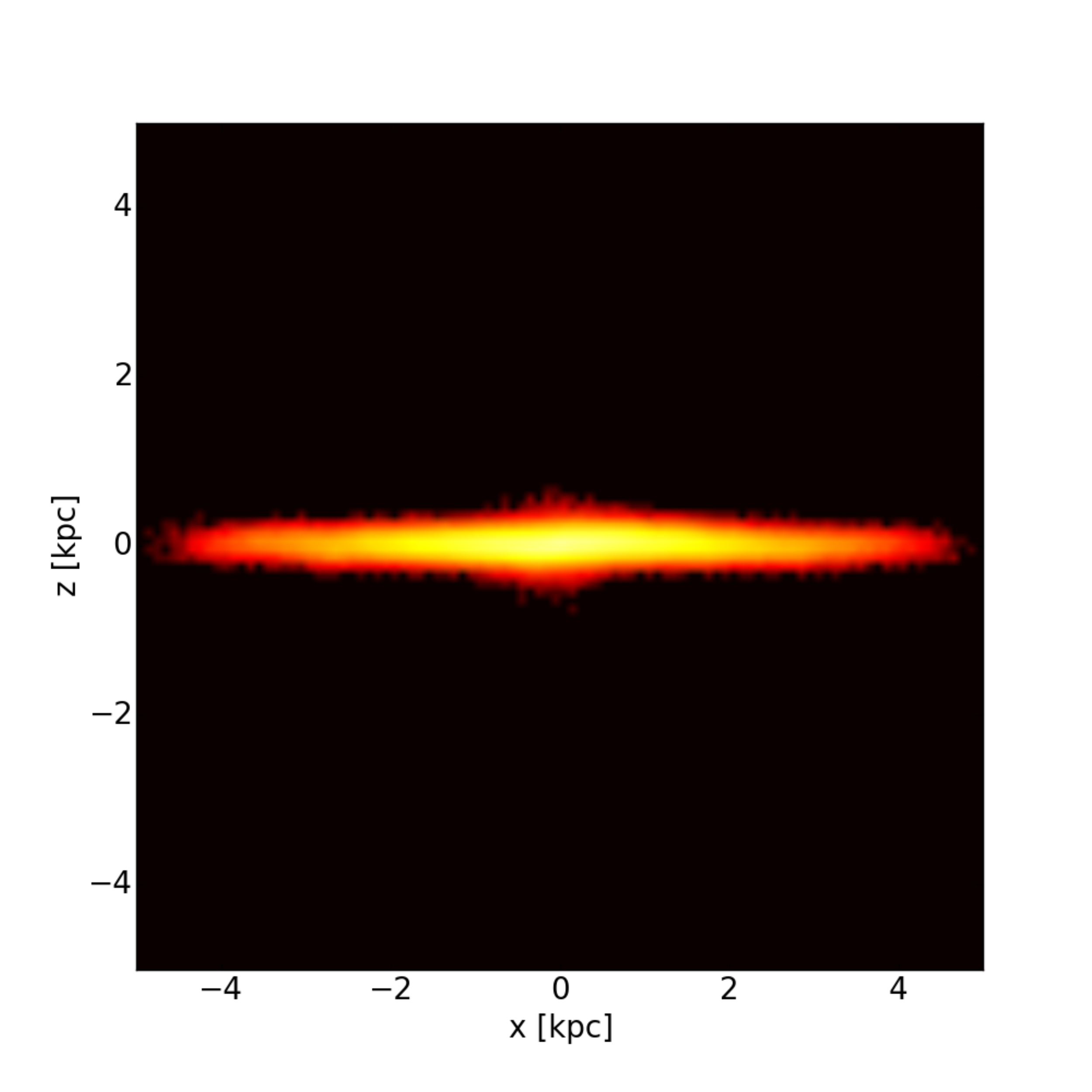}  
\includegraphics[width = .3\textwidth]{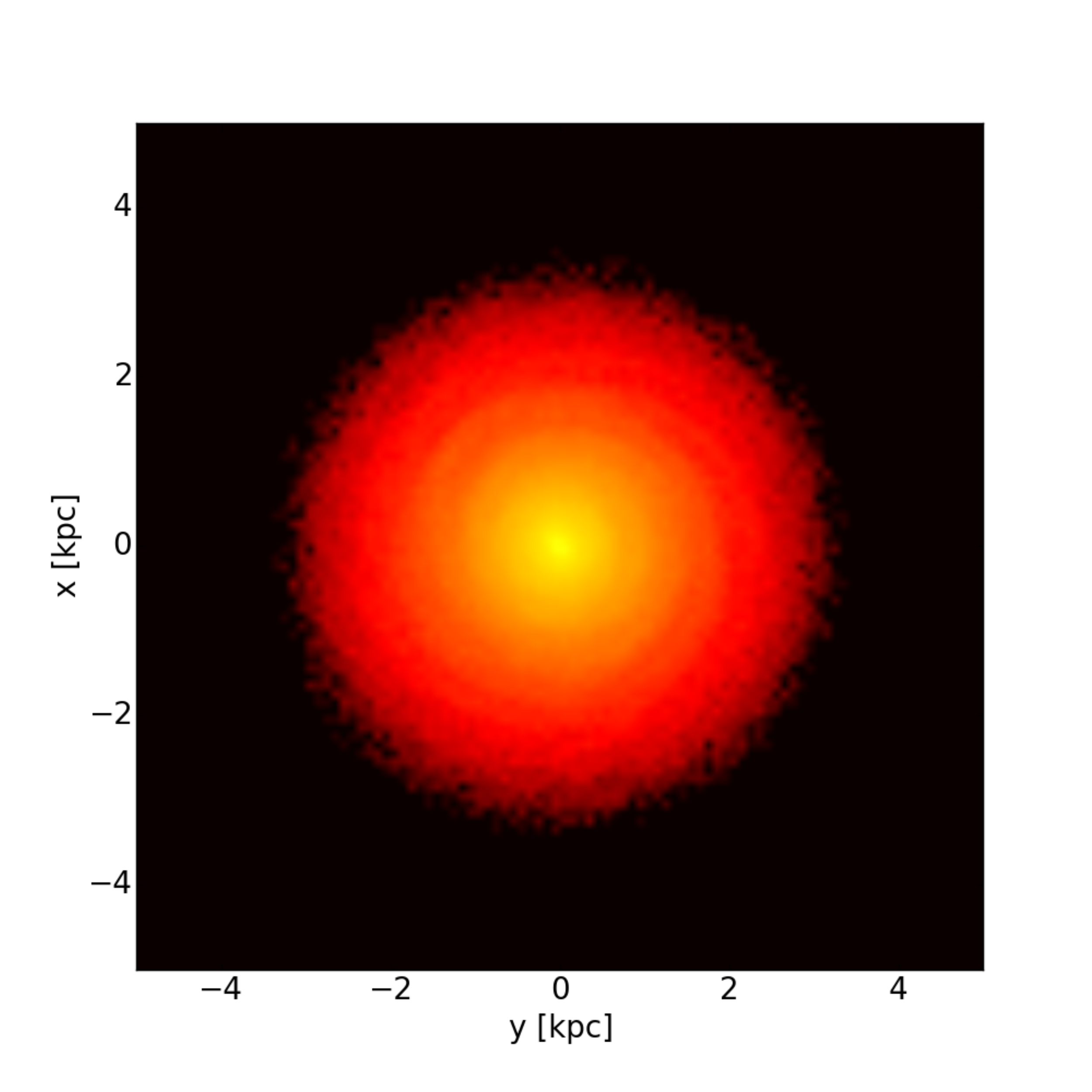}  
\includegraphics[width = .3\textwidth]{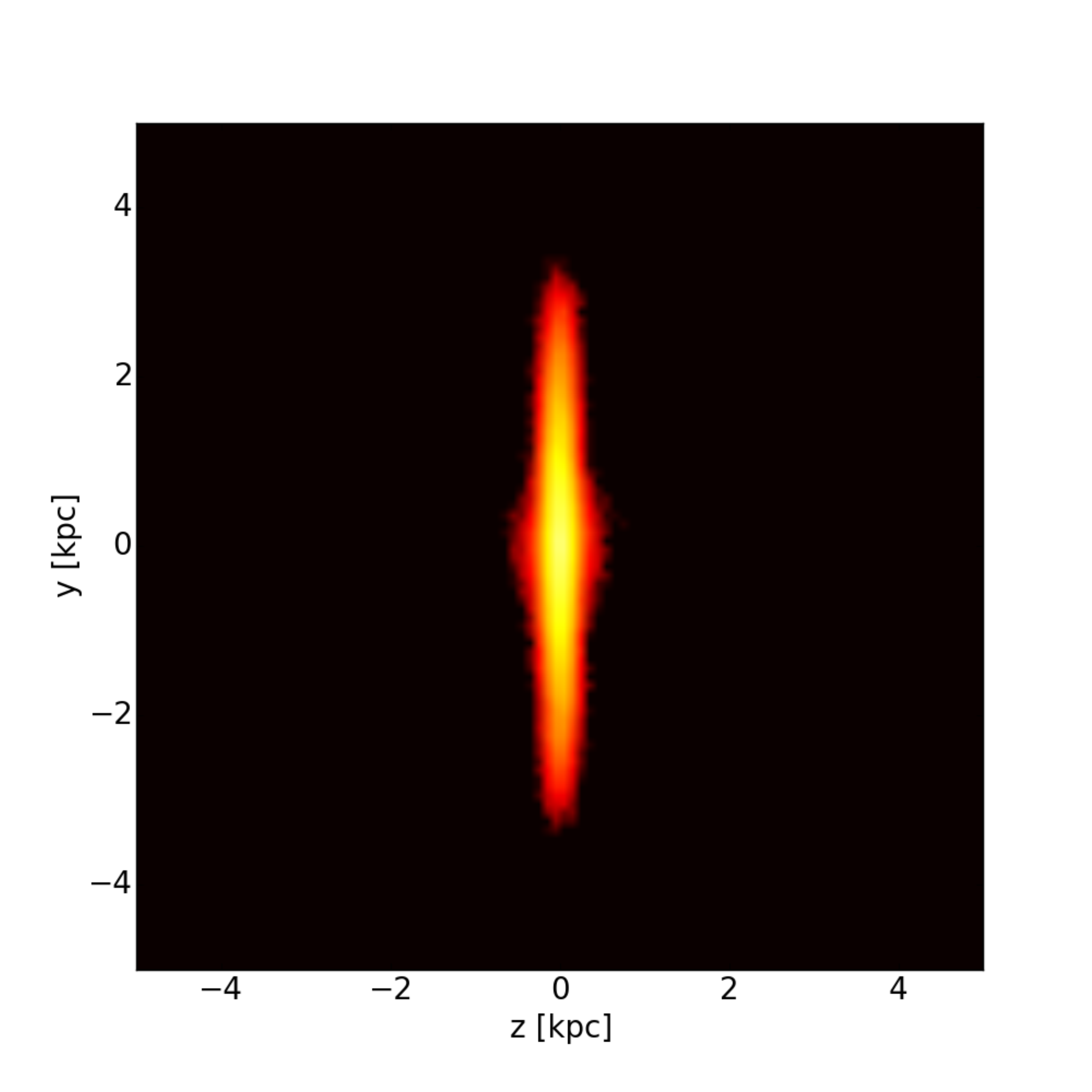}  
\includegraphics[width = .3\textwidth]{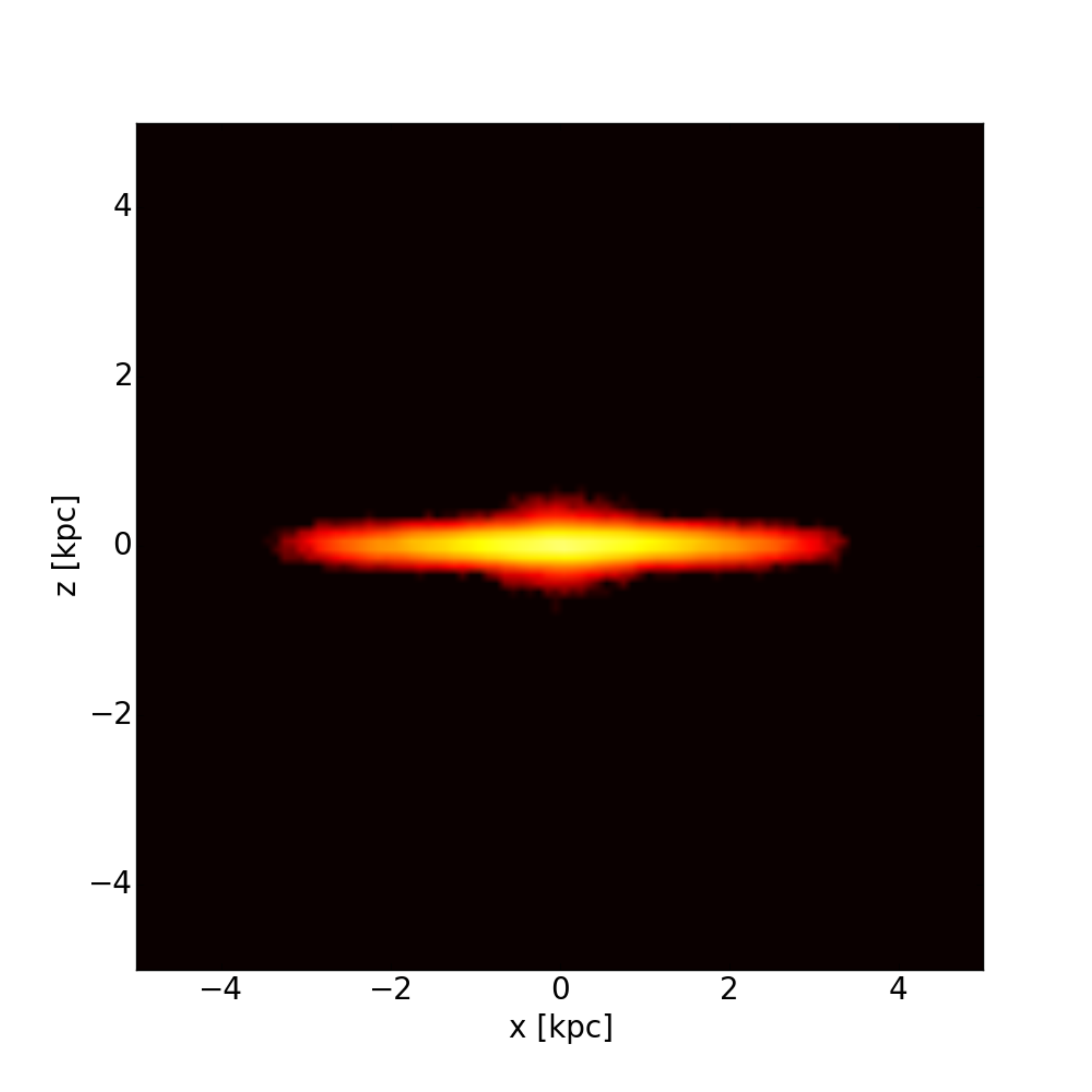}
\includegraphics[width = .7\textwidth]{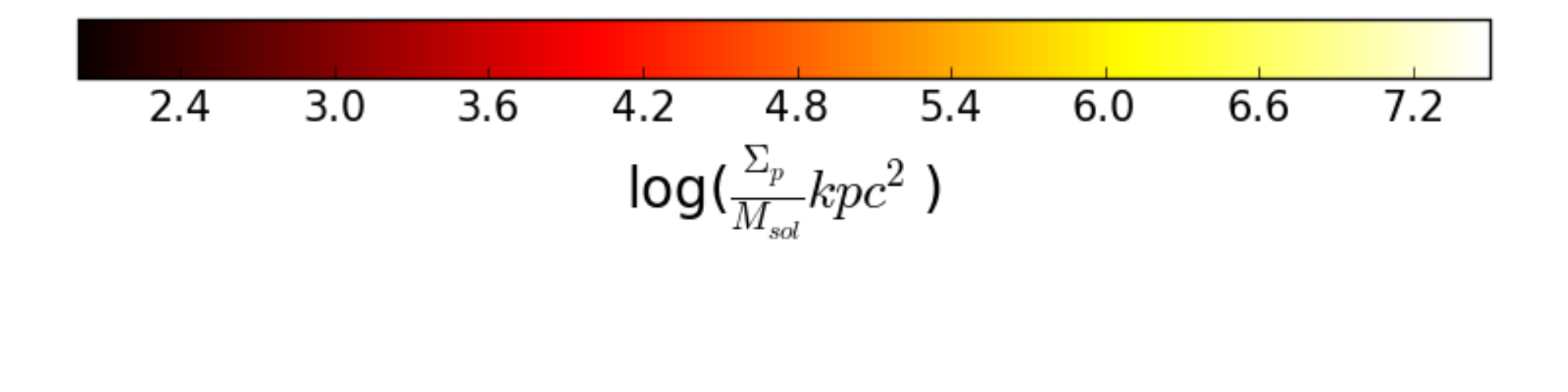}

\caption{\label{fig:figure2} The collage of images describes the stellar distribution of the high resolution version of satellites A, B, and C prior to the replacement.}
\end{figure}

\begin{figure}
\centering
\includegraphics[width = .3\textwidth]{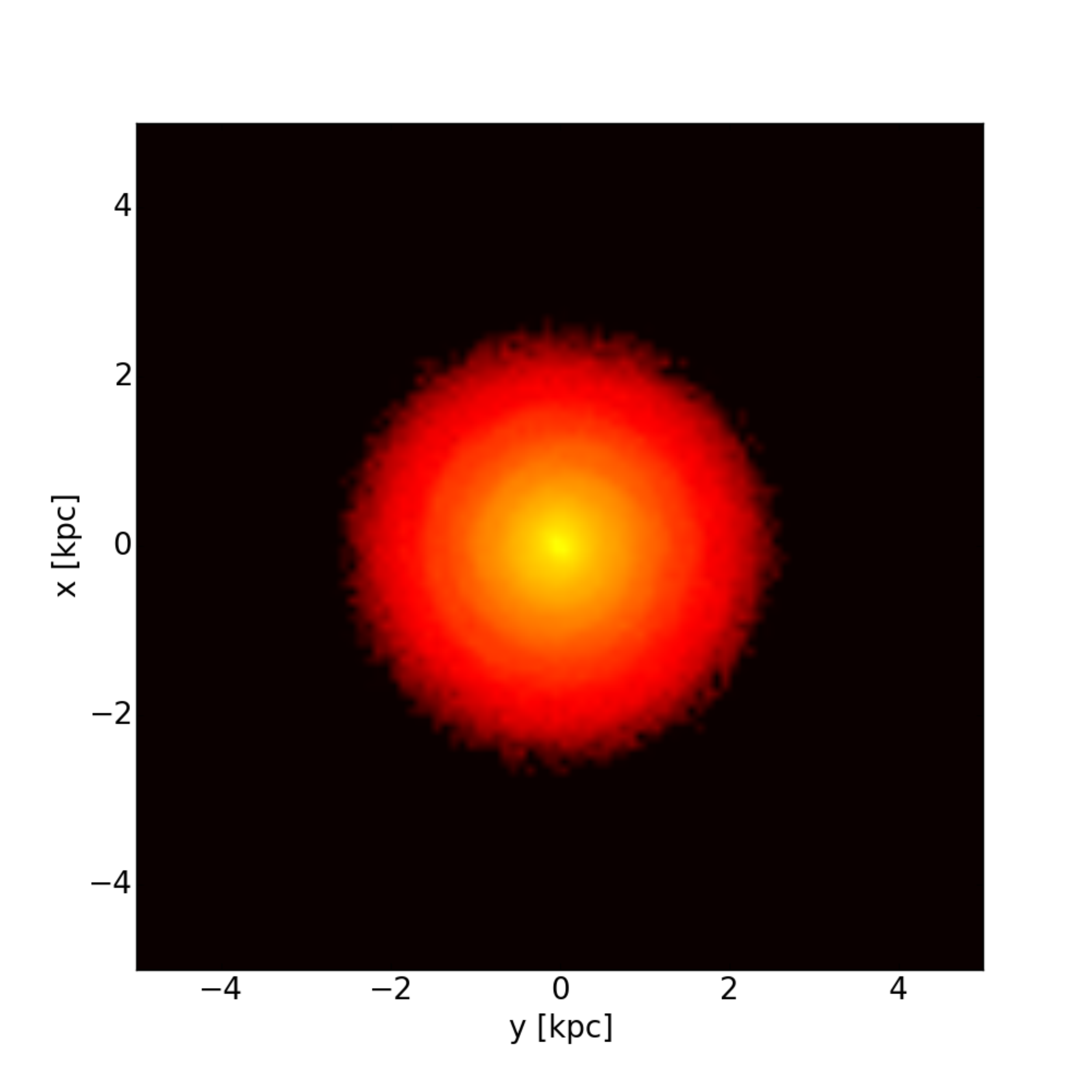}
\includegraphics[width = .3\textwidth]{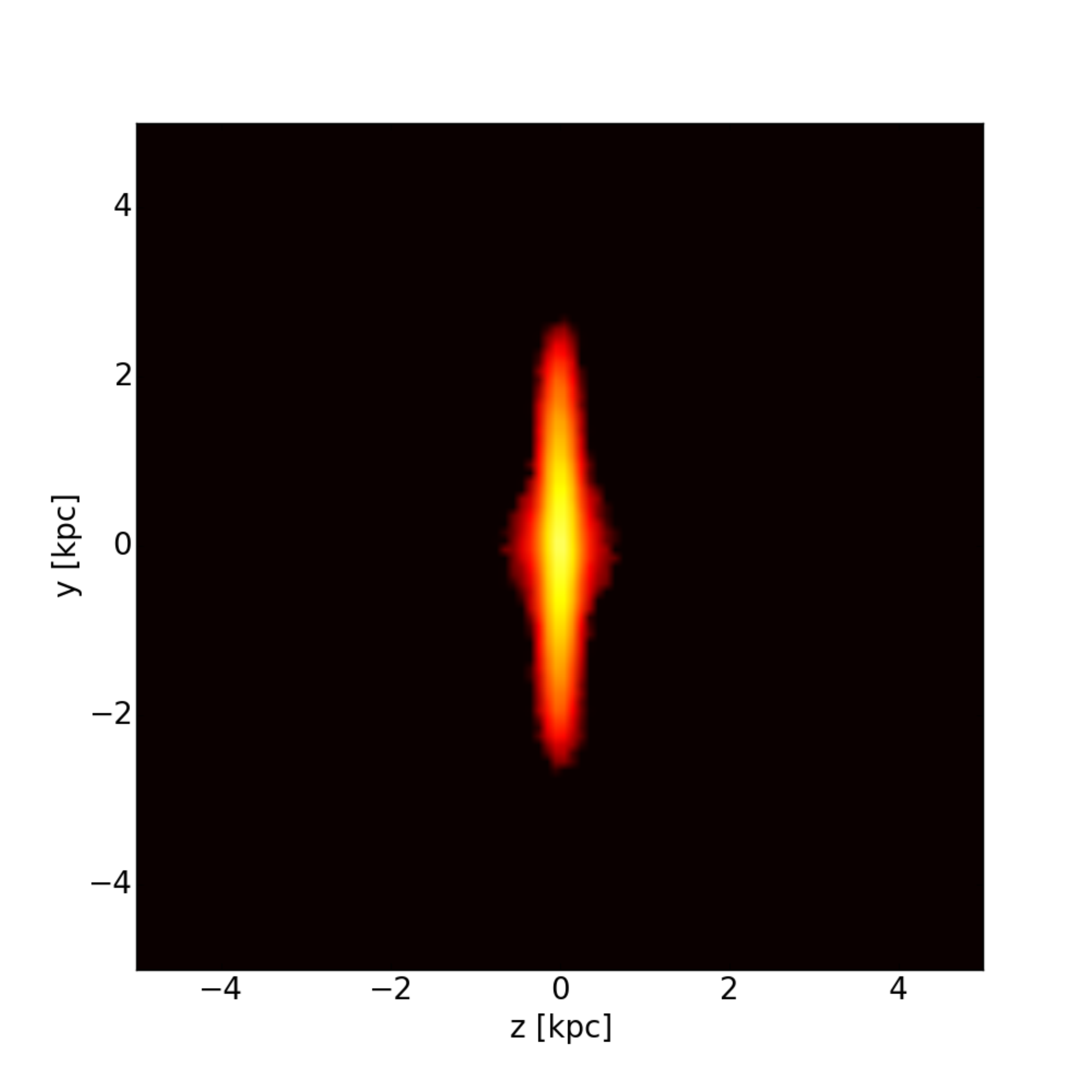}
\includegraphics[width = .3\textwidth]{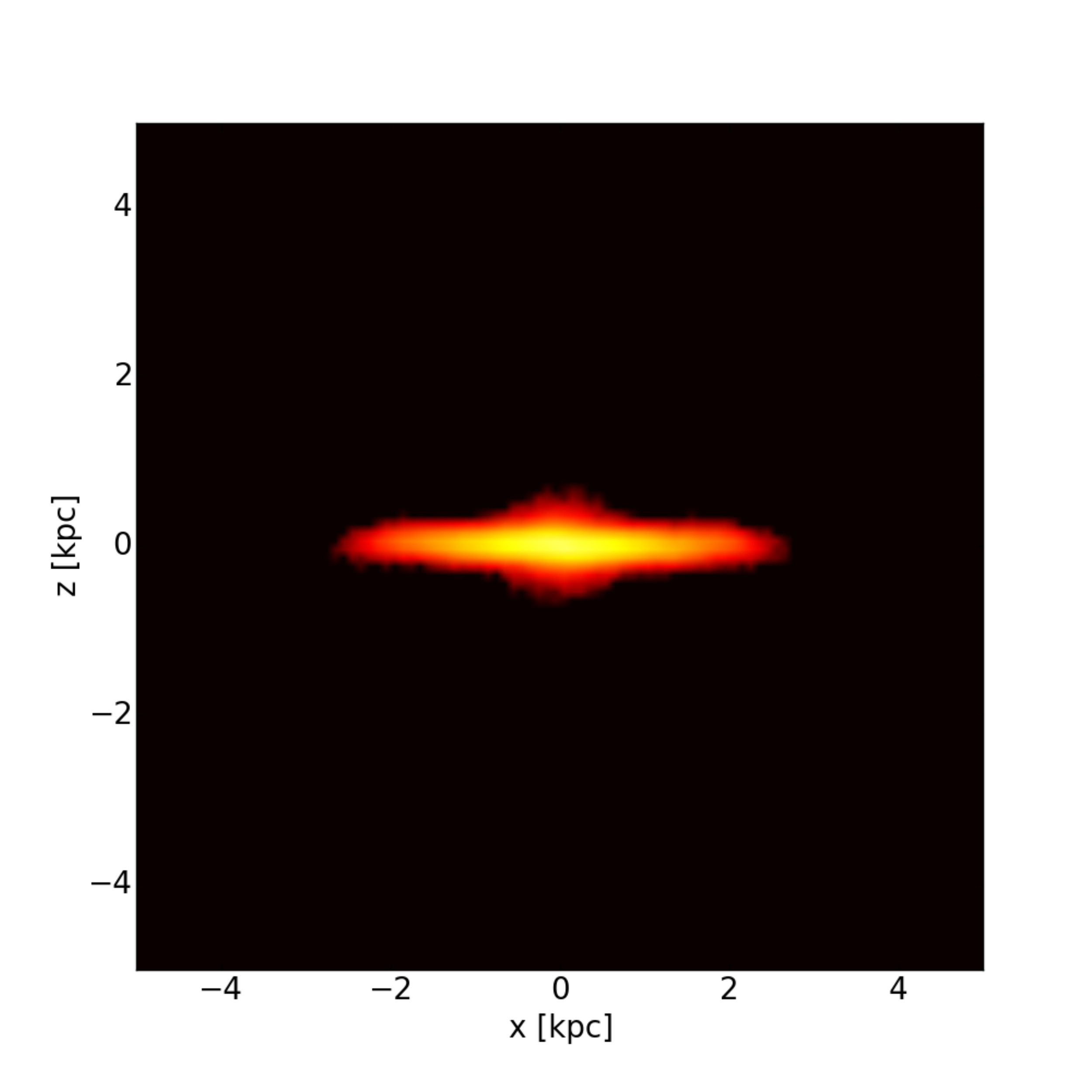}
\includegraphics[width = .3\textwidth]{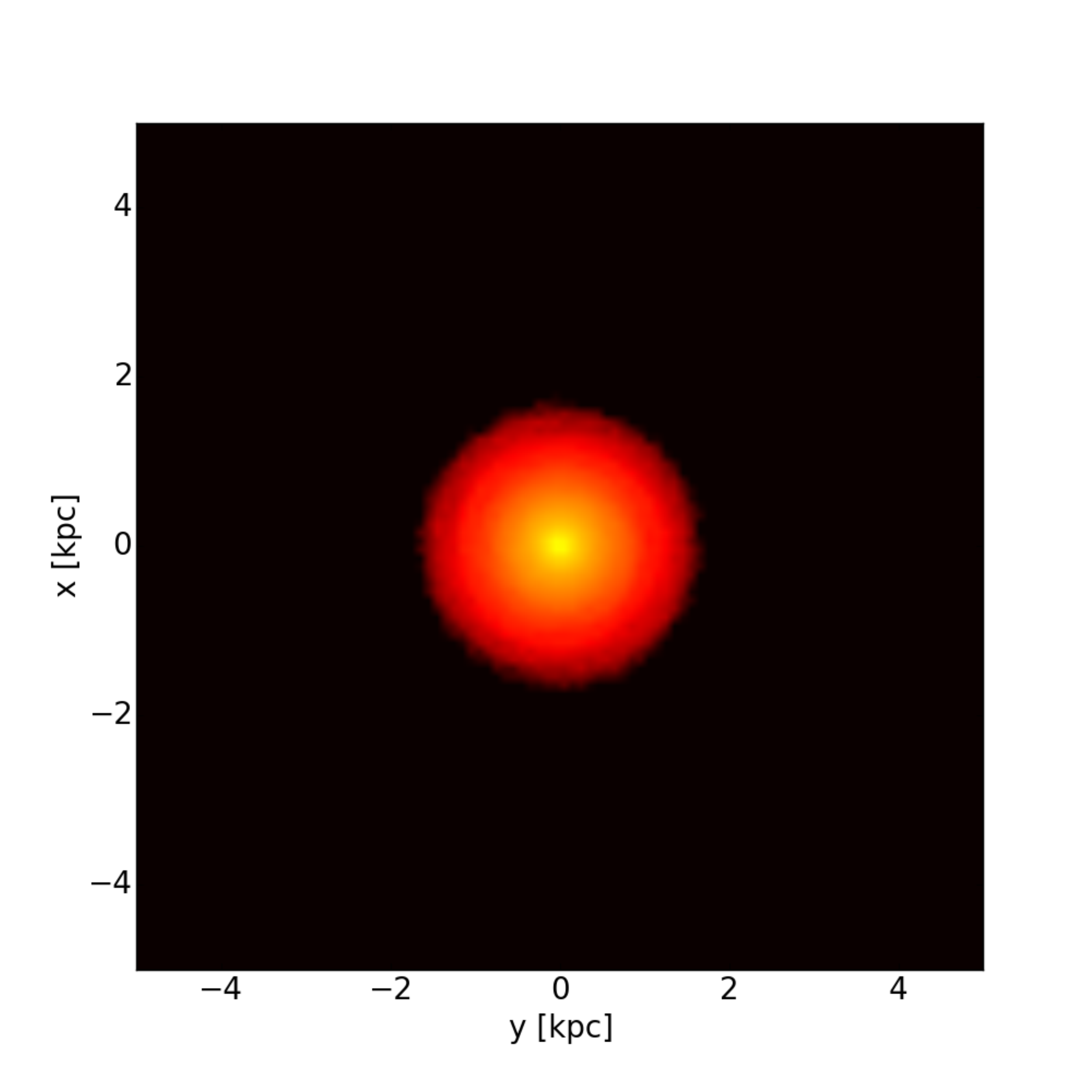}
\includegraphics[width = .3\textwidth]{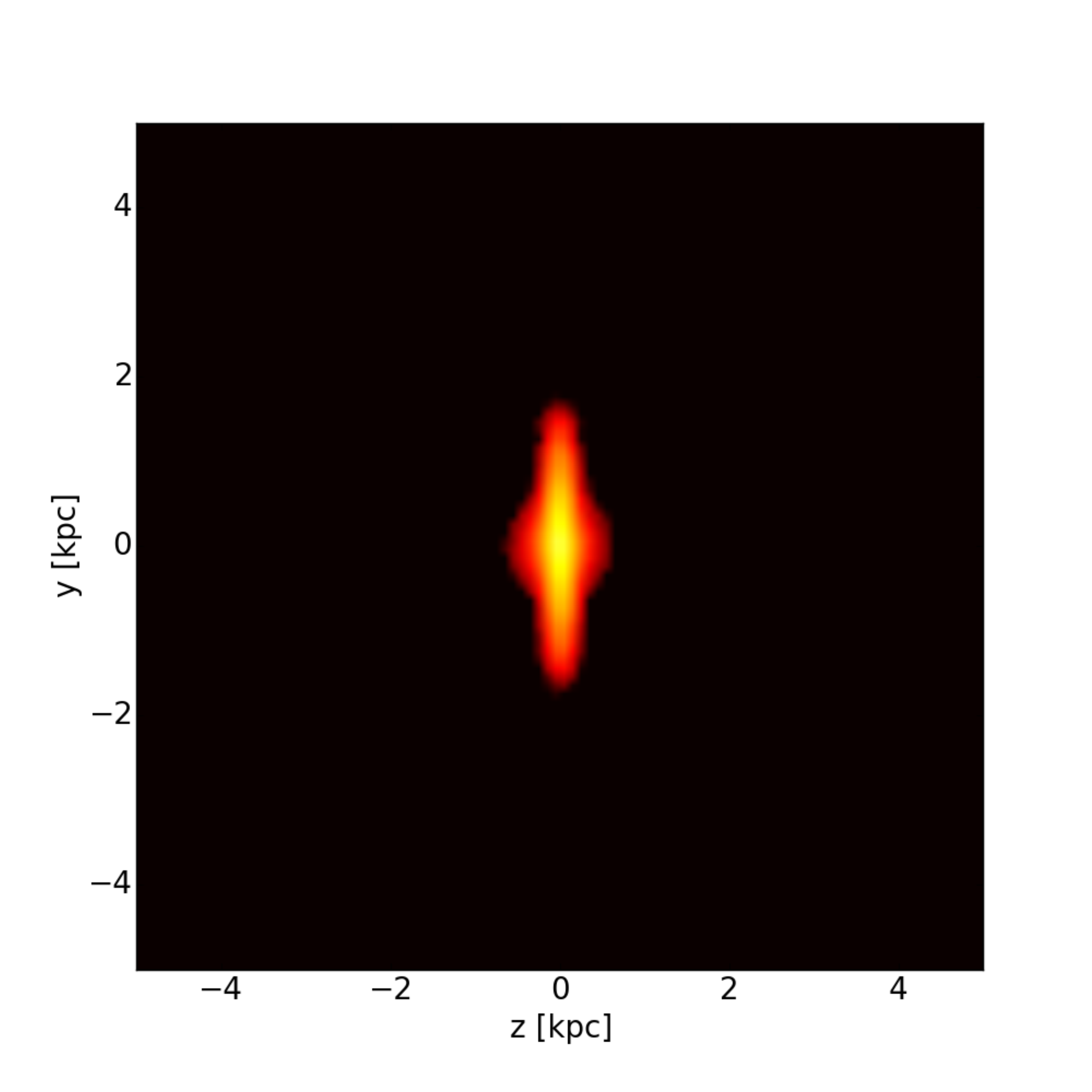}
\includegraphics[width = .3\textwidth]{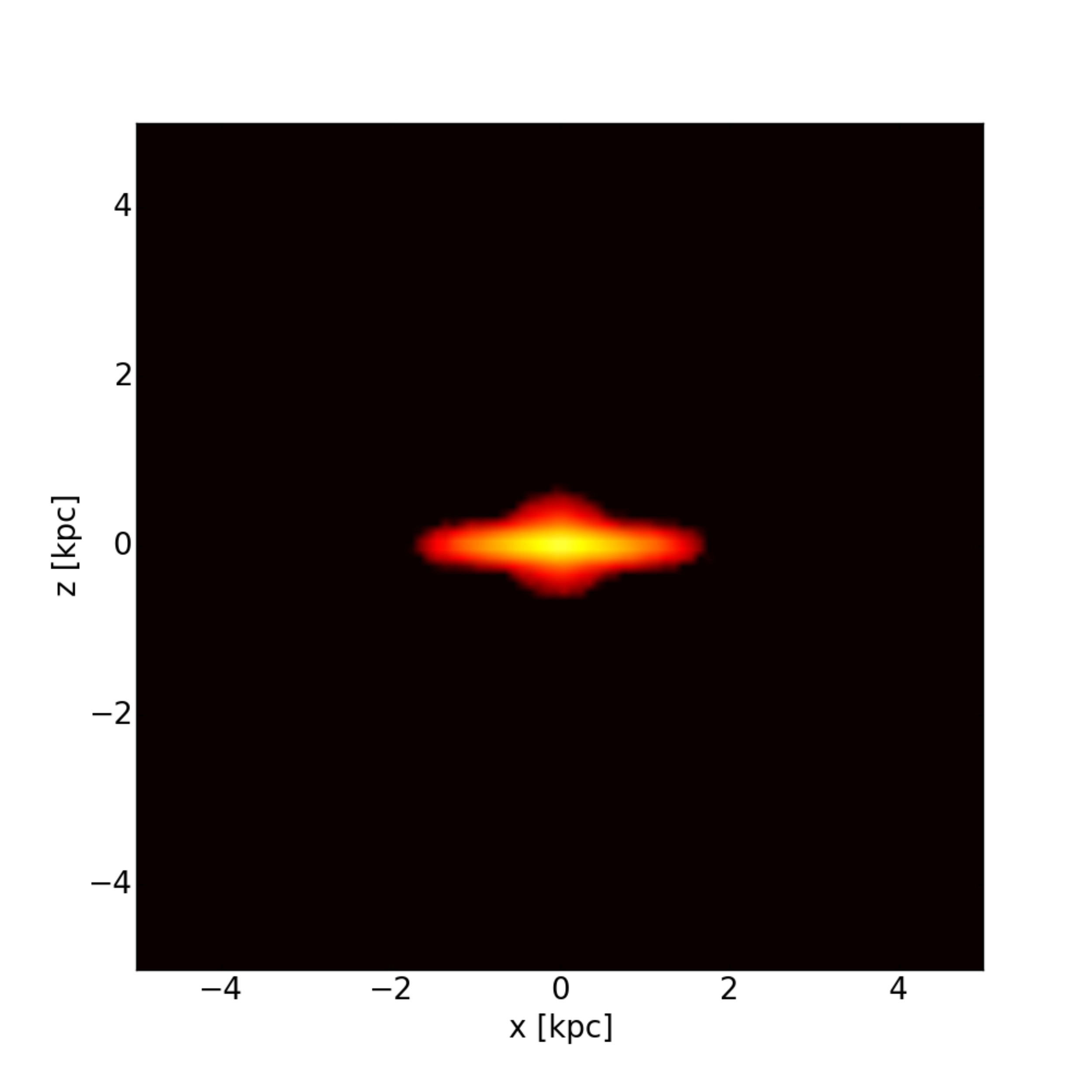}
\includegraphics[width = .3\textwidth]{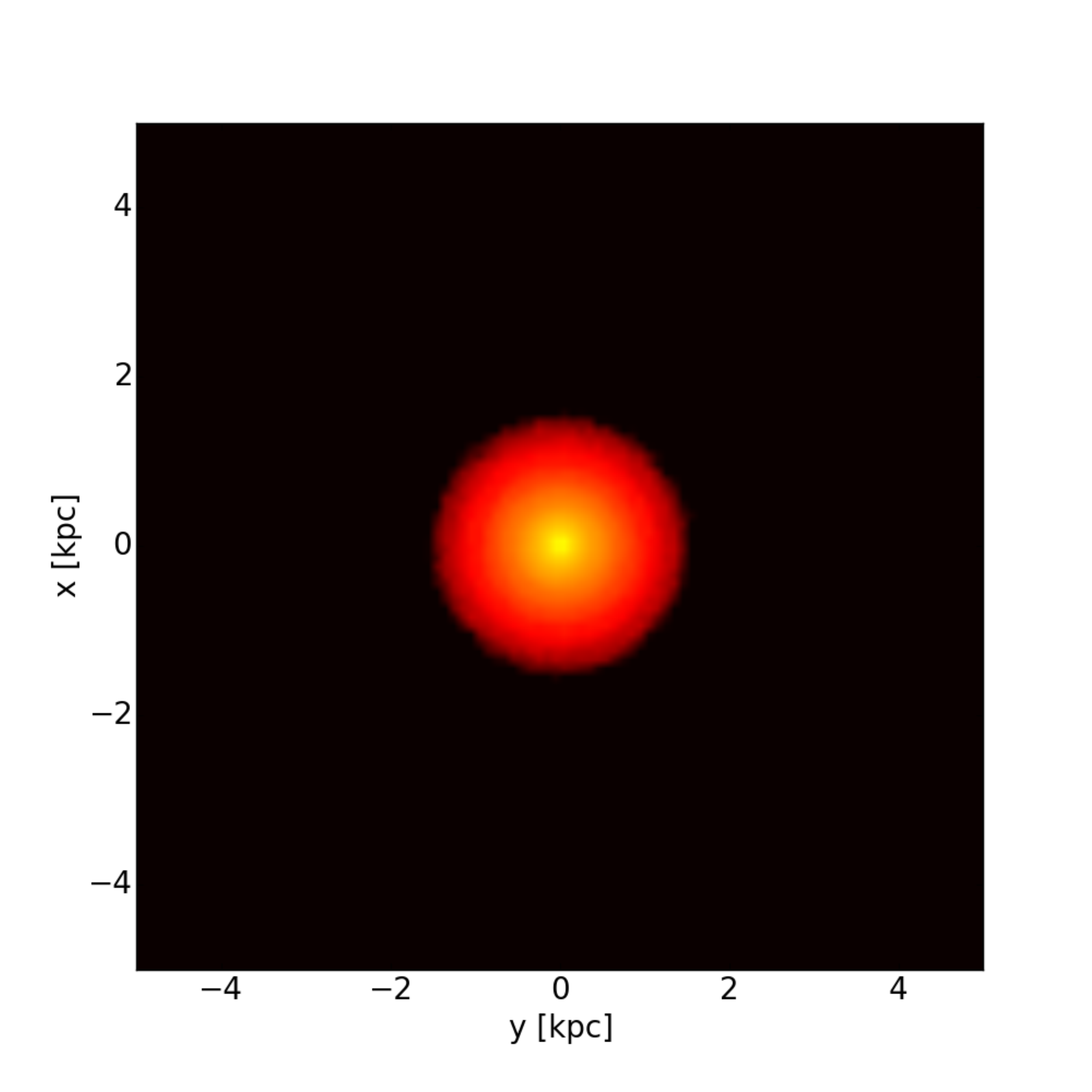}
\includegraphics[width = .3\textwidth]{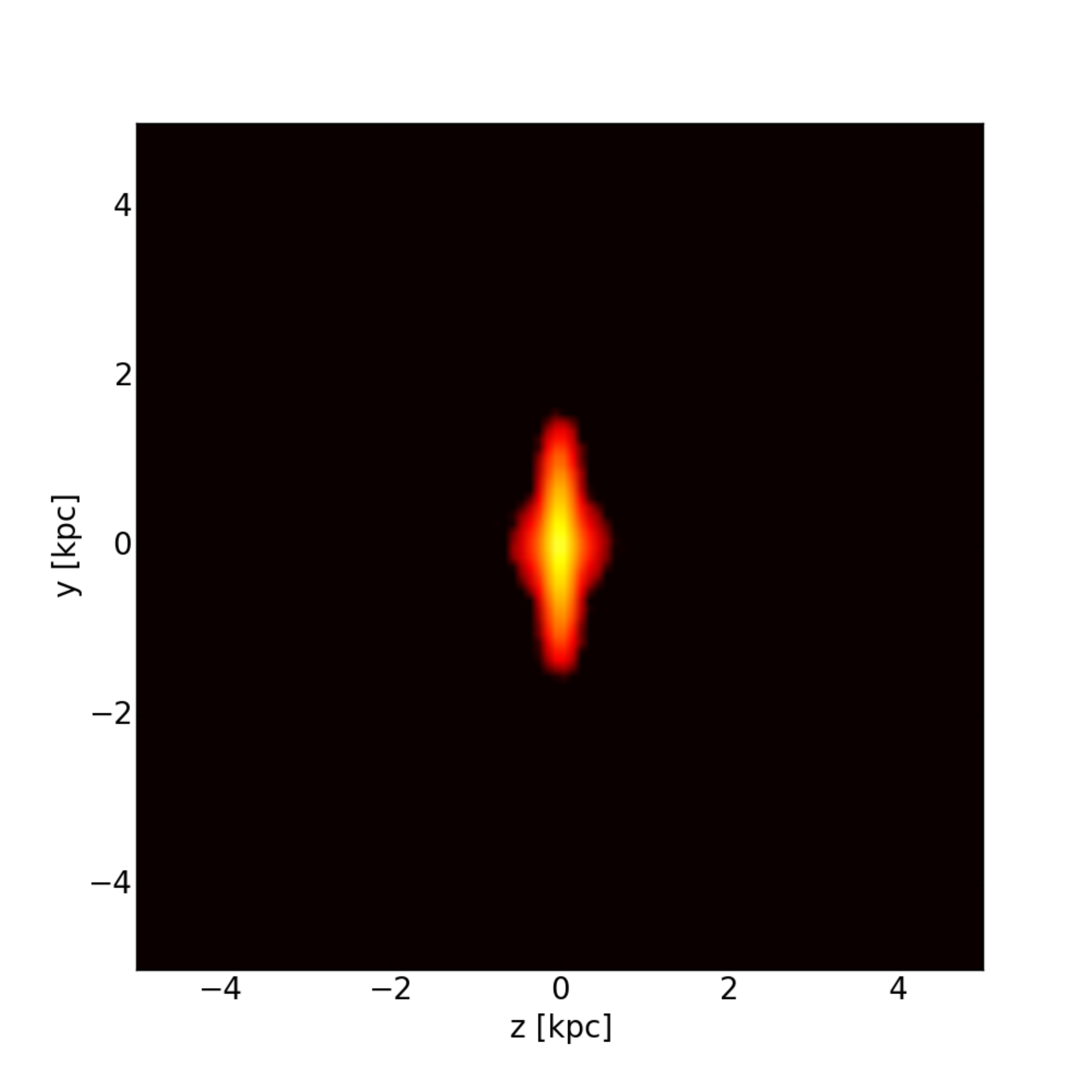}
\includegraphics[width = .3\textwidth]{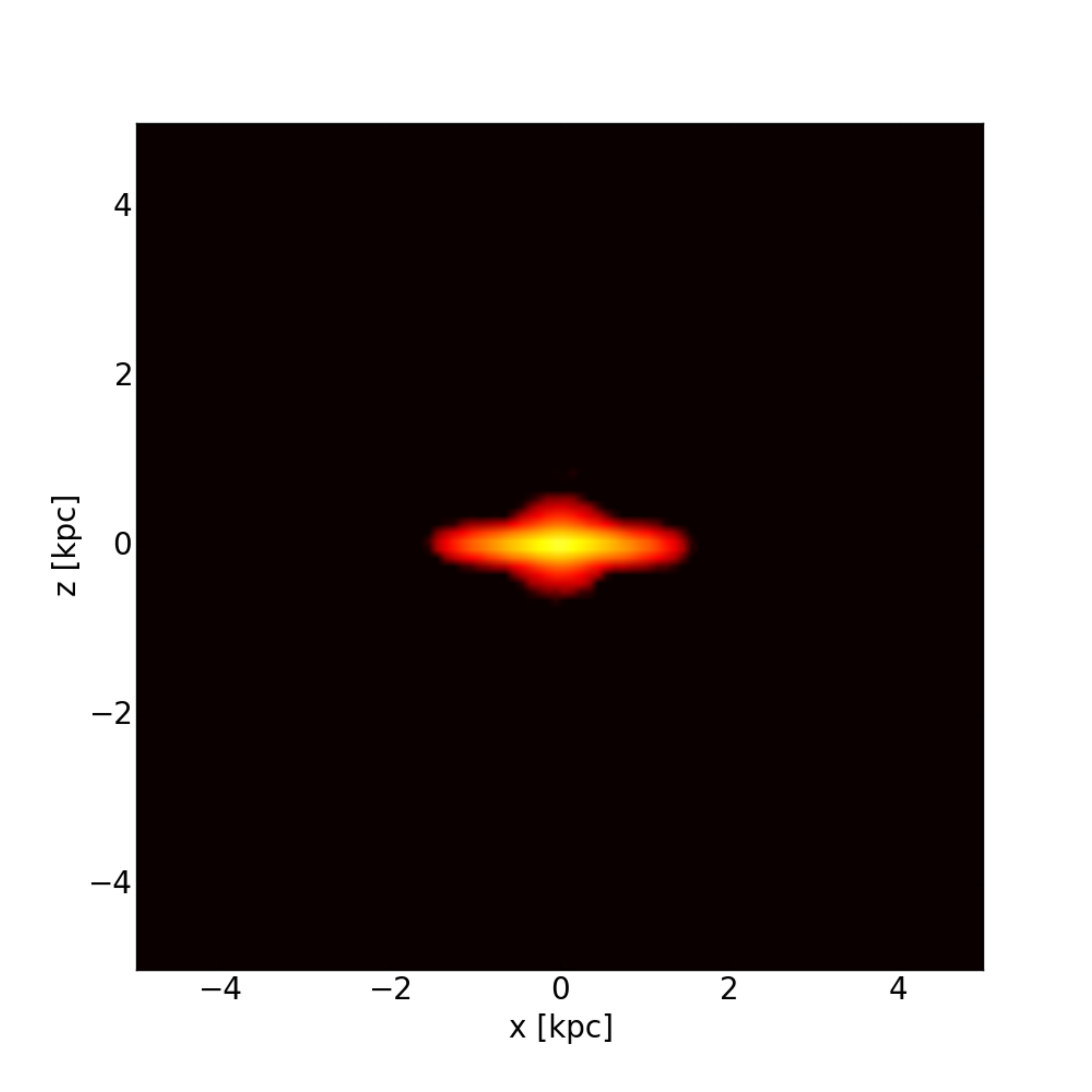}
\includegraphics[width = .7\textwidth]{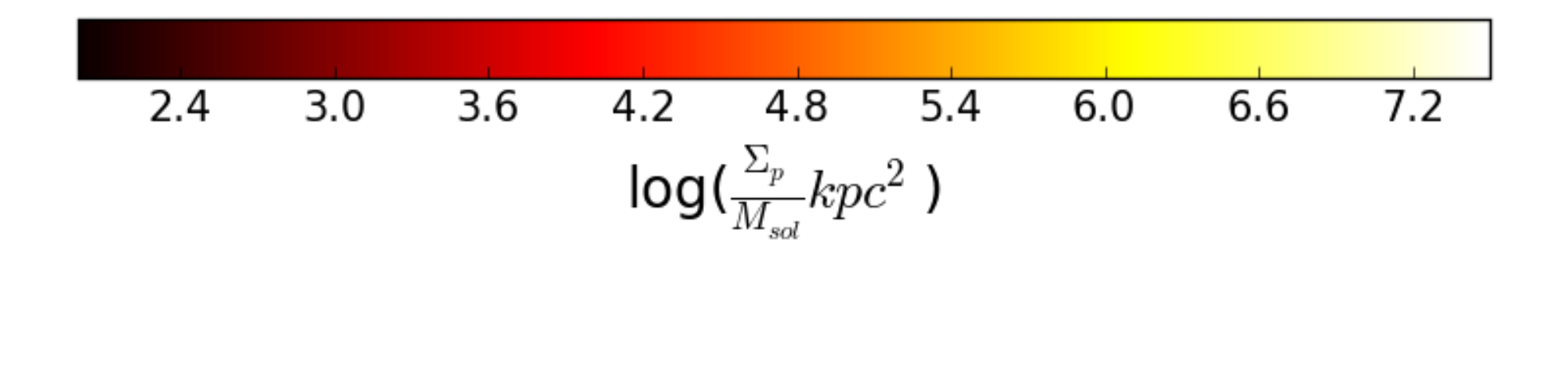}
\caption{\label{fig:figure3} The collage of images describes the stellar distribution of the high resolution version of satellites D, E, and F prior to the replacement.}
\end{figure}

Each of the twelve high resolution generated objects evolve in isolation for 1 Gyr prior to their introduction in the cosmological environment. The state of the objects at the end of the isolation run can be seen in figure \ref{fig:figure2} and \ref{fig:figure3}.  
The purpose of these simulations is to verify the stability of the models and allow the system to remove instabilities prior to replacement. 
Initial discreteness noise is unavoidable, 
and it is known to seed waves in cold disk-like systems which could
lead to mass redistribution (eg \cite{2011EAS....48..369M}). These often manifest themselves as 
transient spiral waves, although the method adopted by our initial condition 
generator combined with the low self-gravity of the light disks embedded in a massive halo
should minimize amplification of the waves. Furthermore, two-body relaxation between more 
massive dark matter particles and lighter stellar particles may also
lead to some spurious evolution (Mayer et al. 2001). 
All isolation runs were performed using the pkdgrav2 code (\cite{2001PhDT........21S}). The softening lengths of the particles were set according to the equation:
\begin{equation}
\epsilon_{dark} = 0.06 \left( \frac{M_{vir}}{10^9 M_{sol}} \right)^{1/3} kpc, 
\end{equation}
\begin{equation}
\epsilon_{star} = 0.02 \left( \frac{M_{stars}}{4 \cdot 10^7 M_{sol}} \right)^{1/3} kpc, 
\end{equation}
And are listed in table \ref{table:table3}.

\begin{figure}
\centering
        \includegraphics[width = 0.4\textwidth]{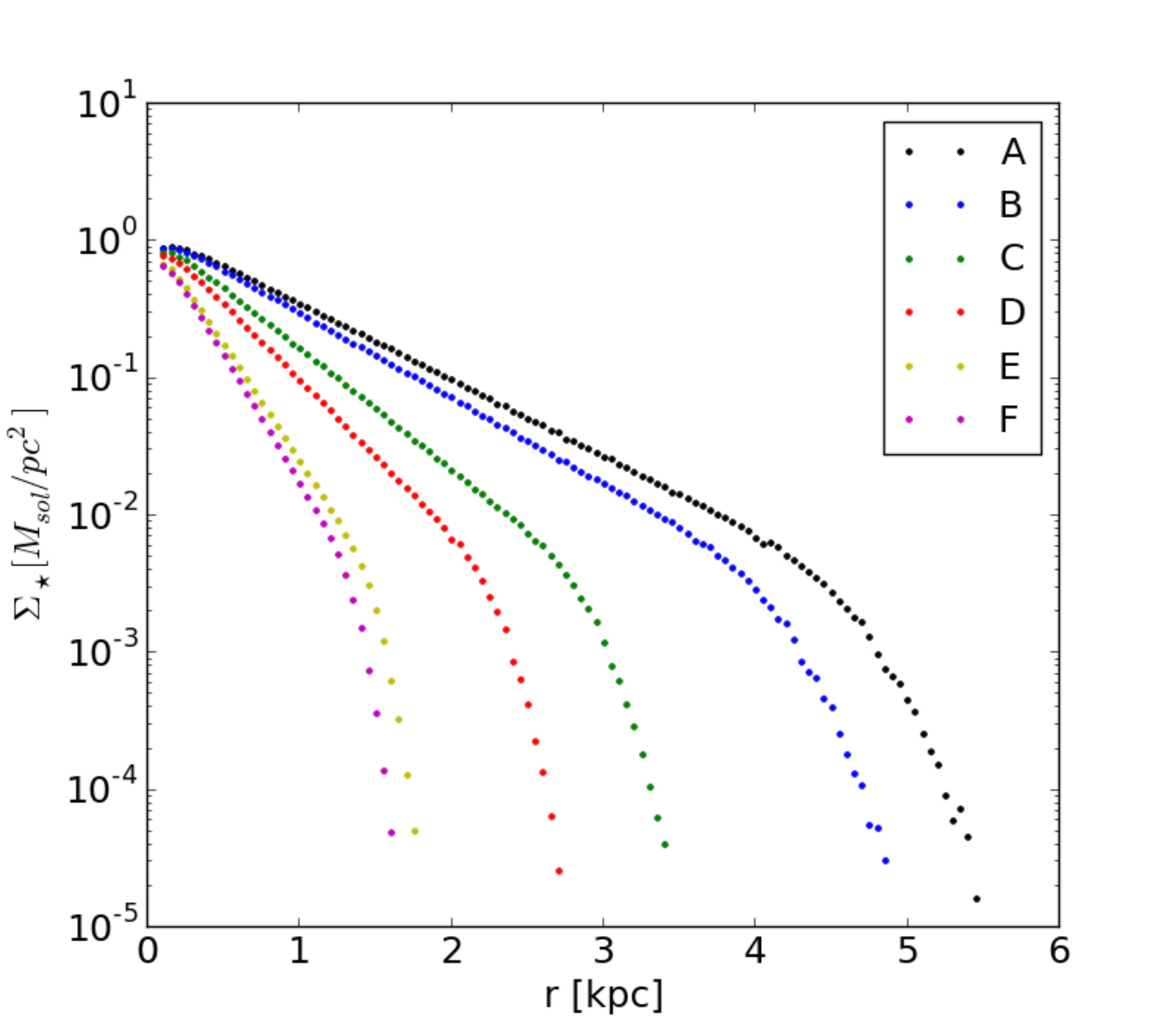}
        \includegraphics[width = 0.4\textwidth]{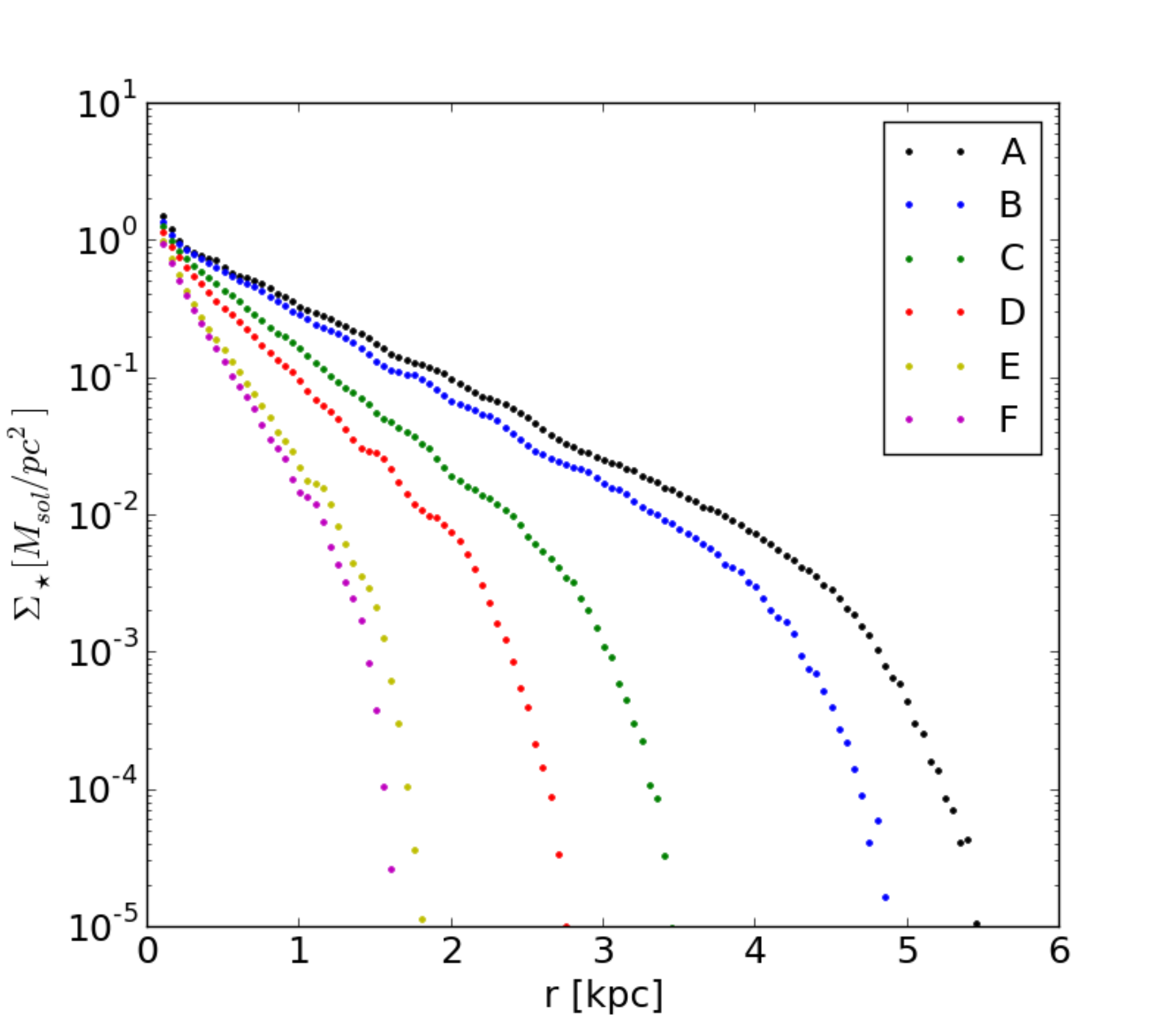}
        \includegraphics[width = 0.4\textwidth]{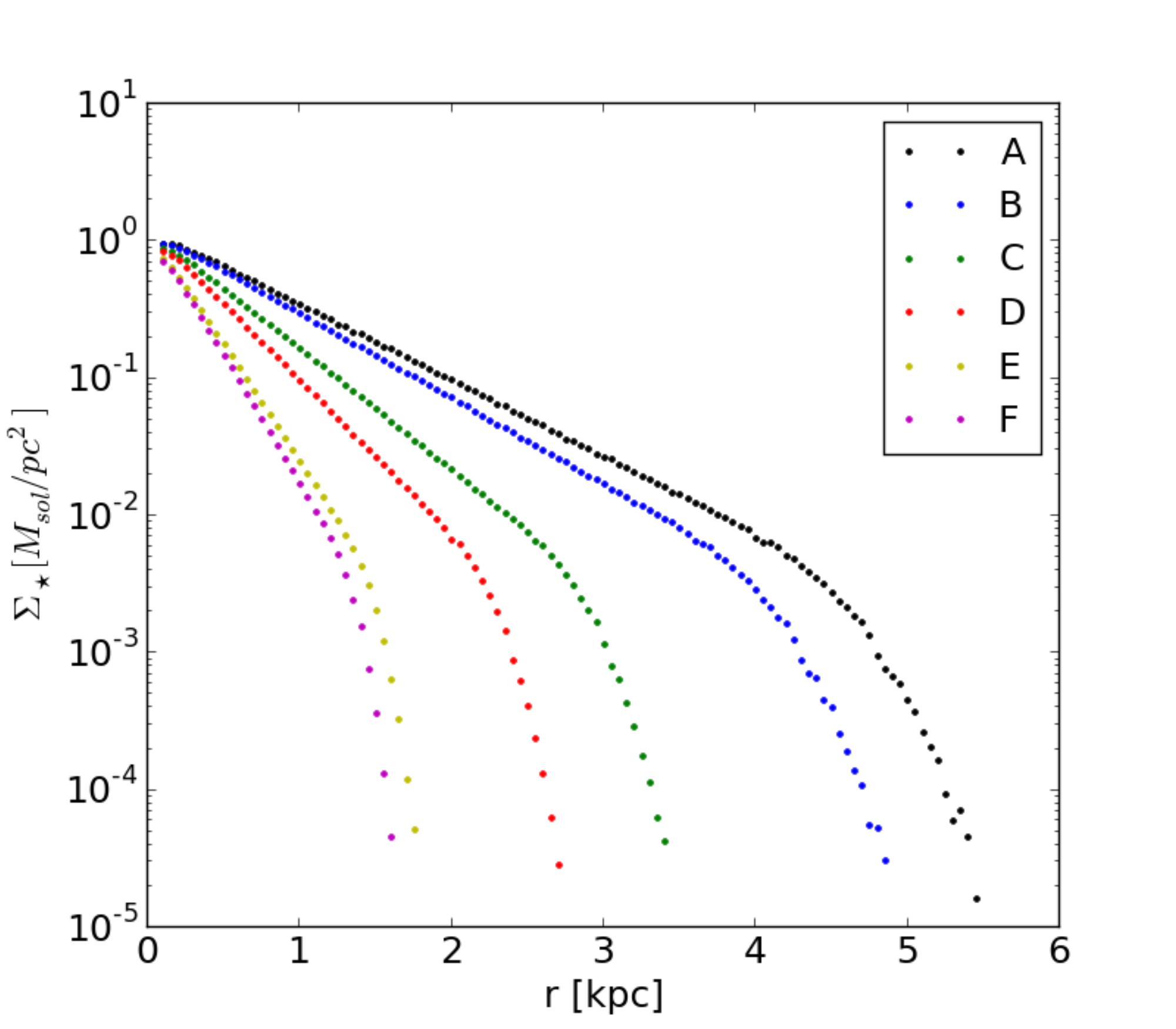}
        \includegraphics[width = 0.4\textwidth]{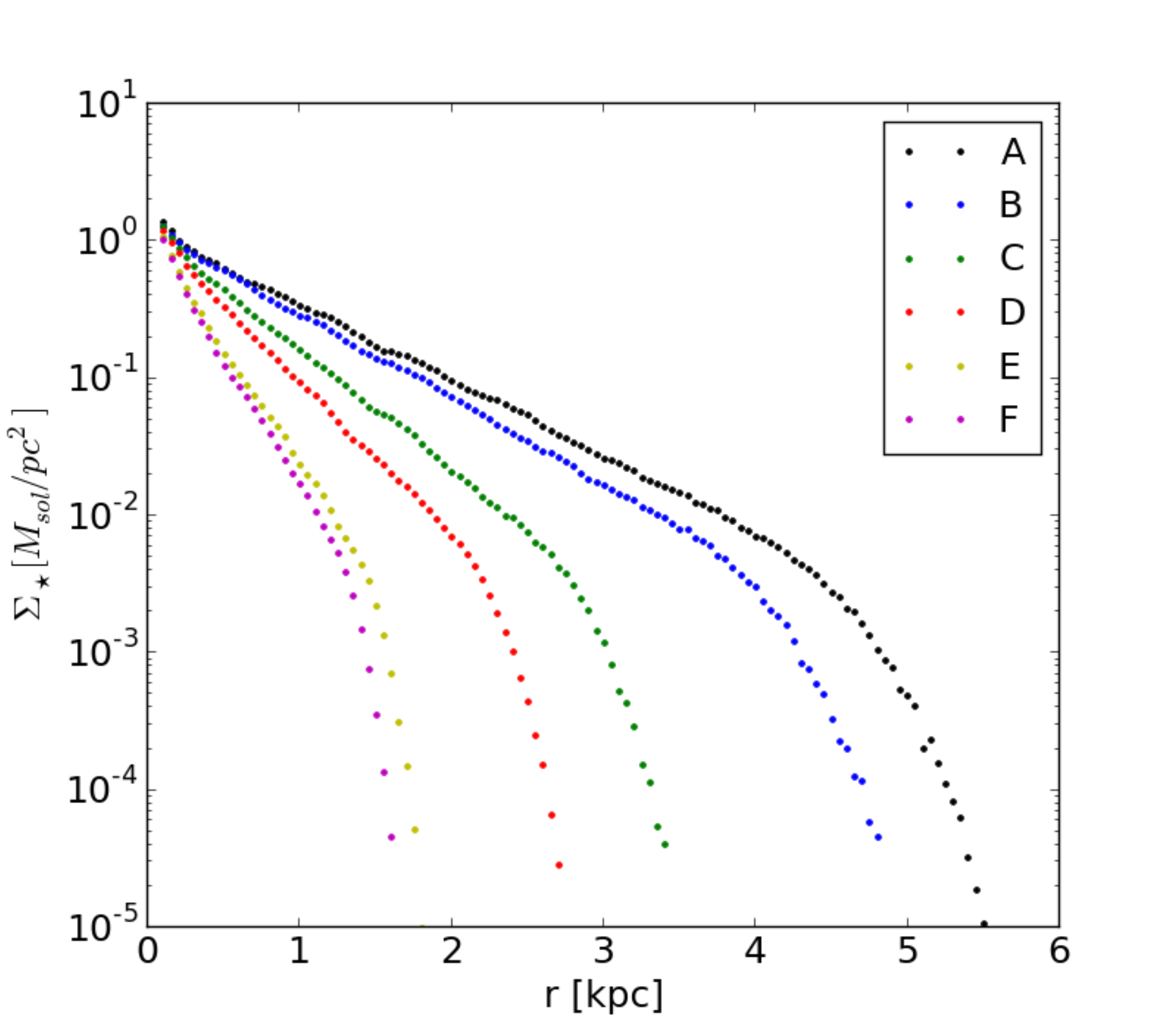}
\caption{\label{fig:figure4}Cylindrically averaged stellar surface densities of the high resolution objects before replacement. 
In the first/second row the halos with $\gamma = 1.0 / 0.6 $ are presented. In the left column the objects before the isolation run, 
while in the right the objects after the isolation runs are presented. The steeper drop in the value of the function for low stellar 
densities is caused by the cuttoff of the stellar disk which is initiated at five scale lengths from the center.}
\end{figure}

\begin{figure}
\centering
        \includegraphics[width = 0.4\textwidth]{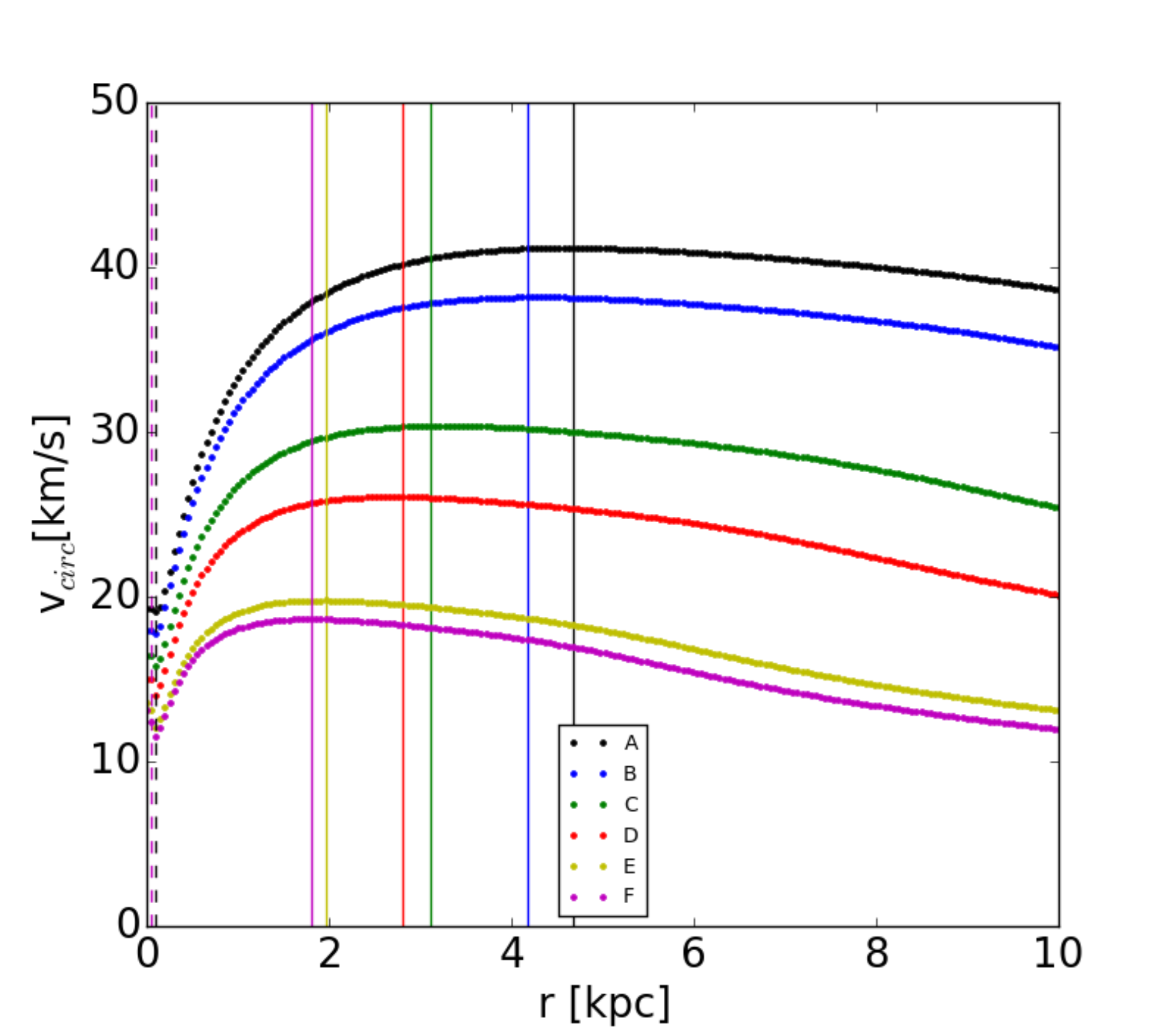}
	\includegraphics[width = 0.4\textwidth]{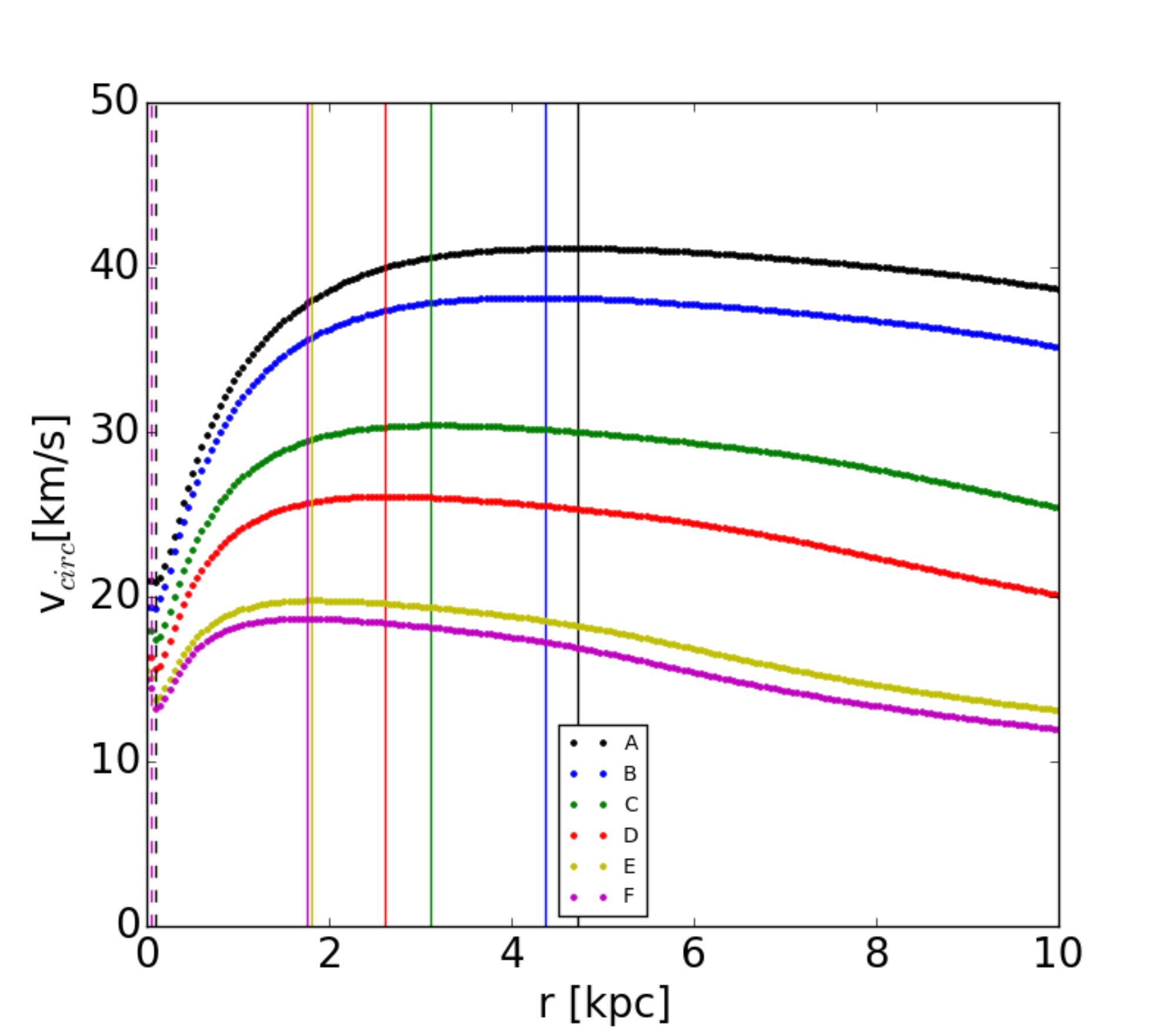}
        \includegraphics[width = 0.4\textwidth]{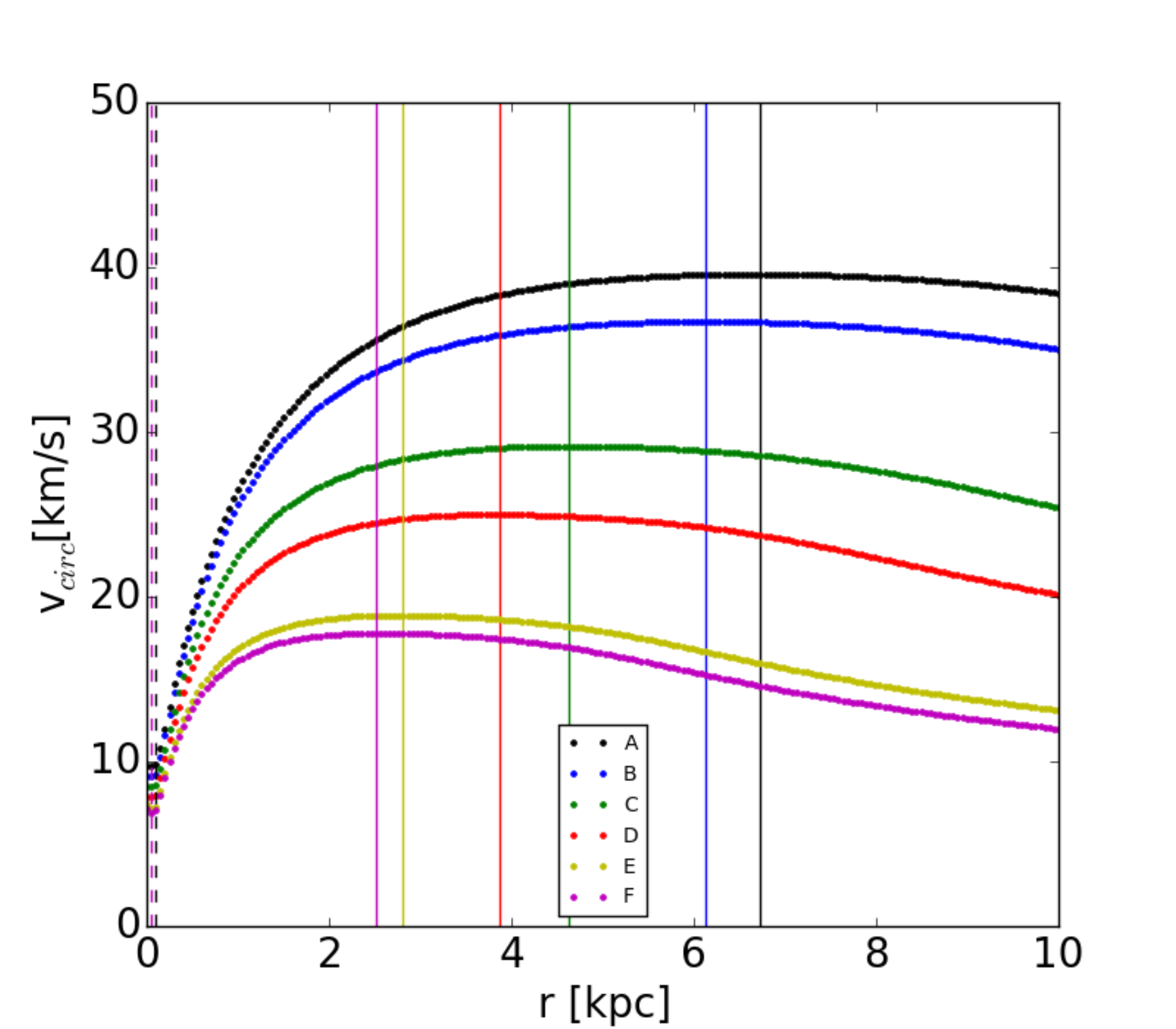}
	\includegraphics[width = 0.4\textwidth]{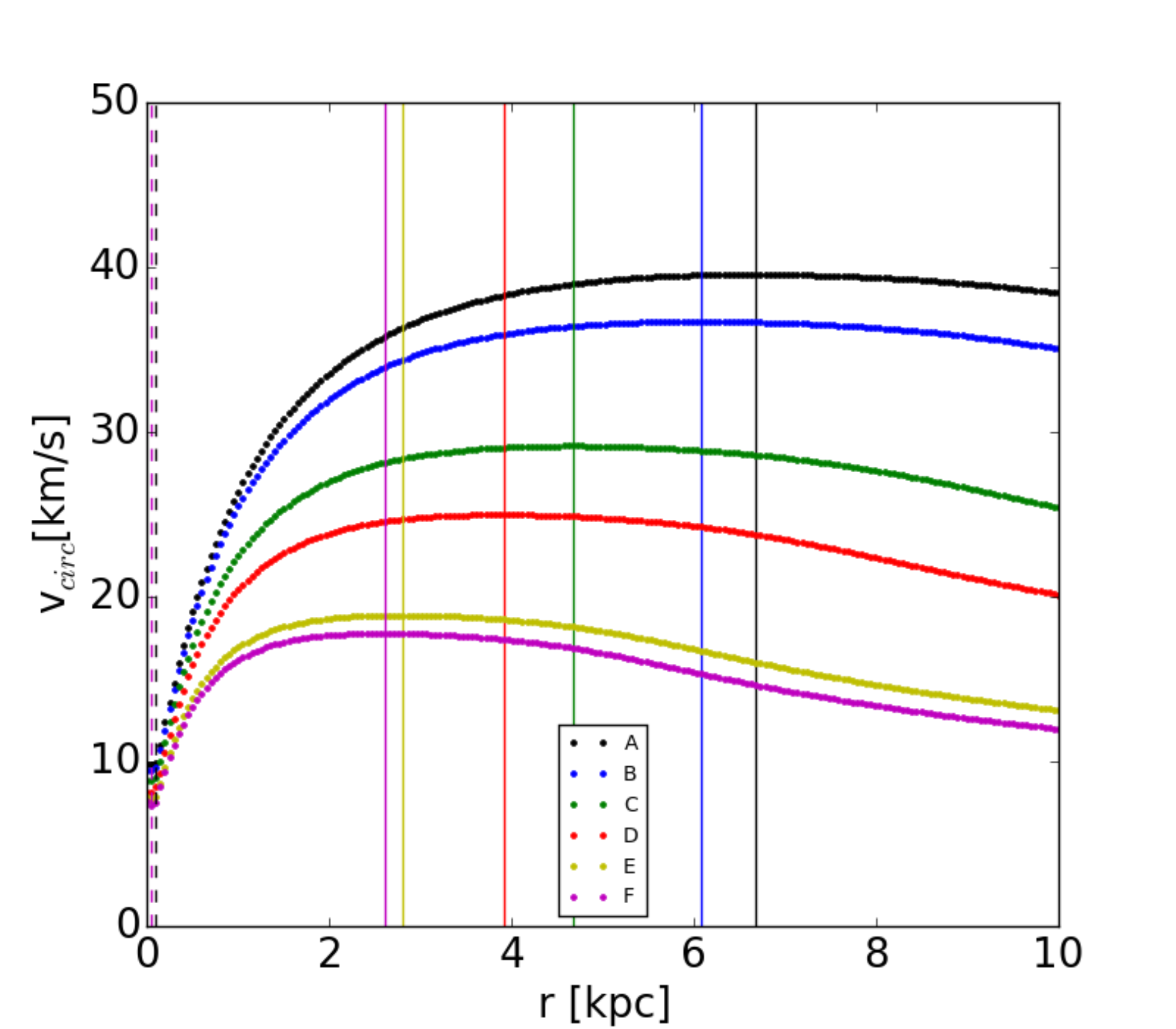}
\caption{\label{fig:figure5}Spherically averaged circular velocities of the high resolution objects as a function of radius. 
In the first/second row the halos with $\gamma = 1.0/0.6$ are presented. 
In the left column the curves corresponding to the objects before the isolation run are shown, 
while in the right the curves corresponding to the objects after the isolation runs are presented. 
The solid color vertical lines mark the radii corresponding to the maximum circular velocities and the 
dashed lines mark the force resolutions of the largest and smallest objects.}
\end{figure}

\begin{figure}
\centering
        \includegraphics[width = 0.4\textwidth]{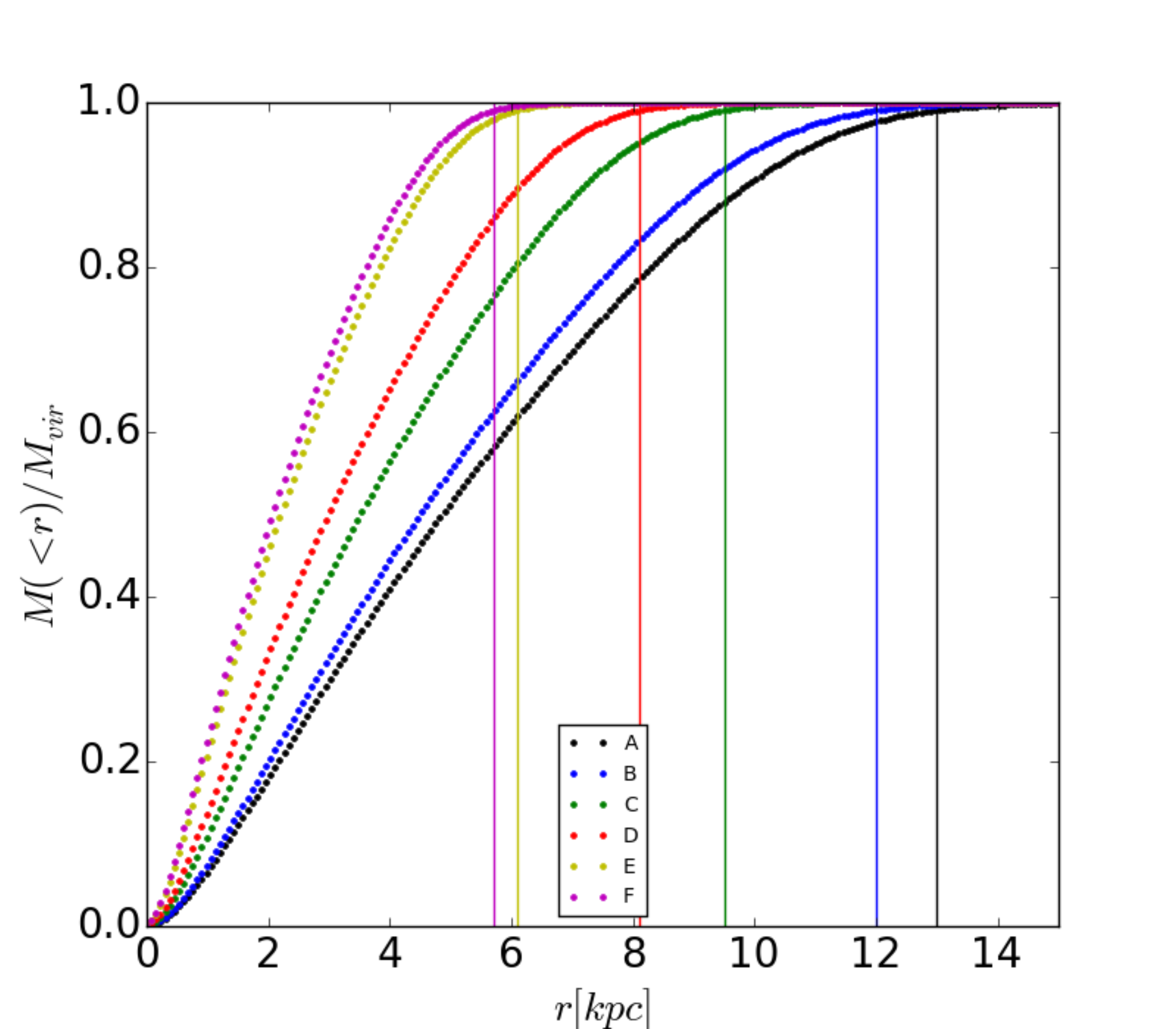}
        \includegraphics[width = 0.4\textwidth]{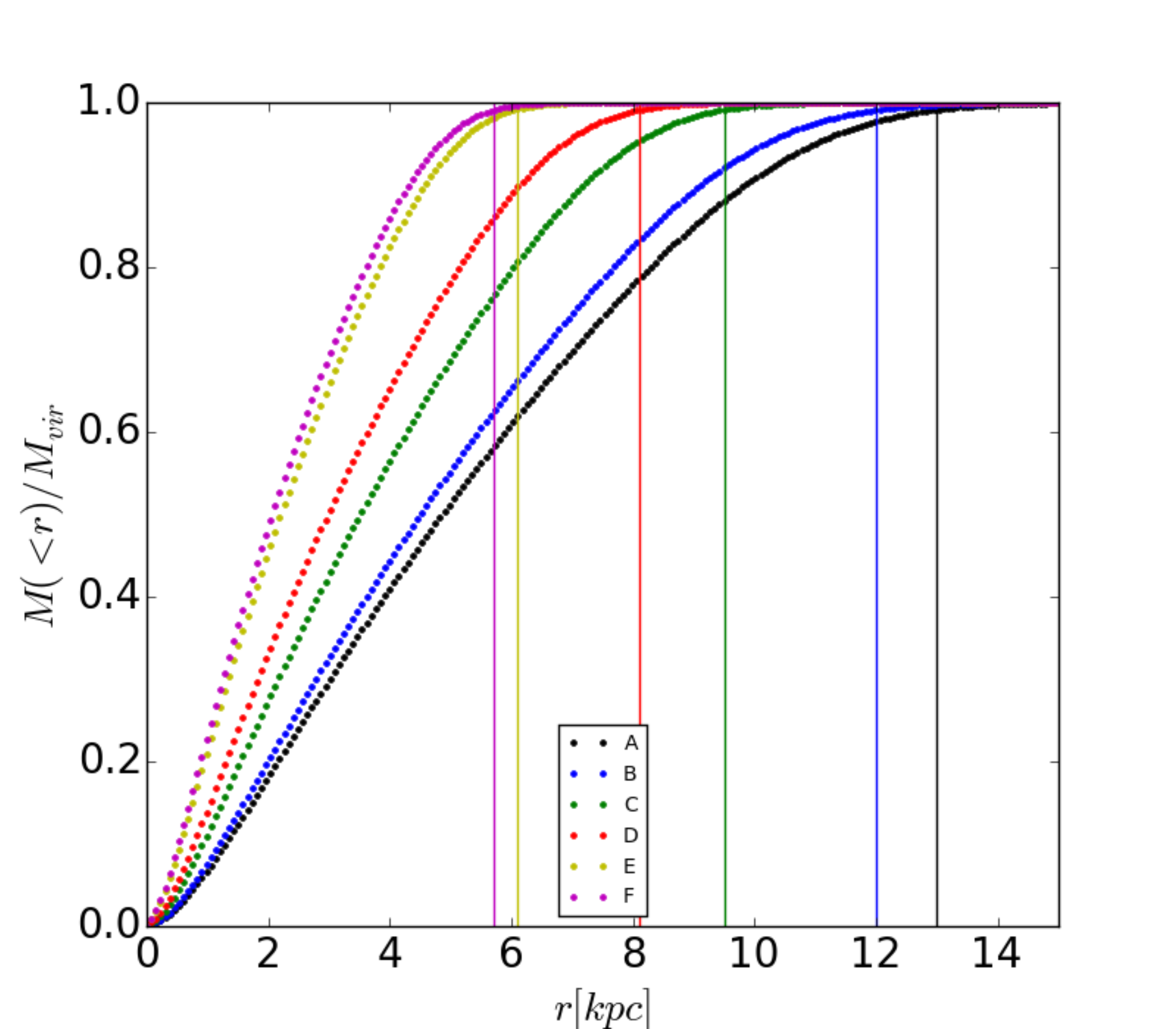}
        \includegraphics[width = 0.4\textwidth]{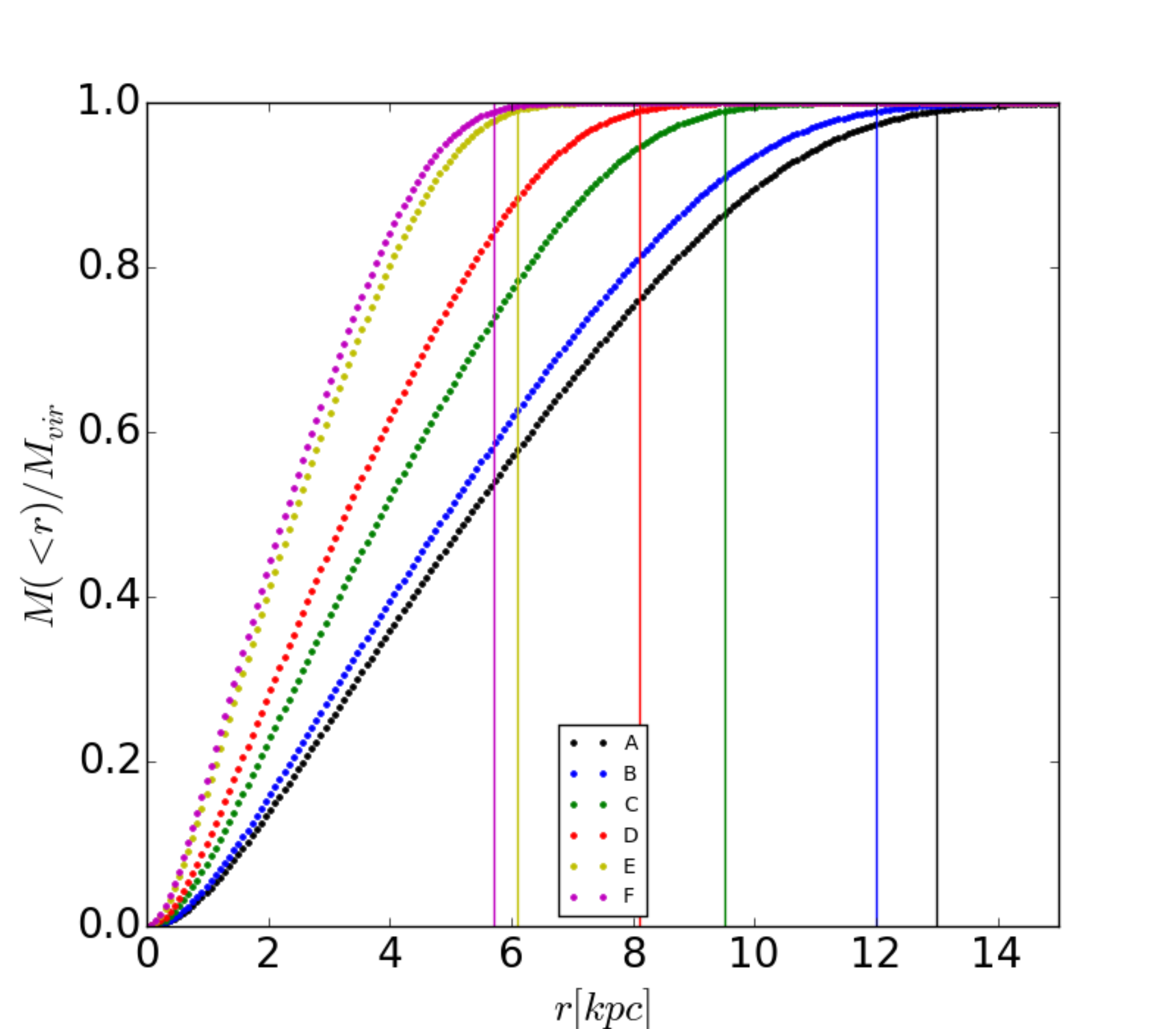}
        \includegraphics[width = 0.4\textwidth]{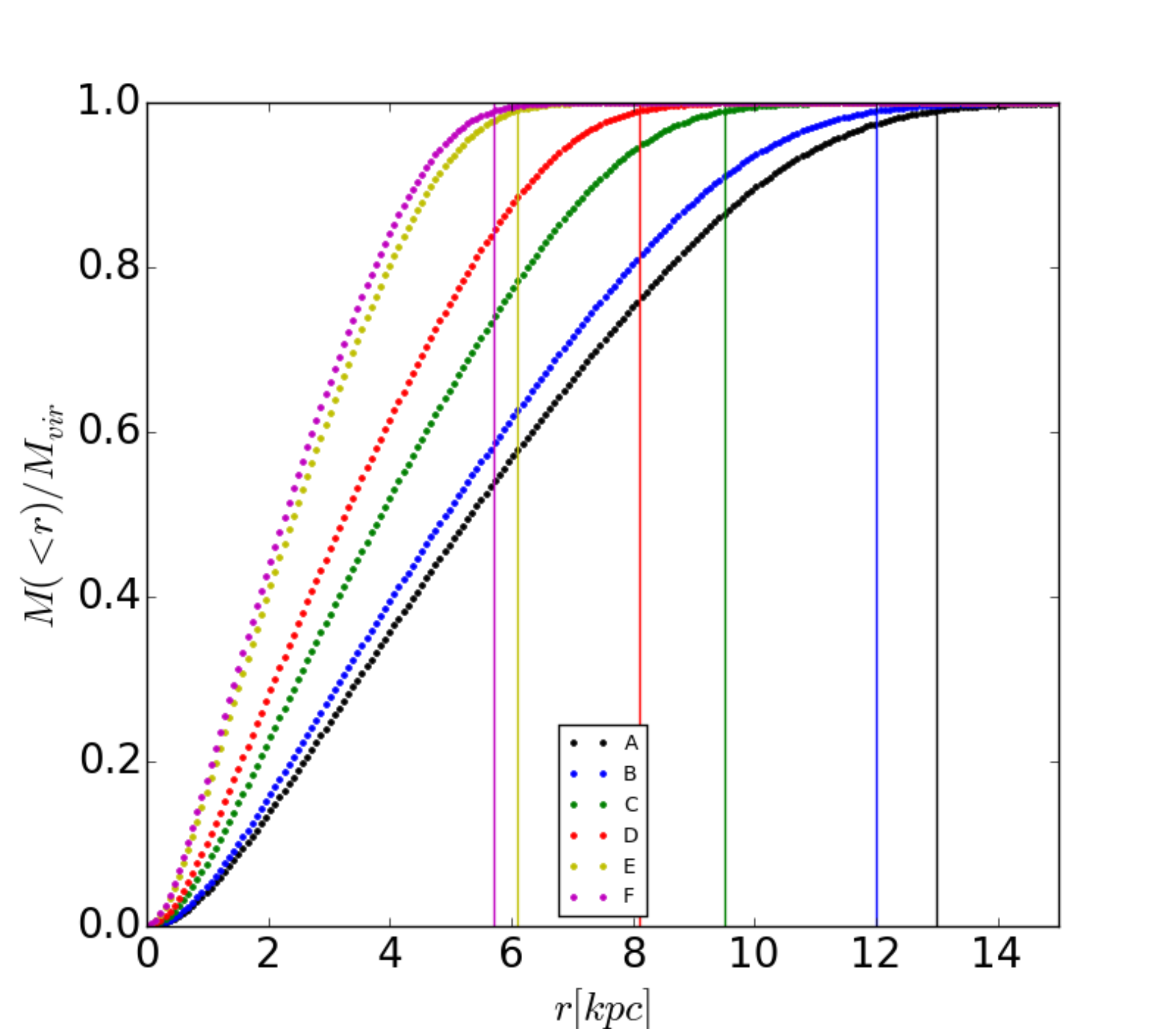}
\caption{\label{fig:figure6}In the left row the cumulative masses of the generated models prior to the isolation run are presented as a function of distance from the center before the isolation run. In the right column the values of same functions after the isolation run are presented. 
The top row corresponds to the objects with $\gamma = 1.0$ while the bottom to the objects with $\gamma = 0.6$. 
The colored vertical lines correspond to the original virial radii of the respective halos.}
\end{figure}

The baryonic distributions of the constructed galaxy models with a cusp coefficient equal to unity are presented in 
figures \ref{fig:figure2} and \ref{fig:figure3}. 
The previously mentioned images correspond to the objects at the end of the isolation run and at the moment of replacement. 
Figure \ref{fig:figure4} shows the cylindrically averaged surface density at the beginning of the isolation run and at the end. 
Small fluctuations are visible as small amplitude waves, which cause departure from perfect axisymmetry.
Despite the existence of the local density fluctuations for both versions there are no 
significant changes happening on scales comparable to the scale length during the isolation run.      
Moreover, the analysis of the global mass distribution including the dominant dark matter component as 
shown by the circular velocity curves in figure \ref{fig:figure5} 
along with the radial distribution of the cumulative mass present in figure \ref{fig:figure6}, 
do not show any appreciable changes. We conclude that the stability of the models is satisfactory.

%%%%%
\subsection{Replacement method and numerical simulations}

At the end of their evolution in isolation the galactic models replace the original objects in four steps. 
All bound particles of the original objects are removed after measuring their centre of mass (COM) velocity and position 
from the cosmological box at redshift 2.87. The particles of the newly created pairs of galaxies have their residual COM velocities and positions removed. 
Afterwards their individual positions and velocities are rotated with random matrices generated with the \cite{arvo1991}. 
Finally the COM phase space coordinates are translated to match the COM phase space coordinates of the original objects that they replace.     
The resulting modified versions of the cosmological volume containing ErisDark are used as initial conditions for the gravity calculation runs down to redshift 0.
ChaNGa was used to carry out these new cosmological runs.
The code ChaNGa uses a fast hierachical oct-tree gravity solver, and is based
on the Charm++ parallel programming infrastructure (\cite{CharmppPPWCPP96}).
It leverages the object-based virtualization and
data-driven style of computation inherent in Charm++ to adaptively overlap communication and computation and achieve high levels of resource
utilization on large systems (\cite{changa2008}, \cite{changa2010}).
The Charm++ load balancing infrastructure is used to distribute pieces of this tree across processors. The leaf
nodes are buckets that contain several particles (usually 8 to 32) whose force calculations are collectively optimized.  At each level of the
tree, multipoles are calculated to speed distant force evaluations.
Time adaptivity is achieved by assigning individual timesteps to particles.  A special load balancer was developed by
the UIUC  Charm++ team to efficiently handle the resulting processor load imbalance (\cite{2015ComAC...2....1M}).

\section{Results and data analysis}
\label{sec3}
The resulting raw simulation data is analized using the AHF in the same manner as performed with the original ErisDark simulation data. 
From the set of halos generated by the AHF code, the six pairs of satellites are identified through the indices of bound star particles. 
As output of the AHF code, we use in our satellite analysis only the list of bound dark matter and star particles. 
There are cases when the halo finder does not detect one of the objects at a particular epoch either due to the pericenter shock or 
temporary interference with other large subhalos. Such events will appear as unavoidable interruption of the flow of data points on the analysis plots. 
As soon as the object is identified again by the AHF code, the flow of data points in the analysis plots  resumes.   \\

\begin{figure}
\centering
        \includegraphics[height = 0.4\textwidth]{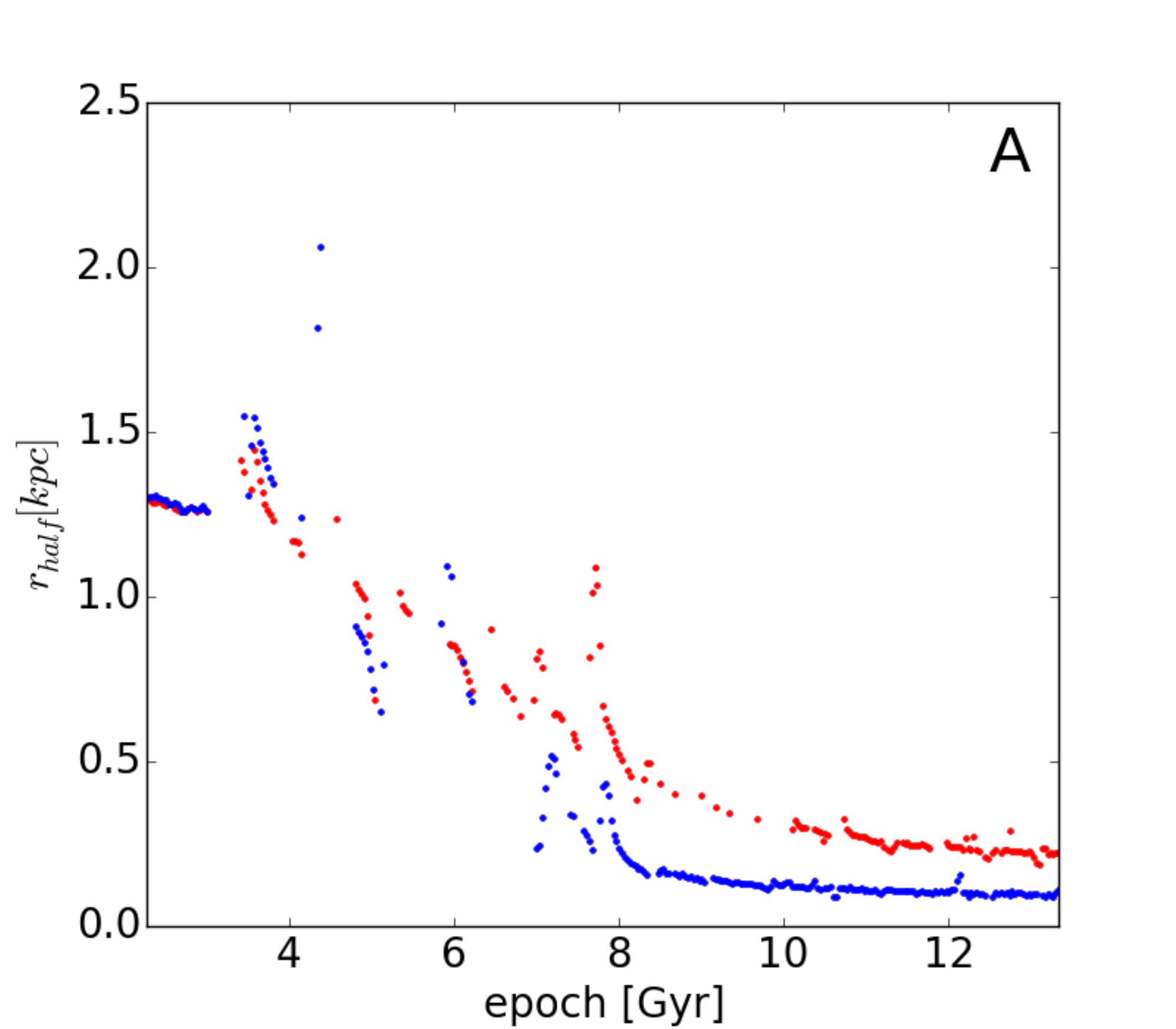}
        \includegraphics[height = 0.4\textwidth]{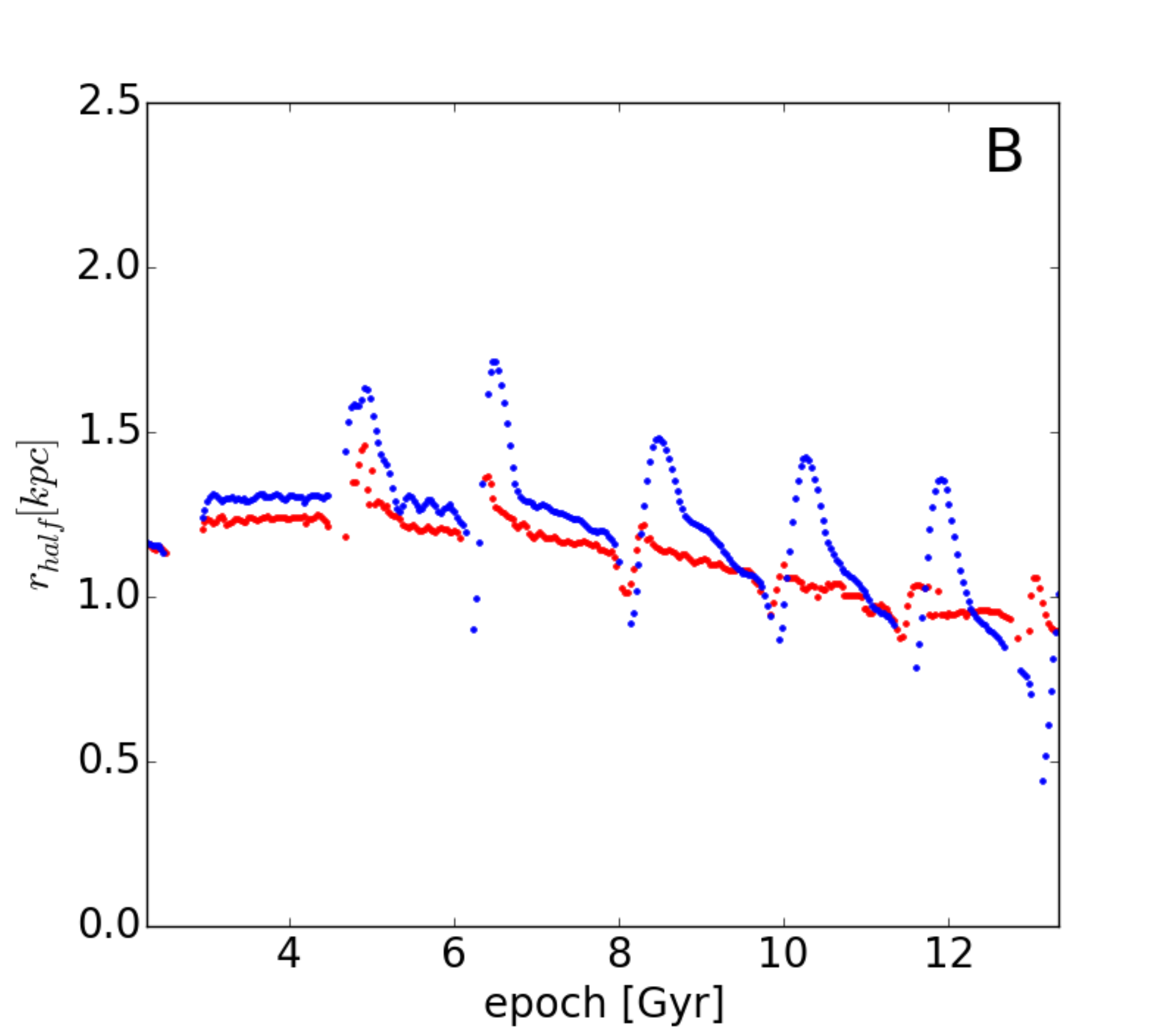}
        \includegraphics[height = 0.4\textwidth]{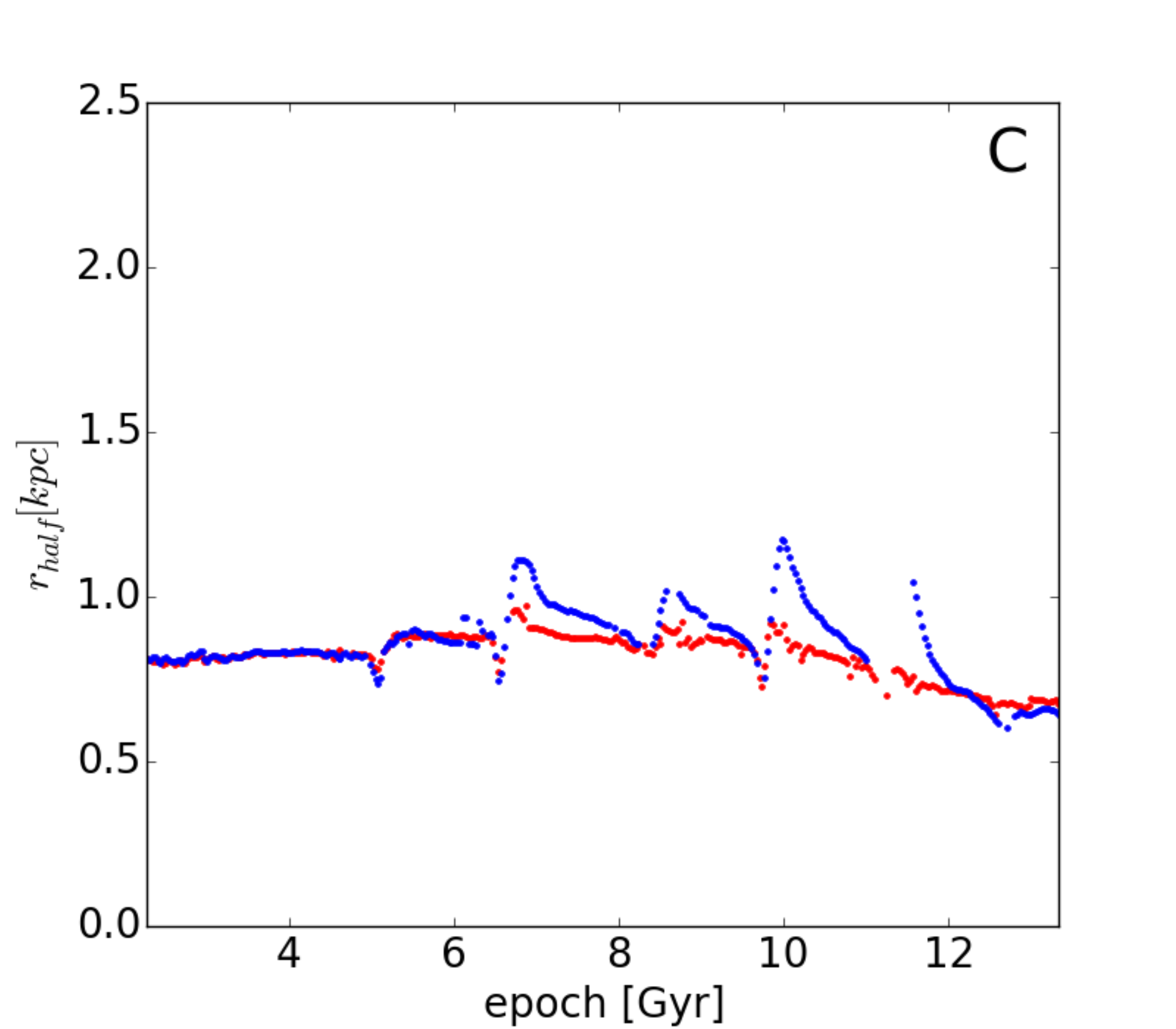}
        \includegraphics[height = 0.4\textwidth]{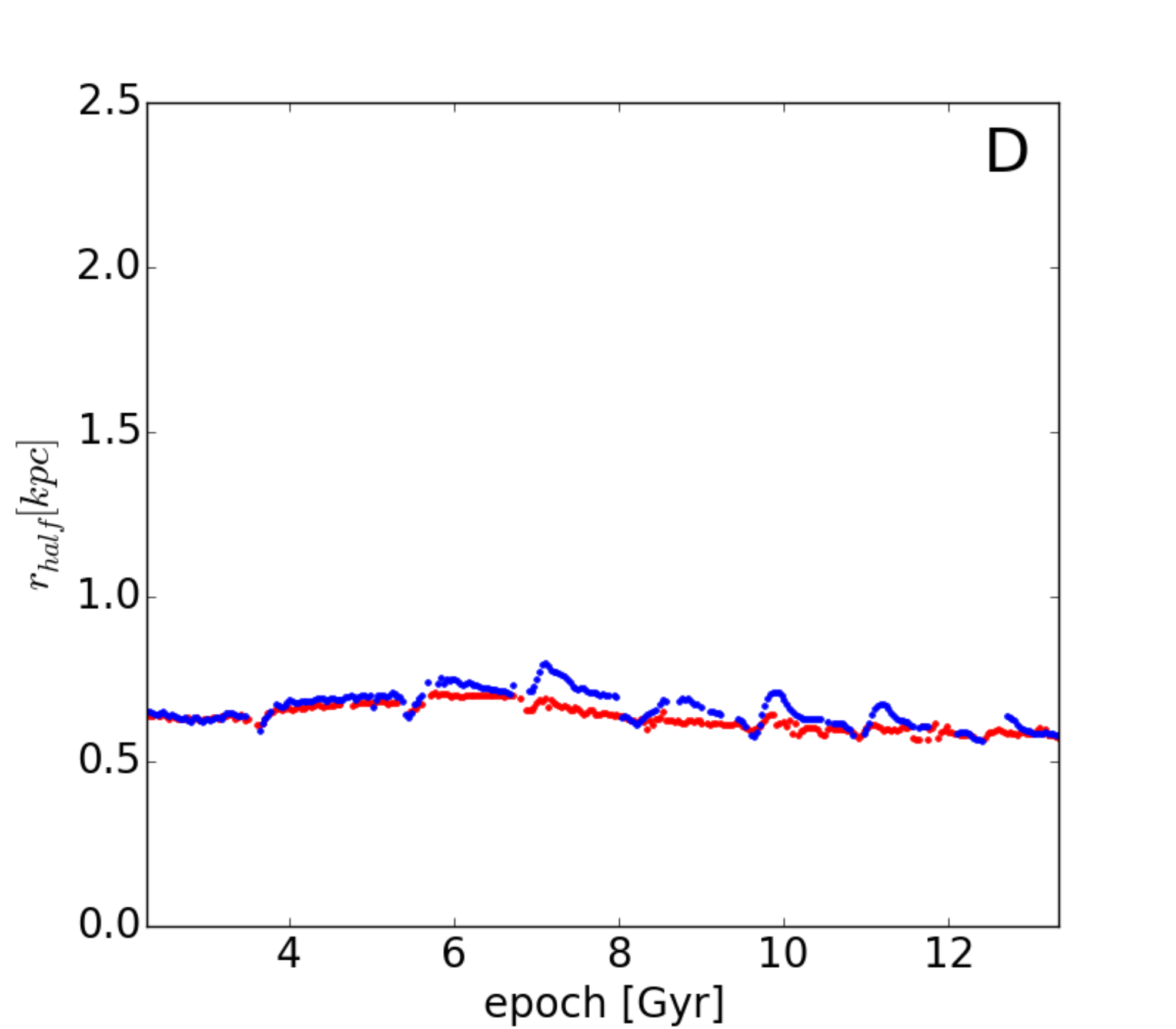}
        \includegraphics[height = 0.4\textwidth]{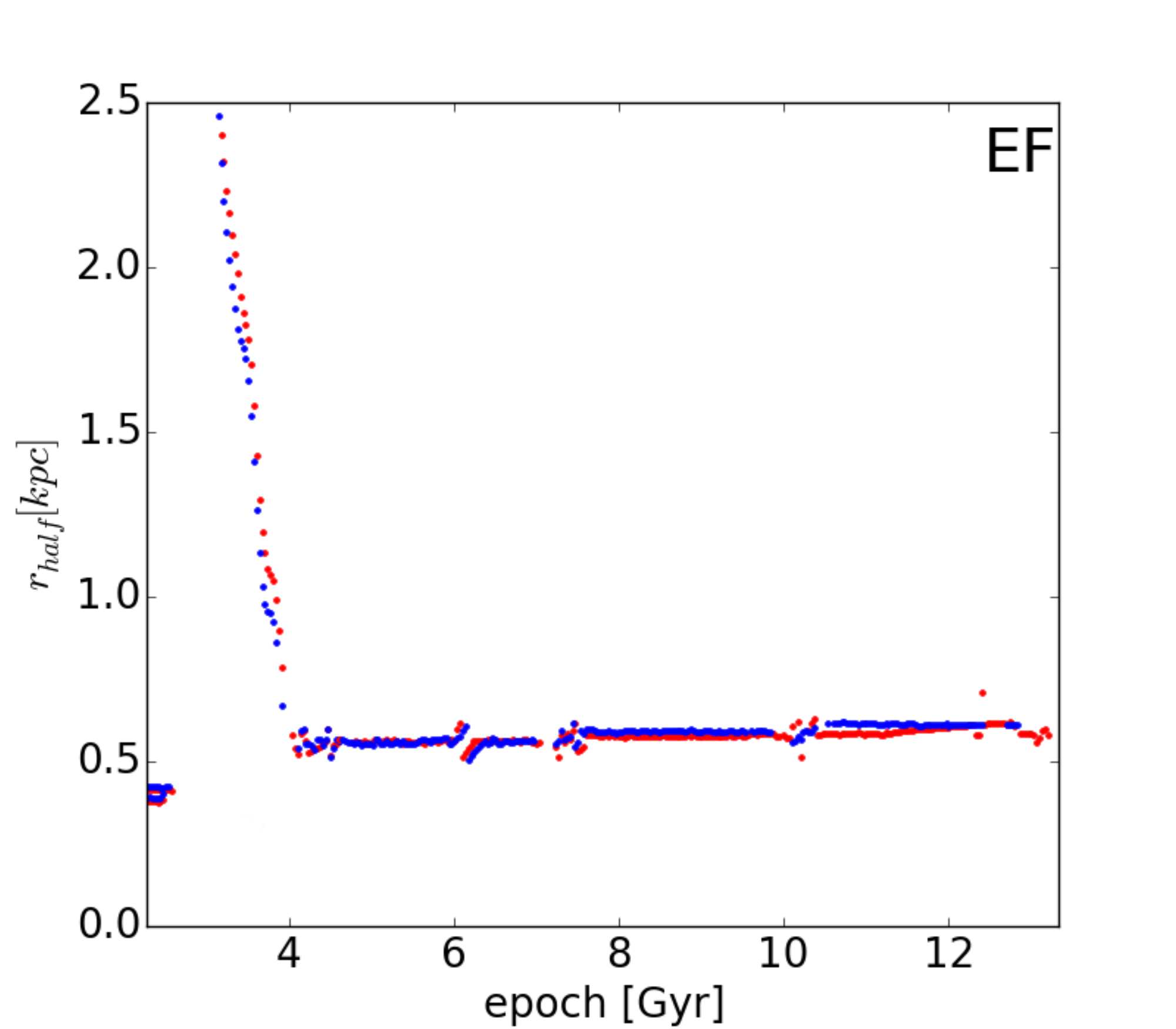}
\caption{\label{fig:figure7} The collage presents the time evolution of the 3D half light/mass radii 
for the five resulting pairs of satellites. The data points corresponding to galaxies with shallow density 
profiles are marked in blue and the points corresponding to galaxies with steep density profiles are marked in red.}
\end{figure}

The center of the satellite is defined as being the position of the stellar density peak of the respective object. 
From this point a 3D half light/mass radius can be defined as the radius of the sphere centered on the previously 
mentioned density peak which includes and excludes half of the mass amount of bound stellar particles. 
All the values of the half light radius associated with each object at the available evaluation epochs are presented in figure \ref{fig:figure7}.  
The next step in the preparation of the satellite data for analysis is the alignment of the object. 
This is performed by defining a set of particles that include both dark matter and stars inside a sphere centered 
on the stellar density peak and with a radius equal to two half light radii. 
All the bound particles of the object including the ones exterior to the two half light 
radii sphere are rotated such that the moment of inertia tensor, 
corresponding to the set of stellar particles interior to the previously mentioned sphere, is diagonalized. 
This process is repeated for each individual object at each epoch of measurement.

Once the satellite is identified and analysed there are a number of diagnostics that are routinely measured, 
as described in the following subsections. In addition to dynamical masses, dark matter masses and stellar masses 
(see tables \ref{table:table4} and \ref{table:table5} for the results at the starting and final time), the key diagnostics used to quantify the morphology of the model galaxies are the shape
as measured by the axis ratios, the ratio between 1D velocity dispersion $\sigma_*$ and rotational velocity $v_{rot}$ of the stars 
(note that we refer here to the actual stellar rotational velocity rather than to the circular velocity), 
and the surface density profiles of the stars. The characteristic velocities are measured inside ($\sigma_*$) and around ($v_{rot}$)
the half-mass radius, which is defined for the whole 3D mass distribution.

Our classification scheme to decide whether or not the final remnant is a dSph is more stringent than the one used in (\cite{2011ApJ...726...98K};\cite{2013ApJ...764L..29K}). 
There two criteria were used simultaneously, namely that the object had to have $c/a > 0.5$ and $v_{rot} / \sigma_*  < 1$. 
With the latter criterion outliers from the typical structural properties of dSphs, such as the Tucana dSph, which has $v_{rot}/\sigma_* \sim 1$, 
would be included. Here instead we use only the criterion $v_{rot}/\sigma_* < 0.5$, which is indeed the case for nearly all observed 
dSphs and that, as we will see, automatically guarantees also that $c/a < 0.5$ in our simulated remnants at $z=0$.
Finally, for brevity we will refer to "cored models" when we talk about the models with inner slope of the dark halo $\gamma=0.6$,
and to "cuspy models" when we discuss models with an NFW profile.

\subsection{Dimensions along the principal axes}
With the object aligned and principal axes identified three maps of the projected stellar  mass corresponding to 
the three pairs of principal axes can be generated. Perpendicular to each of the vectors defining the axes directions 
a square grid with size $4 R_{1/2} \cdot 4 R_{1/2}$ and a million square cells is defined. 
For each cell a projected mass is defined that equals the mass of all bound star particles whose projection on 
the grid is included in the respective cell.\\
Using the information in the projected mass maps, the dimensions of the object along the principal axes can be determined. 
In the process of dimensions calculations ellipses containing cells with different levels of mass density are being fitted. 
The distribution of these ellipses around the center of the image allows for the calculation of the ratio of the dimensions. 
For each of the three planes ${(x,y), (y,z), (z,x)}$ the ratio between the shortest and longest dimensions $c$ and $a$ is measured:
${ \frac{c}{a}_{x,y}, \frac{c}{a}_{y,z}, \frac{c}{a}_{z,x} }$. \\

After the previously presented analysis is performed the first available scalar indicator of the transformation of the object can be determined:
\begin{equation}
\frac{c}{a} = min \left( \left( \frac{c}{a}\right)_{x,y}, \left( \frac{c}{a}\right)_{y,z}, \left( \frac{c}{a}\right)_{z,x} \right).
\end{equation}              
The time evolution of the previously defined quantity is presented in figure \ref{fig:figure8}.

\subsection{Velocity dispersion and tangential velocity}

With the directions of the principal axes determined, the velocity dispersion is computed in the direction of each axis. 
For the previously mentioned quantities the stars considered for the calculations reside in a spherical volume of a radius $r_c$ of interest, centered on the satellite. 
The velocity dispersion of the system associated with radius $r_c$ is:
\begin{equation}
\sigma_{\star}^{r_c} = \left( \frac{ \sigma_x^2 + \sigma_y^2 + \sigma_z^2 }{3} \right)^{1/2} 
\end{equation}      
In order to compute the rotational velocity associated to the system, three cylindrical shell volumes are defined. 
Each of the volumes is defined with a central axis constrained to pass through the maximum density region of the 
object and to be parallel to a principal axis dimension. The inner and the outer radii of the volumes are 0.95 $r_c$ and 1.05 $r_c$.

For all the stars bound to the object and included in the previously defined volume, 
the component of their velocity tangential to any concentric circle centered on the axis of the cylinder is considered for the average. 
The resulting average is the velocity $v_{i,j}^{r_c}$, where $(i,j) \in {(x,y), (y,z), (z_x)}$. \\

With the dispersion and rotation quantities determined, the ratio:
\begin{equation}
\frac{v_{rot}^{r_{c}}}{\sigma_{\star}^{r_{c}}} = \frac{max \left( v_{x,y}^{r_c},v_{y,z}^{r_c},v_{z,x}^{r_c} \right)} { \sigma_{\star}^{r_c}}
\end{equation}
can be used as another indicator of the degree of transformation from rotationally supported satellite to a dispersion supported galaxy (\ref{table:table6}). 
In figure \ref{fig:figure9} the decrease of the ratio due to the effect of the tides can be observed.

\subsection{Projected mass surface densities and cylindrically averaged surface mass densities}

In addition to the quantities determined based on the bound amount of particles for the respective dwarf galaxies, 
the projected surface mass densities (PSMD) and cylindrically averaged surface mass densities (CASMD) have been determined. 
Unlike with the previous quantities where the nearby unbound particles are excluded from the calculations, in the PSMD and CASMD case they are included. 
The plots present in figures \ref{fig:figure2}, \ref{fig:figure3}, \ref{fig:figure10}, \ref{fig:figure11}, 
\ref{fig:figure12} and  \ref{fig:figure13}, are generated by projecting along the principal axes of the bound satellite all stellar particles 
inside a spherical volume of 10 kpc radius centered on the stellar density maximum of the galaxy. In the case of the CASMD calculations, 
the depth of the projection was reduced in order to include the stars with distances from the plane of projection smaller than 2.5 kpc. 
The resulting plots are available in figures \ref{fig:figure4}, \ref{fig:figure14}.

\subsection{Evolution of satellites}
\subsubsection{Galaxy A}
Originally the most massive of the studied objects, satellite A enters the host halo on an orbit which brings it very close to the ErisDark's center. 
Soon after the first pericenter passage, owing to the effect of dynamical friction
 the dwarf galaxy settles on an eccentric orbit with pericenter distances smaller than 5 kpc and apocenter distances less than 50 kpc. 
Before the end of the simulation the cuspy version of the object completes 15 orbits, while the cored version completes 14 orbits, as it can be seen from the top left panel of 
figure \ref{fig:figure15}. In both cases the satellite is found at z = 0 close to its maximum distance from the host. 
The three dimensional half mass radius decreases from the original 1300 pc values to 220 pc and 110 pc for the NFW version and nonNFW version respectively. 
The time evolution of the half-mass radius can be inspected in figure \ref{fig:figure7}. \\

The tides reach the stellar disc and unbinds 10$\%$ of the stellar mass before the second pericenter is reached. 
90$\%$ of the stellar mass of the cored dwarf is lost prior to completion of the fourth orbit (figure \ref{fig:figure16}). 
In the case of the counterpart with steep density profile, seven orbits have to be completed before loosing the same amount of stars.\\

\begin{figure}
\includegraphics[width = .4\textwidth]{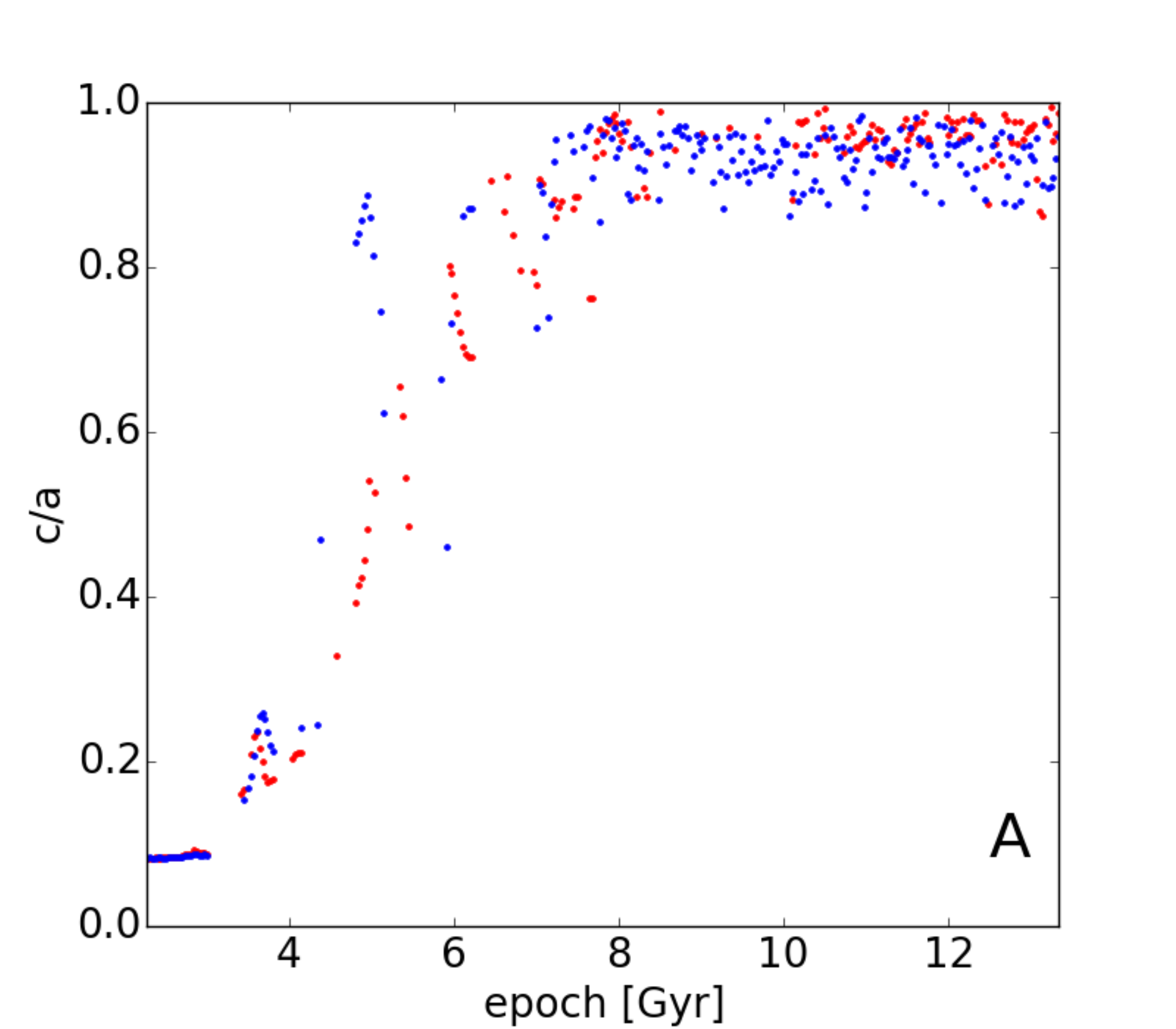}
\includegraphics[width = .4\textwidth]{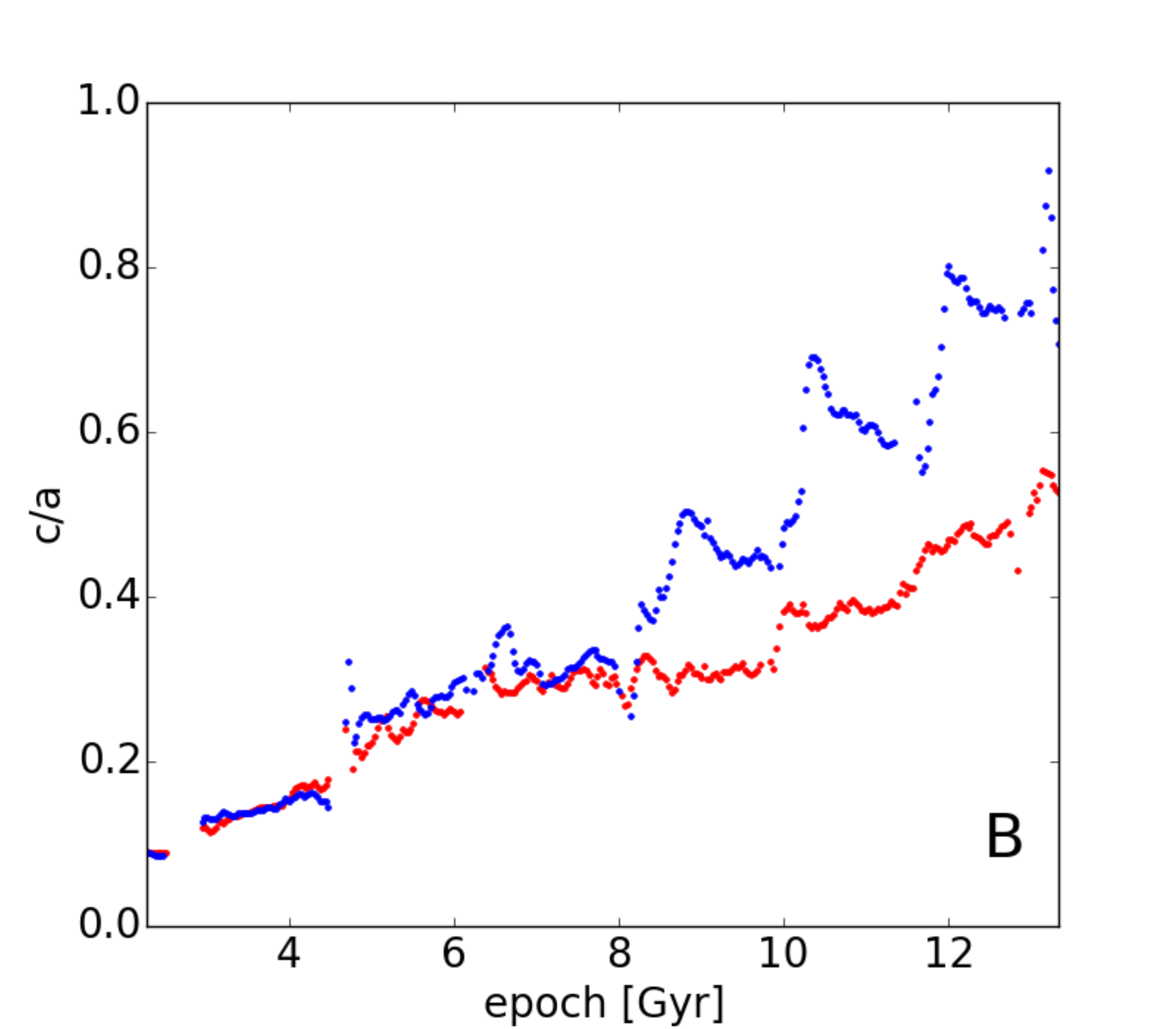}
\includegraphics[width = .4\textwidth]{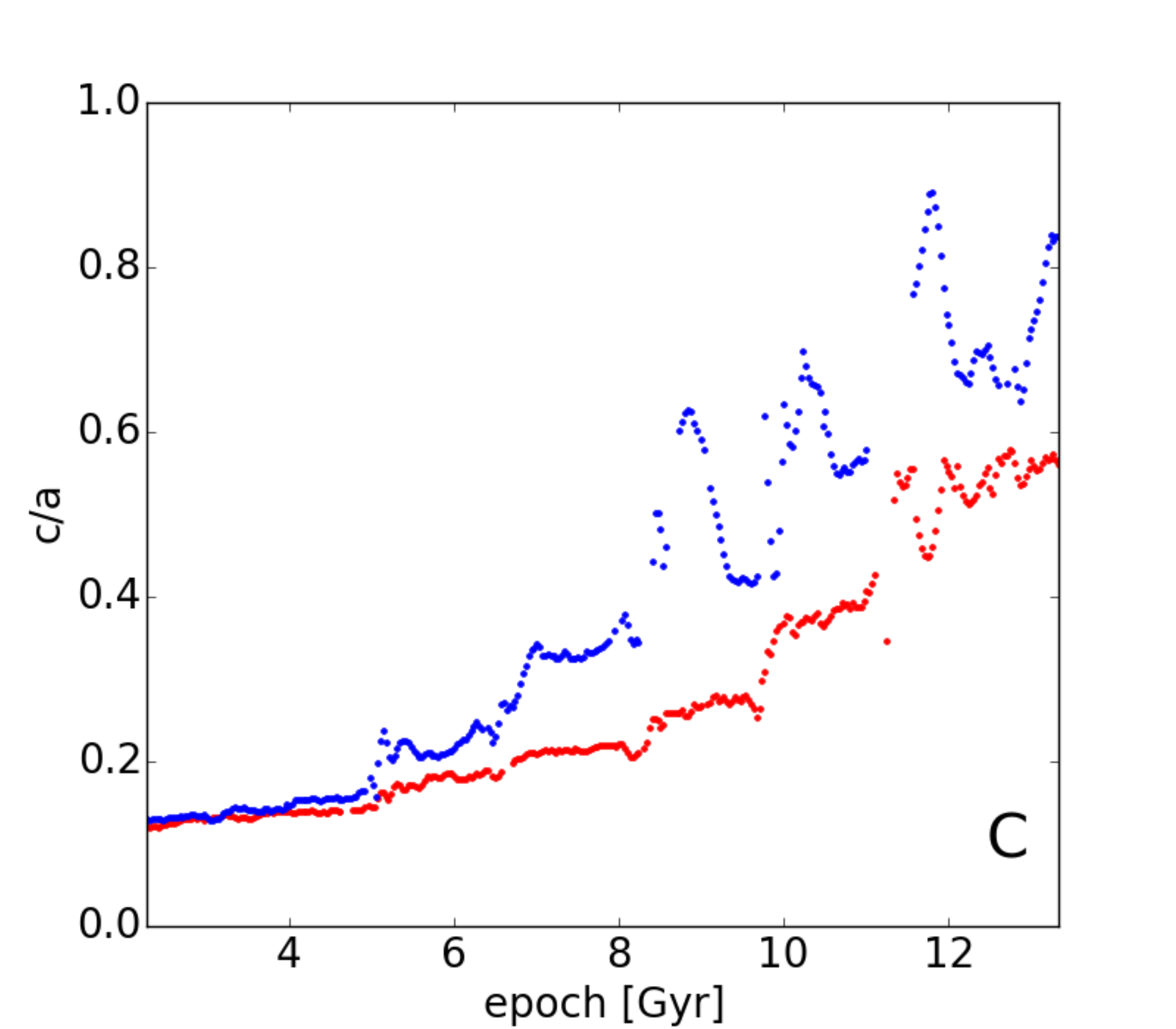}
\includegraphics[width = .4\textwidth]{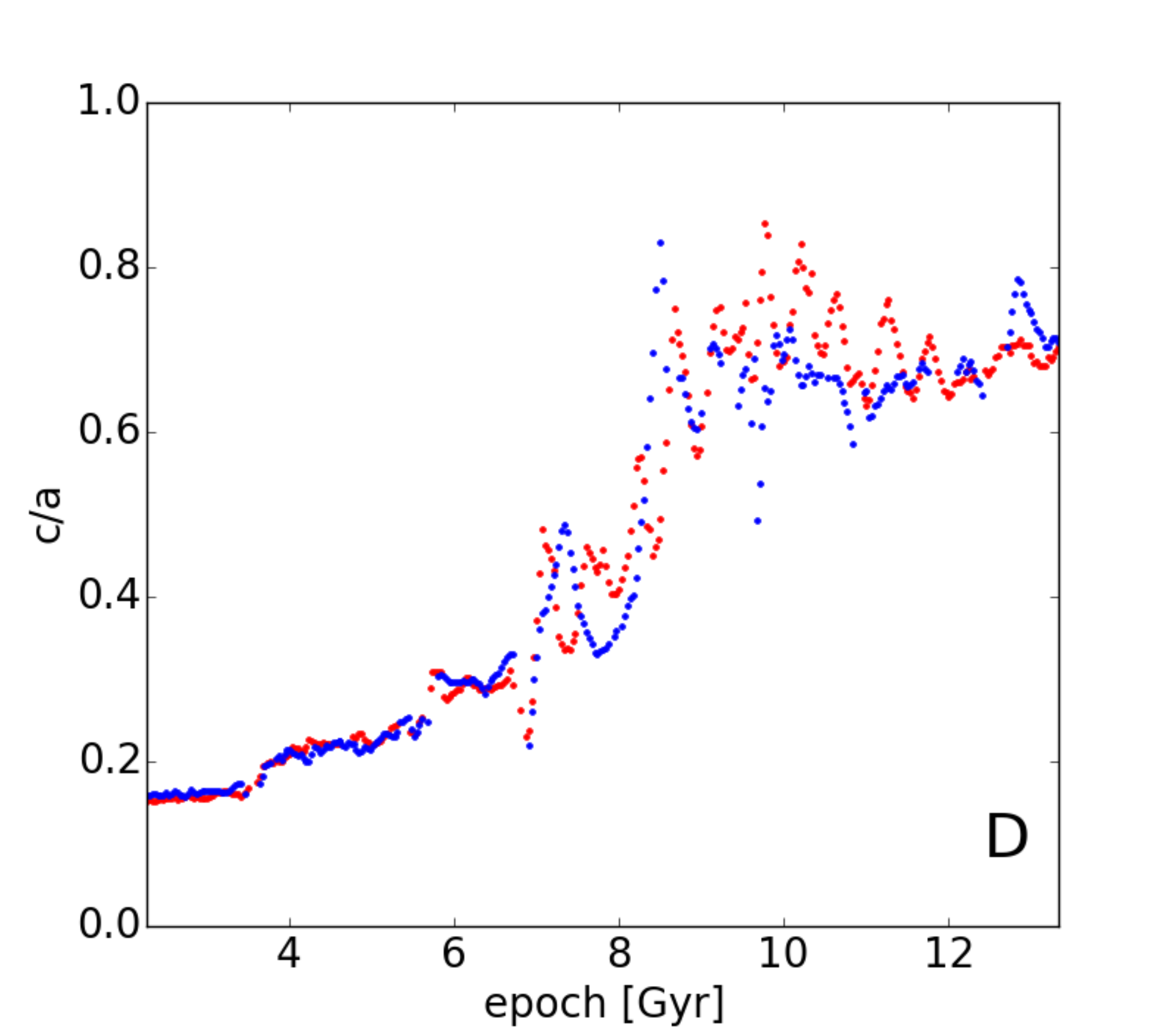}
\includegraphics[width = .4\textwidth]{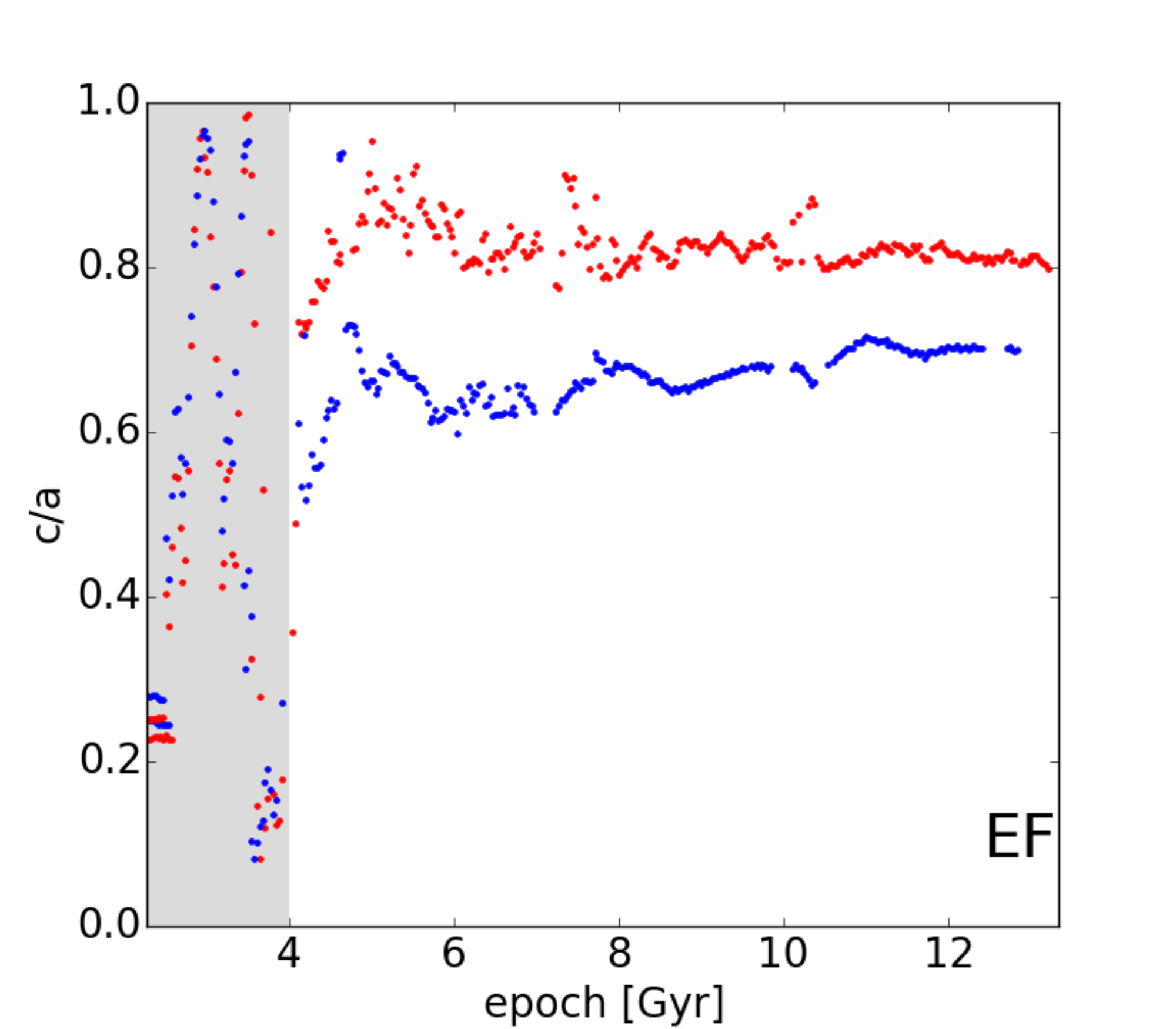}

\caption{\label{fig:figure8} The evolution of the dimension ratio c/a of the objects during the simulation. 
Red/blue dots represent the satellites with steep/shallow central density profile. 
The gray region of the last subplot marks the epochs when the merging process is 
in progress according to the half light radius indicator of \ref{fig:figure7}}
\end{figure}

Based on the projected mass distribution of the satellite, it can be observed that the difference in the halo profile has 
no clear impact on the evolution of the shape of the satellite (figures \ref{fig:figure10} and \ref{fig:figure11}). The ratios between the shortest and longest dimension 
corresponding to both versions reaches the value 0.5 during the third pericenter passage and remains above this value throughout the simulation (figure \ref{fig:figure8}).

\begin{figure}
\centering
\includegraphics[width = .4\textwidth]{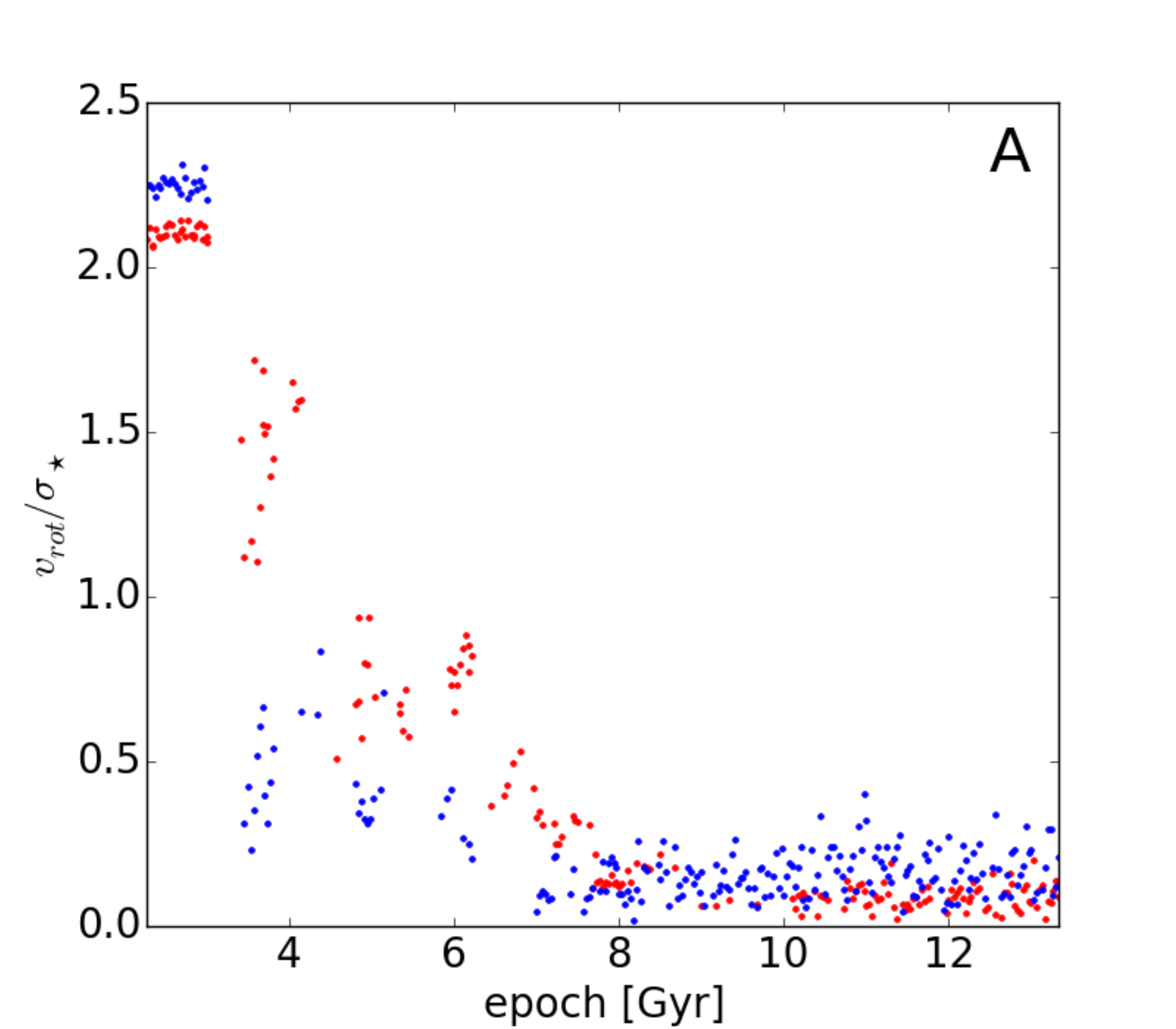}
\includegraphics[width = .4\textwidth]{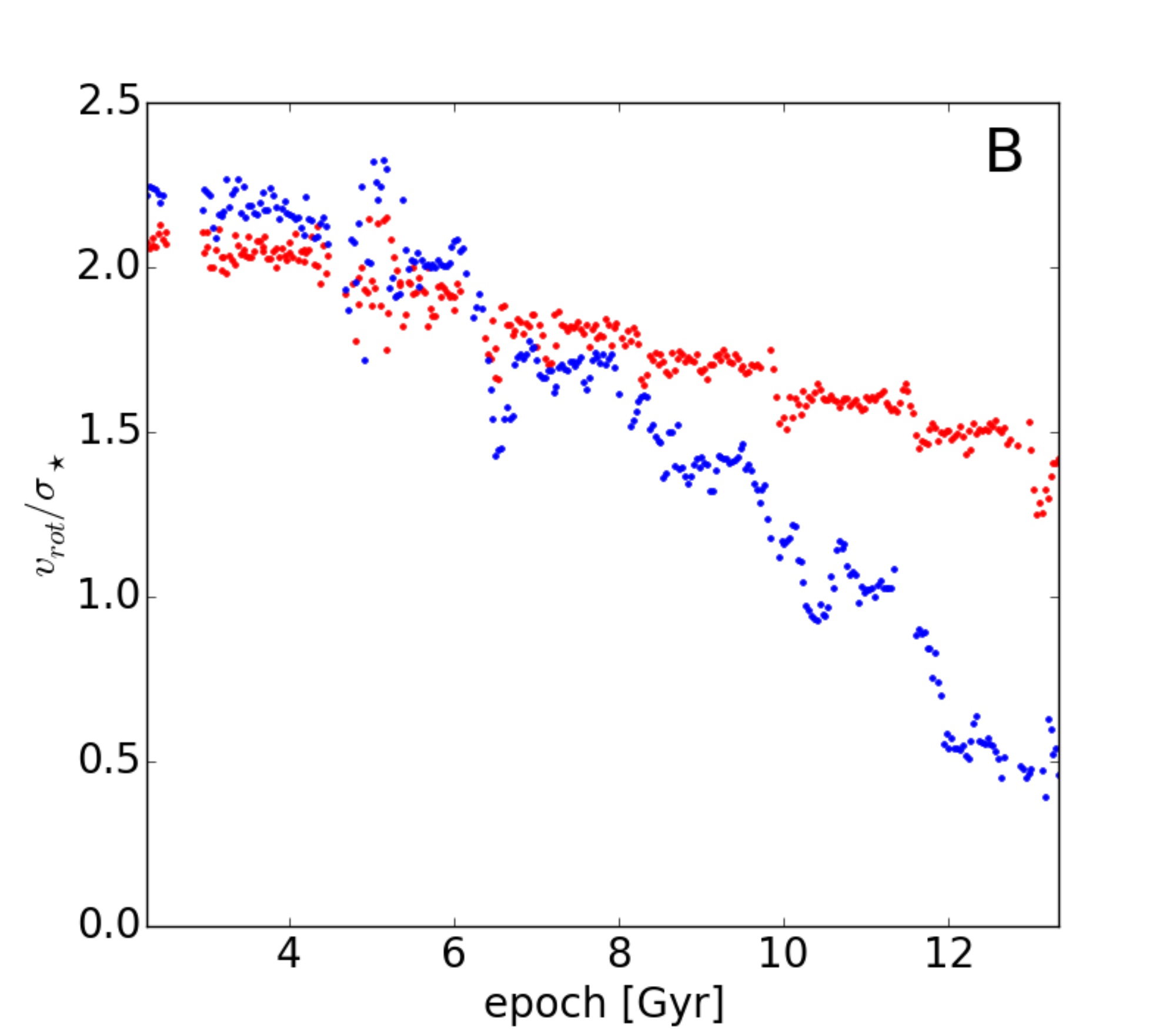}
\includegraphics[width = .4\textwidth]{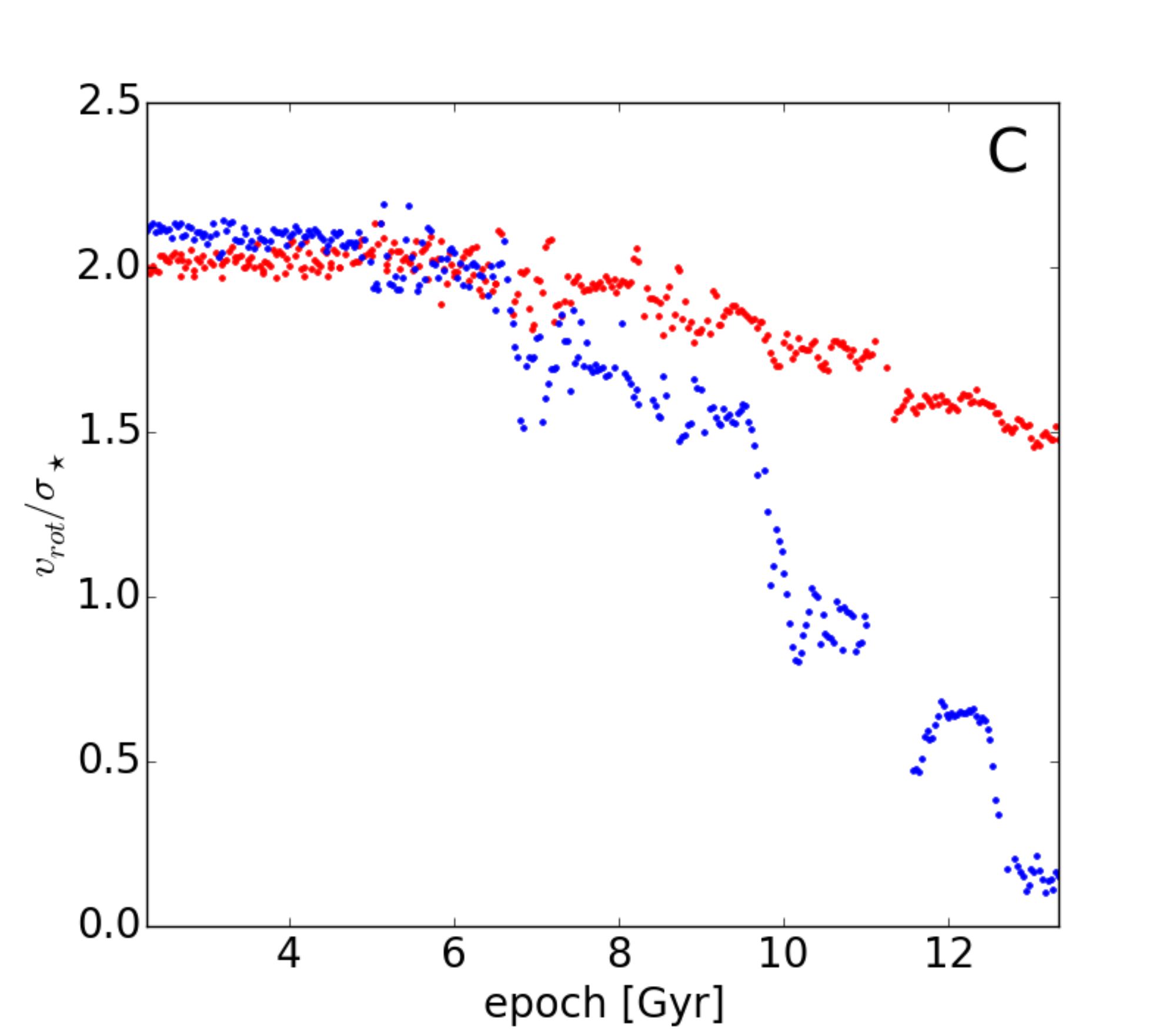}
\includegraphics[width = .4\textwidth]{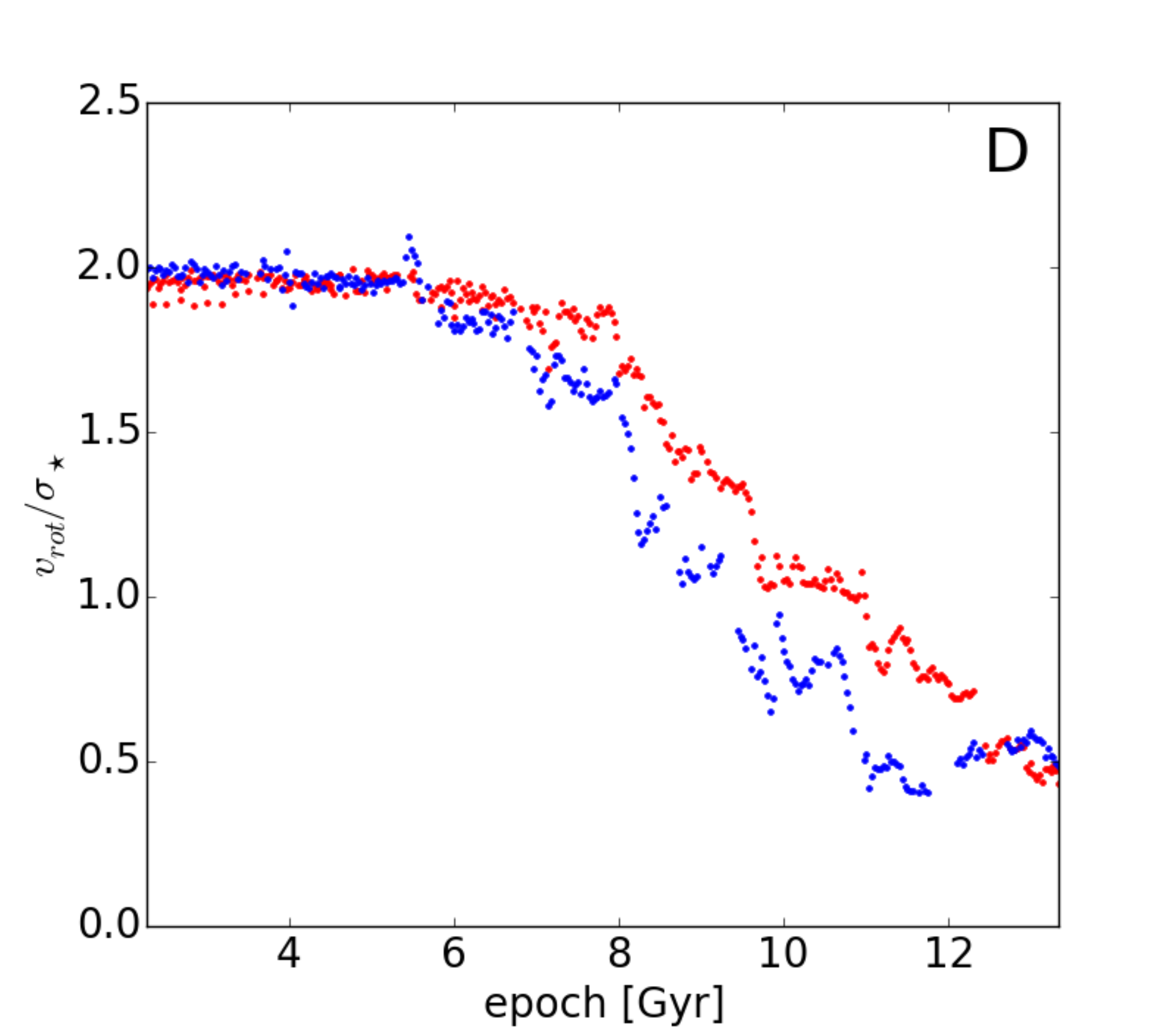}
\includegraphics[width = .4\textwidth]{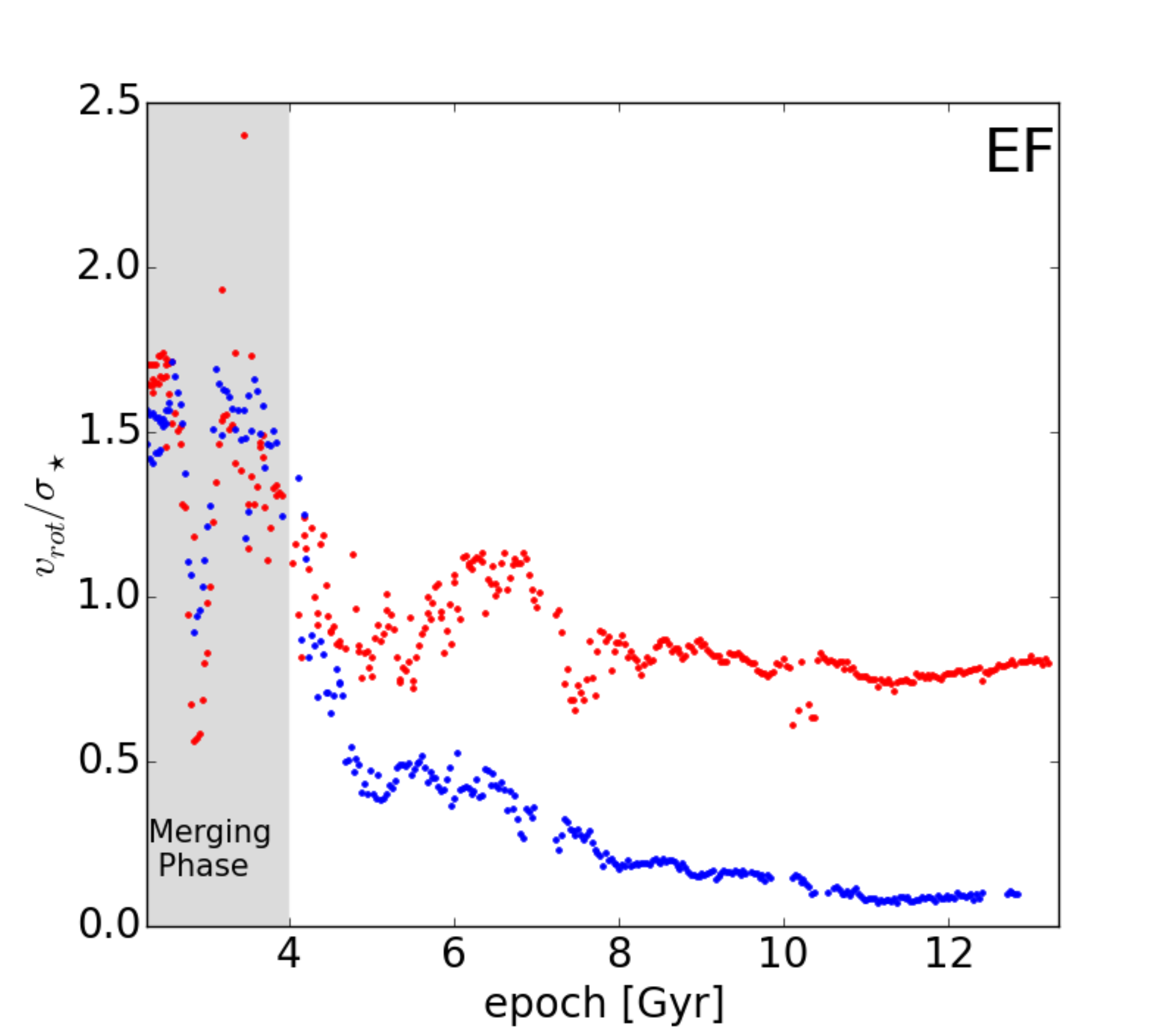}

\caption{\label{fig:figure9} The evolution of the ratio between the circular velocity and the velocity dispersion, 
both estimated for the 3D half-light radius.The data points corresponding to galaxies with shallow density profiles are 
marked in blue and the points corresponding to galaxies with steep density profiles are marked in red.}
\end{figure}

\begin{figure}
\centering
\includegraphics[width = .3\textwidth]{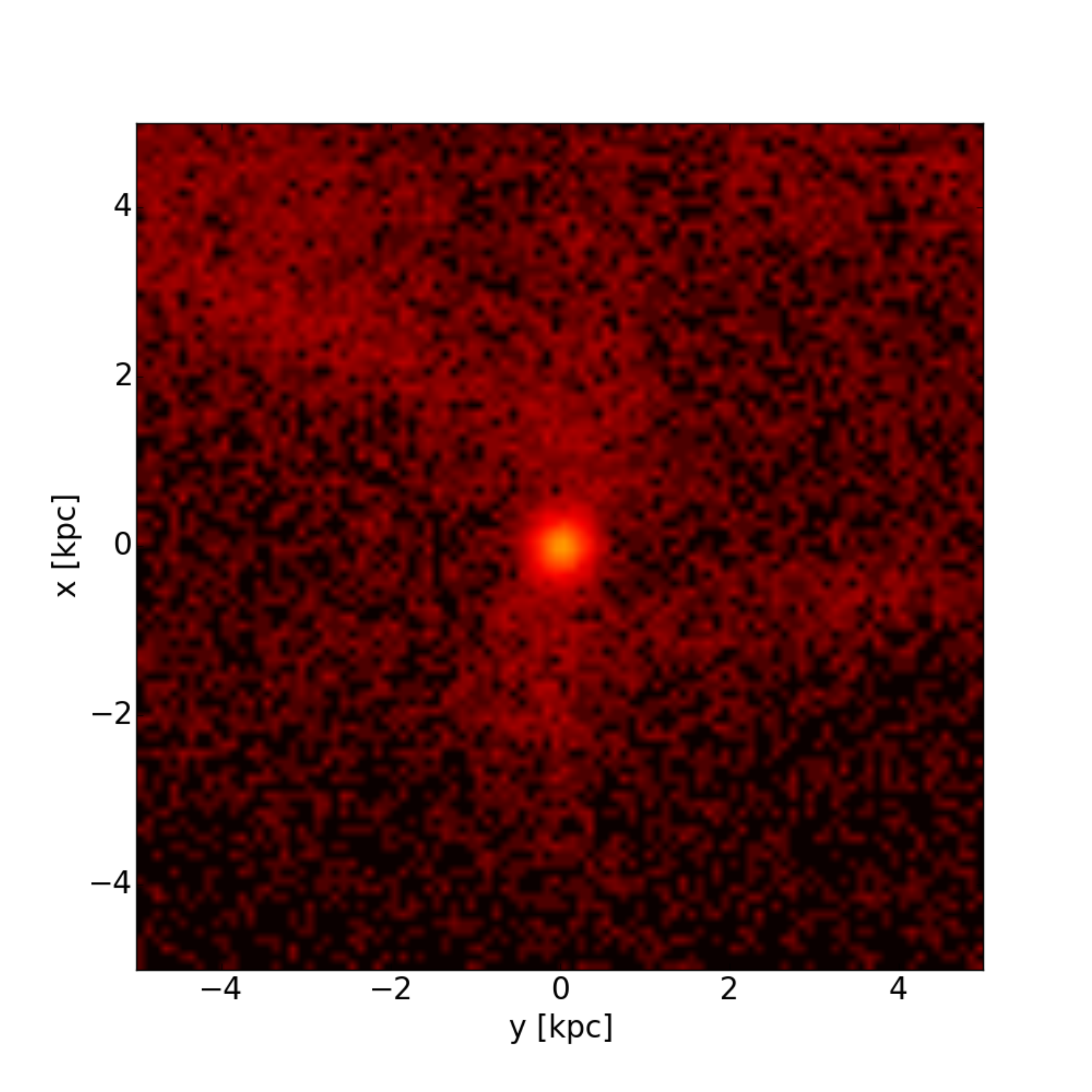}
\includegraphics[width = .3\textwidth]{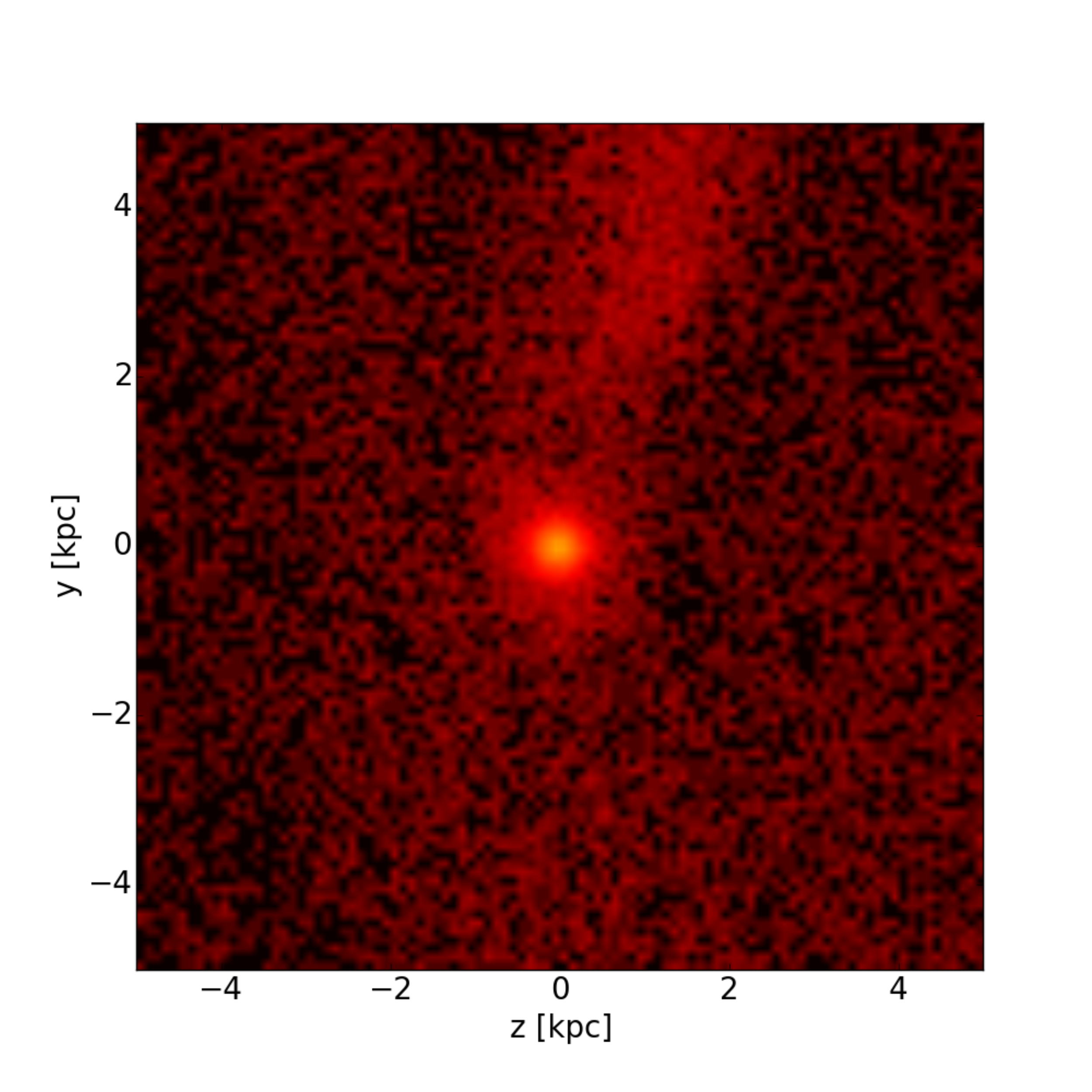}
\includegraphics[width = .3\textwidth]{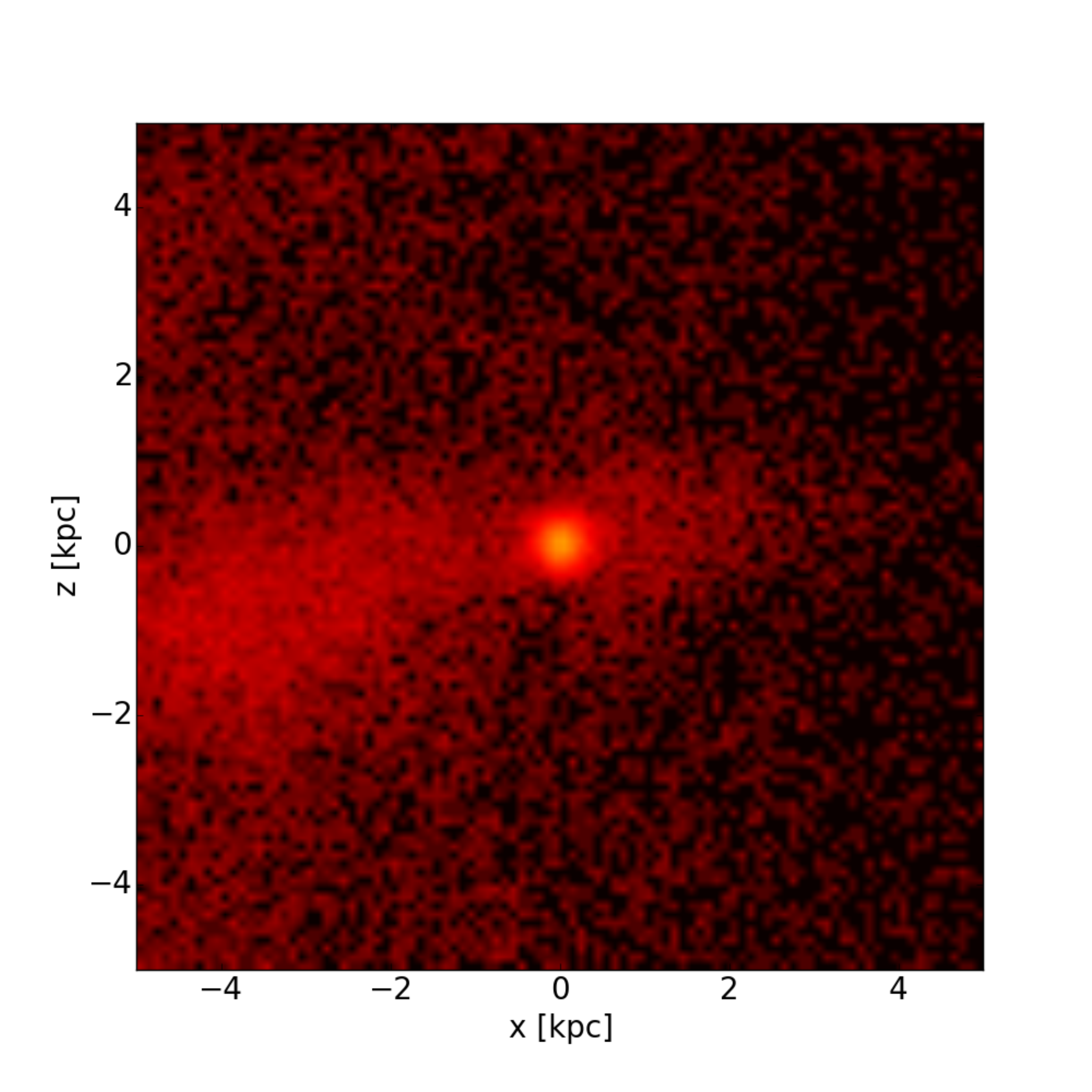}
\includegraphics[width = .3\textwidth]{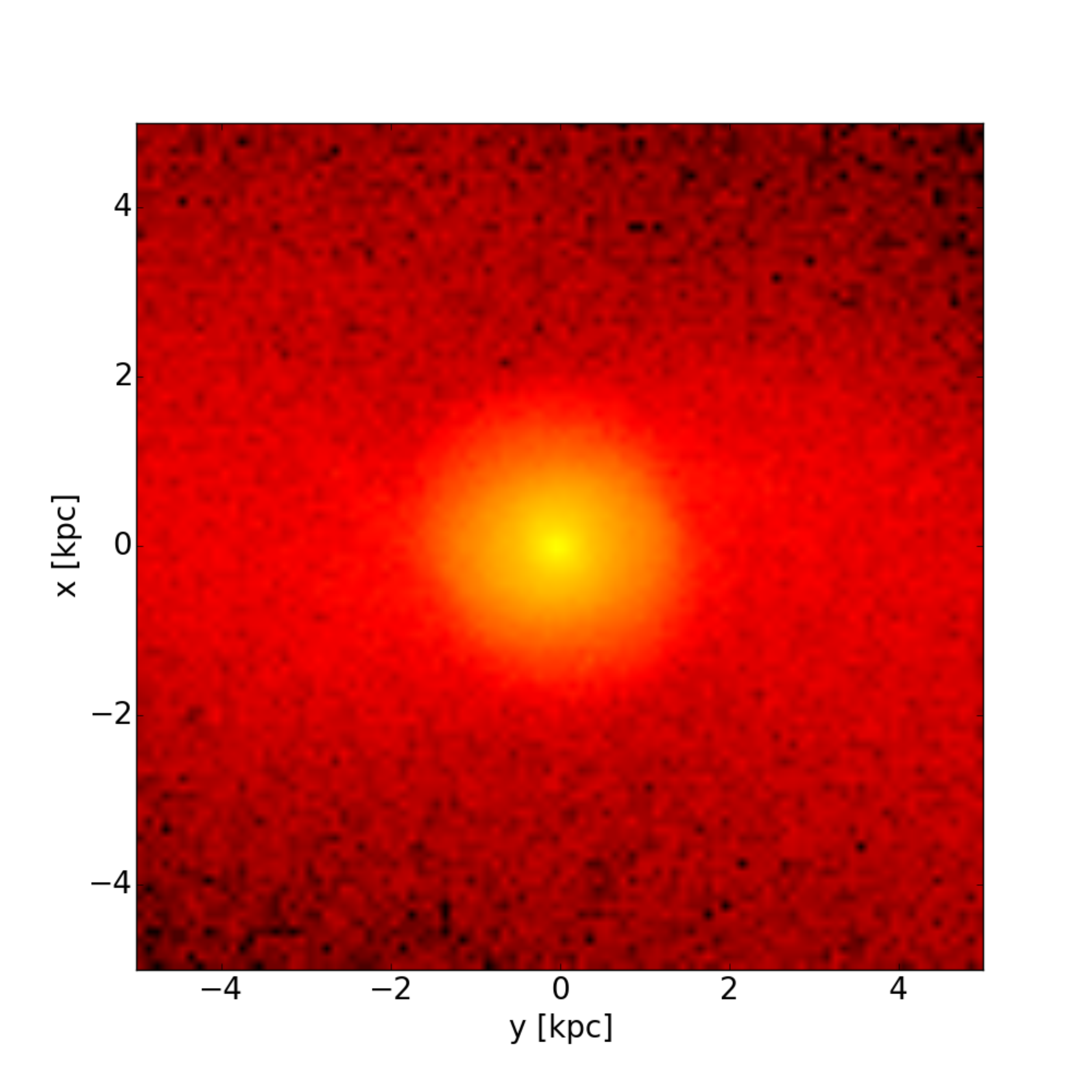}
\includegraphics[width = .3\textwidth]{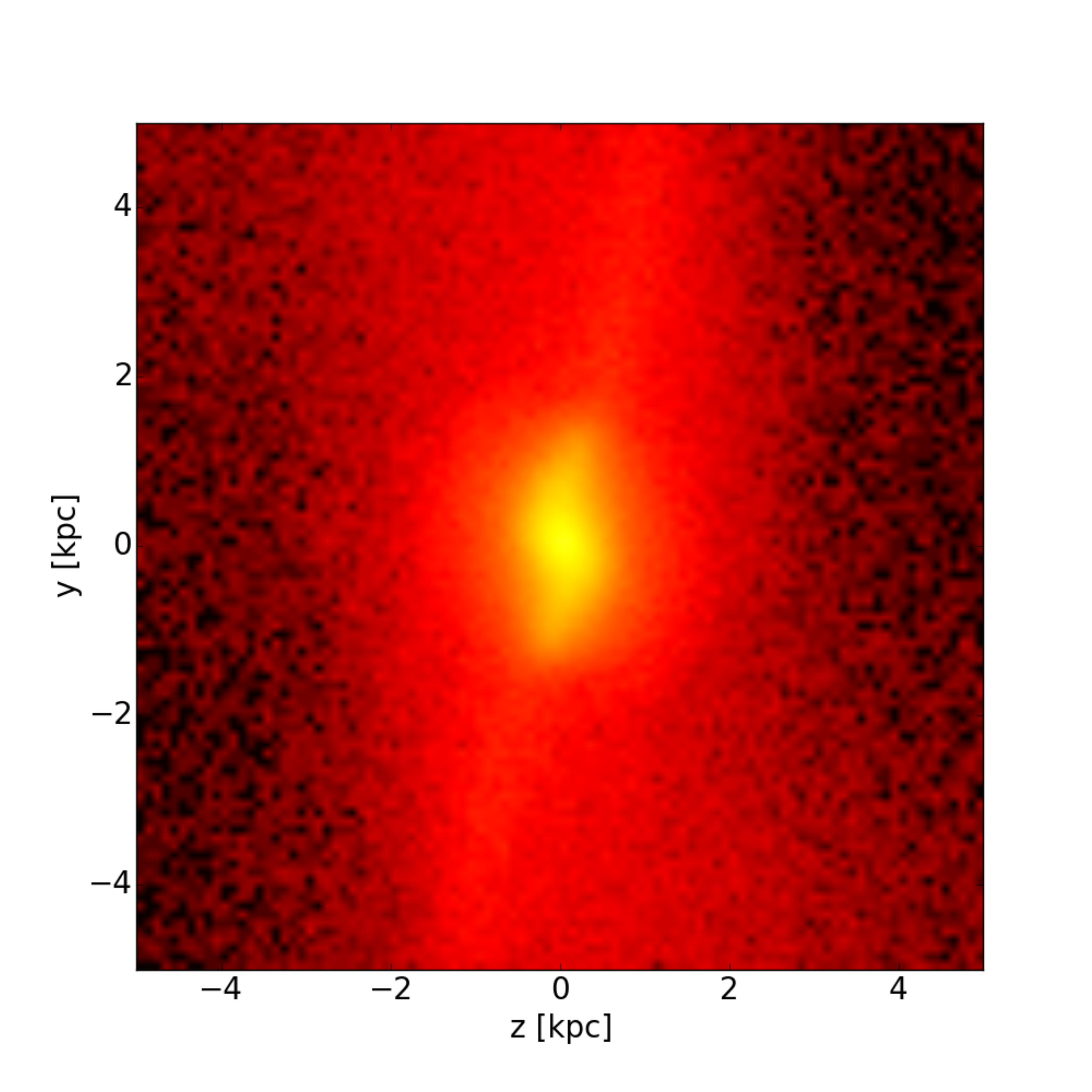}
\includegraphics[width = .3\textwidth]{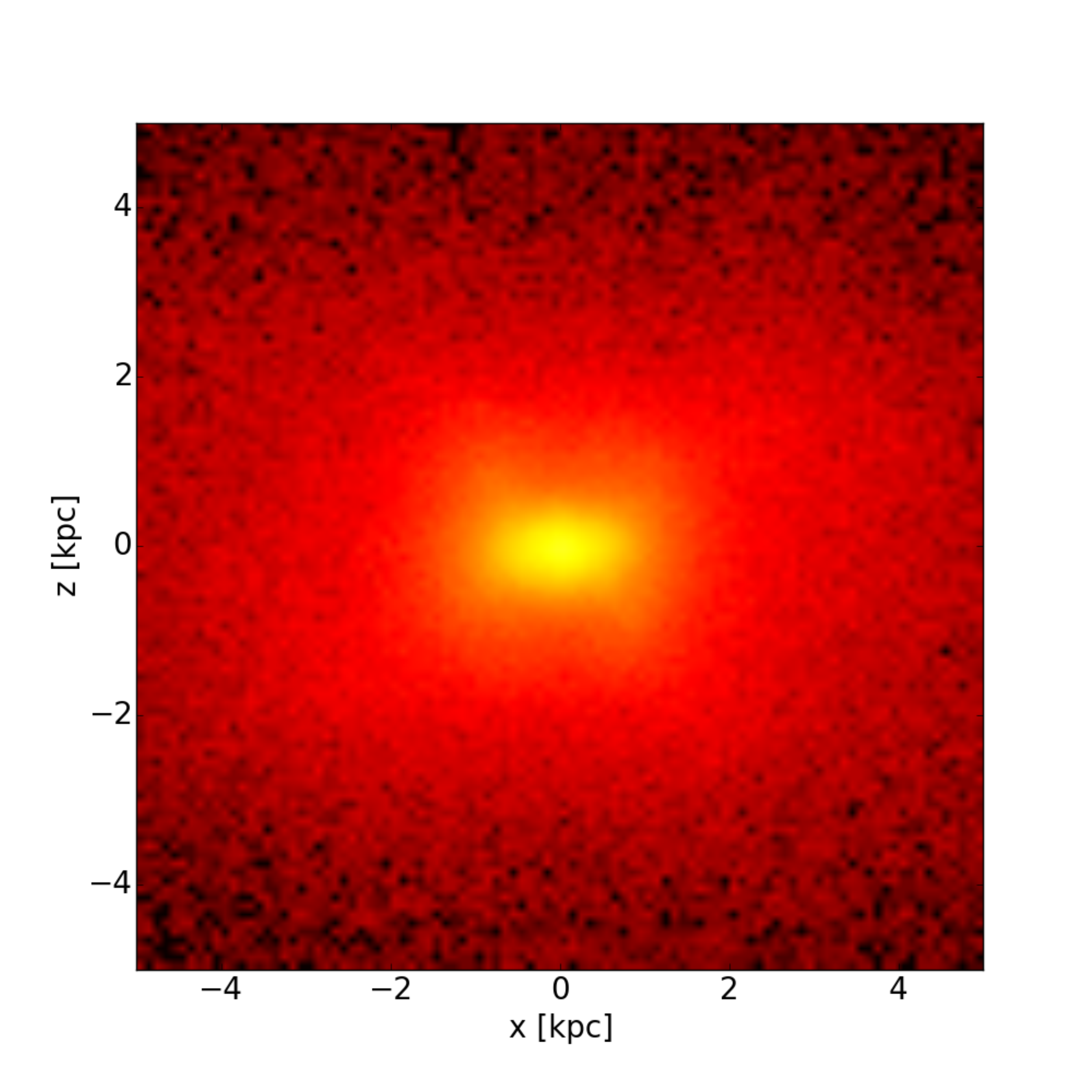}
\includegraphics[width = .3\textwidth]{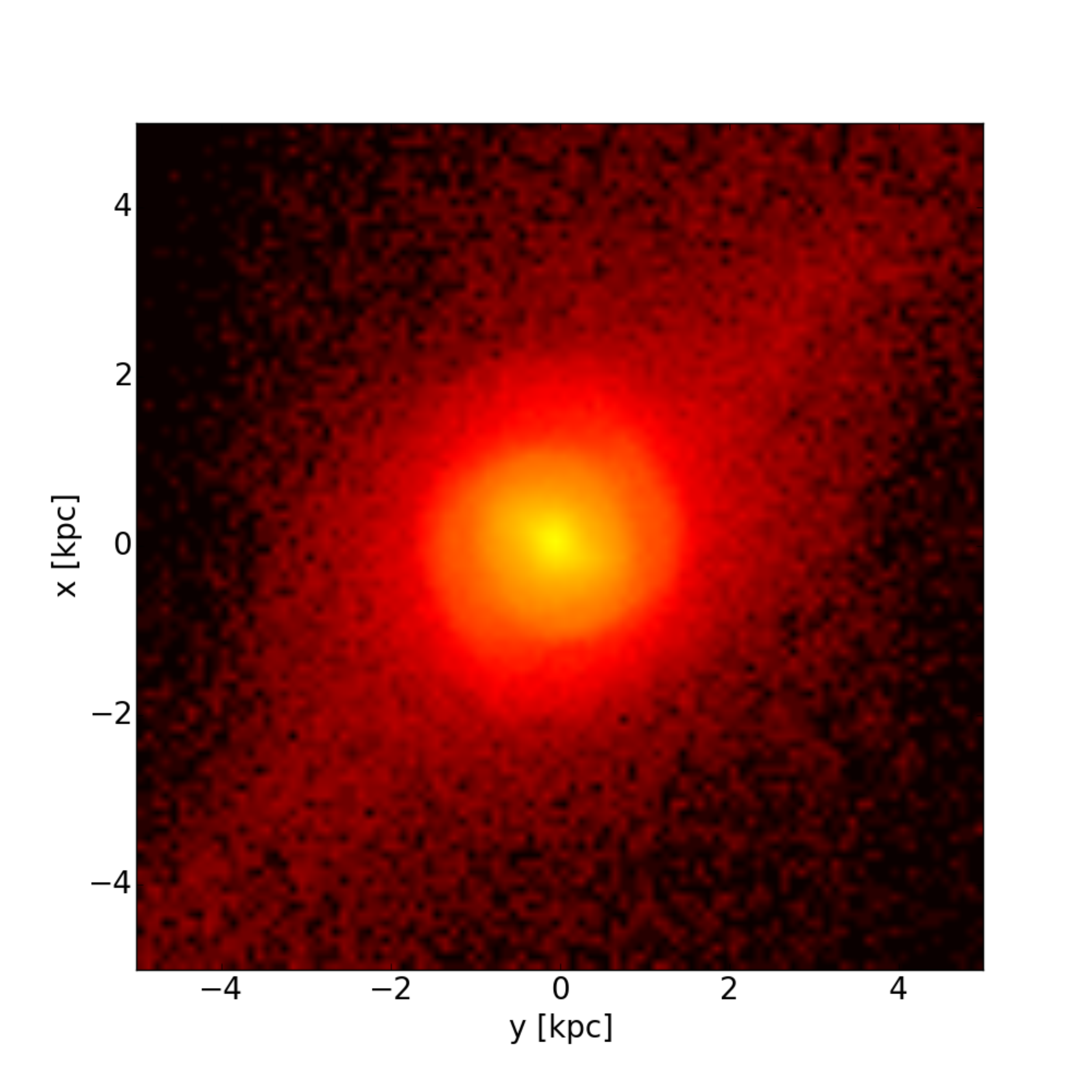}
\includegraphics[width = .3\textwidth]{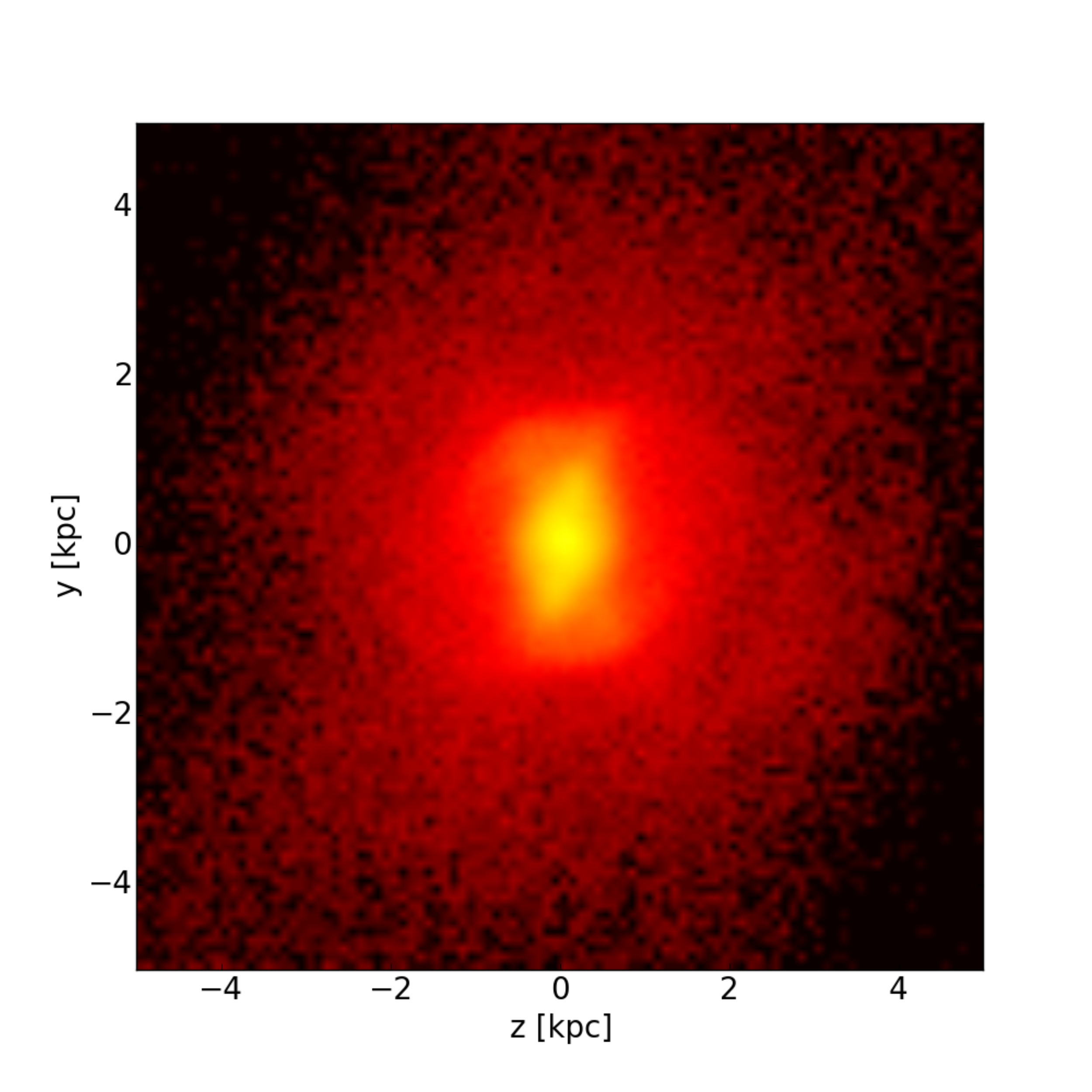}
\includegraphics[width = .3\textwidth]{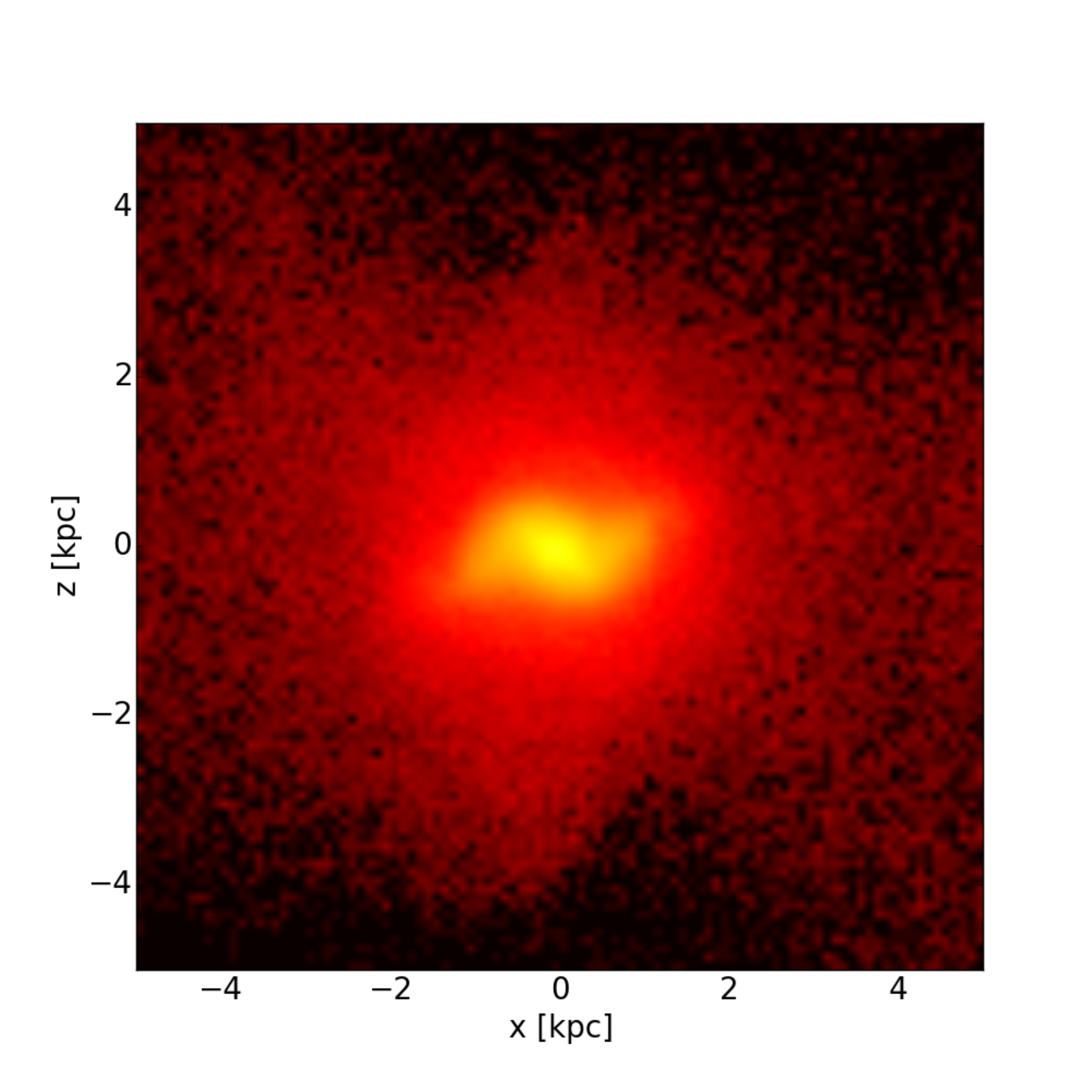}
\includegraphics[width = .7\textwidth]{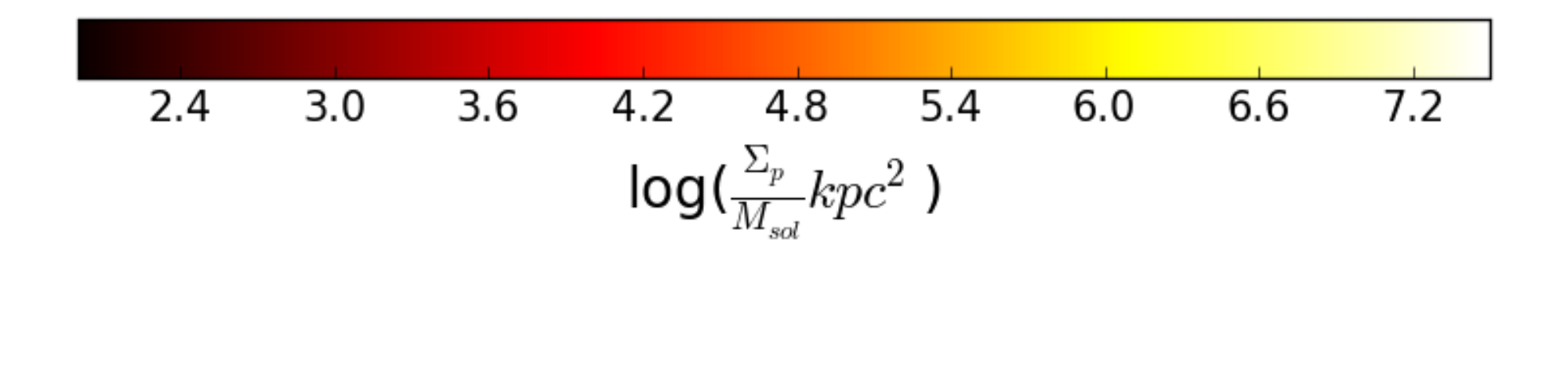}

\caption{\label{fig:figure10} The projected surface mass density maps of objects A (first row), B (second row), and C (last row) 
with $\gamma = 1.0$ at redshift 0. The first/second/third column corresponds to the projections along the smallest/middle/largest dimension.}
\end{figure}

\begin{figure}
\centering
\includegraphics[width = .3\textwidth]{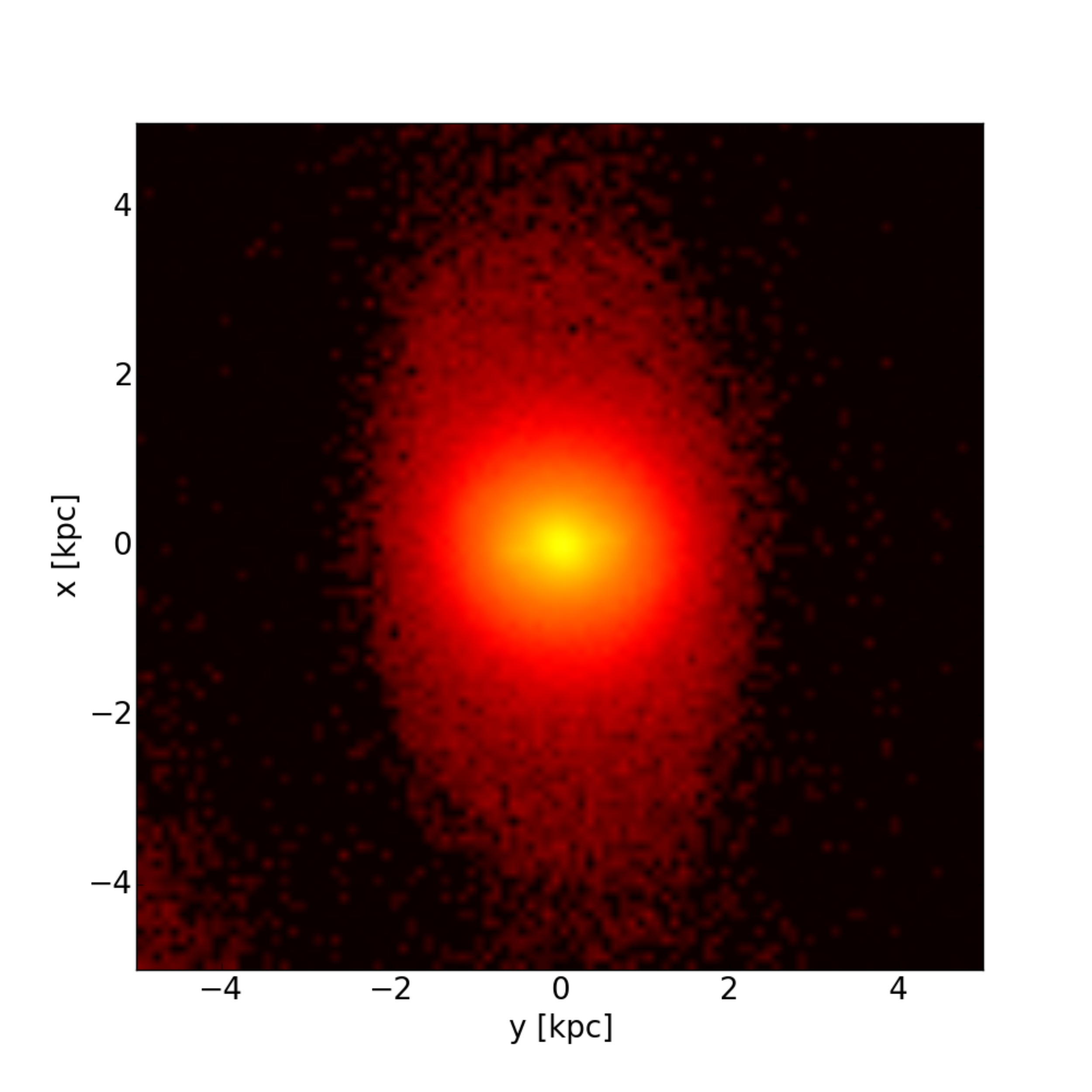}
\includegraphics[width = .3\textwidth]{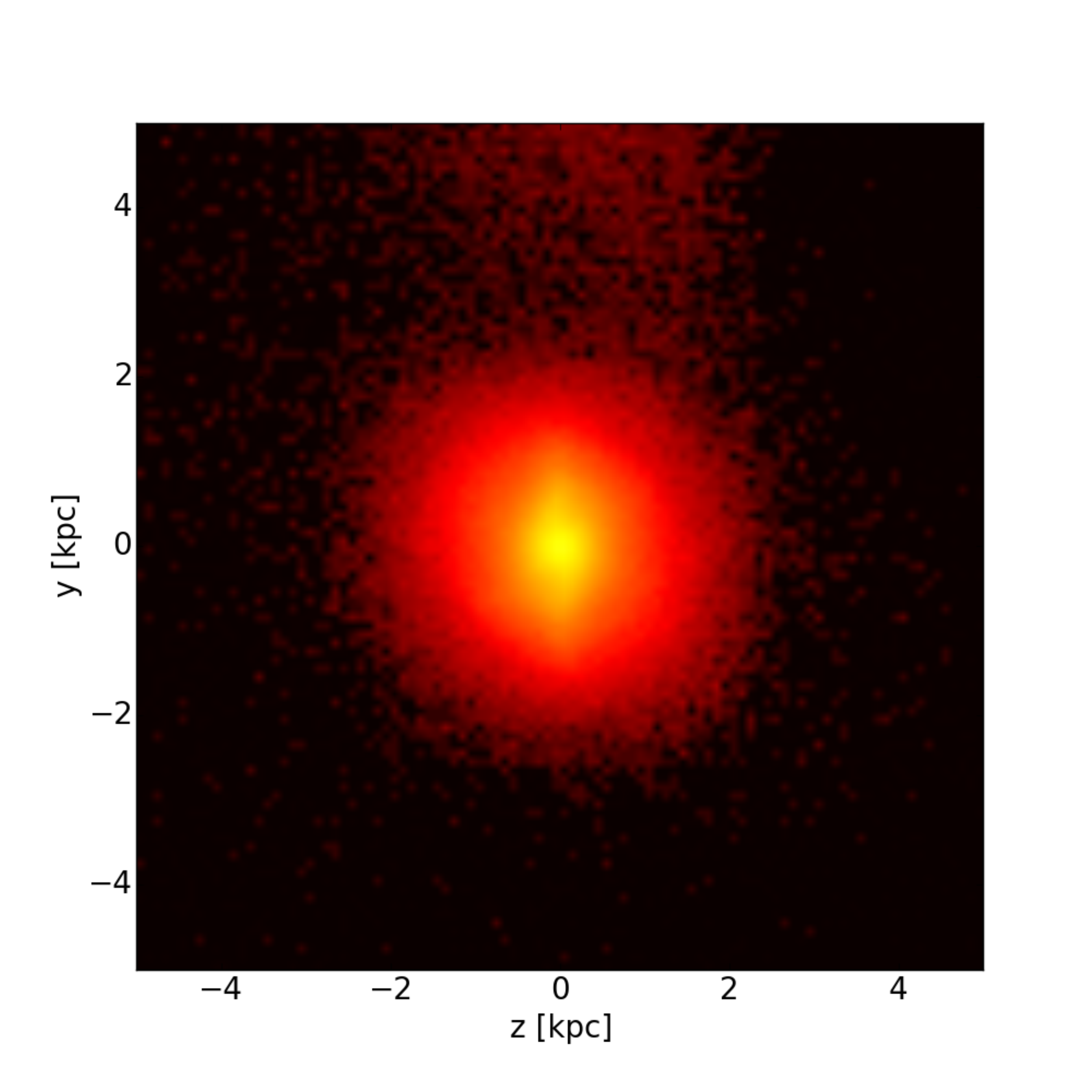}
\includegraphics[width = .3\textwidth]{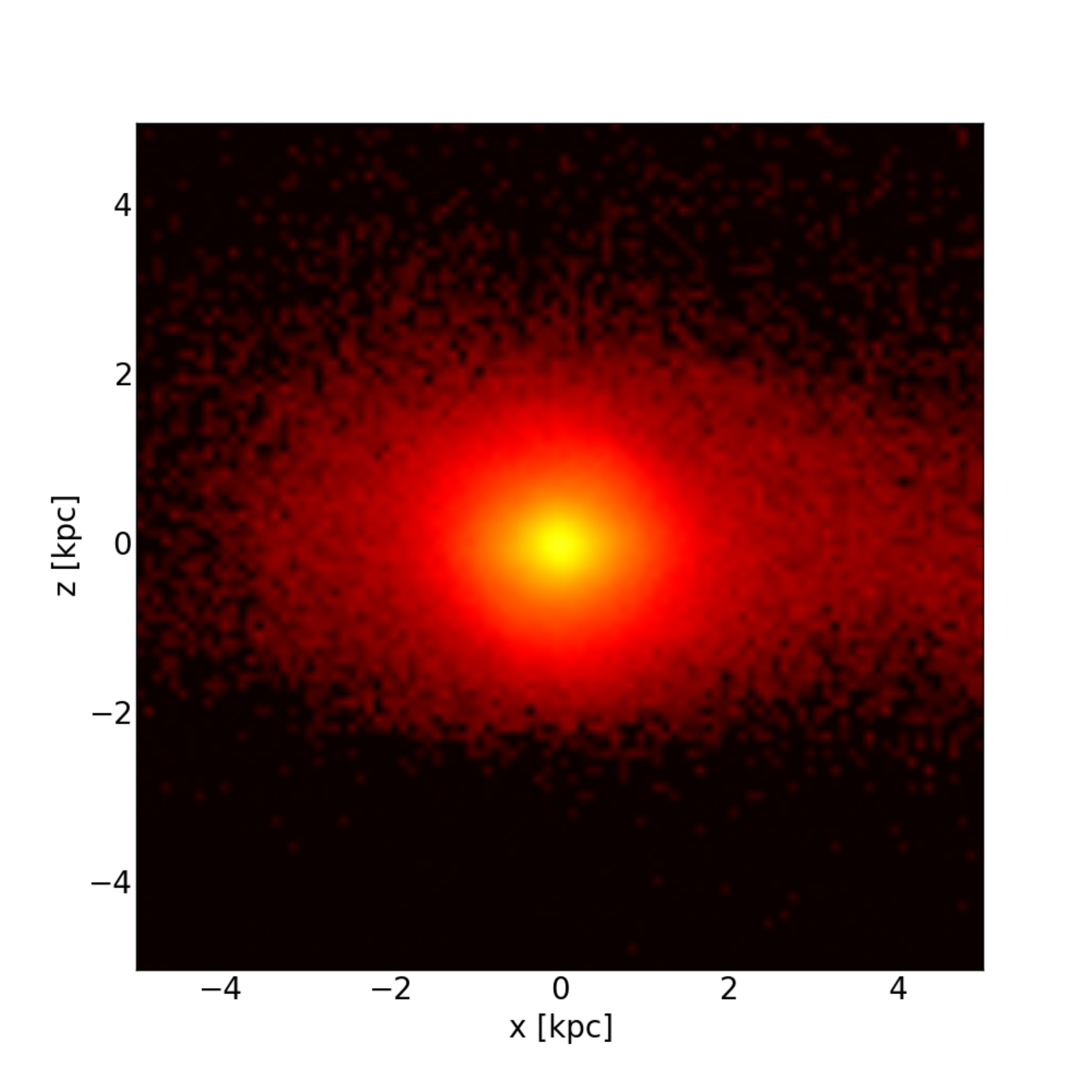}
\includegraphics[width = .3\textwidth]{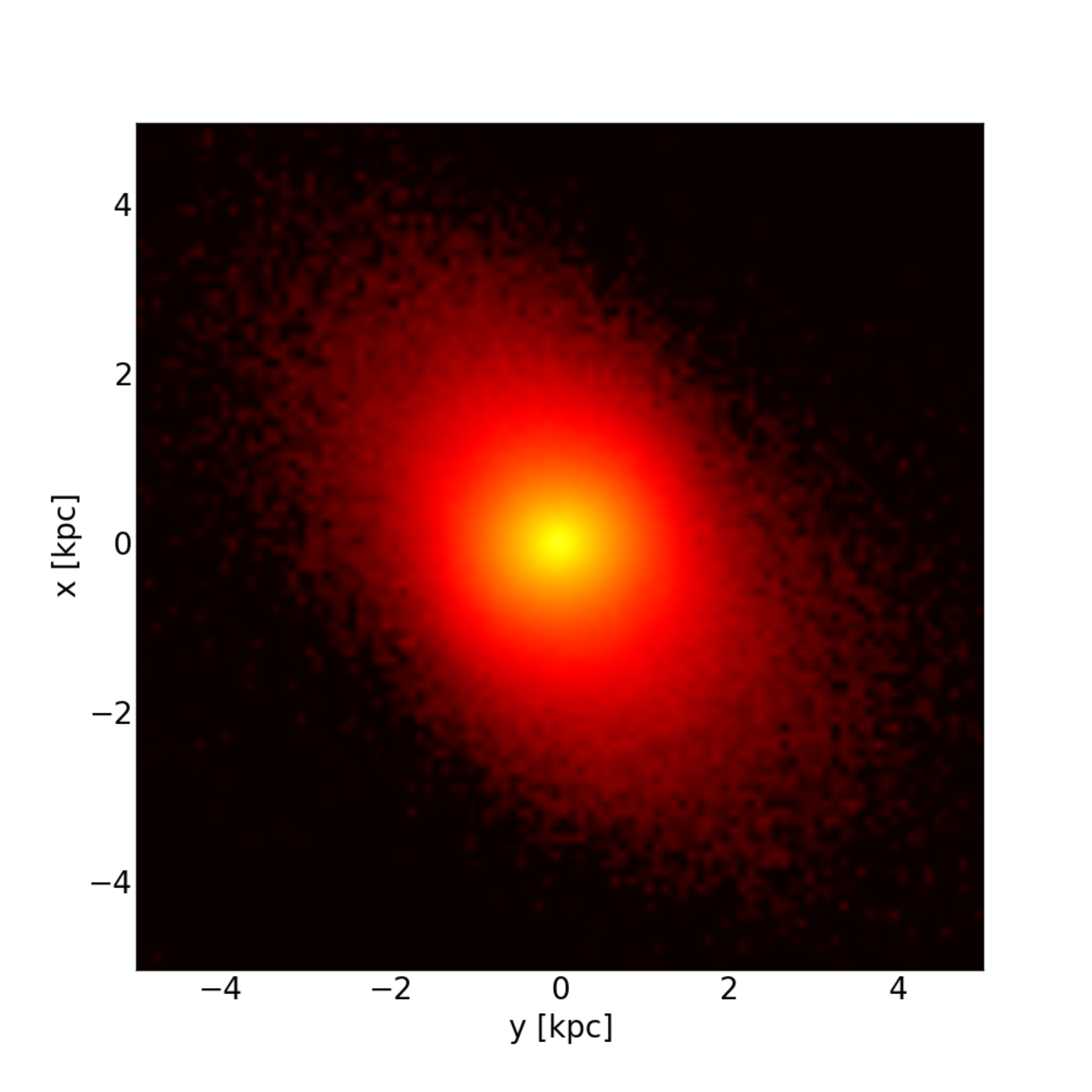}
\includegraphics[width = .3\textwidth]{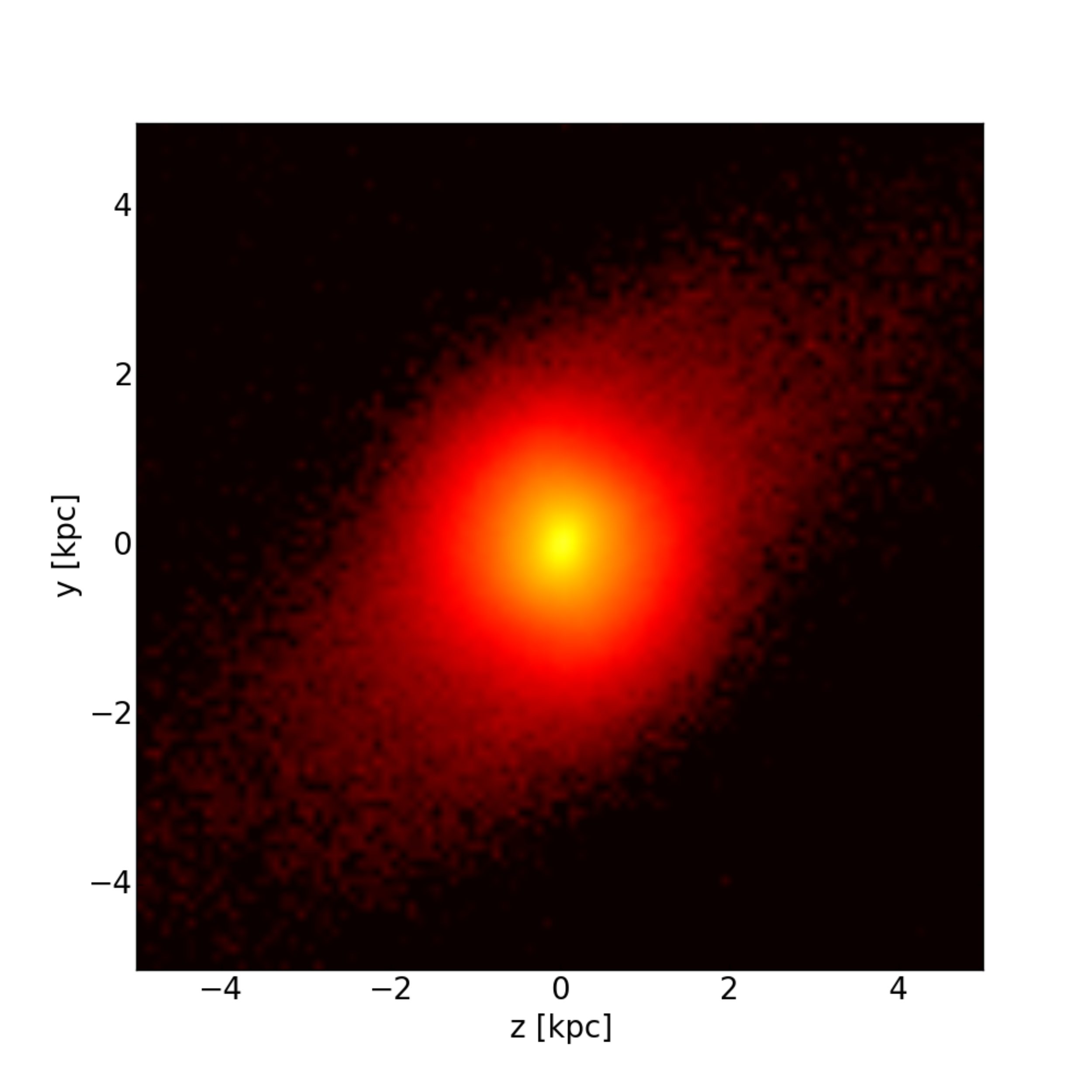}
\includegraphics[width = .3\textwidth]{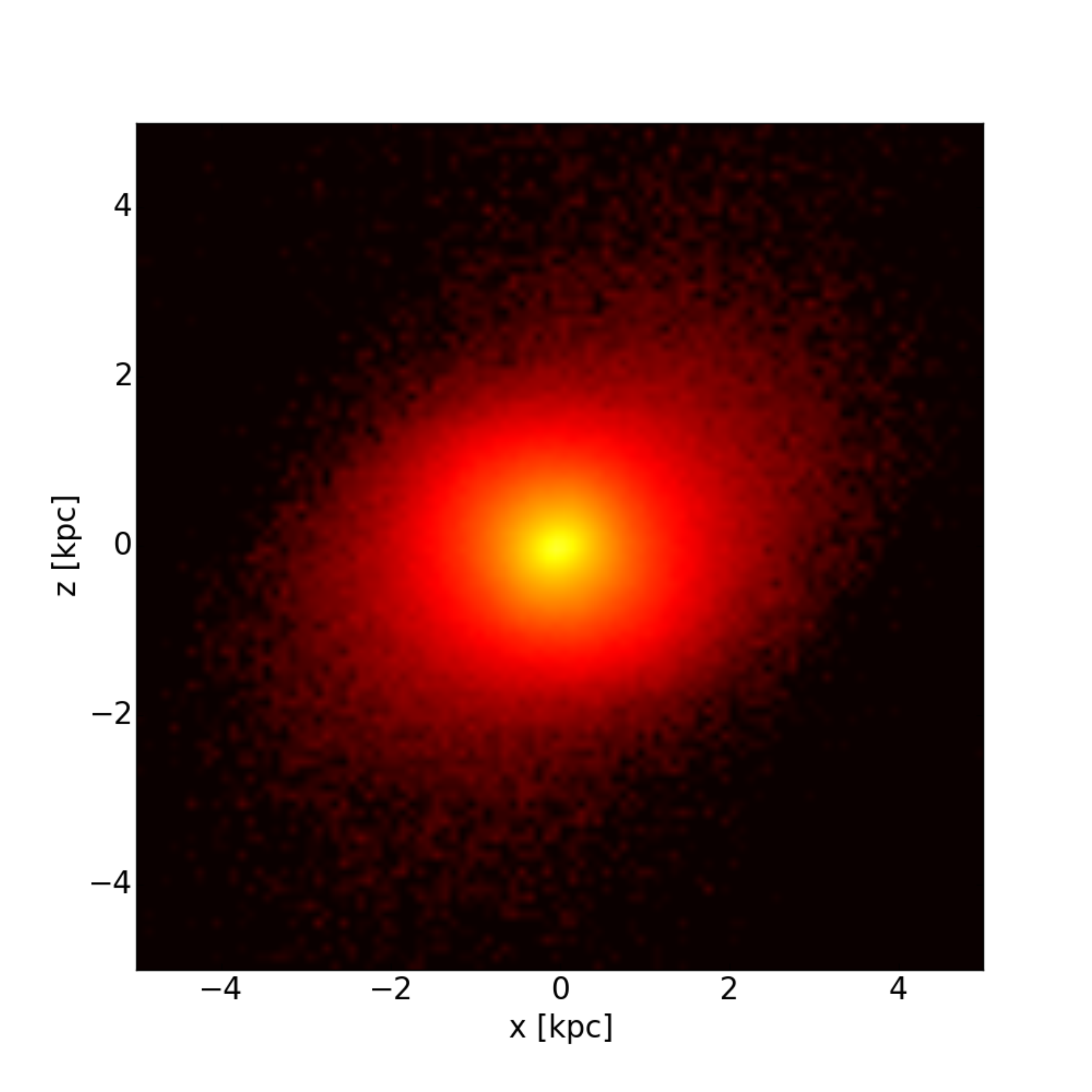}
\includegraphics[width = .7\textwidth]{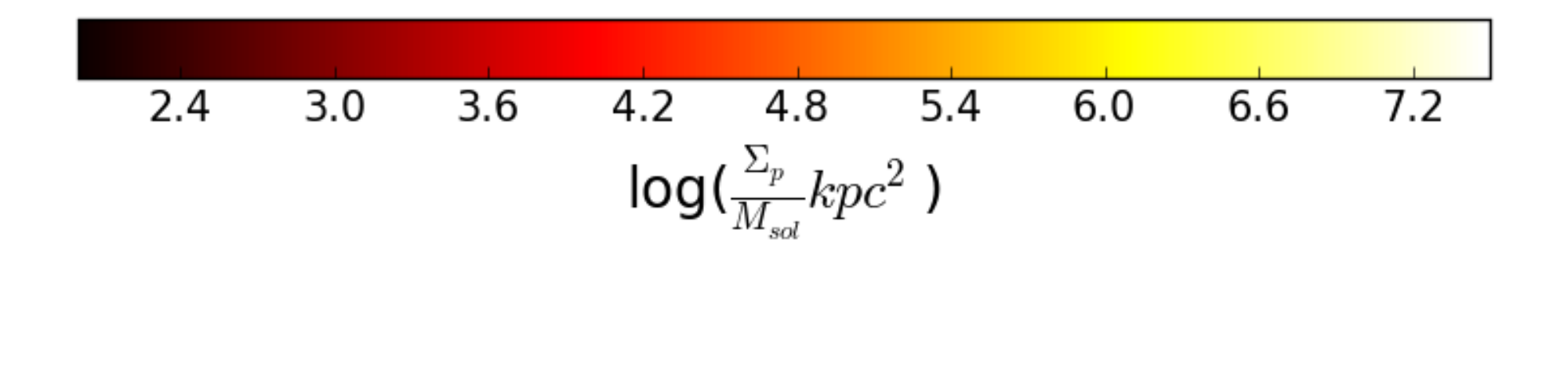}

\caption{\label{fig:figure11} The projected surface mass density maps of the objects D (first row) and EF with $\gamma = 1.0$ at redshift 0. 
The first/second/third column corresponds to the projections along the smallest/middle/largest dimension.}
\end{figure}

\begin{figure}
\centering
\includegraphics[width = .3\textwidth]{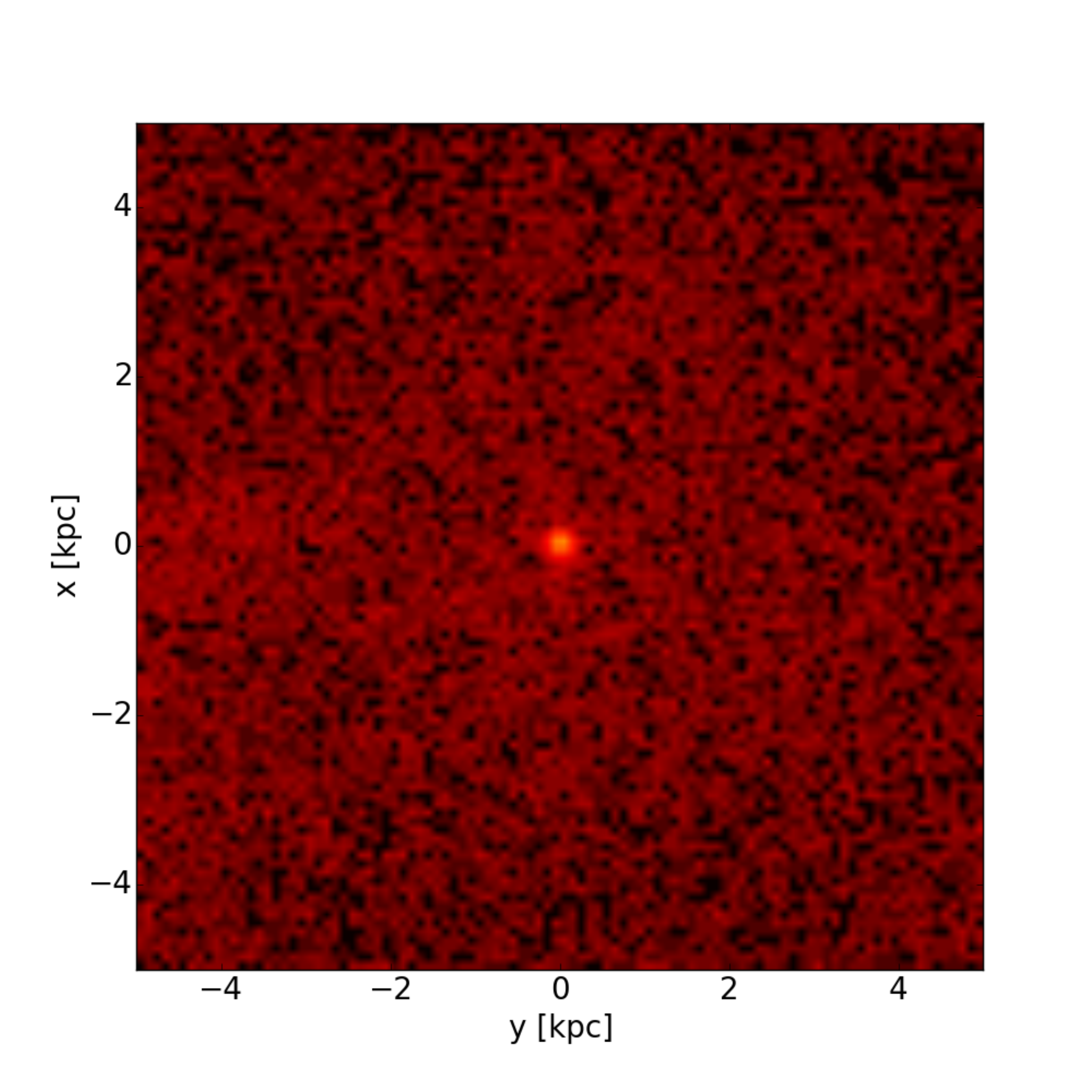}
\includegraphics[width = .3\textwidth]{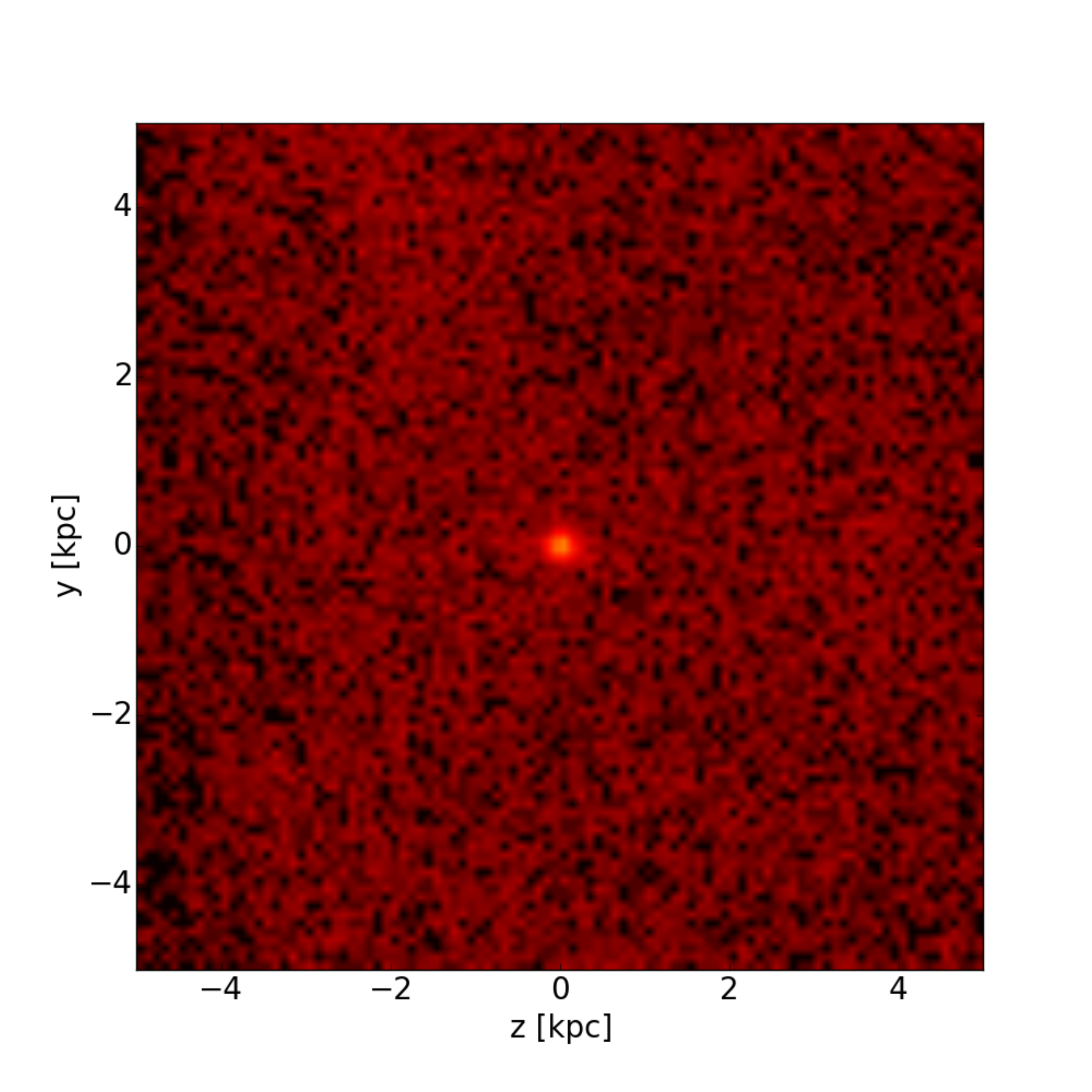}
\includegraphics[width = .3\textwidth]{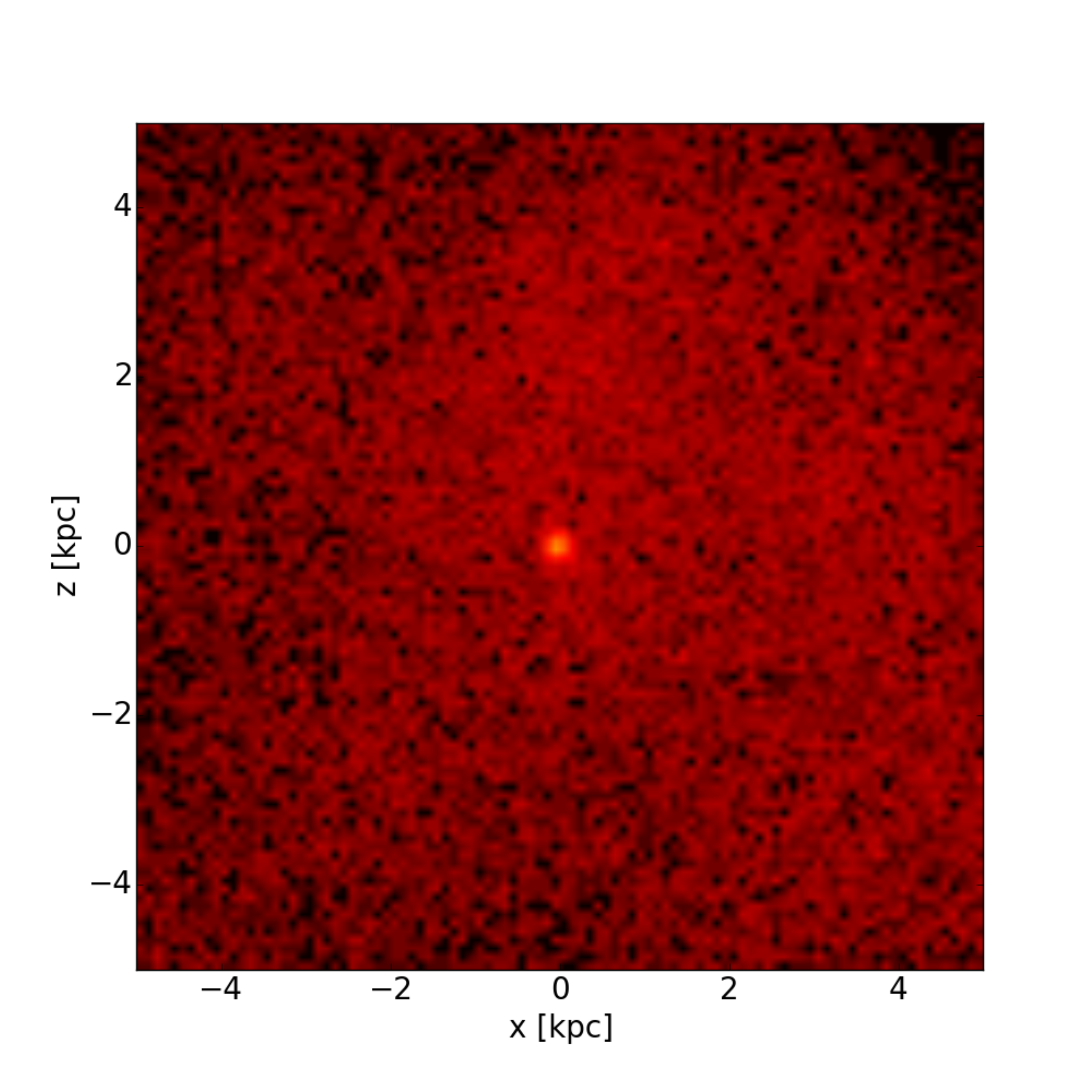}
\includegraphics[width = .3\textwidth]{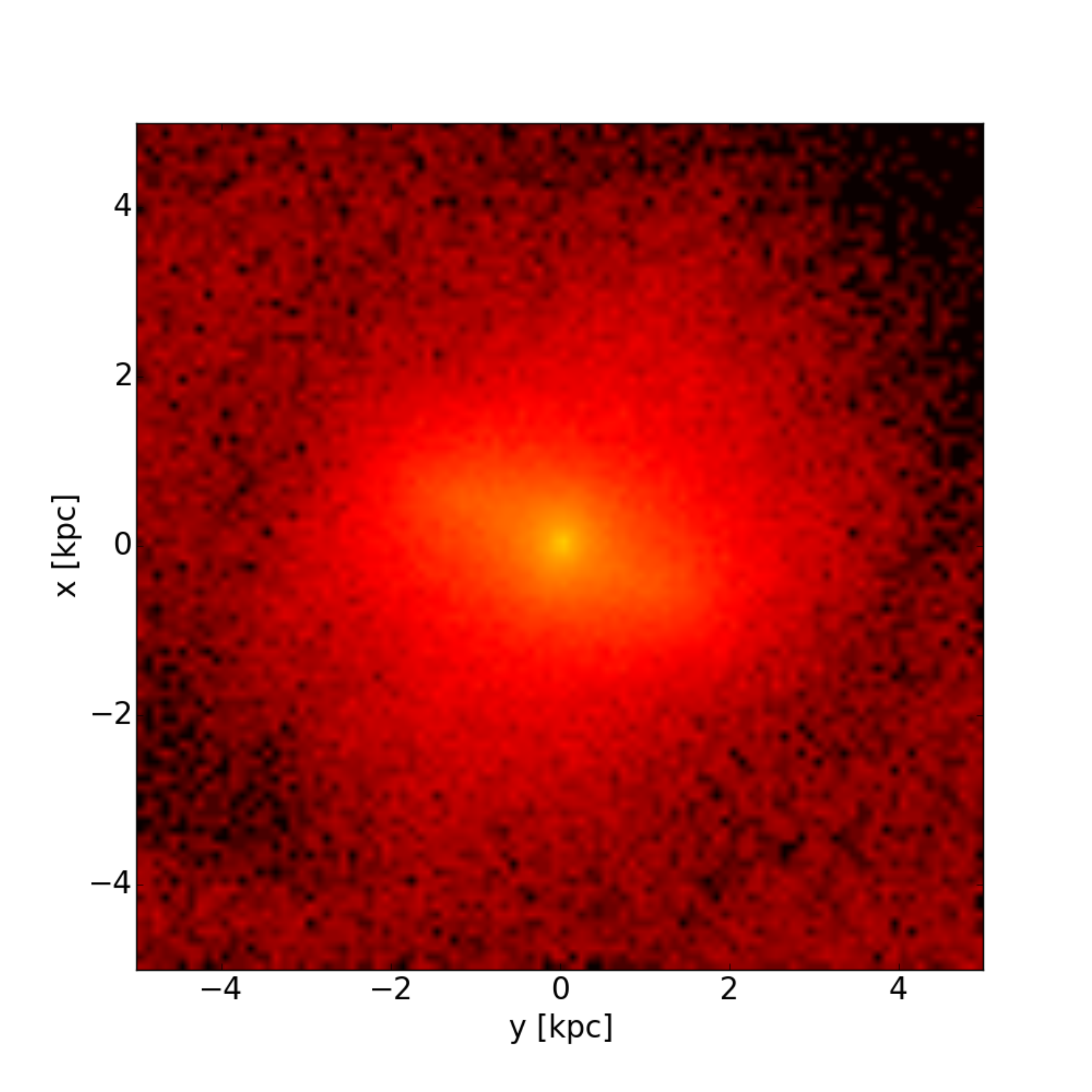}
\includegraphics[width = .3\textwidth]{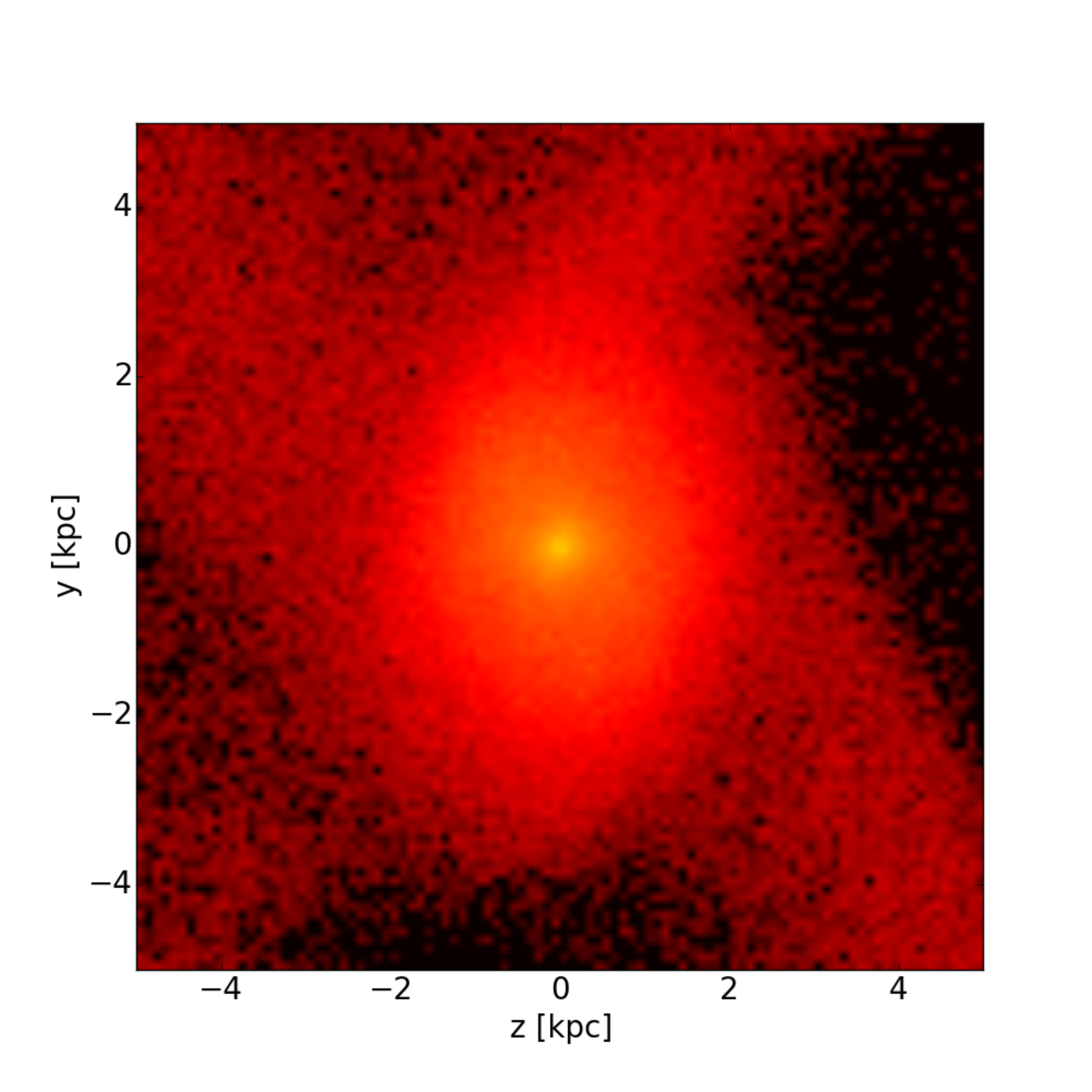}
\includegraphics[width = .3\textwidth]{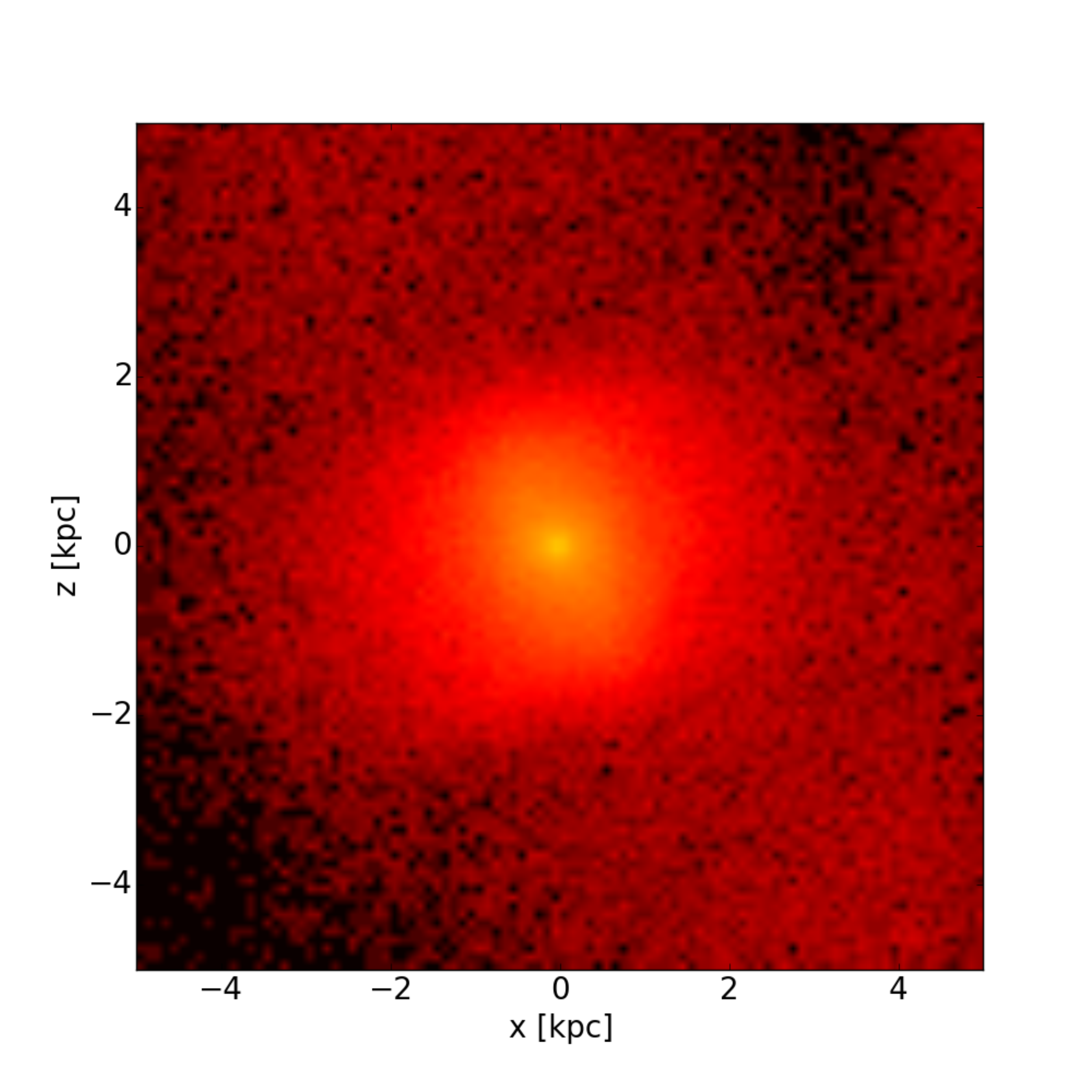}
\includegraphics[width = .3\textwidth]{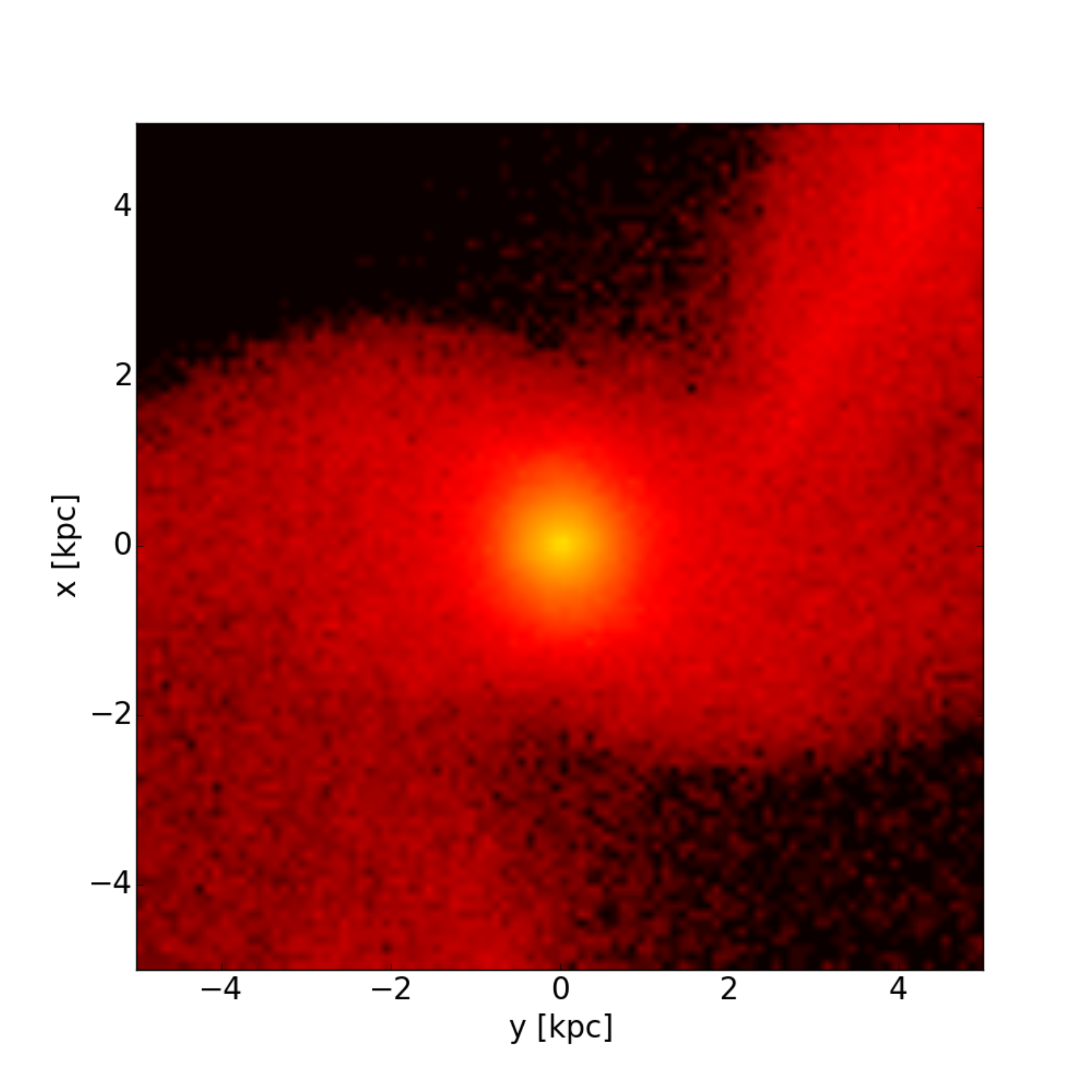}
\includegraphics[width = .3\textwidth]{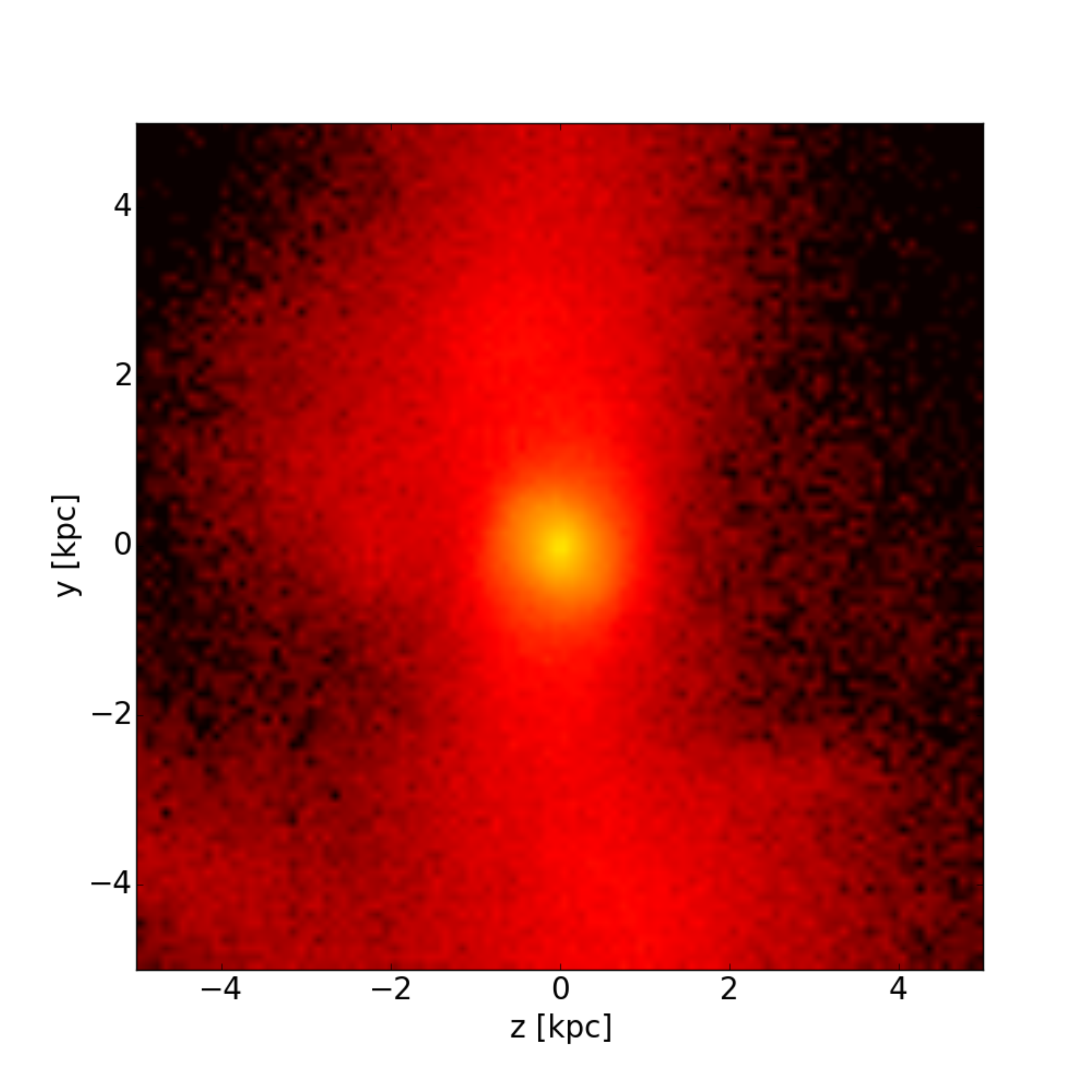}
\includegraphics[width = .3\textwidth]{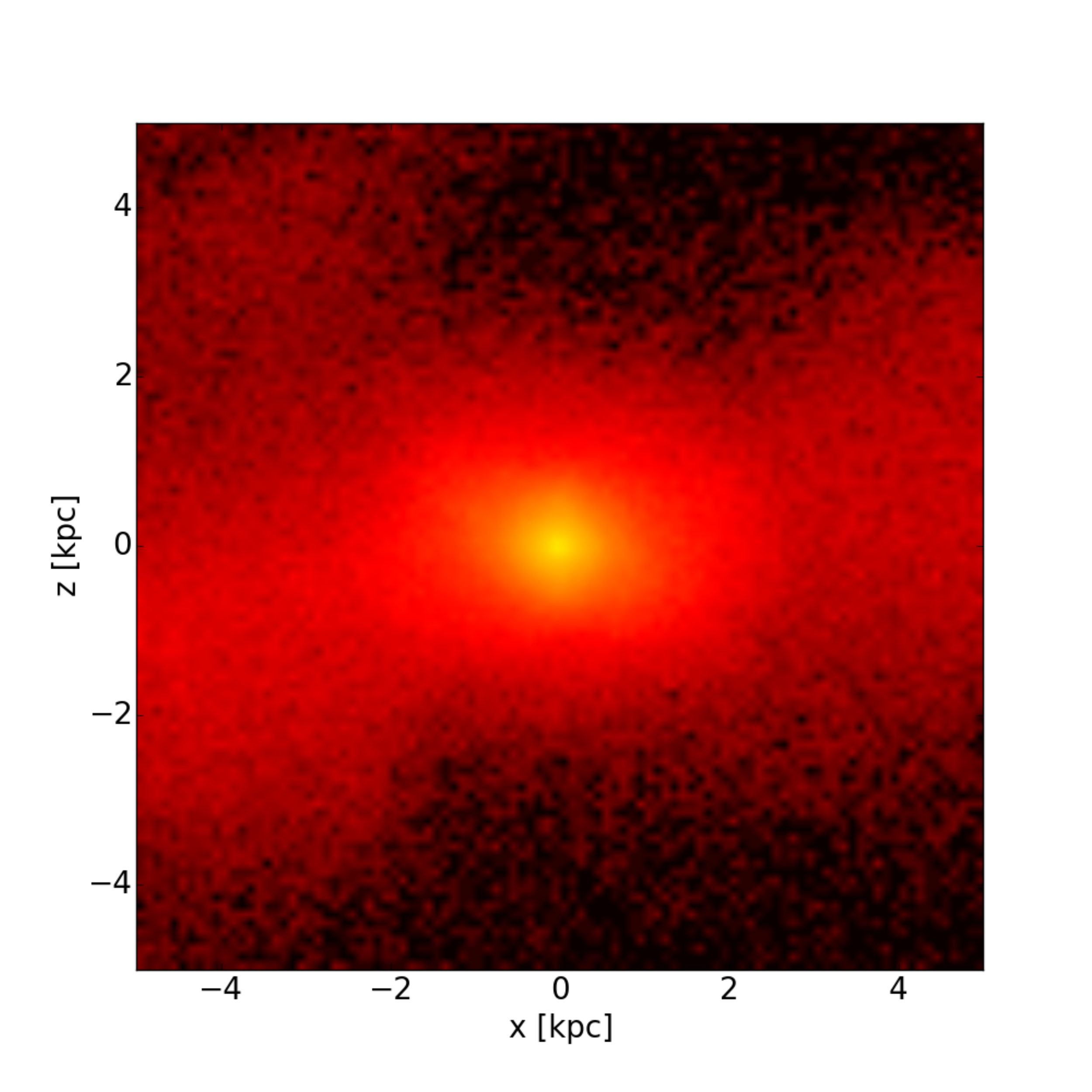}
\includegraphics[width = .7\textwidth]{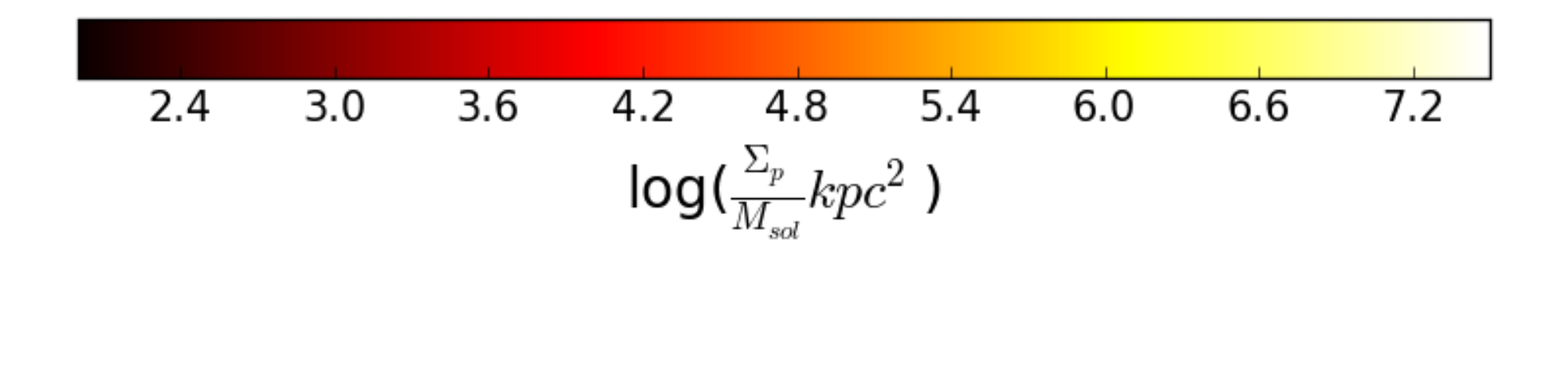}

\caption{\label{fig:figure12} The projected surface mass density maps of objects A (first row), B (second row), and C (last row) 
with $\gamma = 0.6$ at redshift 0. The first/second/third column corresponds to the projections along the smallest/middle/largest dimension.}
\end{figure}

\begin{figure}
\centering
\includegraphics[width = .3\textwidth]{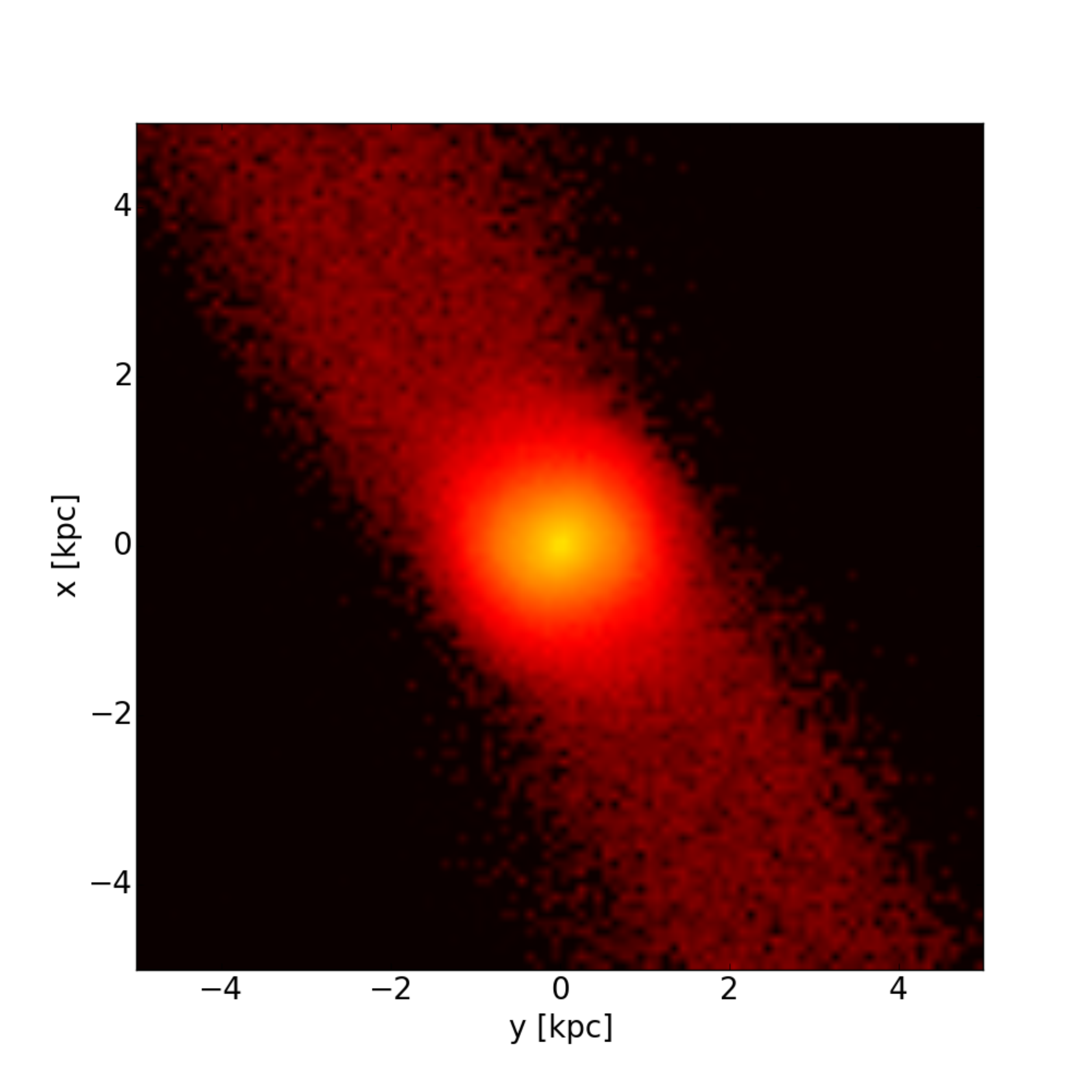}
\includegraphics[width = .3\textwidth]{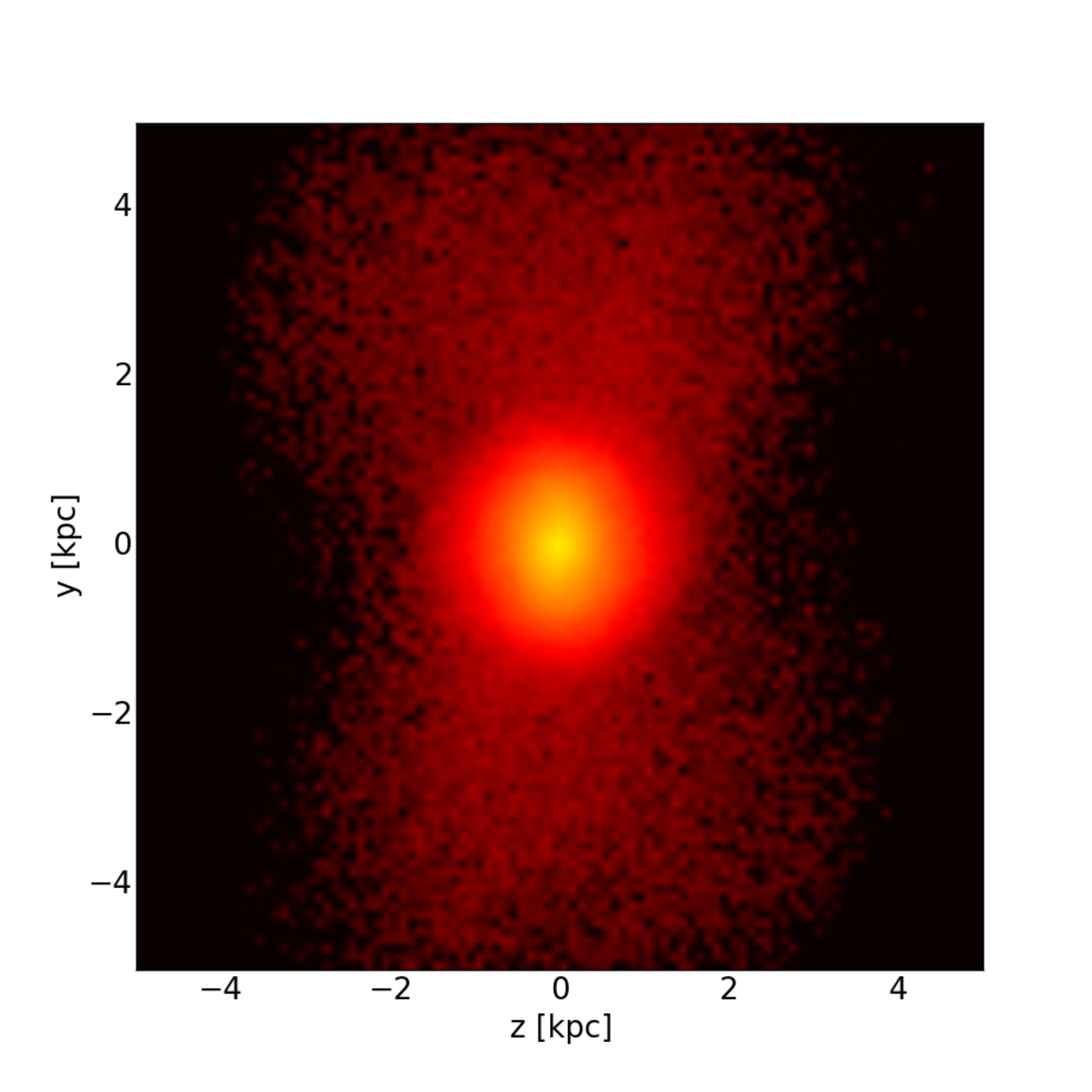}
\includegraphics[width = .3\textwidth]{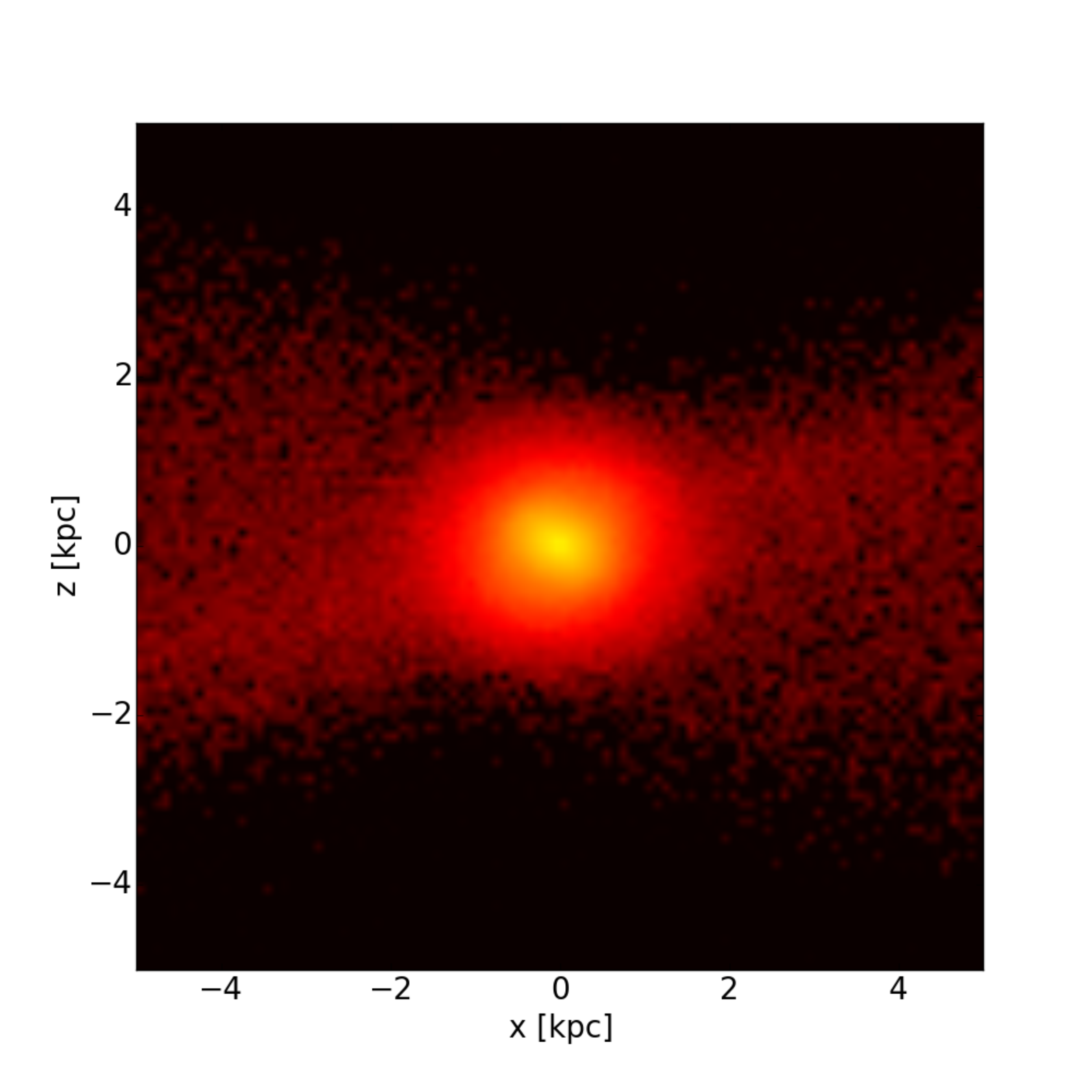}
\includegraphics[width = .3\textwidth]{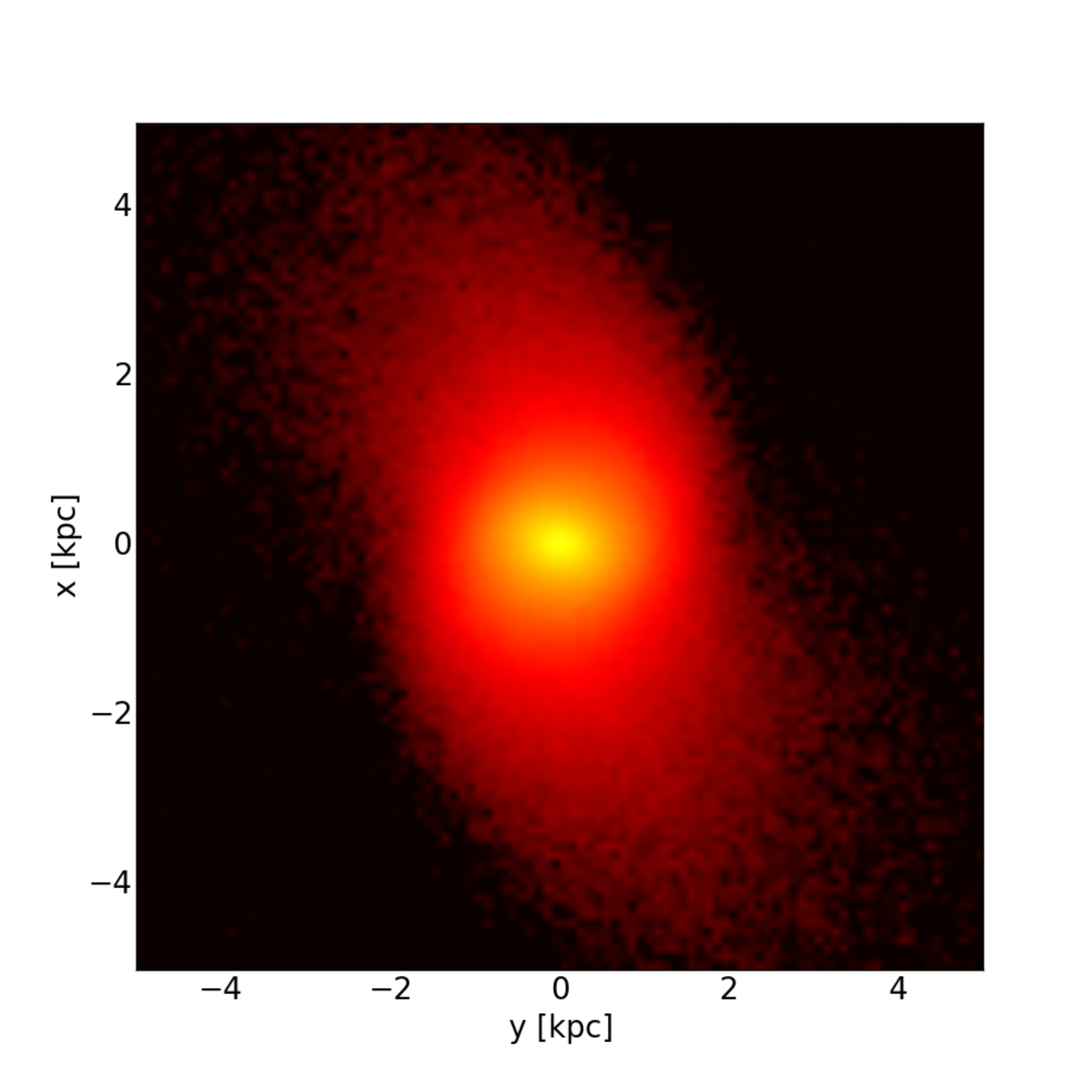}
\includegraphics[width = .3\textwidth]{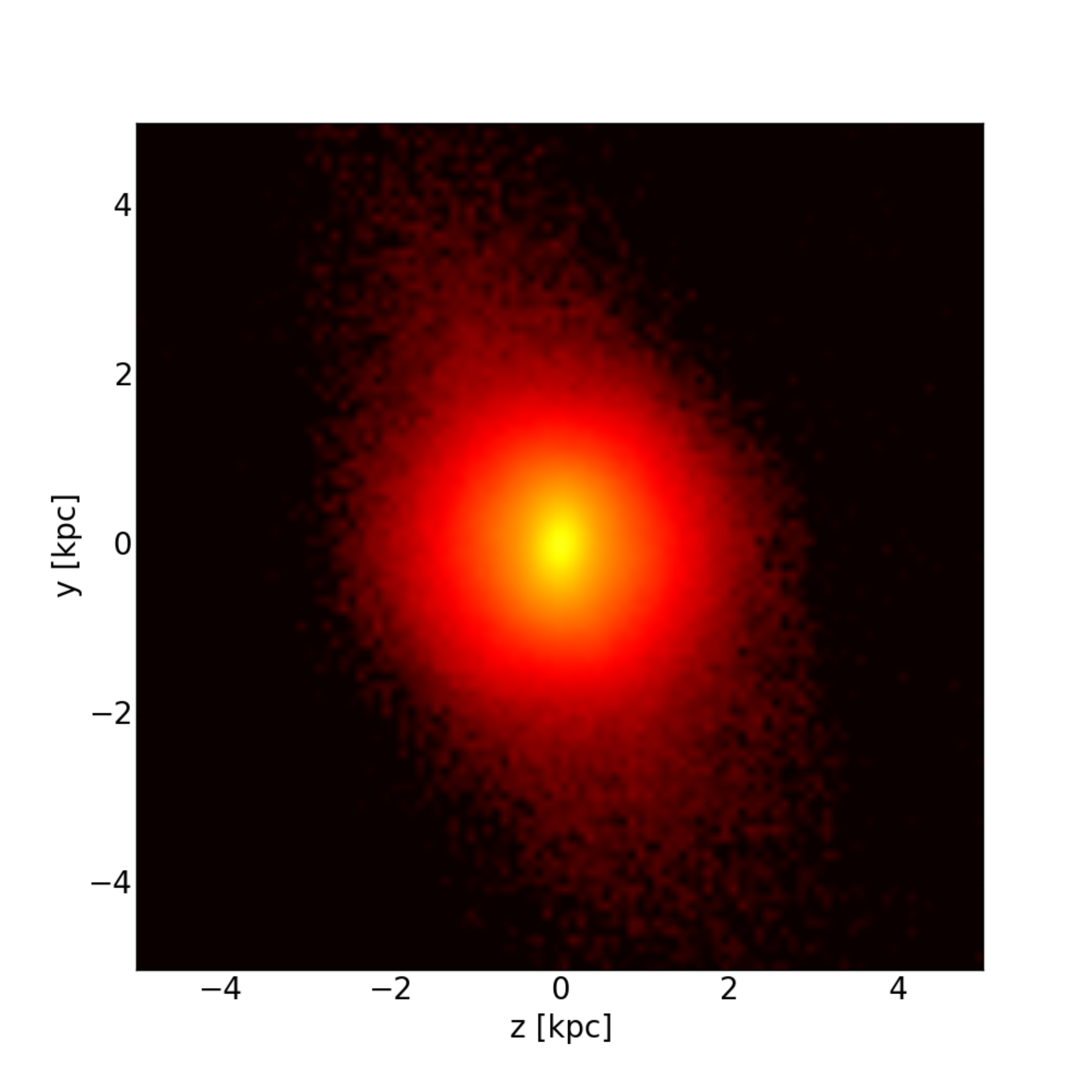}
\includegraphics[width = .3\textwidth]{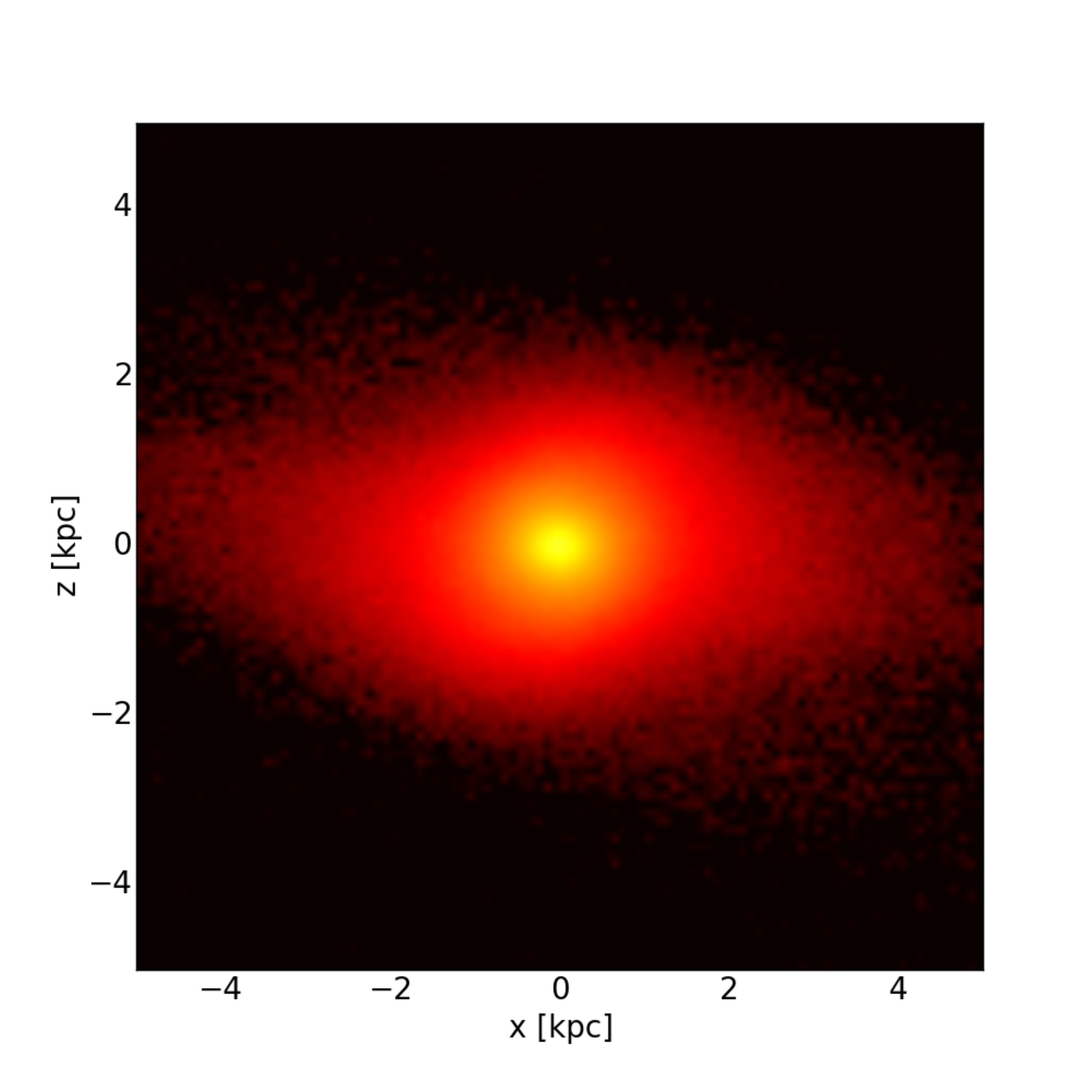}
\includegraphics[width = .7\textwidth]{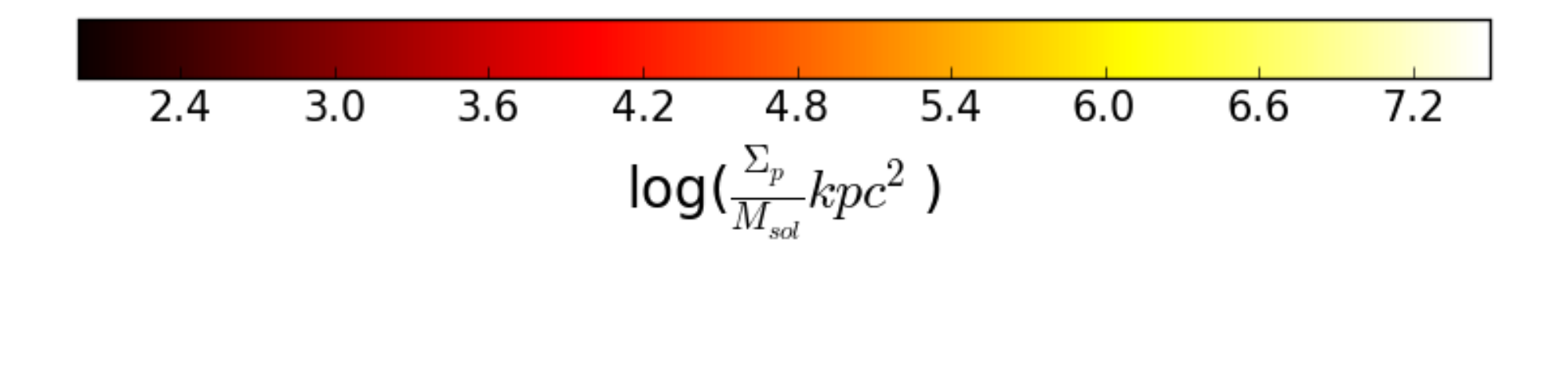}

\caption{\label{fig:figure13} The projected surface mass density maps of the objects D (first row) and EF with $\gamma = 0.6$ at redshift 0. 
The first/second/third column corresponds to the projections along the smallest/middle/largest dimension.}
\end{figure}

From a kinematic perspective, the ratios between the rotational velocity and the dispersion associated with the respective 
half-light radius decrease to the 0.5 level differently for the two versions (figure \ref{fig:figure9}). Before the first orbit is completed the 
cored satellite is described by a $c/a$ of 0.5. At the same time the ratio of the NFW variant decreases to values in the 1.0 - 1.7 range. 
Later during the third orbit, the value 0.5 is reached as well. It should be noted that the final size of the object with $\gamma$ = 0.6 is comparable 
to the softening length of its dark matter particles. Therefore the very final stages in the evolution of the object should be
regarded with caution as the response of the object to tides is affected by the limited resolution (\citep{1996ApJ...457..455M}).

The evolutionary tracks of the descriptive parameters $c/a$, $v_{rot}/\sigma_{\star}$, and $M_{\star}$ are available in figures 
\ref{fig:figure8}, \ref{fig:figure9}, and \ref{fig:figure16} respectively. 

\subsubsection{Galaxy B}
Unlike the previous dwarf galaxy, the second object has a significantly wider orbit. 
The apocenter distance varies around the 100 kpc value with a pericenter distance less than 10 kpc. 
During the 11 Gyr timespan of the simulation the satellite completes six orbits, and at the epoch corresponding to z = 0 it is aproaching its 7th apocenter (figure \ref{fig:figure16}). 
The value of half mass radius shows large fluctuations in the case of the variant with shallow dark matter density profile, with large variations when the 
object is in the proximity of the periapsis. The half light radius of the alternative presents smaller fluctuations. 
Each of the versions of dwarf galaxies start with a half light radius of 1160 pc and end with a value of 1010 pc in the case of the cored dwarf and 900 pc
in the case of the cuspy dwarf (figure \ref{fig:figure7}).     \\
The stellar discs of the two versions start to lose significant amounts of stars during the second pericenter passage. 
By the end of the simulations the version with the higher central density has lost half of the initial stellar mass. 
In the case of the other variant this level of loss is reached before the 4th apocenter epoch. Moreover at z=0 90$\%$ of the original stellar mass is lost (figure \ref{fig:figure16}).\\
The velocity ratio corresponding to the NFW profile reaches the level of 1.5 at redshift 0. At the same epoch the cored dwarf reaches a value of 0.5 (figure \ref{fig:figure8}).\\
Spatially,  the two versions of the object, maintain similar ratios of the minor and major dimensions until epoch 8 Gyr. 
Afterwards in the case of the variant with a cusp coefficient of 0.6, the ratio grows faster. 
The divergence in the velocity ratio appears at the same time; namely after the epoch corresponding to the 4th pericenter. 
Finally at z =0 , c/a fluctuates around 0.8 for the non-NFW variant and around 0.5 for the NFW variant (figure \ref{fig:figure9}). 
When studying the projected density maps it is evident that the cupsy object maintains  a two-armed spiral pattern while the 
cored dwarf approaches spherical symmetry (figures \ref{fig:figure10} and \ref{fig:figure12}).

\subsubsection{Galaxy C}
The third most massive object is originally on a trajectory with a pericenter distance of 40 kpc and an apocenter distance of 110 kpc. 
Soon after one orbit is completed it decays to a trajectory closer to the host. The following orbits are characterised by pericenter 
distances in the range 10-20 kpc and apocenter distances in the 90-100 kpc range. As the previous satellite before the last epoch it completes six orbits (figure \ref{fig:figure15}).\\

Kinematically, the characteristics of the two versions diverge after epoch 7.5 Gyr which roughly corresponds to the third apocenter passage (figure \ref{fig:figure8}). 
The indicator of the spheroidal transformation c/a clearly reflects the large difference in the transformation level (figure \ref{fig:figure9}). 
While the cuspy object maintains a ratio around 1.5,  the cored object reaches values below 0.2. 
The shape indicators of the two variants diverge earlier during the third pericenter passage and eventually converge to 0.7 values at z=0, 
roughly the time of the last apocenter passage.\\

A visual inspection of the density map discloses a more complex case (figures \ref{fig:figure10} and \ref{fig:figure12}). The shape of the stellar distribution in the inner regions resemble case by case, 
the distribution of the satellite B pair. At the same time the outer, low density and spatially more extended regions are very different. 
Particularly they are more isotropically distributed in the case of object C than in the case of object B for both versions of the respective object.

The stellar masses of the two versions of satellite C remain approximately constant until the 7 Gyr epoch. 
From this moment on the stellar mass decreases from 1.7 $10^6 M_{\odot}$ to 1.0 $10^6 M_{\odot}$  in the case of 
the cuspy dwarf, and to 4 $10^5 M_{\odot}$ in its cored counterpart.
The original value of the half light radius for both versions was 810 pc, while the final values are 680 pc and 650 pc in the cuspy
and cored dwarf, respectively (figure \ref{fig:figure8}).
Fluctuations similar to ones observed with object B are present, although with smaller amplitudes. 
Moreover they start to be visible at later epochs than in the case of object B.

\subsubsection{Galaxy D}
The object D is bound to ErisDark on an orbit with a pericenter distance of about 20 kpc and an apocenter distance in the 70-80 kpc range. 
During the simulation time frame the dwarf galaxy passes through 7 pericenters, and at z = 0 it is approaching the 8th (figure \ref{fig:figure15}). 
The tides of the host galaxy start to significantly affect the stellar component just before 7 Gyr ago. 
From this epoch onwards the cuspy object loses $3 \times 10^5 M_{\odot}$ mass. During the same period its cored
counterpart loses 8 $10^5 M_{\odot}$ (figure \ref{fig:figure16}). 
Similarly to the previous cases the stellar component in the cored model begins to lose substantial mass
when  the onset of large fluctuations in the values of the $c/a$ ratio is also observed, 
reflecting the more impulsive response to tidal shocks relative to the cuspy case (figure \ref{fig:figure9}).
The fluctuations are particularly enhanced for the cored version. 
Close to the present epoch $c/a$ reaches $\sim 0.7$ for both variants of the model.\\
Similarly, the velocity ratio starts to diverge at the 7 Gyr epoch and converge to similar values of 0.5 at z = 0 (figure \ref{fig:figure8}). 
Shapewise the final state of the two versions appear to be similar (figures \ref{fig:figure11} and \ref{fig:figure13}). 
Moreover the evolution of the half light radius is similar for both cases, showing small fluctuations and a small decrease of the value 
from the original 640 pc values to 570 pc and 580 pc for the cuspy and cored object, respectively (figure \ref{fig:figure7}).

\subsubsection{Galaxy EF}

The Dwarf galaxy EF, which is the result of a merger between two objects comparable in size, evolves on a wide orbit with the pericenter distance varying 
from 30 to 50 kpc and apocenter distances with values as high as 160 kpc (figure \ref{fig:figure15}). Since infall the satellite completes three orbits. 
After the merger, 
the stellar mass of the object remains virtually constant for both variants (figure \ref{fig:figure16}). 
The ratio $v_{rot}$/$\sigma_{\star}$  stabilizes around $0.8$ for the object resulting from the merging of two cuspy dwarfs. 
For the other variant, $v_{rot}$/$\sigma_{\star}$ continues to decrease until it reaches values close to 0.1 at 13 Gyr epoch (figure \ref{fig:figure9}). 
In the case of the shape ratio c/a the object resulting from the merger of the cuspy subhalos has on average a higher value;
after 12 Gyr the average c/a is around  0.8 for the cuspy pair and 0.7 for the cored pair (figure \ref{fig:figure8}).

The original values of the half mass radius prior to the merging of the two objects were 420 pc for object E and 390 pc for object F. 
Immediately after the merger the radius of the resulting objects is 550 pc and during the remaining simulation time it increases to 610 pc,
with no significant differences between the cuspy and core variants (figure \ref{fig:figure7}).  
This highlights a clear difference with tidally stirred dwarfs, in which the half-light radius always decreases. 

\section{Discussion}
In the current work the evolution of five pairs of satellites entering a MW-sized halo has been presented. 
Each pair consisted of two objects differentiated by a single parameter, namely the cusp coefficient of the dark matter halo. 
In each pair, each of the objects was composed of a stellar disk component and a dark matter halo. 
The virial mass of the objects spanned one order of magnitude from 3.3 $10^8 M_{\odot}$ to 3.83 $10^9 M_{\odot}$. 
The simulation commences at z =2.87 when all dwarf galaxies are in the proximity of their future host halo, yet outside its virial volume. 
The original distances to the host center ranged from 75 to 165 kpc. Since the last major merger experienced by the host halo
takes place before z = 3 the described population of objects is representative of the earliest infall objects and thus the most tidally affected 
population. Galaxies with later infall epochs will be affected by the tides on average to a lesser extent.\\

All of the objects with cored density profiles match our definition of dSphs before z=0. 
At the same time only two out of four objects with cuspy density profiles satisfy the criterion. 
The more efficient transformation of cored models is expected as their response to tides is more impulsive owing to their
lower gravitational binding profile for comparable mass (\cite{2013ApJ...764L..29K}). 
In particular the binding energy decreases much faster towards the center allowing the tides to disturb the stars found closer to the center
of the satellite (\cite{2012ApJ...751L..15L}).

Figures \ref{fig:figure17}, \ref{fig:figure18}, \ref{fig:figure19} and \ref{fig:figure20} show the time evolution of various structural properties of all models, highlighting how the cored models in general fit 
better the range of values that classical dSphs and UFDs span for such properties (\cite{2012AJ....144....4M}).
We caution, though, that there is a lack of objects 
with small half light radii, which may reflect the relatively small sample of initial conditions adopted here (see next section). 

The half mass radius, velocity and shape ratios have significant fluctuations on orbital timescales.
Their fluctuations appear to be in phase with orbital position and grow in amplitude as more stellar mass is unbound from the respective satellite.
In general the objects with an original $\gamma = 0.6$ show significantly greater amplitudes in the fluctuations than the ones with $\gamma = 1.0$,
again reflecting the different binding energy profiles which place cored models in the more impulsive regime for the response to tides
(which will be maximal near pericenter).
The minimum of the fluctuations in half mass radius is reached when the object is near its periapsis, while the maxima is reached a few
hundred million years afterwards.

In general the satellites with $\gamma=0.6$ are also transformed faster into dwarf spheroidals than their counterparts.
The final dark matter to stellar mass ratio are significantly higher in the cuspy remnants (see table \ref{table:table5}), yet the final values in cored remnants are
still high enough to match those of classical dSphs and UFDs. 
Assuming a stellar mass-to-light ratio $(M/L)_* \sim 3-4$, appropriate
for an old stellar population, (table \ref{table:table5}) would yield values in the range $M/L = 90-1000$ for the cored remnants.
Instead velocity dispersions and dynamical masses of cuspy remnants indeed appear to exceed
observed values in model B and D, because of the very high central dark matter density.
Note this result is not coincidental rather it reflects the fact that, while the core remnants
have radii of less than 1 kpc, the initial dark matter-to-stellar 
mass ratio was very high even inside the central kpc in order to
match the constraints from abundance matching with the chosen density profiles of dark matter.

All the tidally stirred remnants are found at $R < 100$ kpc from the host at $z=0$. 
Within such a distance from the MW and M31 all the galaxies in the
mass range of the remnants (stellar masses $10^4 - 10^6 M_{\odot}$) are dSphs 
(or UFDs, which structurally appear similar to classical dSphs, see \cite{2012AJ....144....4M}). 
One is thus tempted to conclude that this is evidence of the fact that the most massive satellites falling early inside the host, which
is the type of objects we are studying here, have predominantly shallow dark matter profiles. In other words, dwarf galaxies
with the properties of the cuspy remnants that did not transform fully into dSphs are not observed in the inner halo of the MW and M31, 
which means cuspy progenitors must be rare (the only irregular galaxies in the inner halo, the LMC and SMC, are 
a few magnitudes more luminous than any of our remnants).
Note that the latter inference is well grounded even with the small sample in this 
paper since specifically this sample contains the objects that have experienced the longest tidal evolution possible inside the main host 
(early infall population). We thus expect a larger number of such
incompletely transformed objects with disk-like features in the much larger sample with varying infall time that will be presented in Paper II. 

Note that idealized simulations already hinted at such a more efficient transformation of the cored models, 
still they never found such a strong effect because the mass ratio adopted between the satellite and the main host was at least an order 
of magnitude higher than in this work. The reason is that here the host halo is relatively low in mass (closer to $\sim 10^{11} M_{\odot}$ 
than to $10^{12} M_{\odot}$ when the most massive satellites fall in at $z \sim 2$, an effect that
only a fully cosmological simulation can account for.
The case for cored progenitors  is suggested by the notion that dwarf galaxy satellites which had the highest star formation rates, 
as required to reach a stellar mass in the range of $10^6-10^7 M_{\odot}$ already at $z > 2$, 
are those in which feedback was most effective to turn cusps into cores (\cite{2010Natur.463..203G}; \cite{2012MNRAS.422.1231G}; \cite{2014ApJ...792...99S}; \cite{2015arXiv150202036O}). These are the objects in which the star formation (SF) efficiency, 
defined as ratio between stellar mass assembled and dark halo mass, was particularly high (\cite{2014ApJ...790L..17M}). 
A high star formation efficiency in classical dSphs versus similarly low mass field
dIrrs is indeed seen in SF histories based on hi-res color magnitude diagrams (eg \cite{2014ApJ...786...44S}; Aparicio et al. in preparation). 

Interestingly, gas-poor dwarfs with disk features have been detected in clusters (\cite{2006AJ....132..497L}) and have been explained 
as harassed galaxies that have accreted onto the cluster core only recently, thus have not been able to transform fully into dwarf spheroidals. 
It is tempting to speculate that such a population would be present in clusters even if the progenitors have dark matter cores before 
infall since a much larger fraction of galaxies accrete at $z  < 1$ in clusters as opposed to $z > 1$  for satellites into galaxy-sized halos. 
This is strongly suggested by the evolution of some of our models, which also in the cored variant transform fully into
dSphs only after nearly 10 billion years (eg see models B and D in figure \ref{fig:figure9}).

\begin{figure}
\includegraphics[width = .4\textwidth]{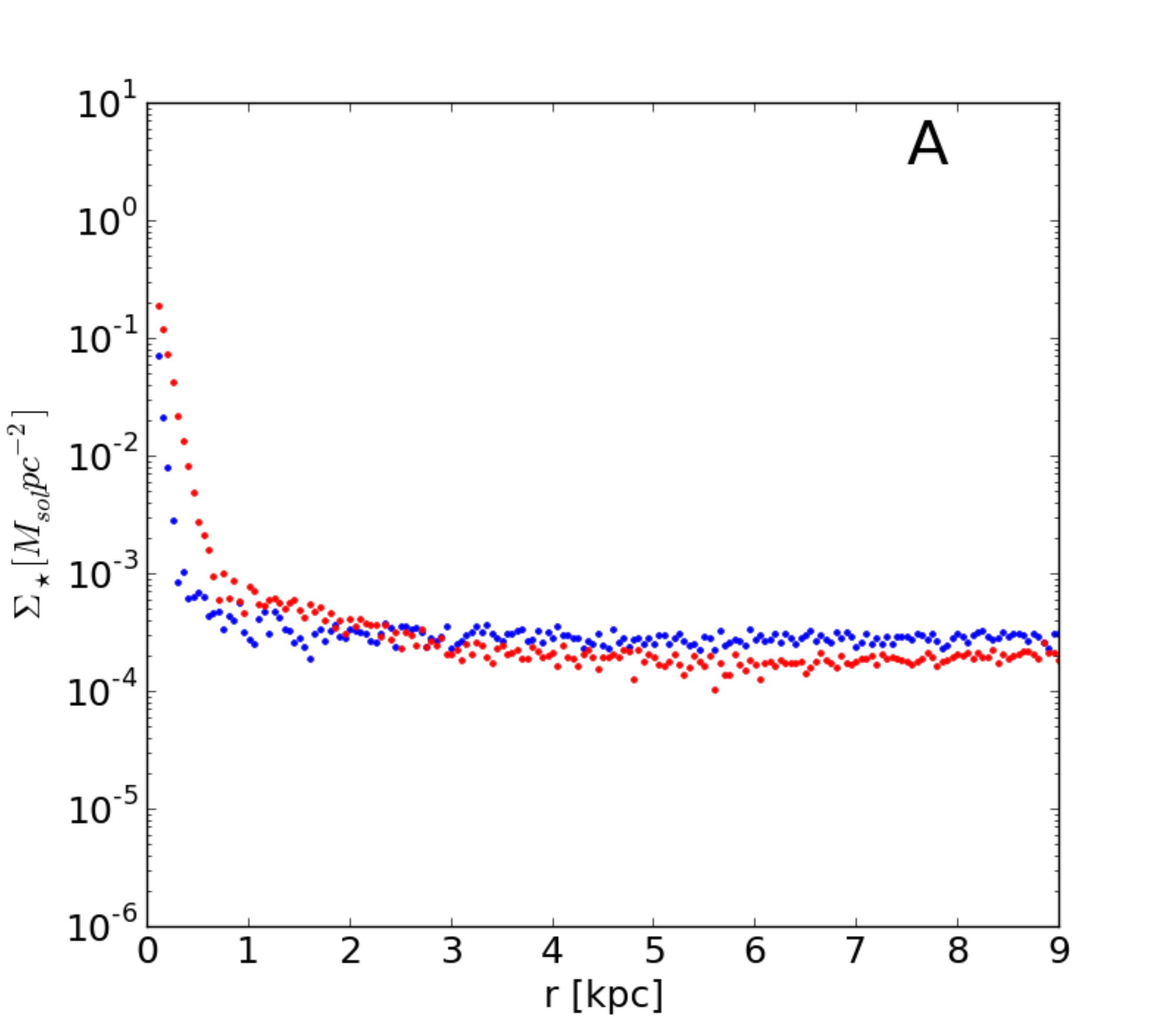}
\includegraphics[width = .4\textwidth]{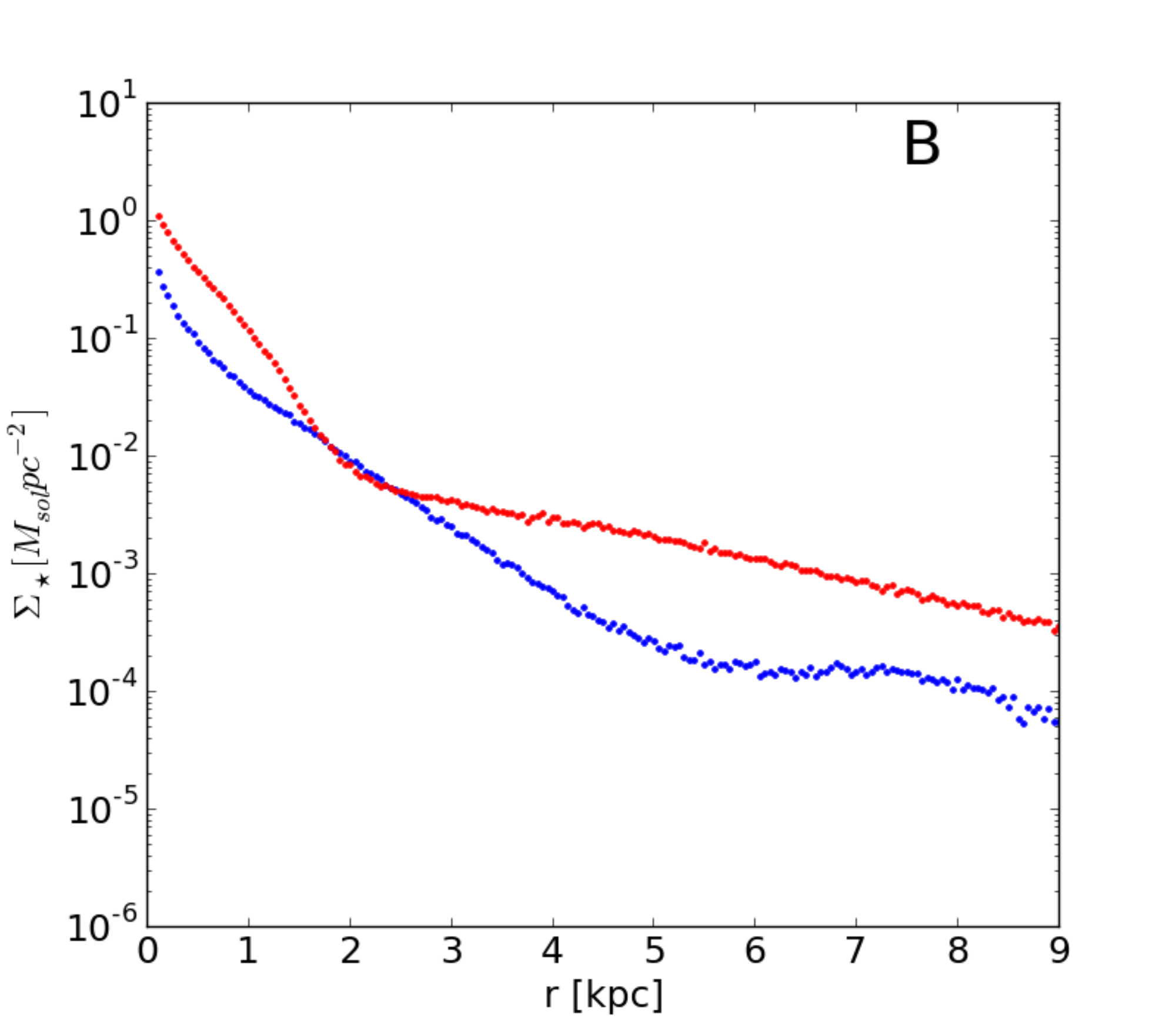}
\includegraphics[width = .4\textwidth]{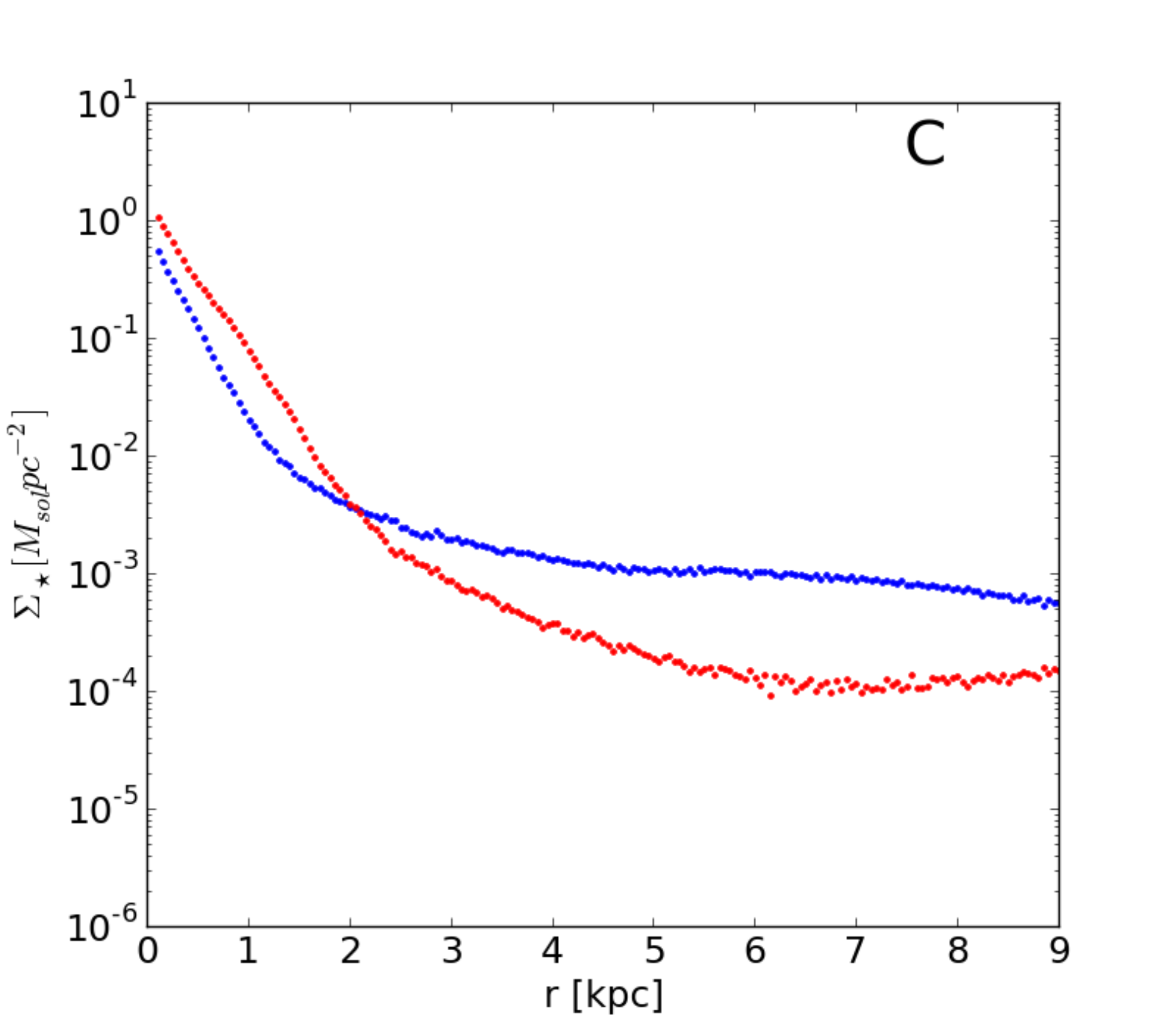}
\includegraphics[width = .4\textwidth]{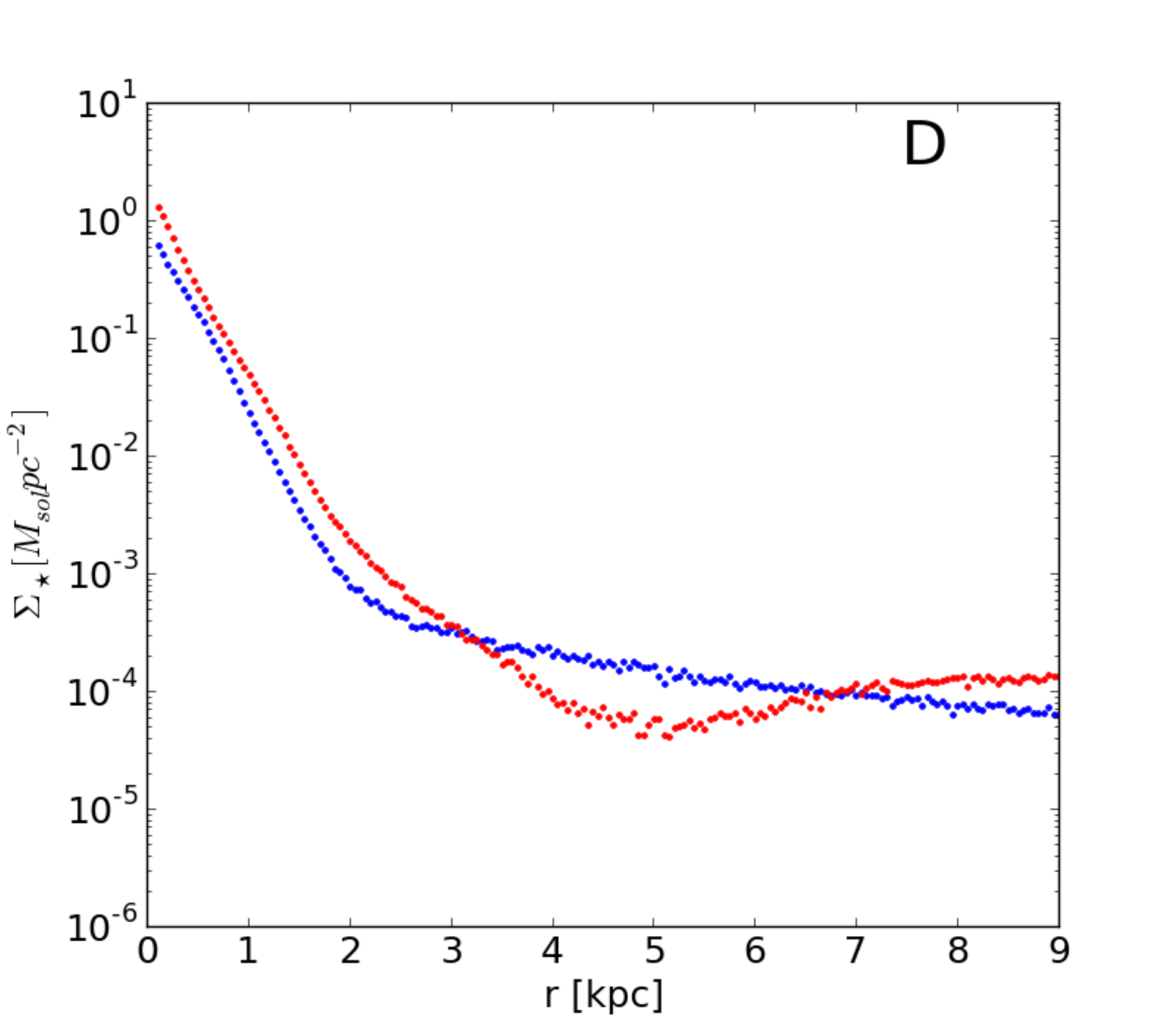}
\includegraphics[width = .4\textwidth]{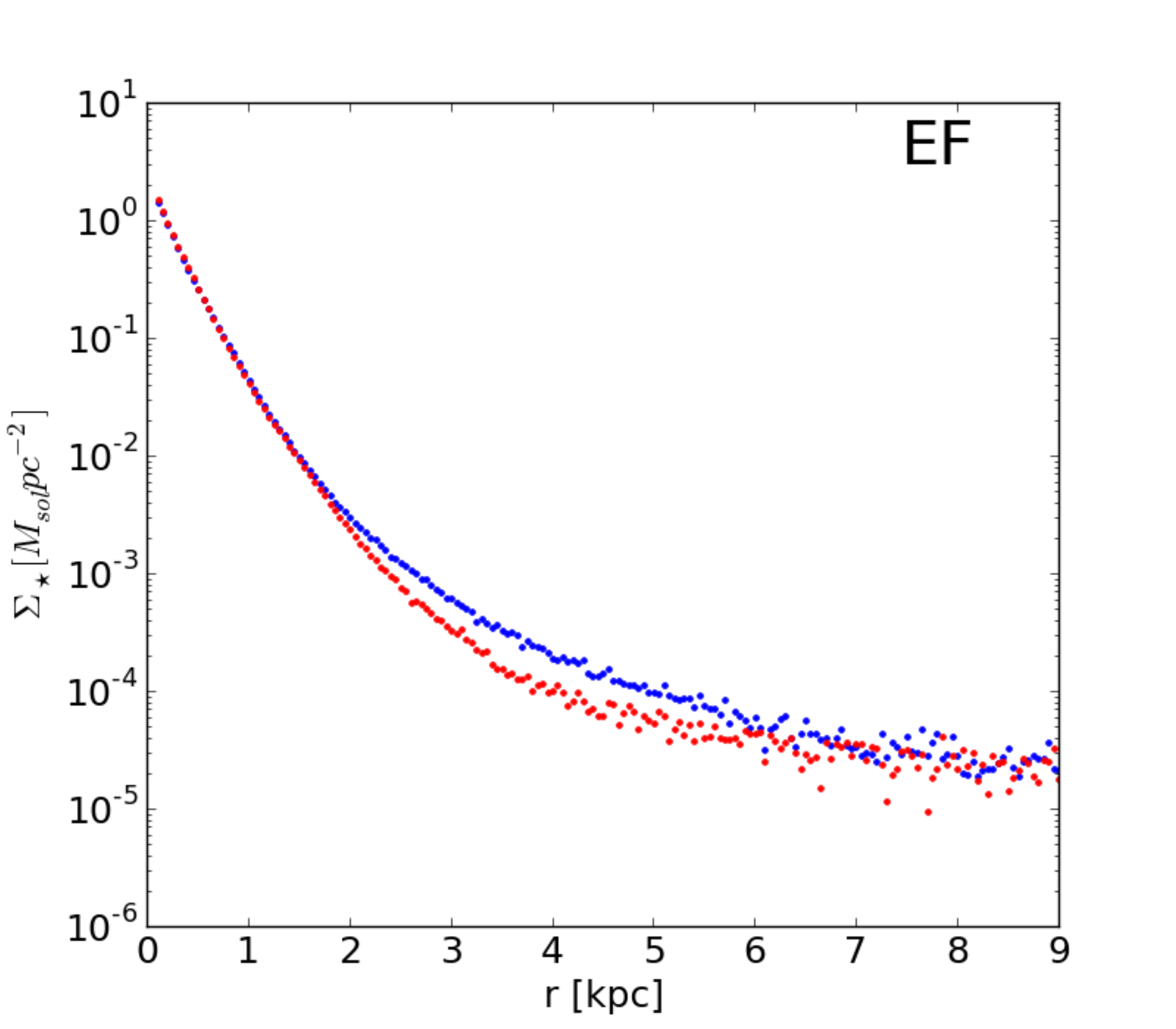}

\caption{\label{fig:figure14} The cylindrically averaged surface mass density profiles are presented. 
All quantities are computed for the direction corresponding to the shortest dimension. 
Red/blue dots mark the satellites with steep/shallow central density profile. 
The top/middle/bottom row illustrates the surface density of satellites A and B / C and D / EF.}
\end{figure}

\begin{figure}
\centering
        \includegraphics[height = 0.4\textwidth]{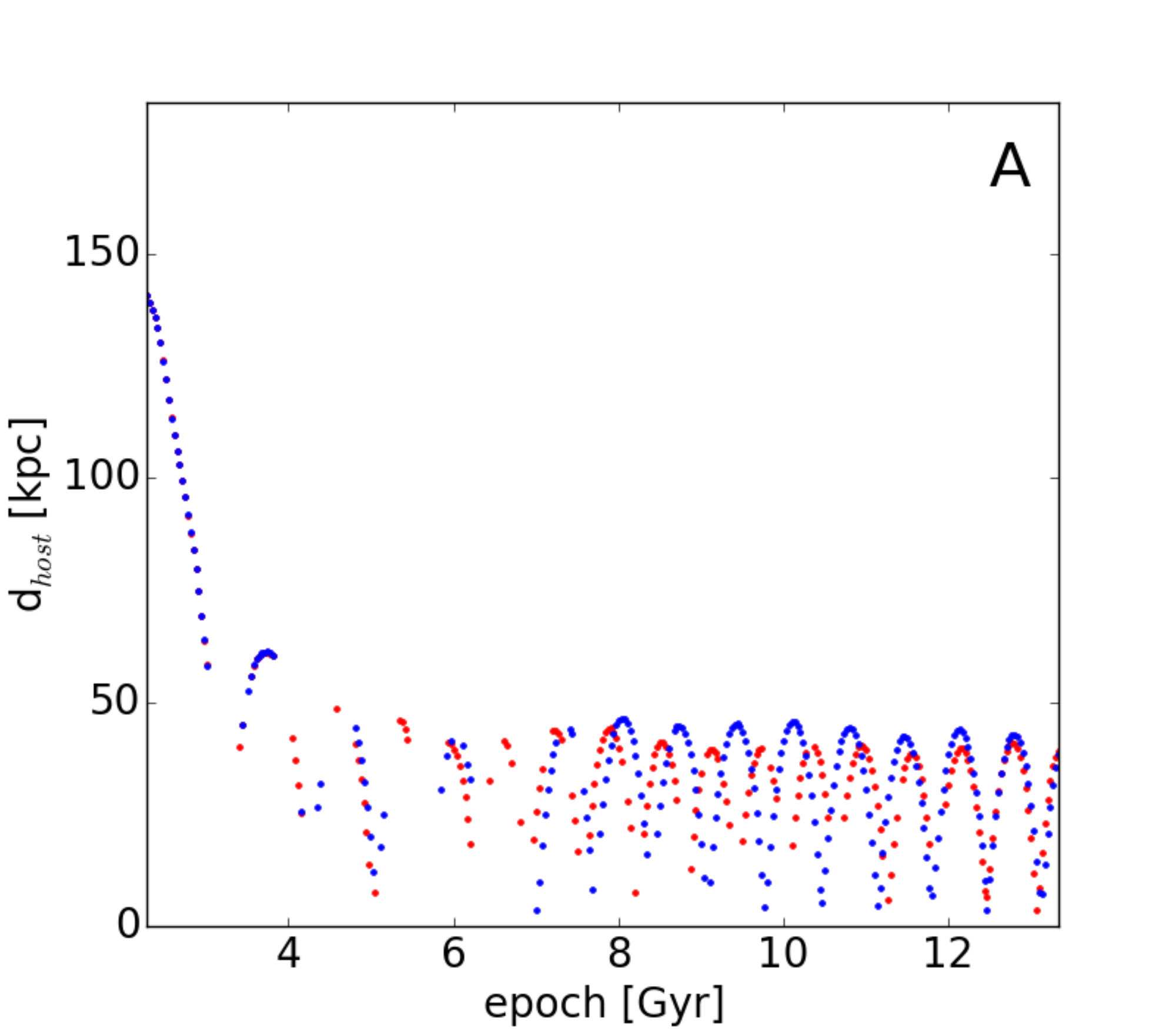}
        \includegraphics[height = 0.4\textwidth]{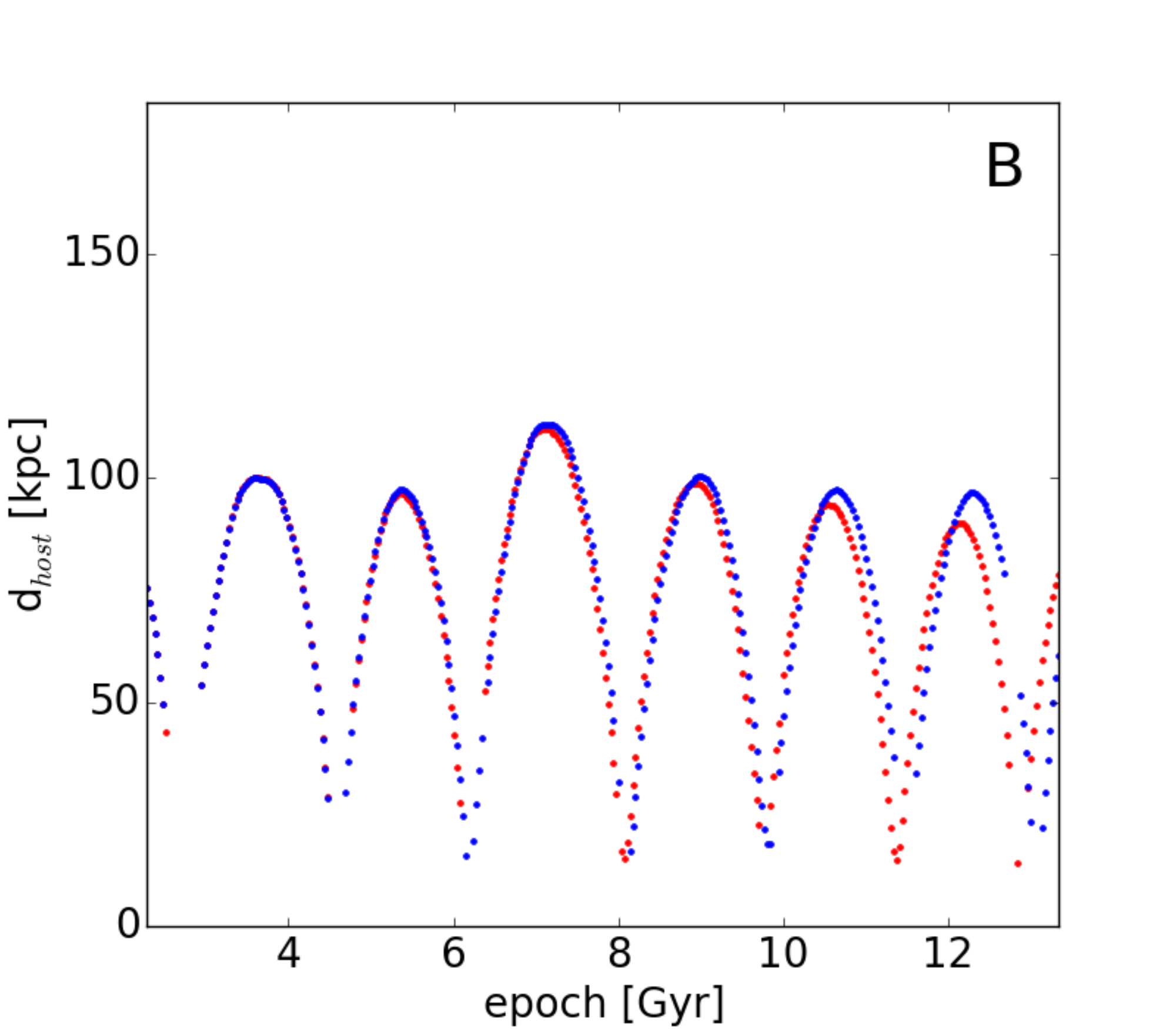}
	\includegraphics[height = 0.4\textwidth]{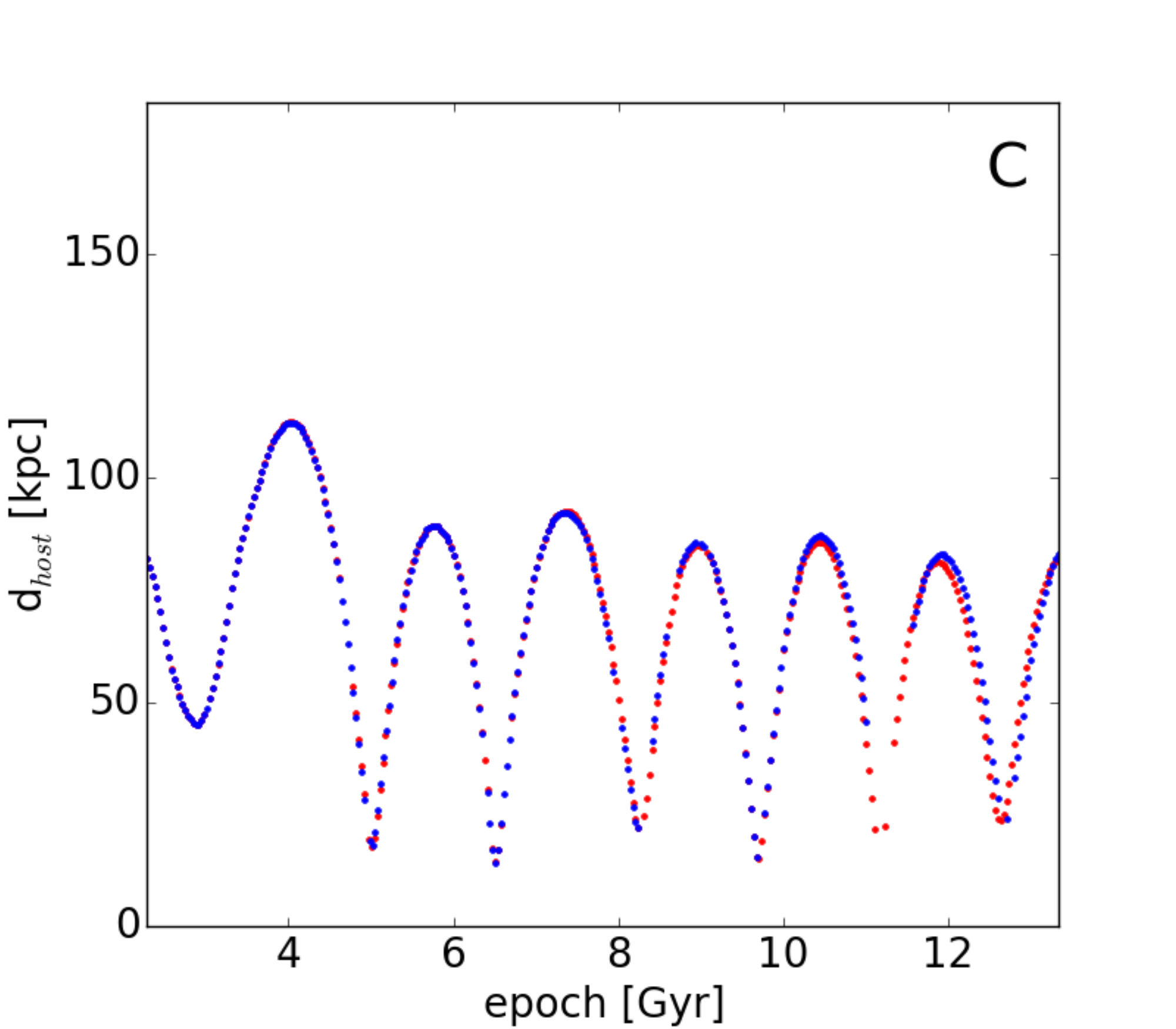}
	\includegraphics[height = 0.4\textwidth]{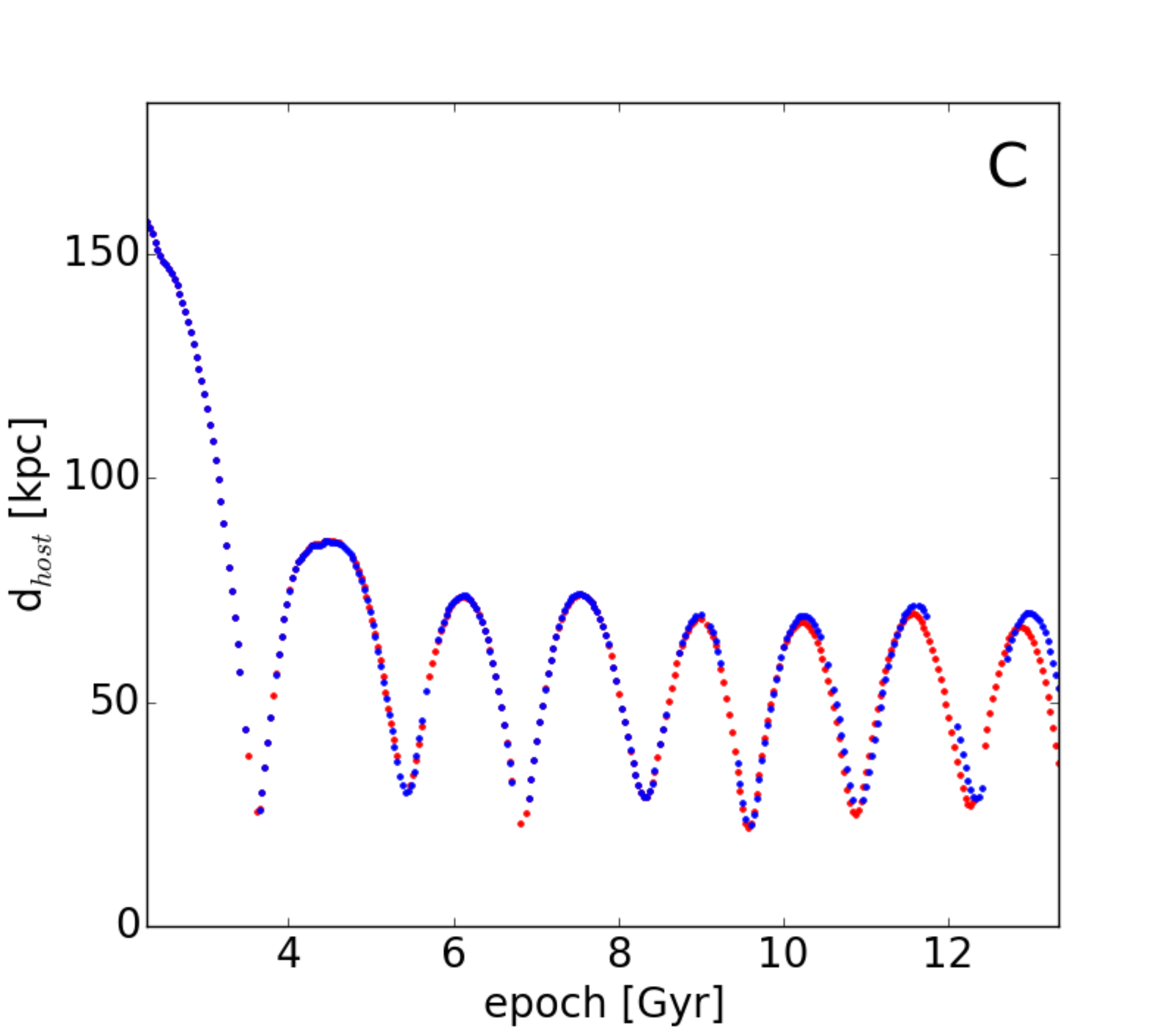}
	\includegraphics[height = 0.4\textwidth]{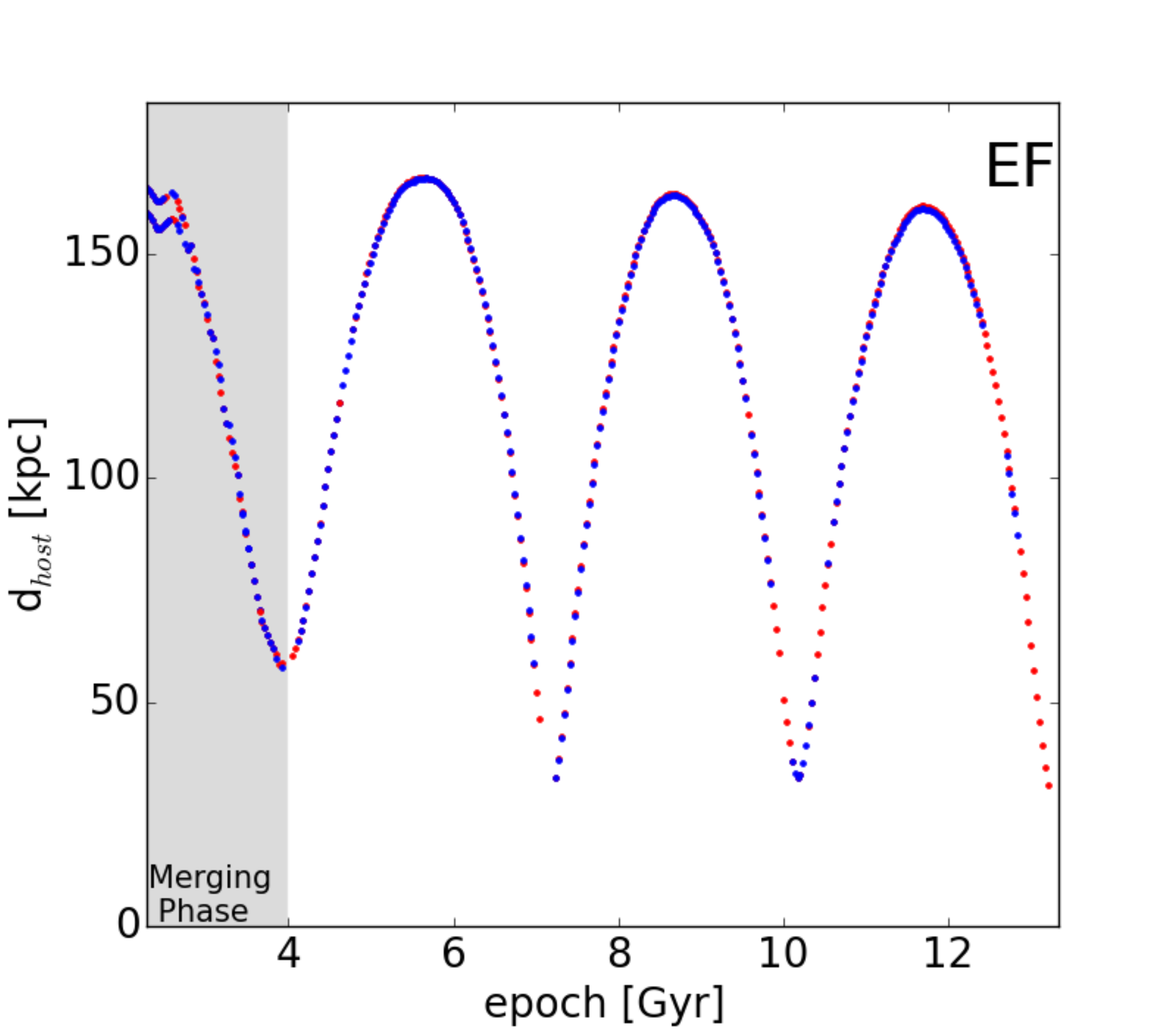}
\caption{\label{fig:figure15} Distances between the satellite galaxies and their host halo. 
The versions of dwarf galaxies with $\gamma =$ 1.0 / 0.6 are marked in red/blue. 
For object EF (the merging case) the gray region marks the epochs before the evolution 
of the half mass radius stabilizes (as can be seen in figure \ref{fig:figure7}).}
\end{figure}

\begin{figure}
\includegraphics[width = .4\textwidth]{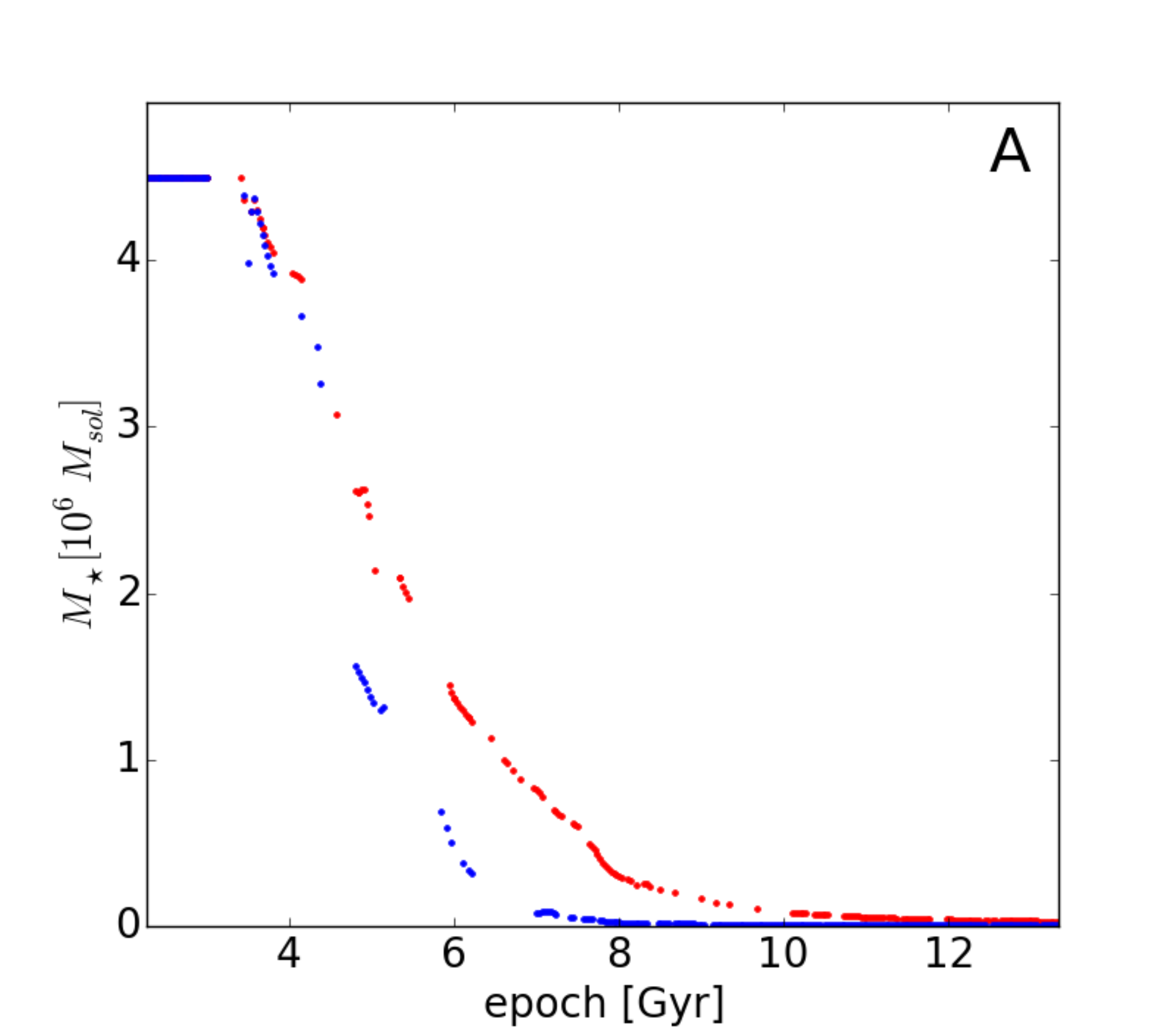}
\includegraphics[width = .4\textwidth]{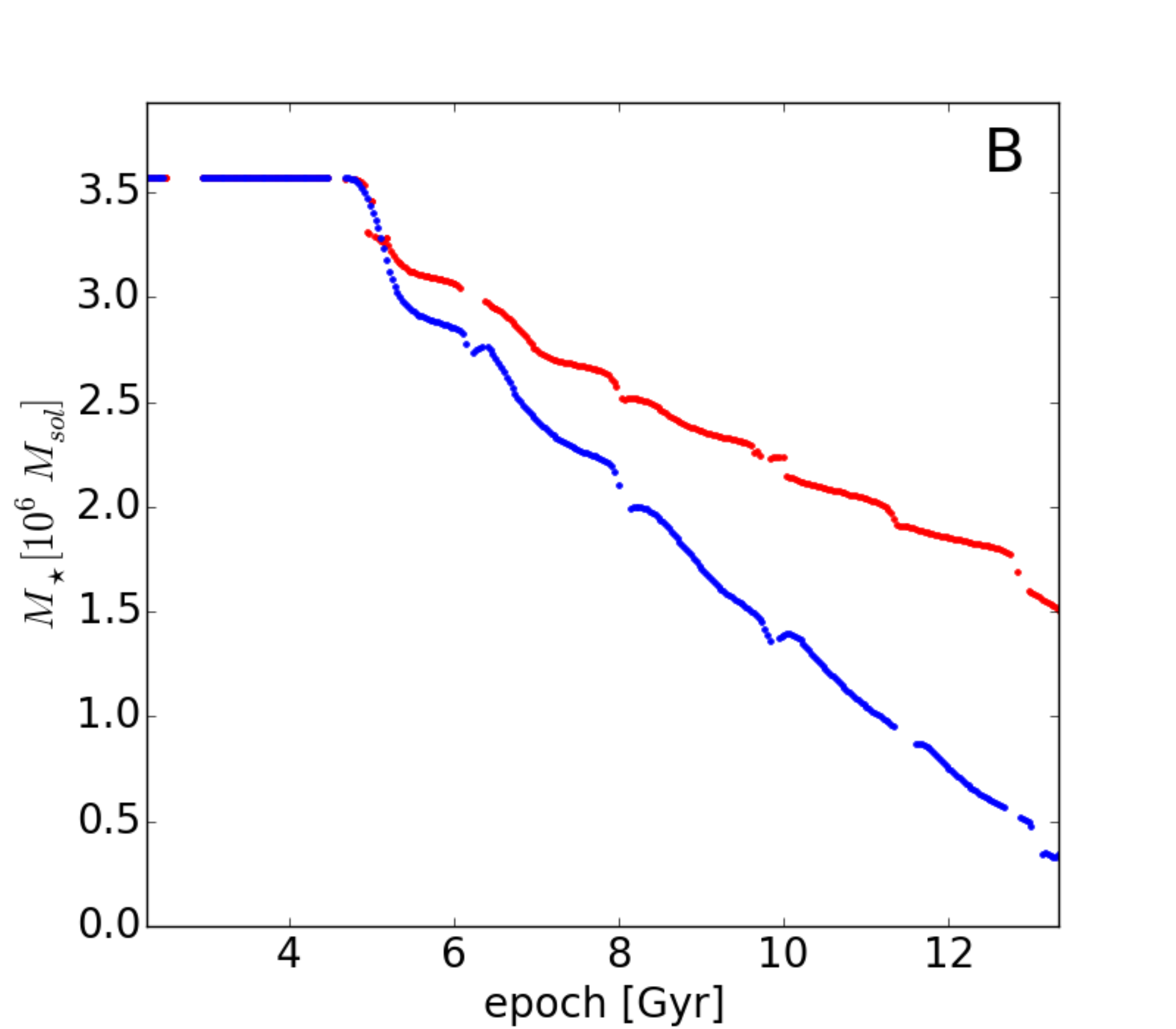}
\includegraphics[width = .4\textwidth]{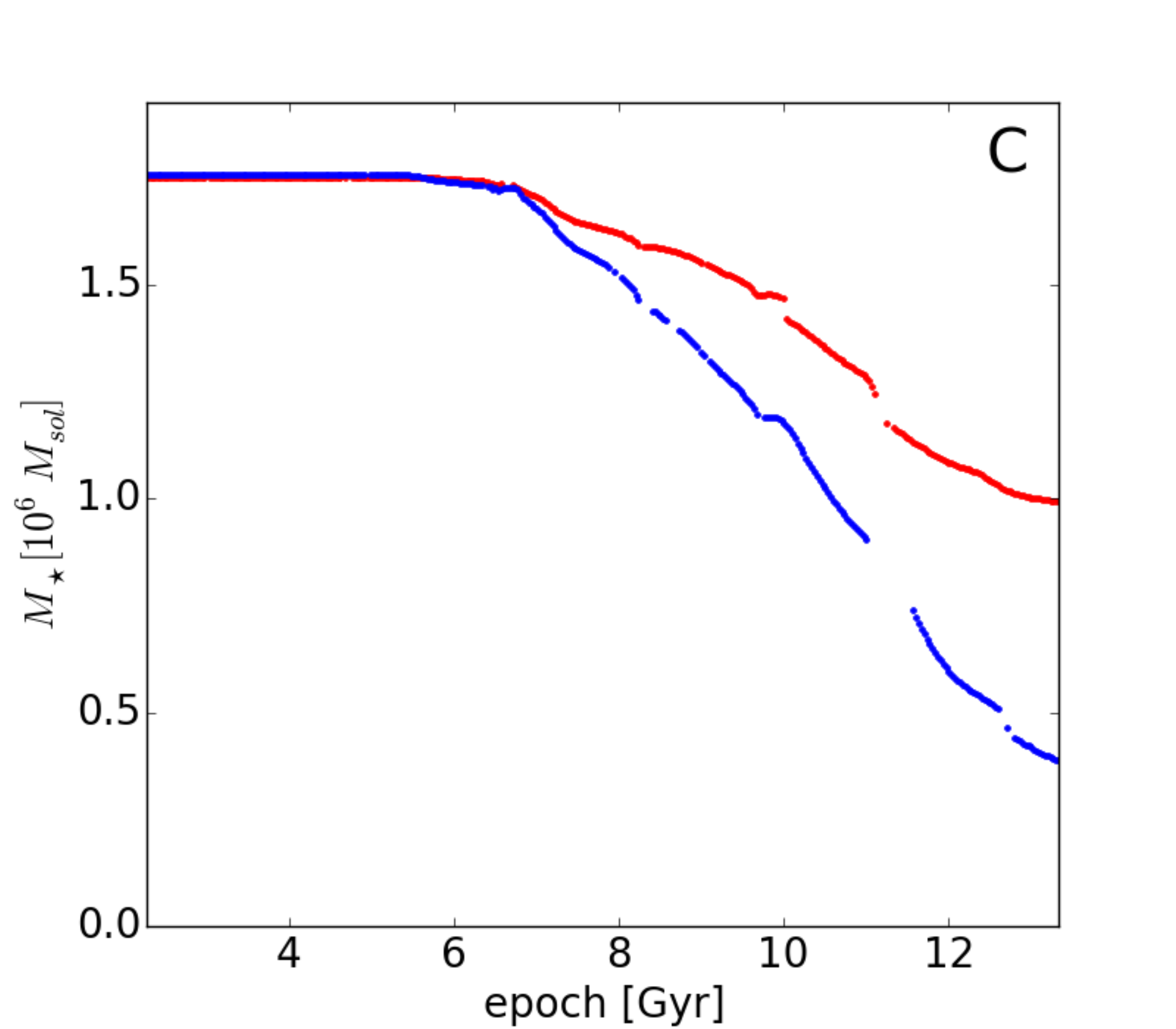}
\includegraphics[width = .4\textwidth]{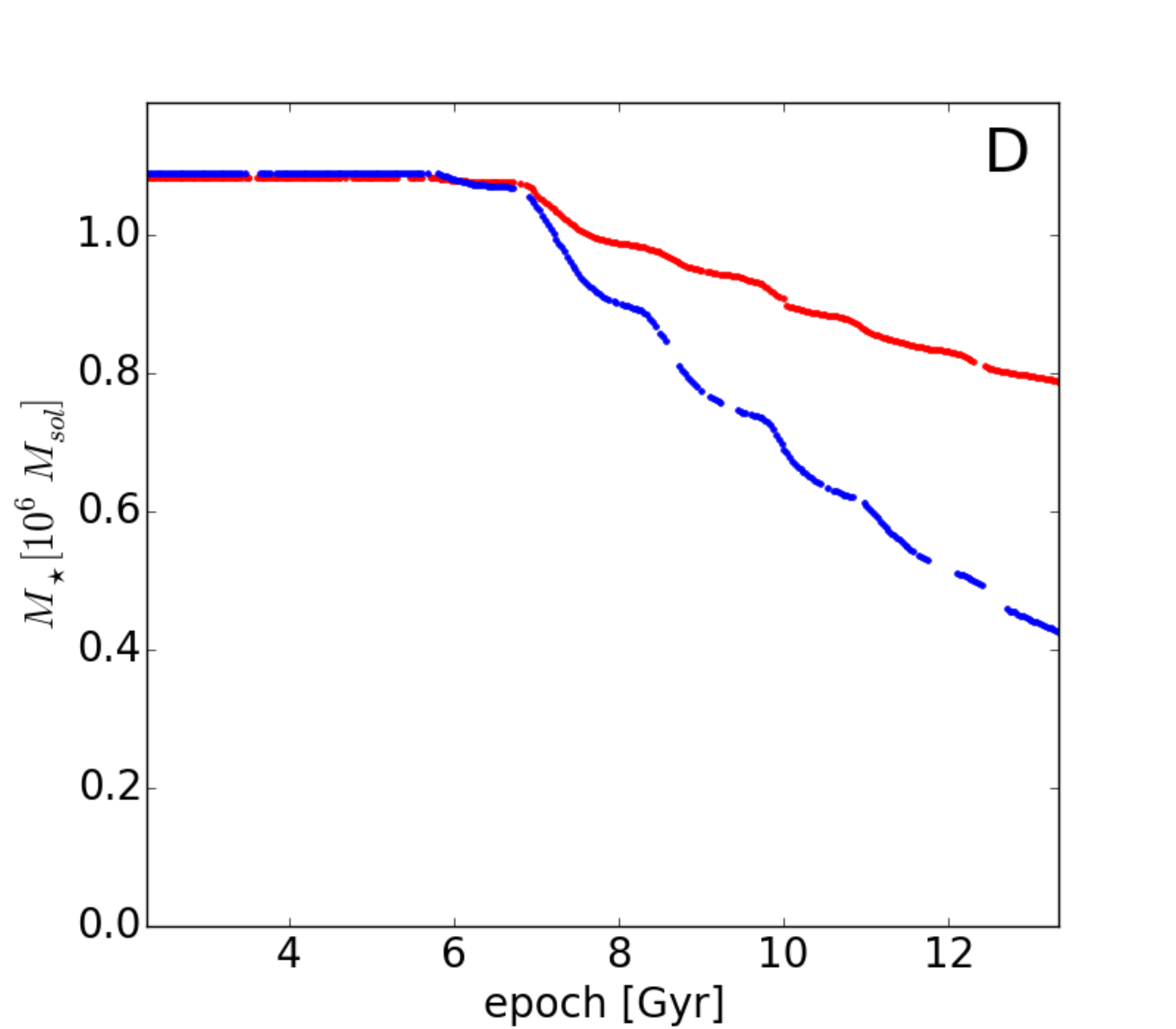}
\includegraphics[width = .4\textwidth]{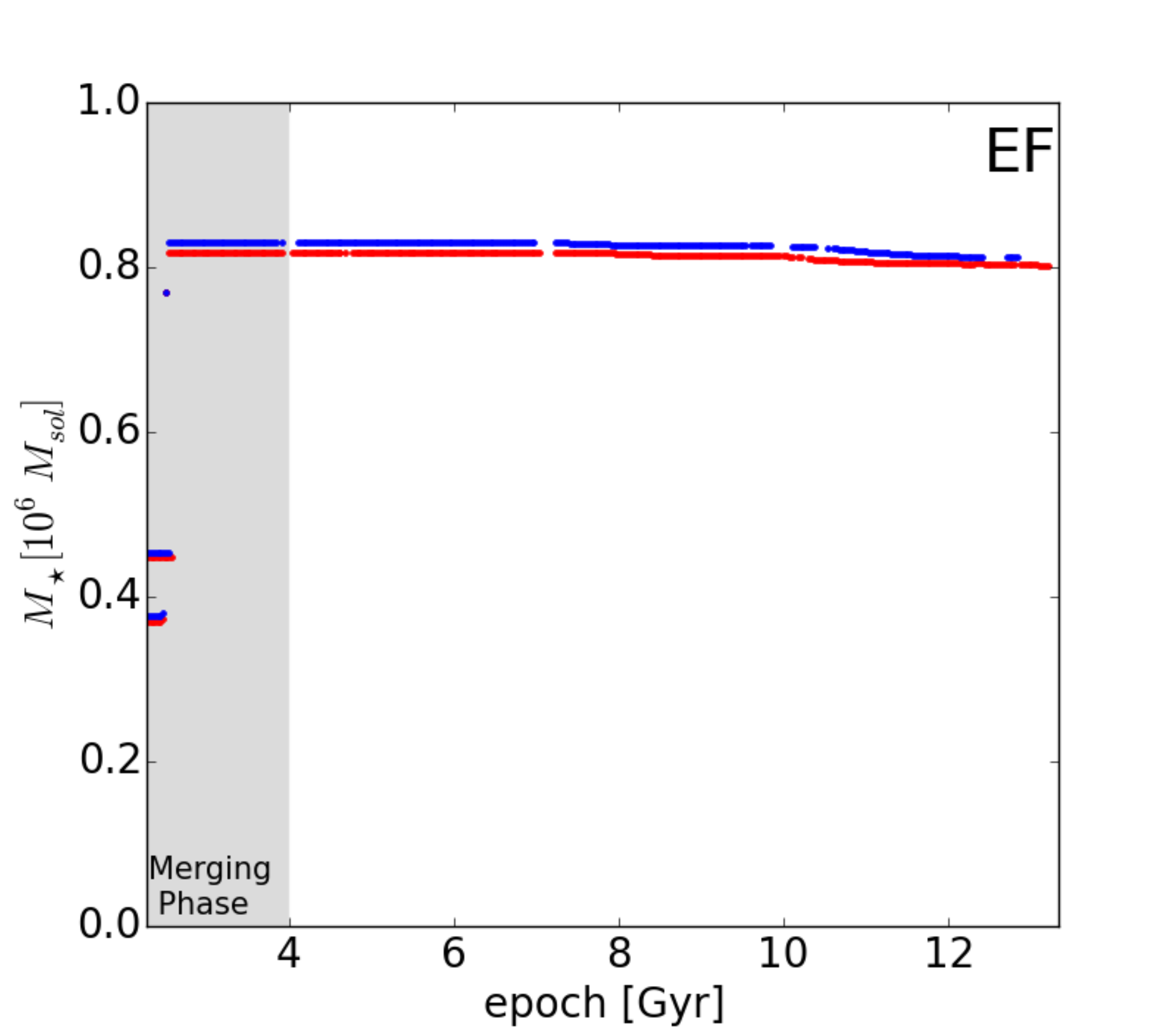}

\caption{\label{fig:figure16} The change in the mass of the bound stellar component over time is presented. 
Red/blue dots represent the satellites with steep/shallow central density profile.}

\end{figure}

\begin{figure}
\includegraphics[width = .4\textwidth]{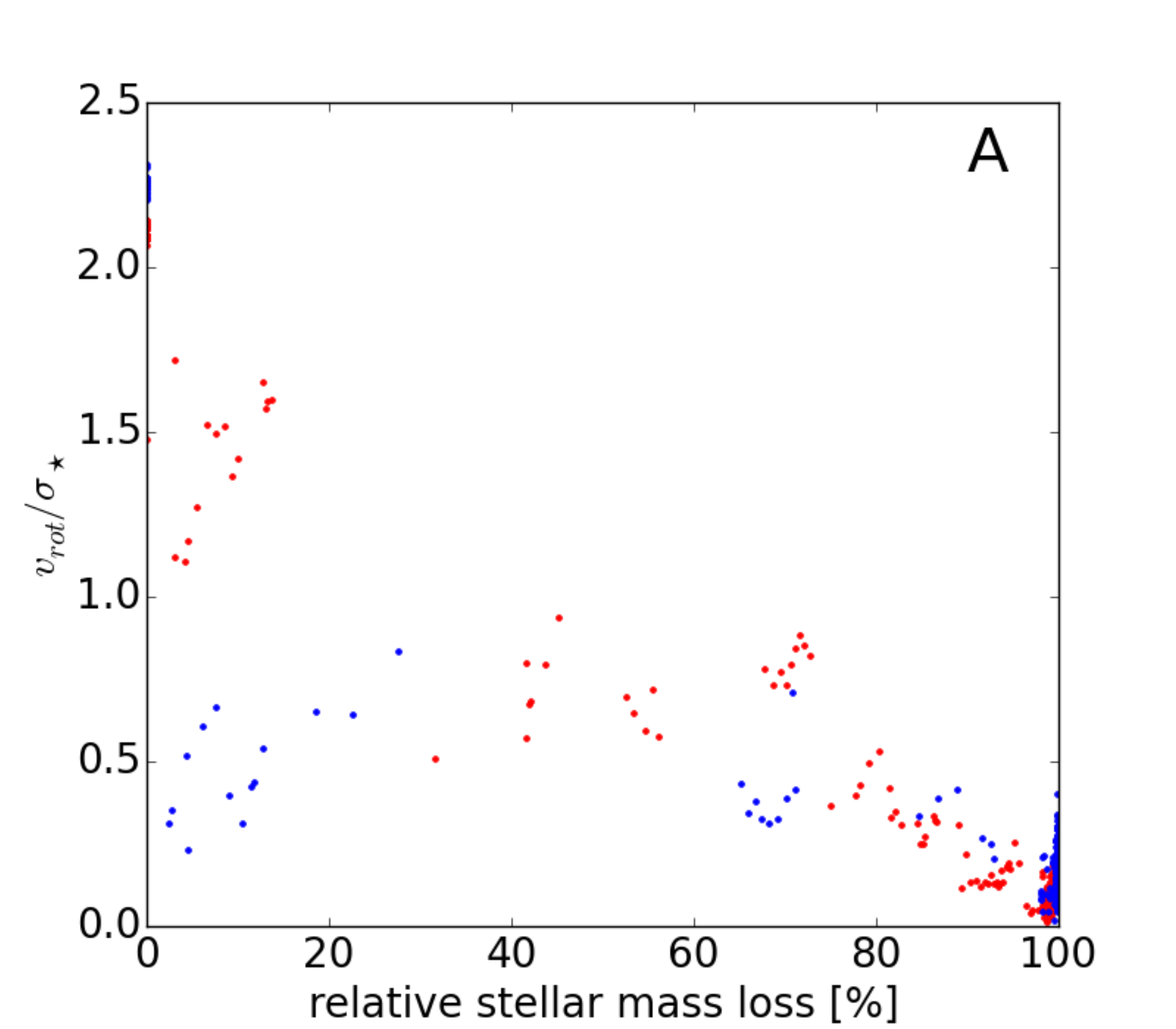}
\includegraphics[width = .4\textwidth]{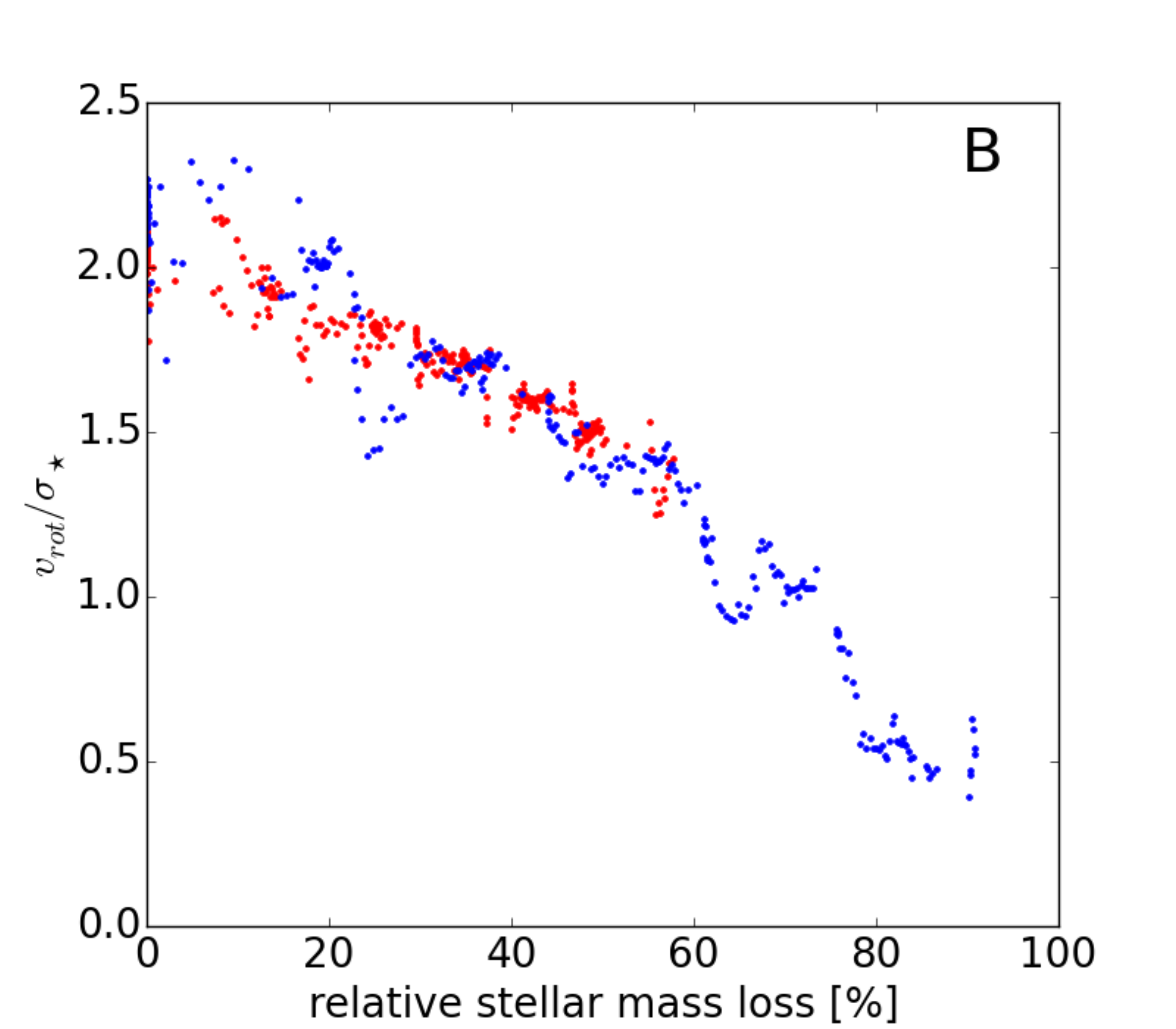}
\includegraphics[width = .4\textwidth]{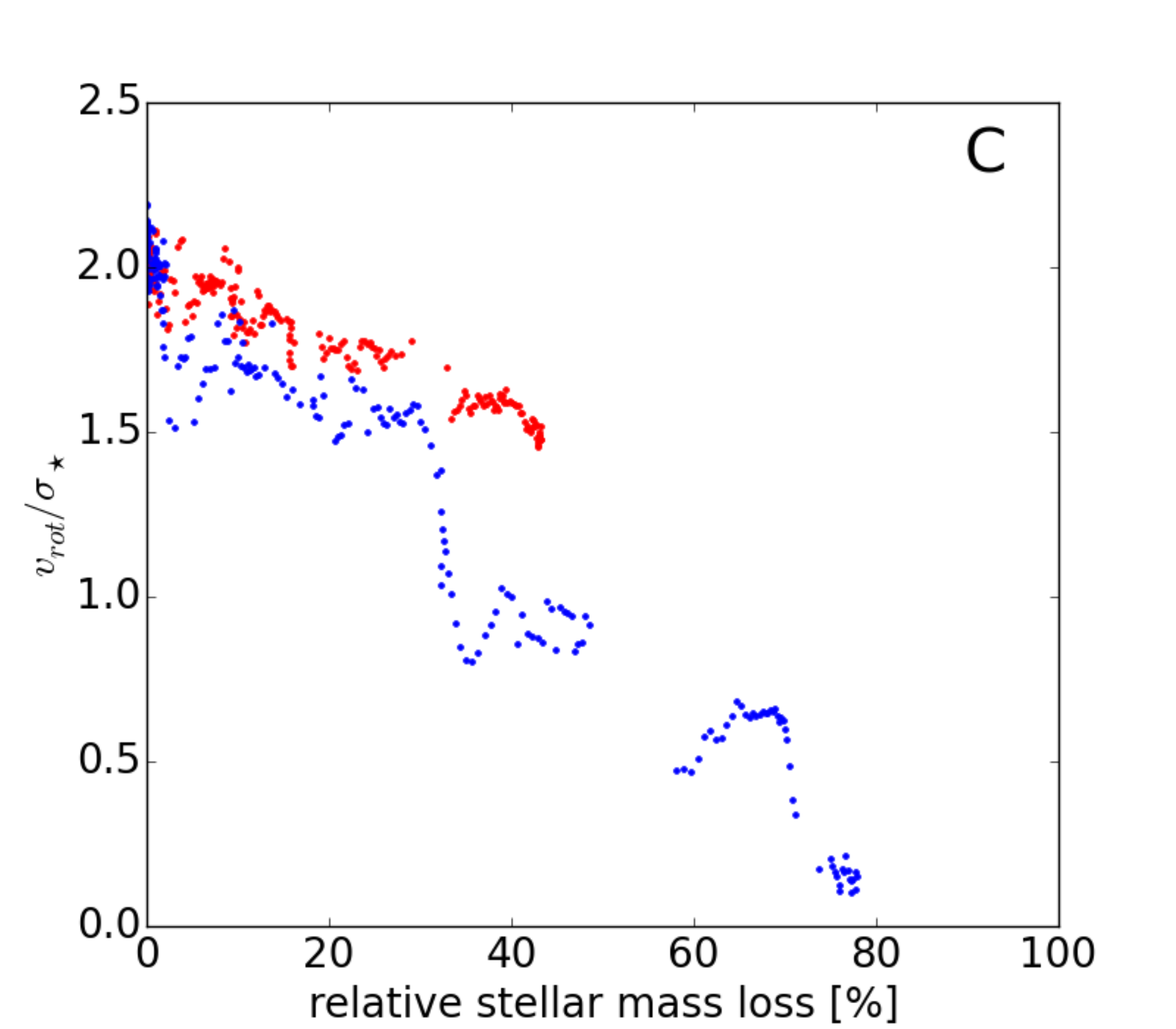}
\includegraphics[width = .4\textwidth]{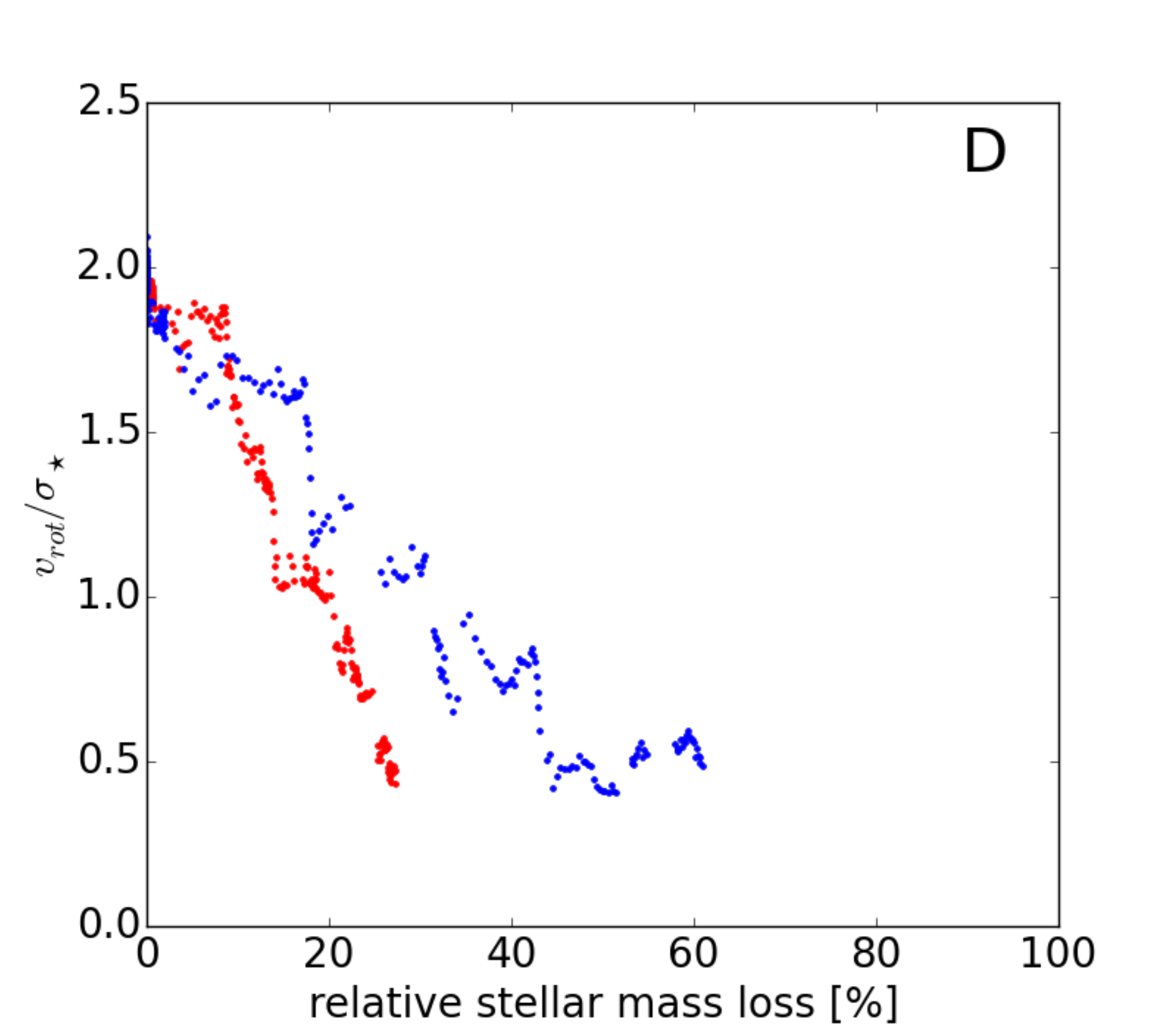}
\includegraphics[width = .4\textwidth]{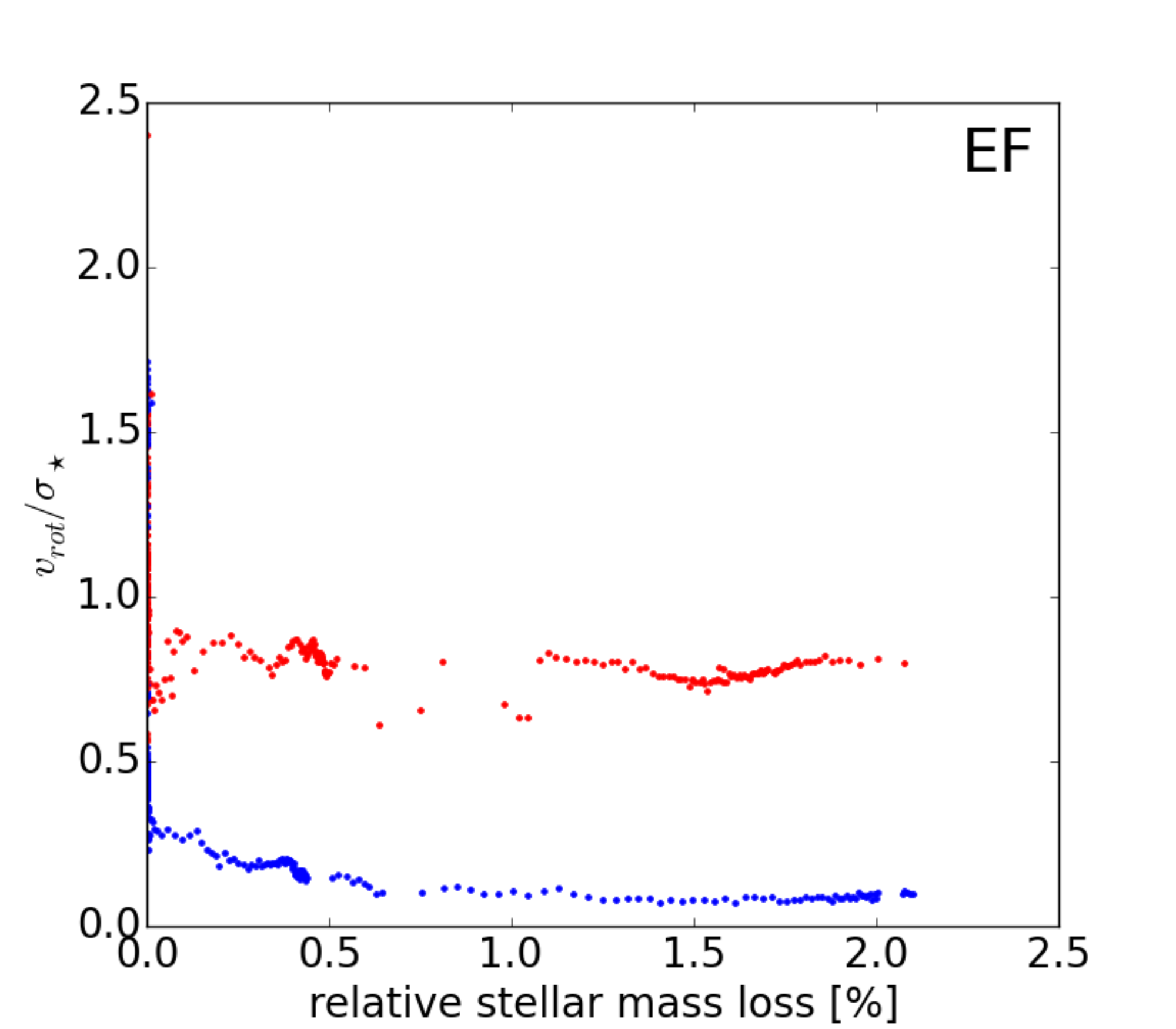}

\caption{\label{fig:figure17} The correlations between the amount of stellar mass lost and velocity ratios. 
Red/blue dots represent the satellites with steep/shallow central density profile. 
From top to bottom the subplots correspond to satellite (A and B), (C and D) and EF.}
\end{figure}

\begin{figure}
\includegraphics[width = .4\textwidth]{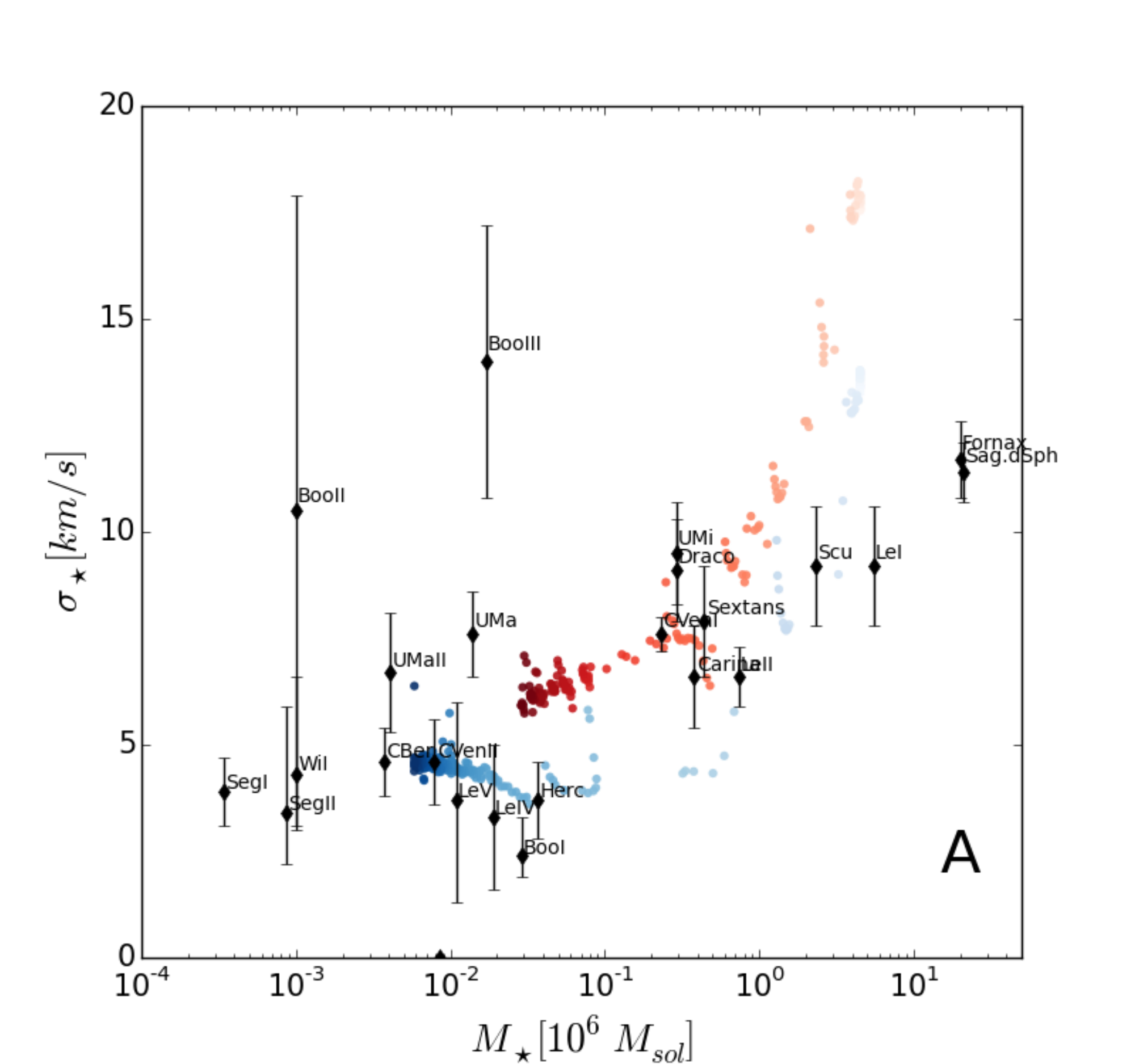}
\includegraphics[width = .4\textwidth]{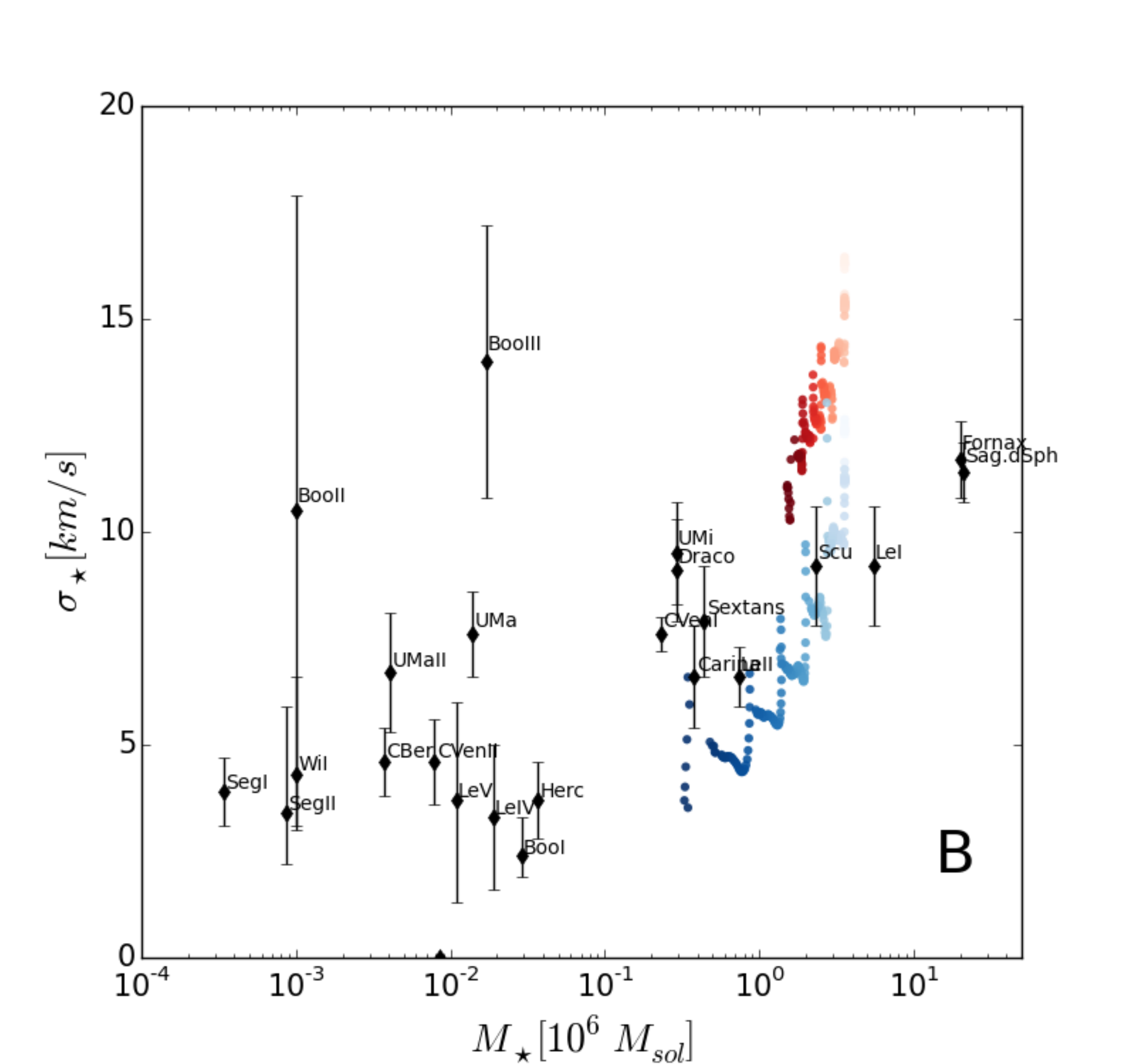}
\includegraphics[width = .4\textwidth]{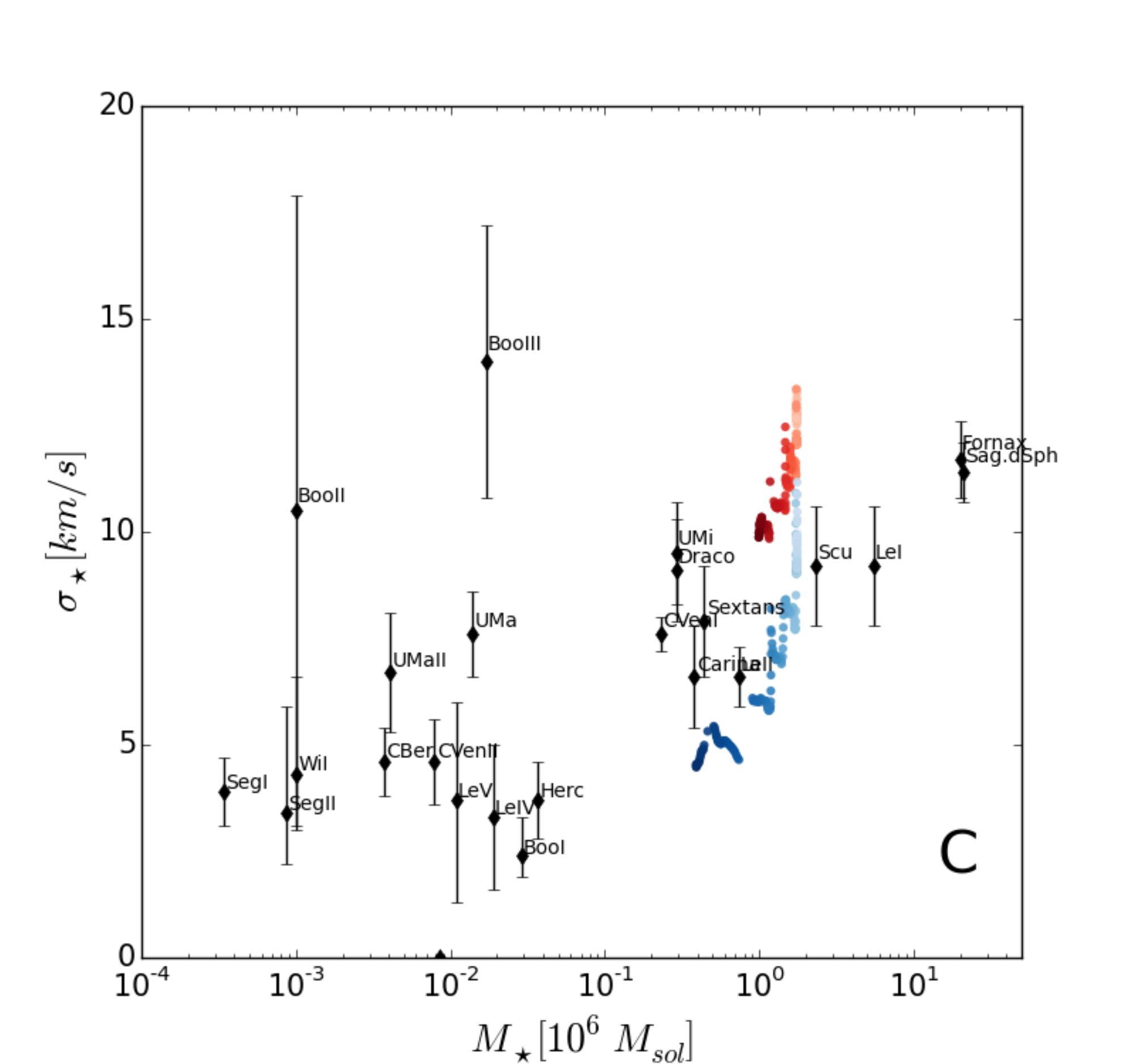}
\includegraphics[width = .4\textwidth]{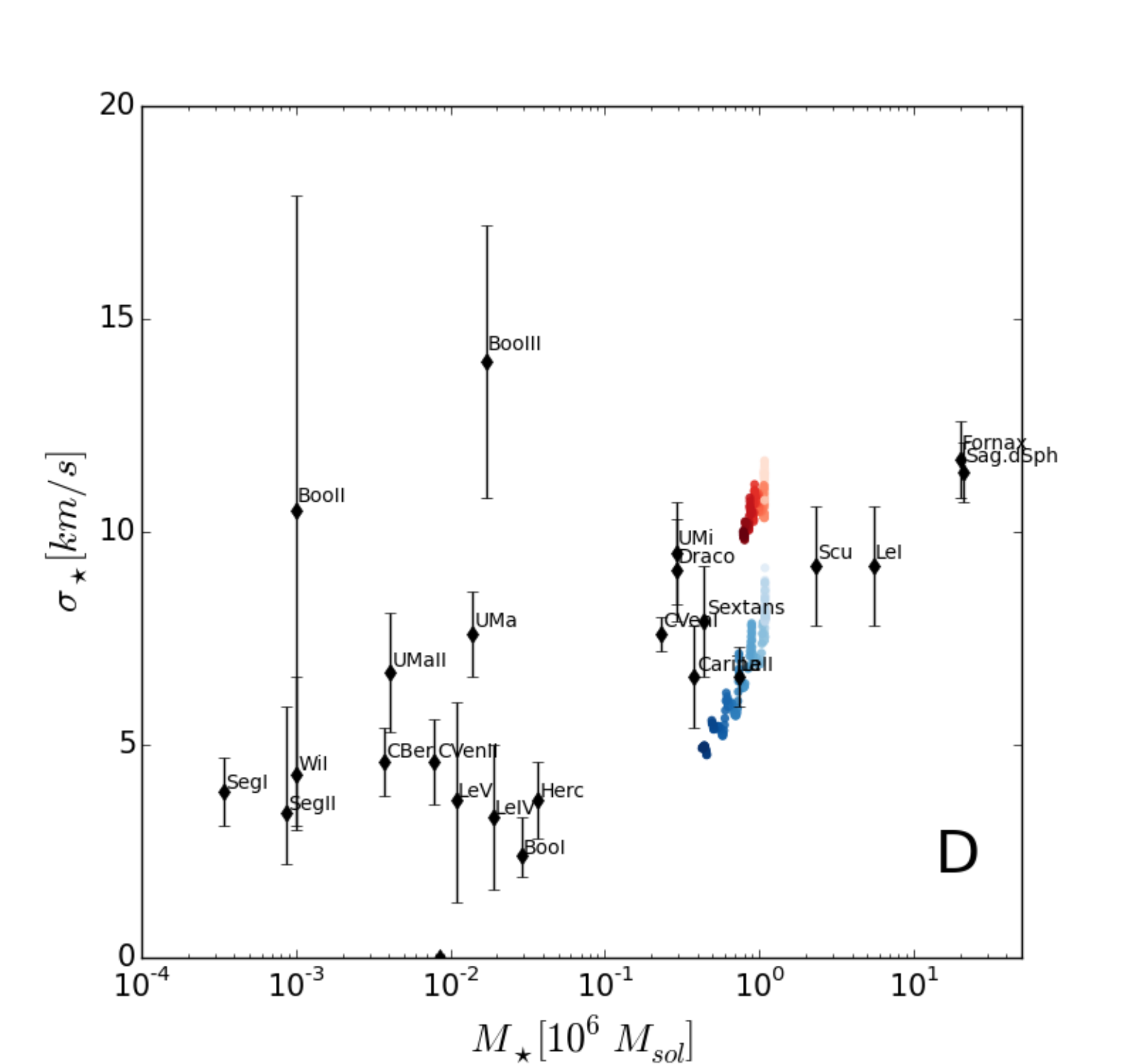}
\includegraphics[width = .4\textwidth]{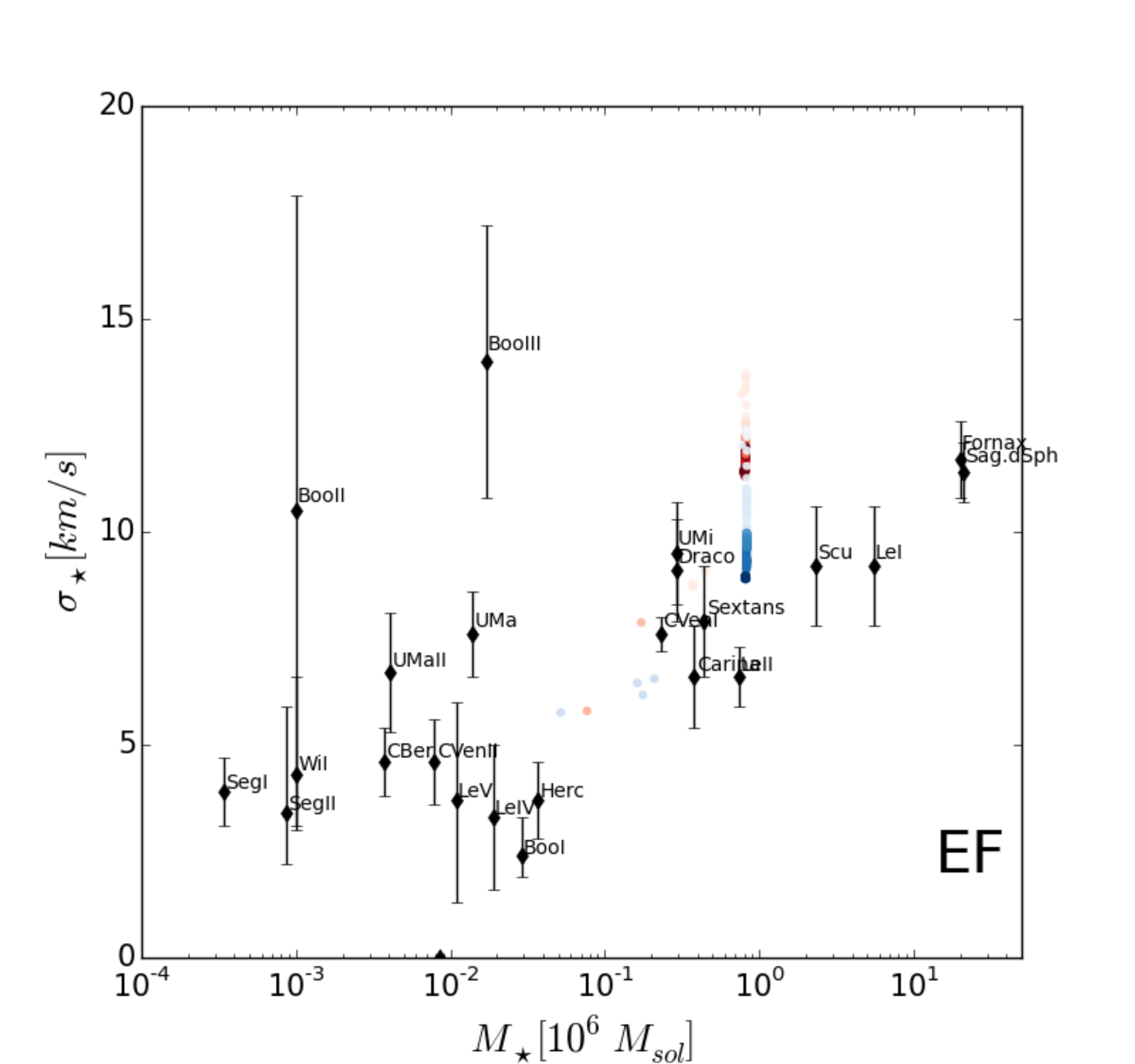}
\includegraphics[width = .4\textwidth]{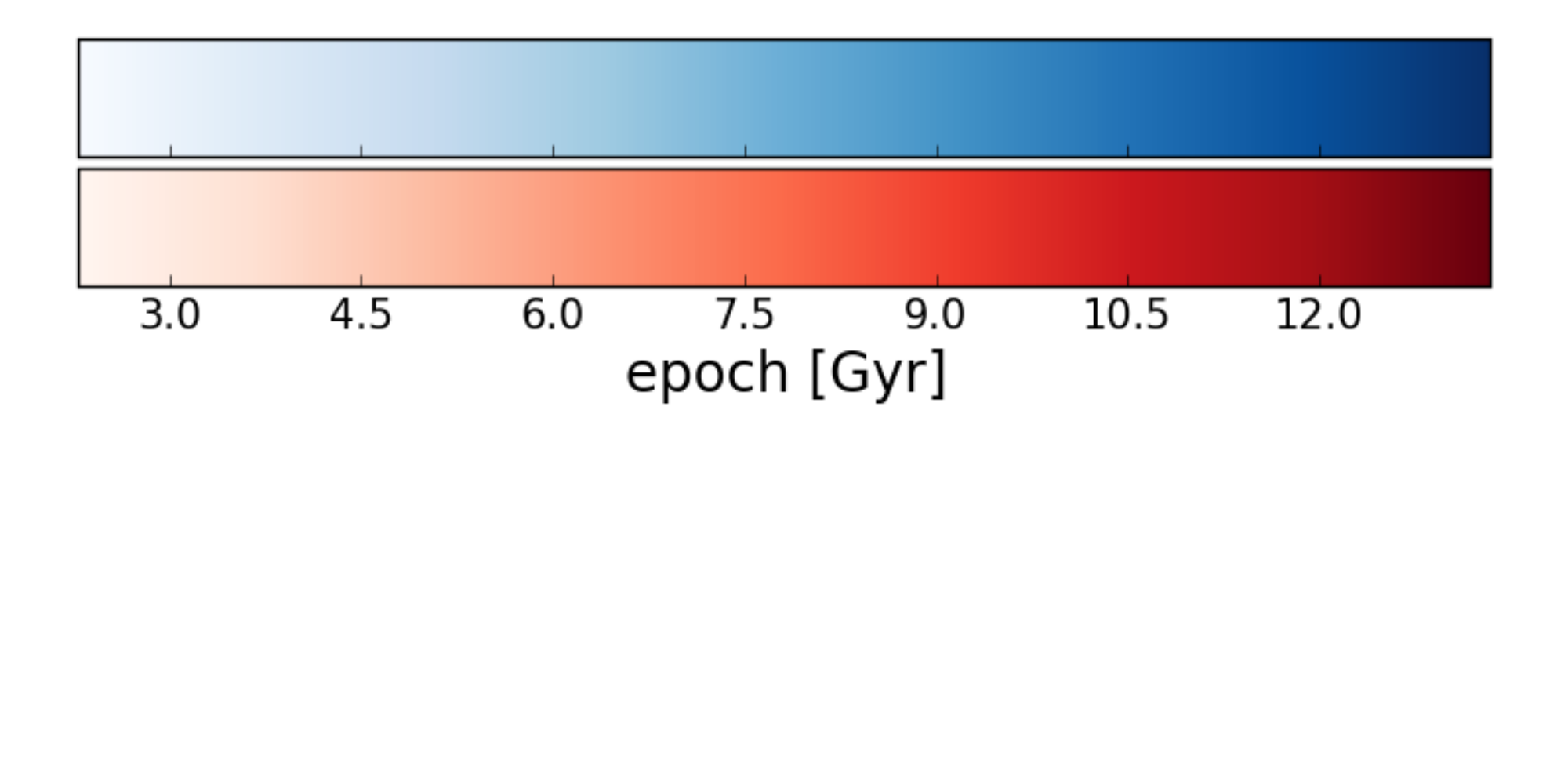}

\caption{\label{fig:figure18} The evolution of the mass of the bound stars and velocity dispersion of 
simulated satellites in comparison to the values corresponding to the local dwarf galaxies as compiled in the paper \cite{2012AJ....144....4M}.  
Red/blue dots represent the satellites with steep/shallow central density profile. 
From top to bottom the subplots correspond to satellite (A and B), (C and D) and EF. 
The black rhombuses mark the local dwarf spheroidals found within 300 kpc from MW's center.}
\end{figure}

\begin{figure}
\includegraphics[width = .39\textwidth]{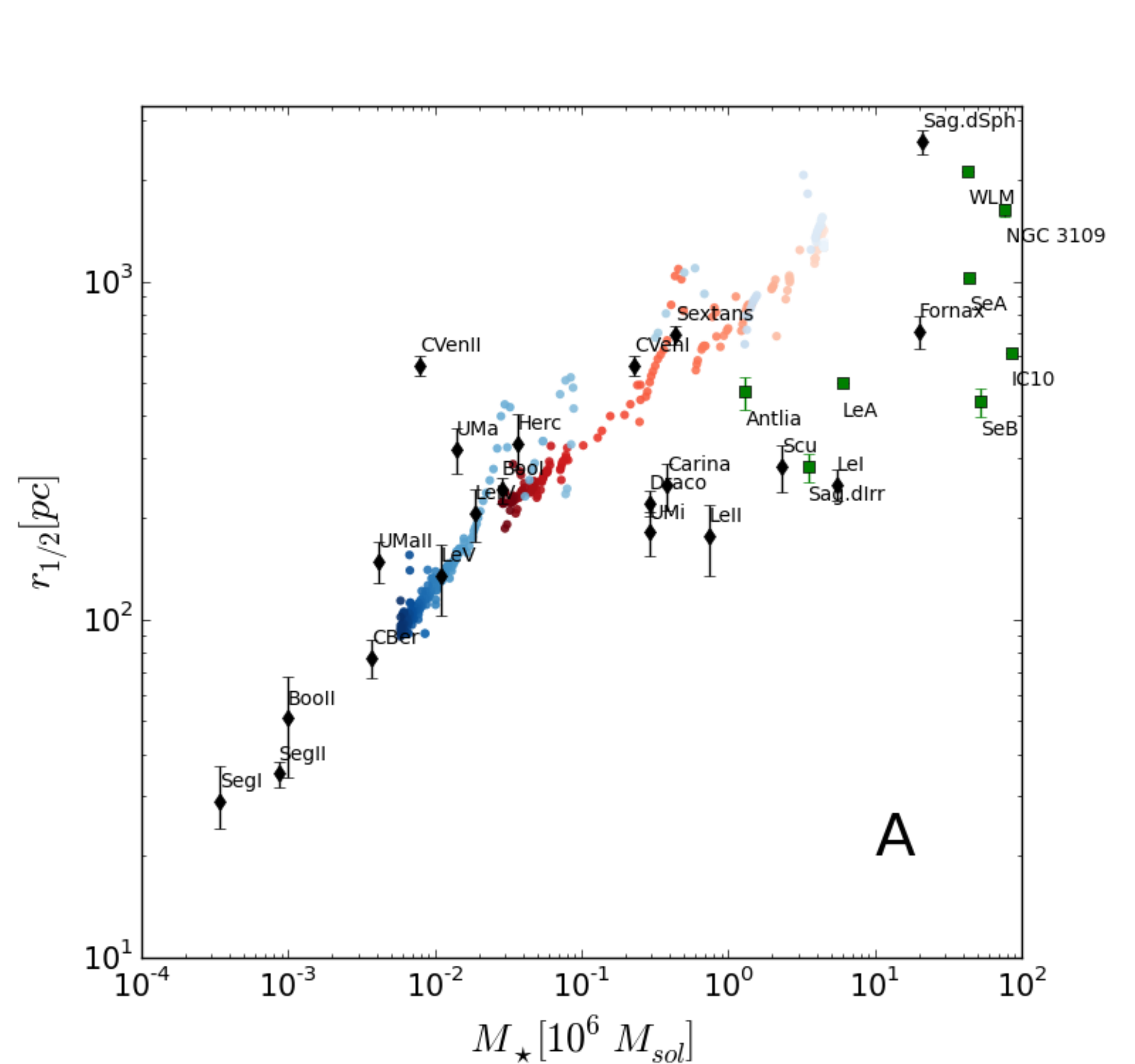}
\includegraphics[width = .39\textwidth]{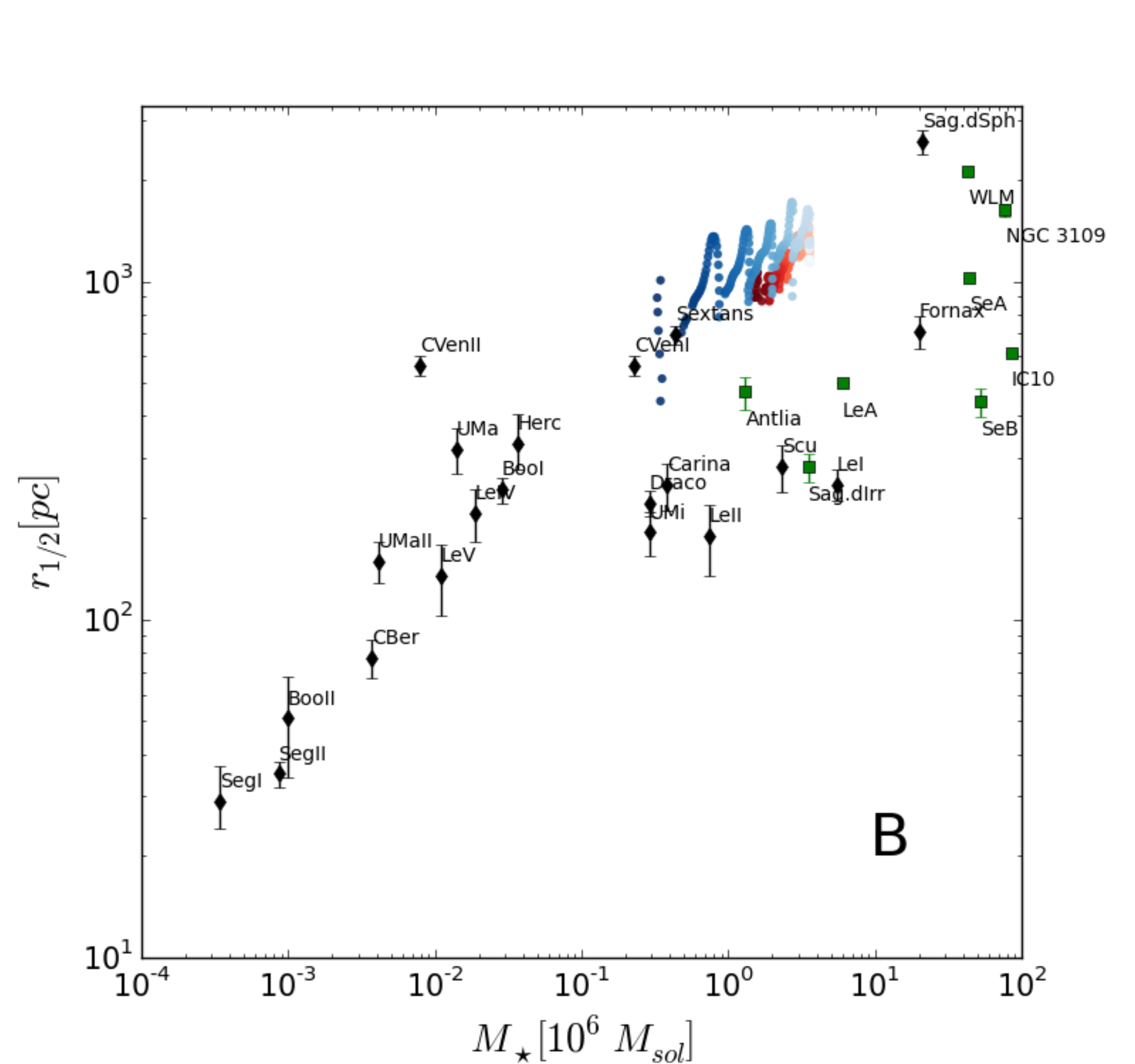}
\includegraphics[width = .39\textwidth]{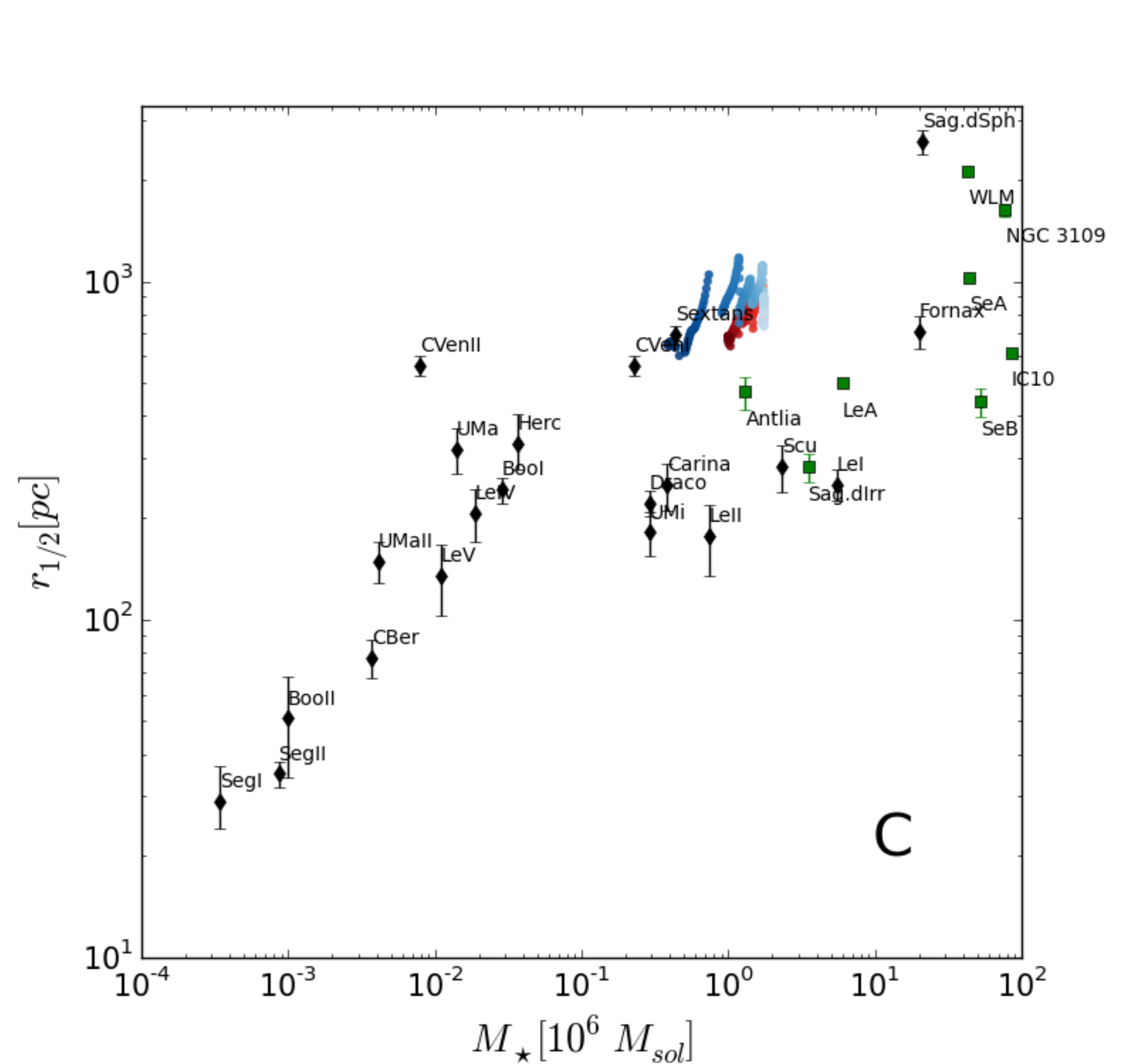}
\includegraphics[width = .39\textwidth]{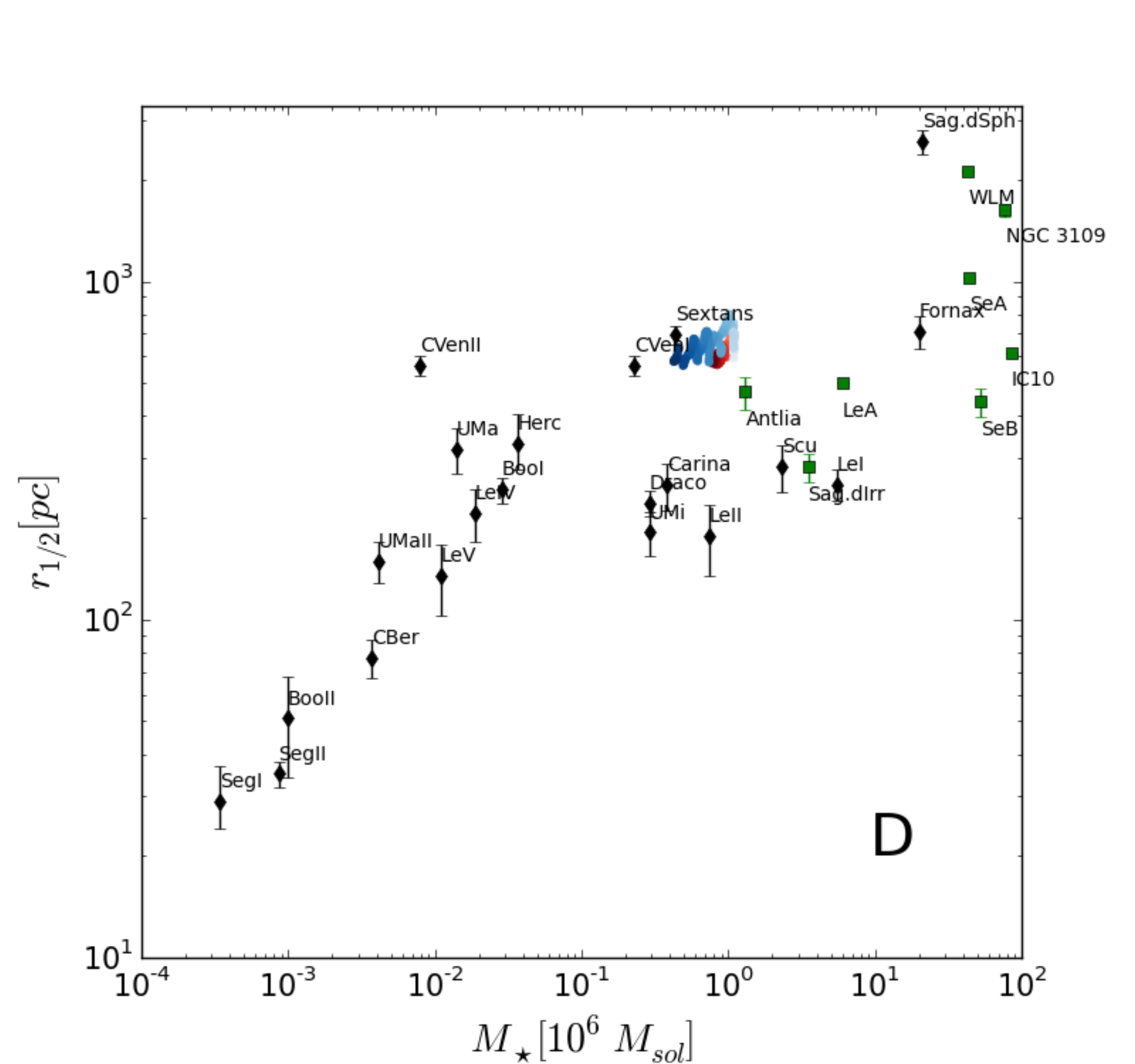}
\includegraphics[width = .39\textwidth]{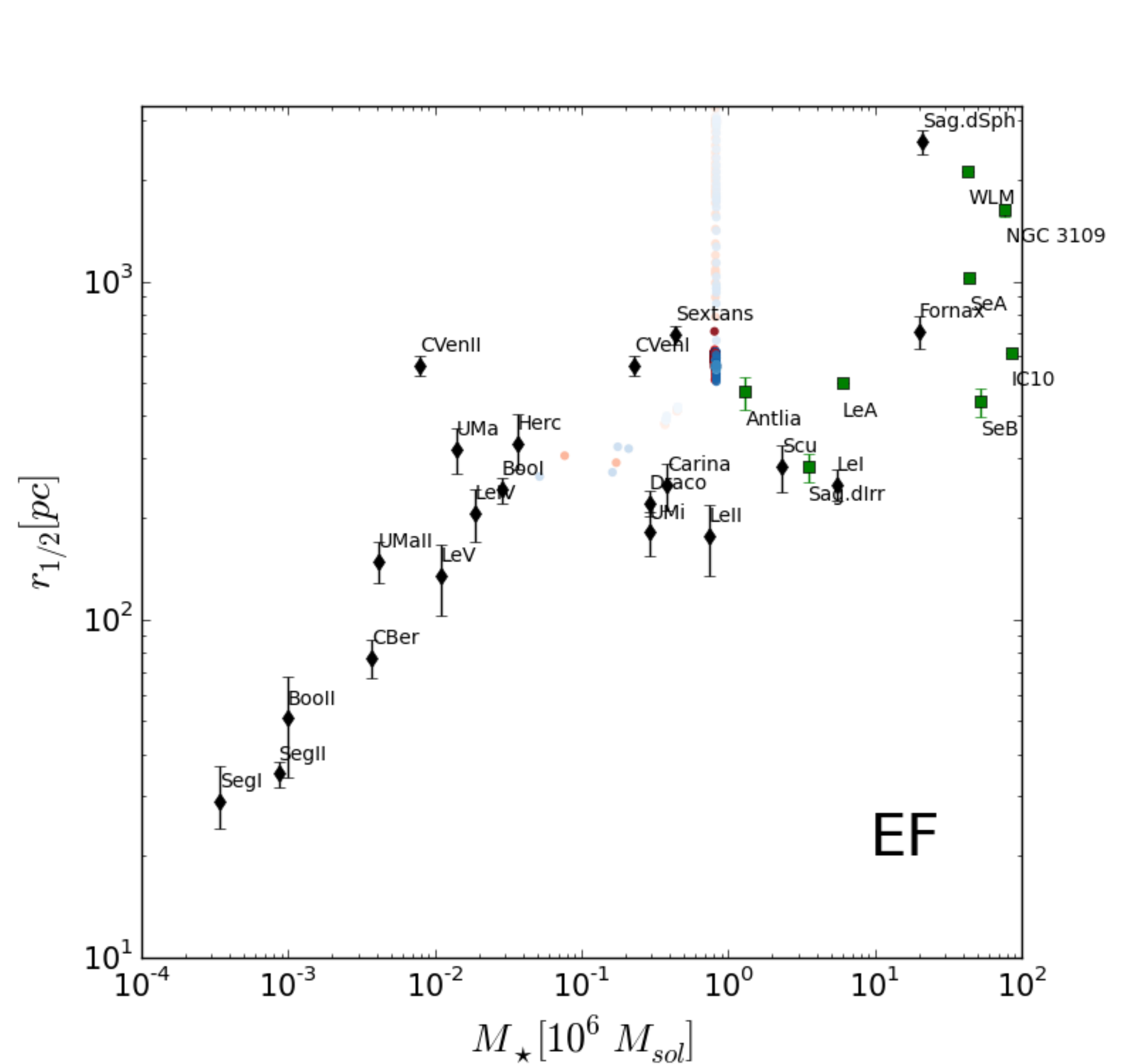}
\includegraphics[width = .39\textwidth]{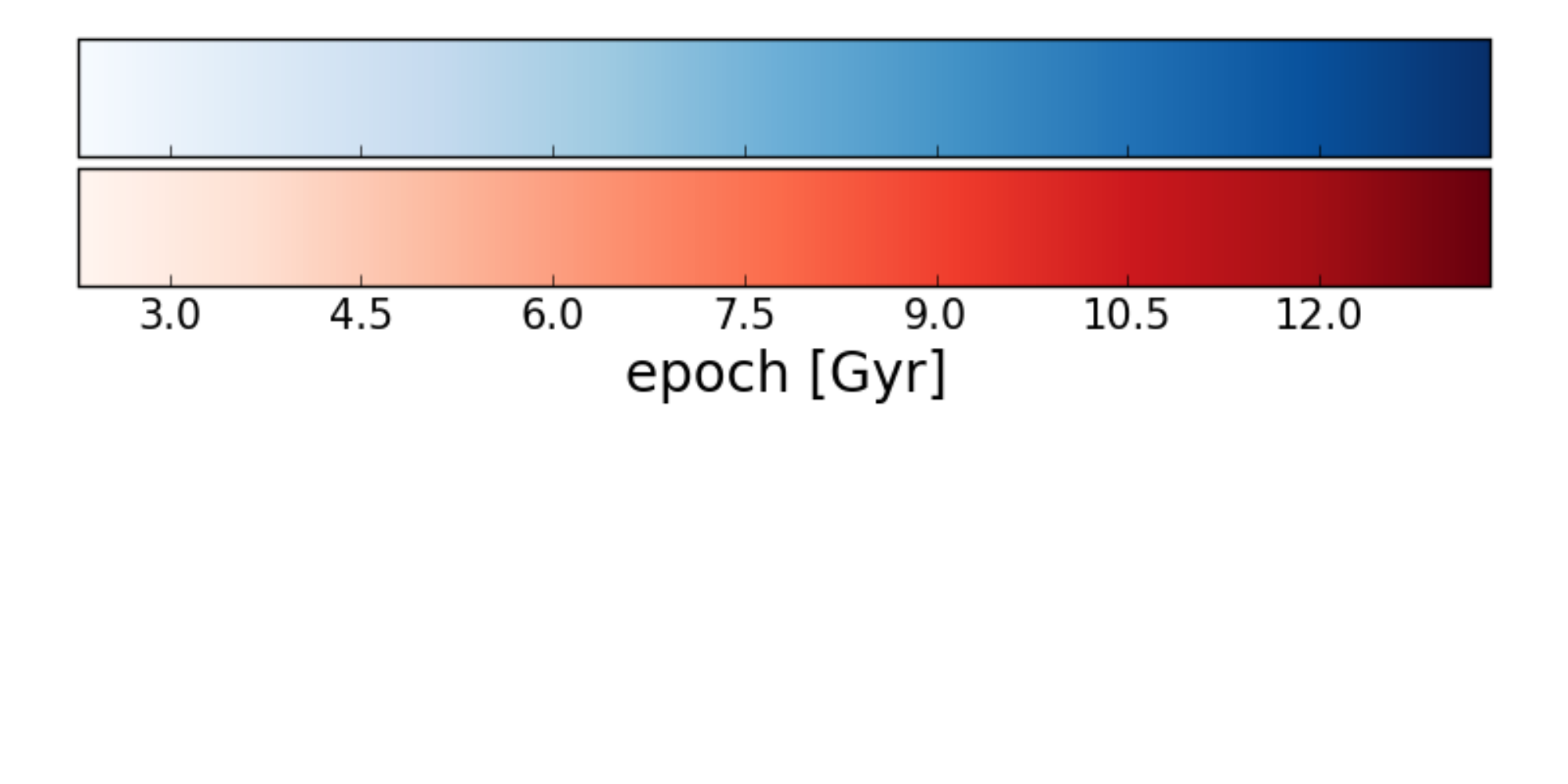}

\caption{\label{fig:figure19} The evolution of the mass of the bound stars and the half light/mass radius of the simulated satellites
in comparison to the values corresponding to the local dwarf galaxies as compiled in the paper \cite{2012AJ....144....4M}.  Red/blue dots represent the 
satellites with steep/shallow central density profile.  
The black rhombuses mark the local dwarf spheroidals found whithin 300 kpc from MW's center. The green squares mark the field dwarf irregulars.}
\end{figure}

\begin{figure}
\includegraphics[width = .39\textwidth]{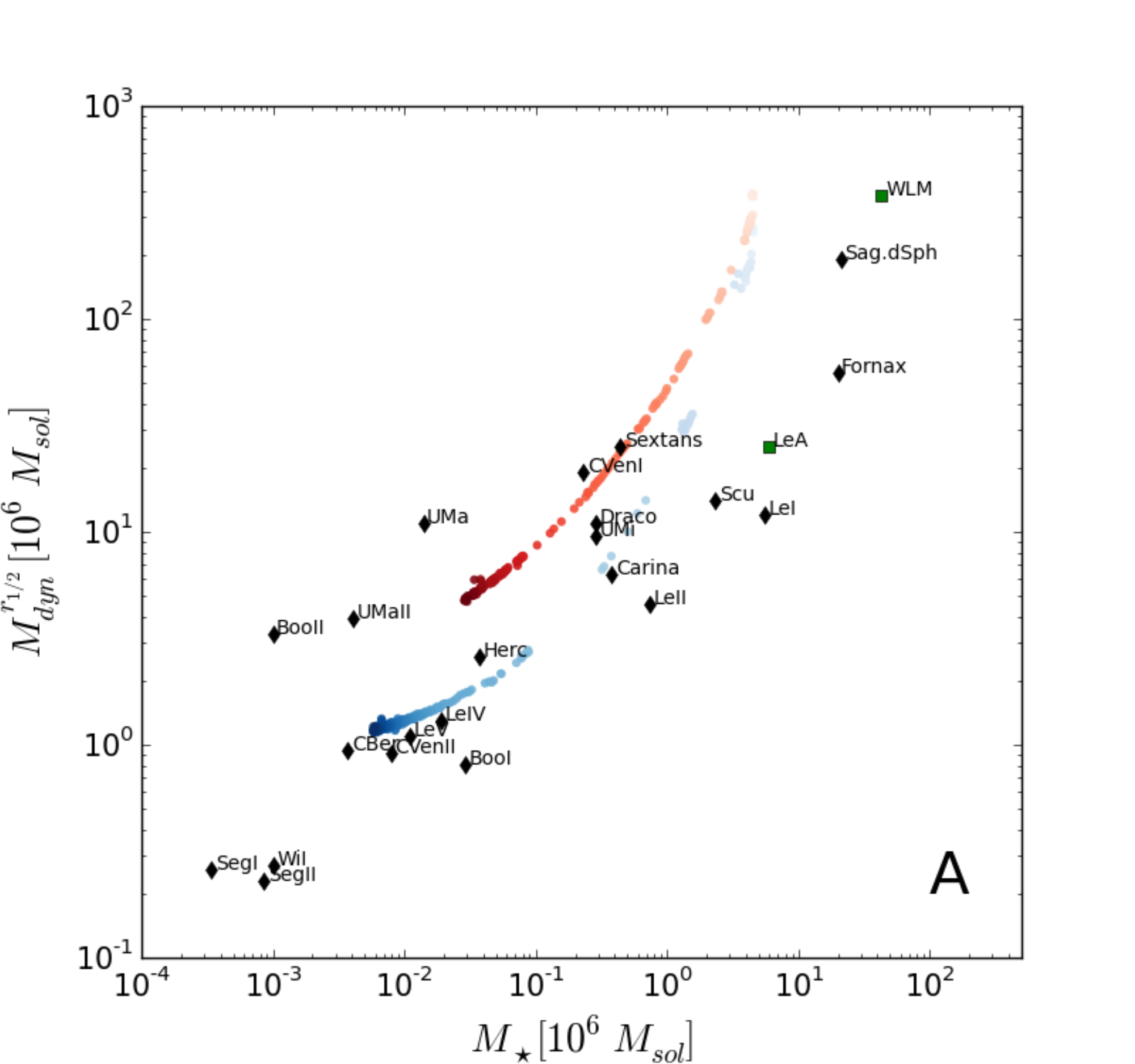}
\includegraphics[width = .39\textwidth]{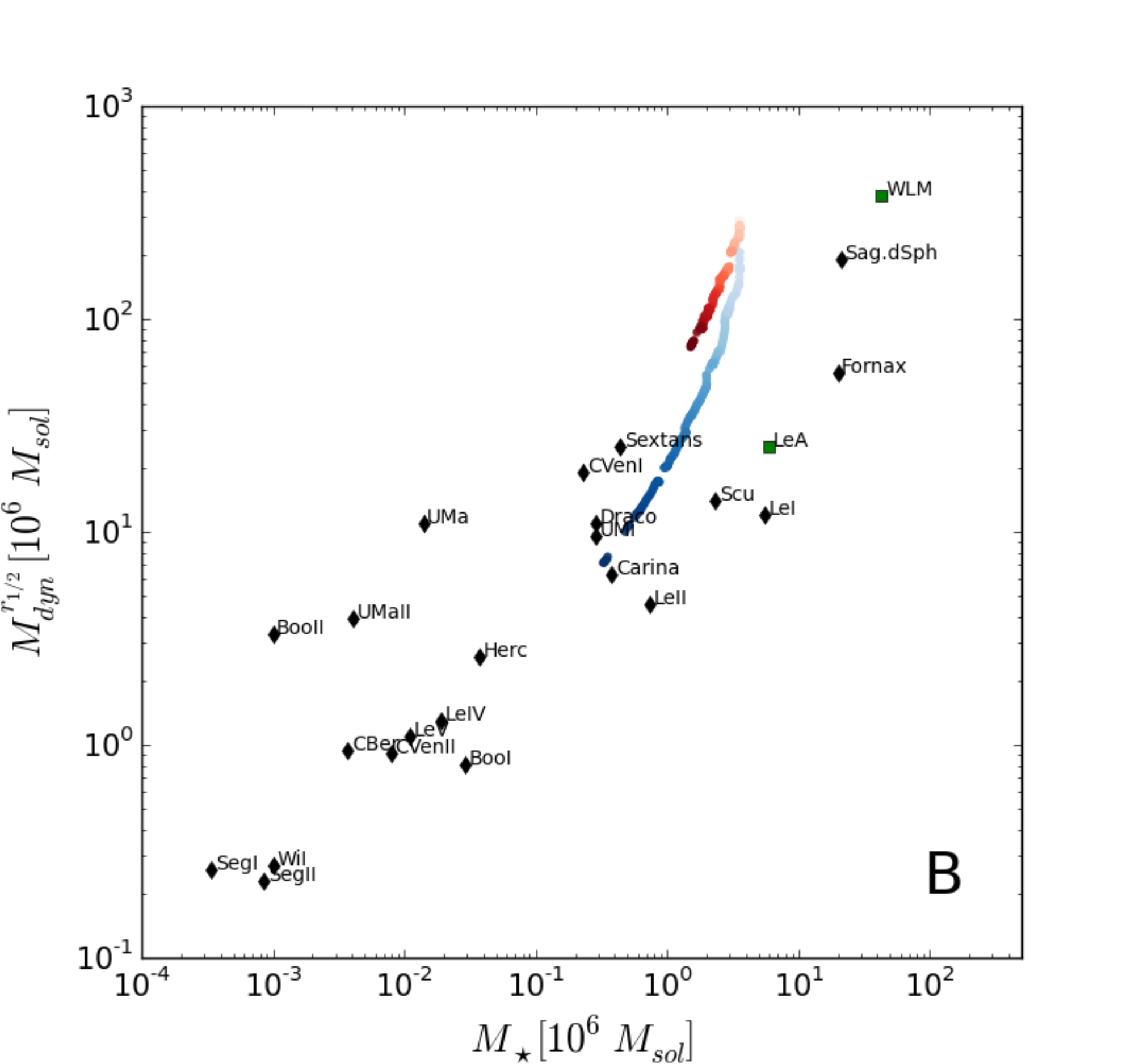}
\includegraphics[width = .39\textwidth]{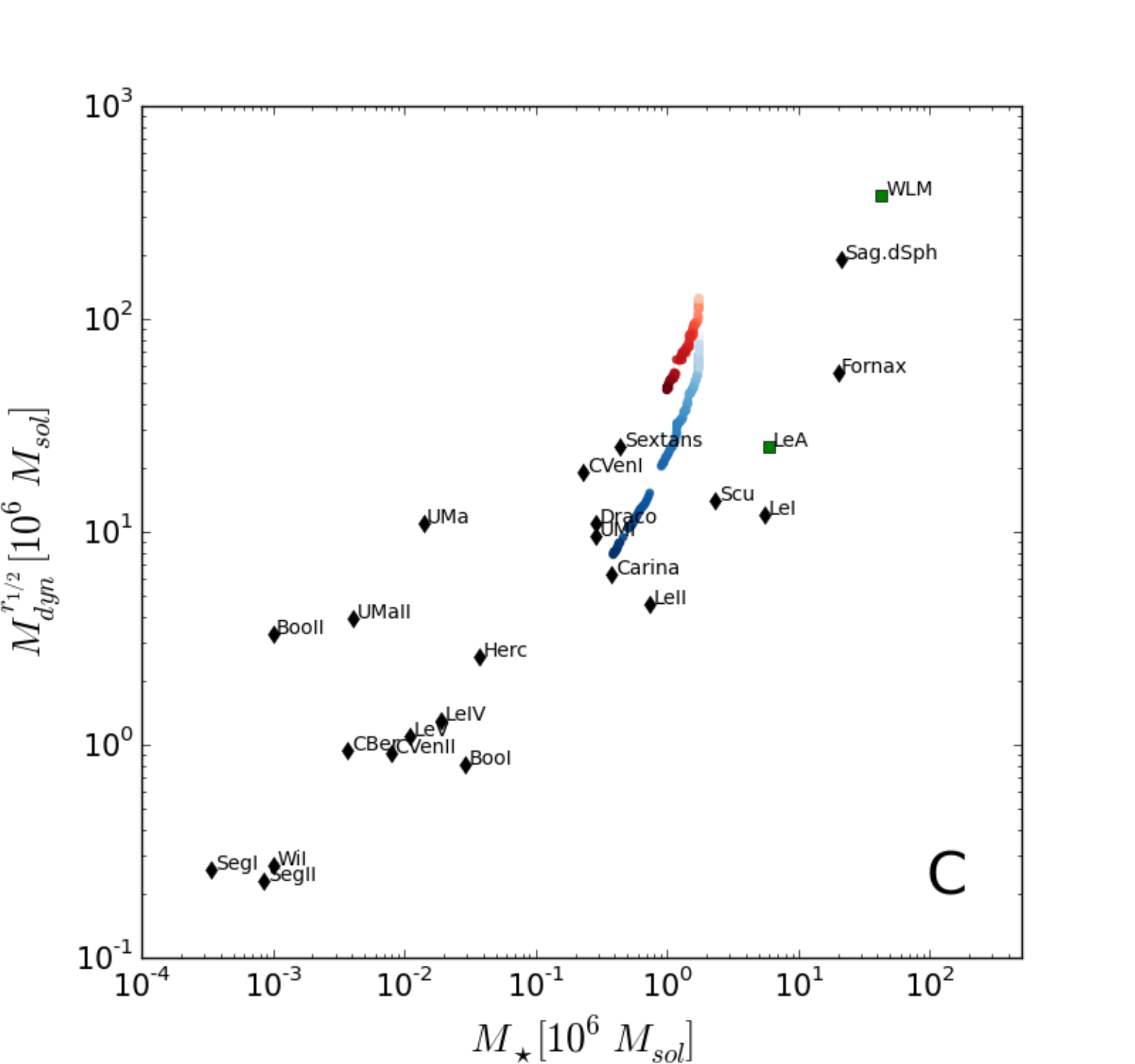}
\includegraphics[width = .39\textwidth]{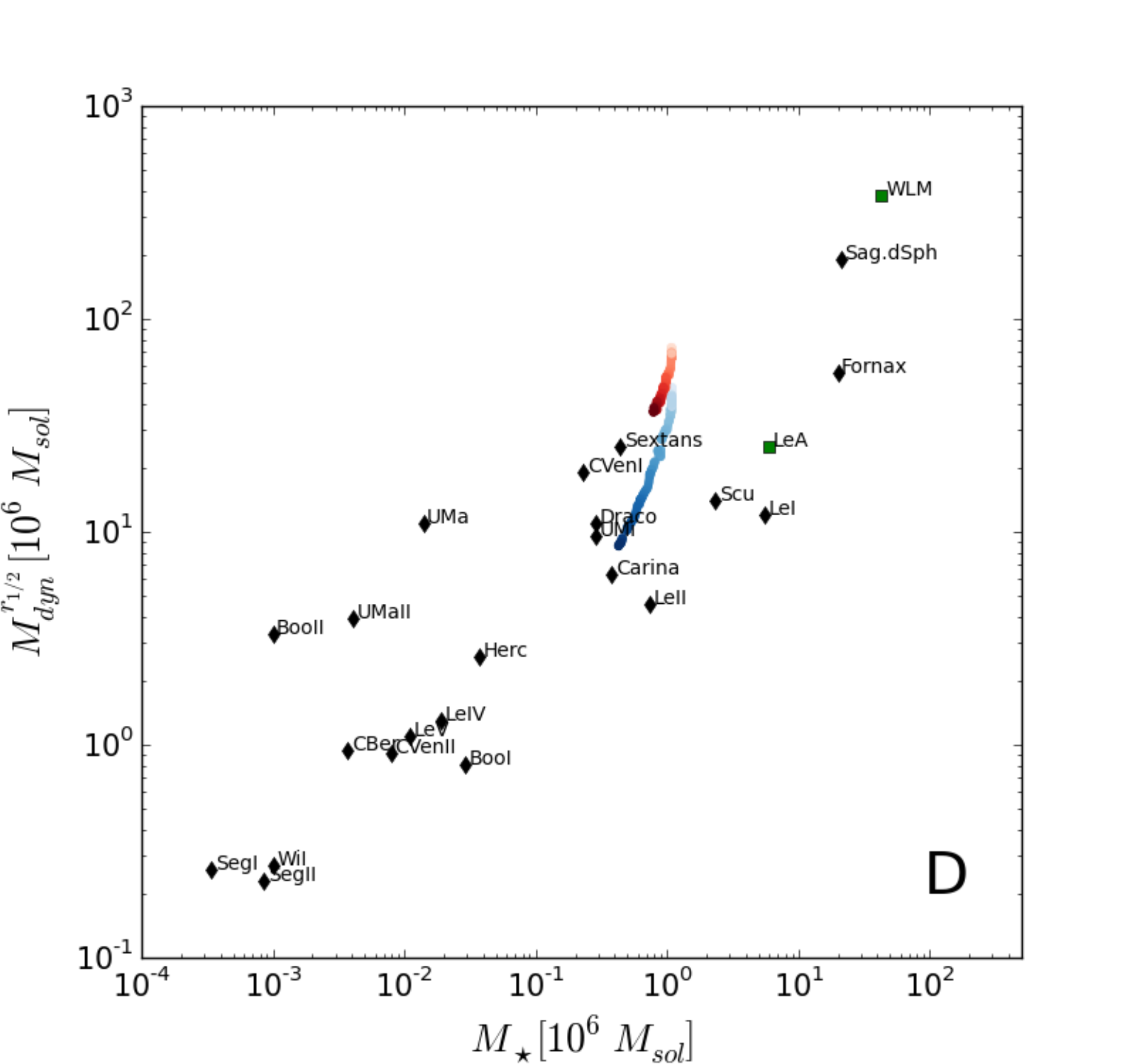}
\includegraphics[width = .39\textwidth]{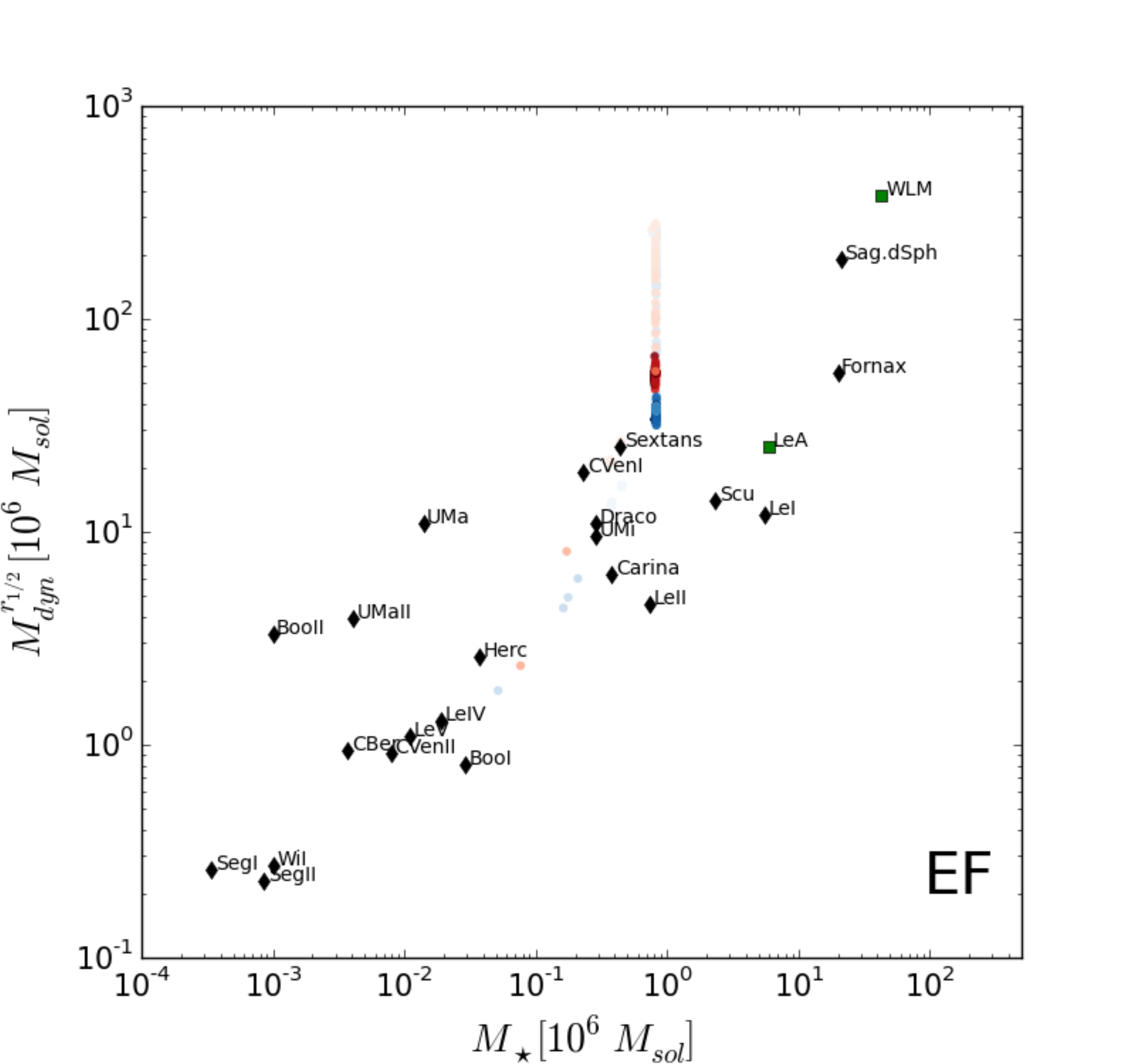}
\includegraphics[width = .39\textwidth]{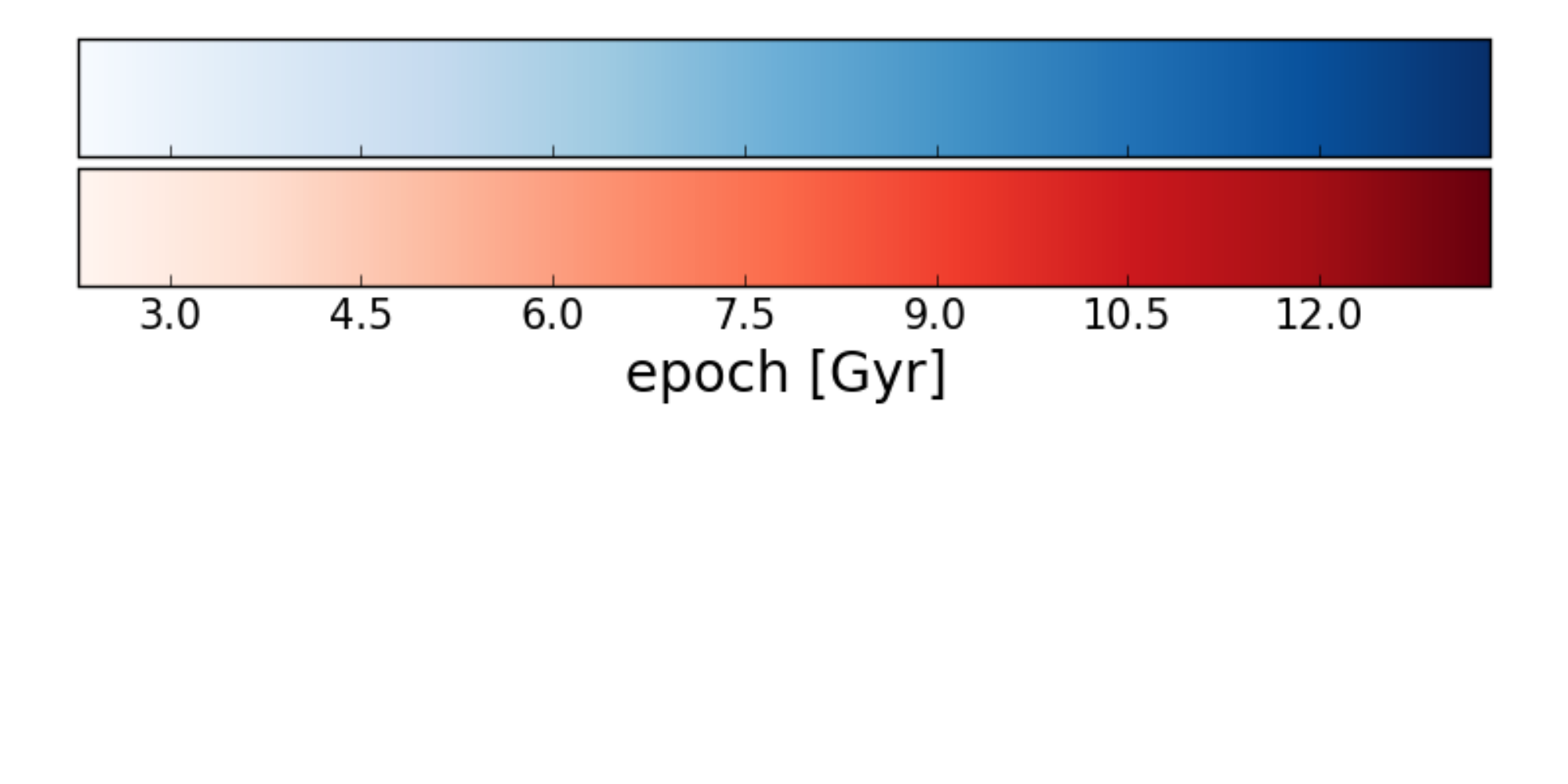}

\caption{\label{fig:figure20} The evolution of the mass of the bound stars and dynamical mass within the half light 
radius of simulated satellites in comparison to the values corresponding to the local dwarf galaxies as compiled in the paper \cite{2012AJ....144....4M}.  
Red/blue dots represent the satellites with steep/shallow central density profile. The black rhombuses mark the local dwarf spheroidals found whithin 300 kpc from MW's center. 
The green squares mark the field dwarf irregulars.}
\end{figure}

\begin{figure}
\includegraphics[width = .4\textwidth]{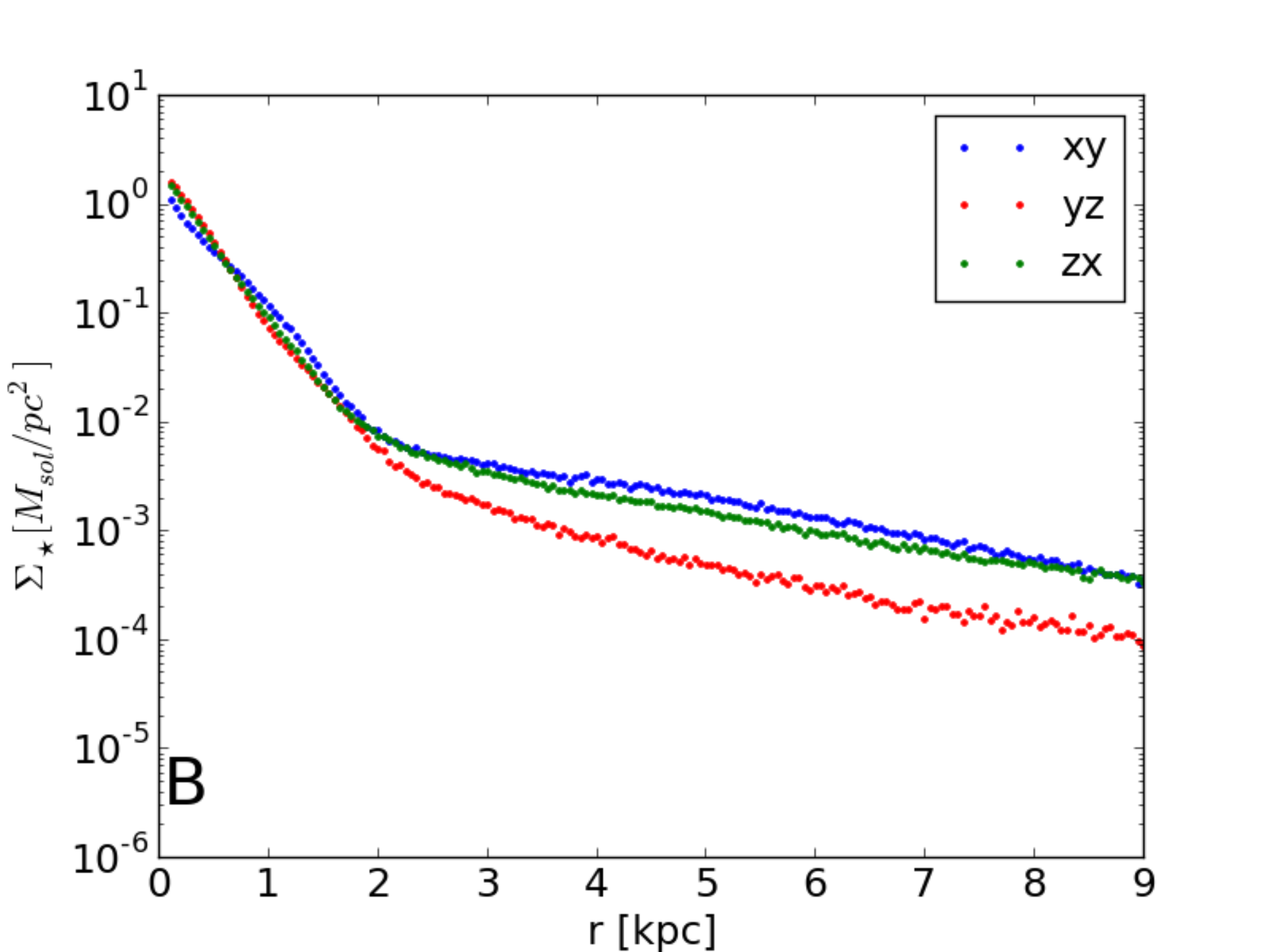}
\includegraphics[width = .4\textwidth]{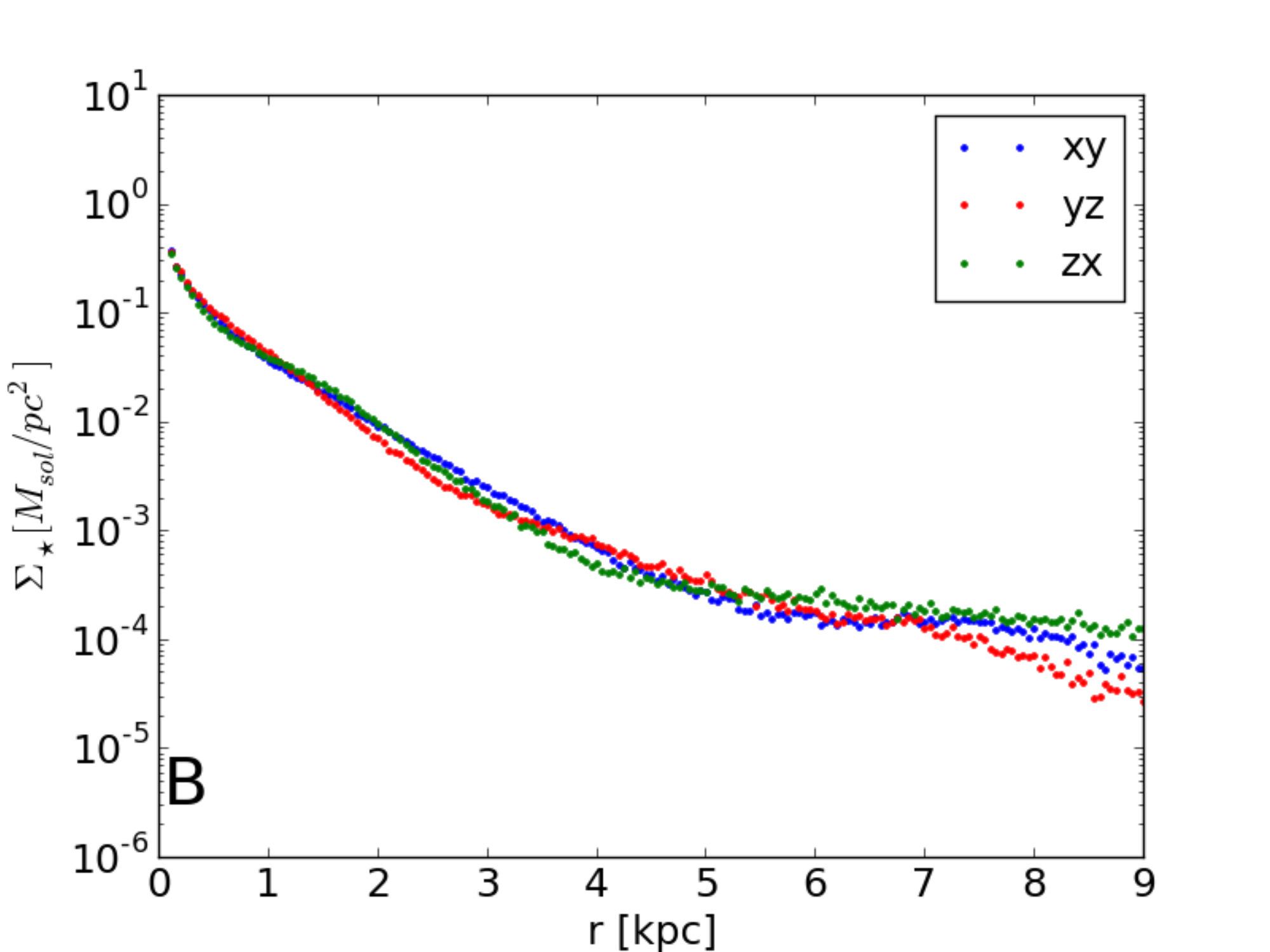}
\includegraphics[width = .4\textwidth]{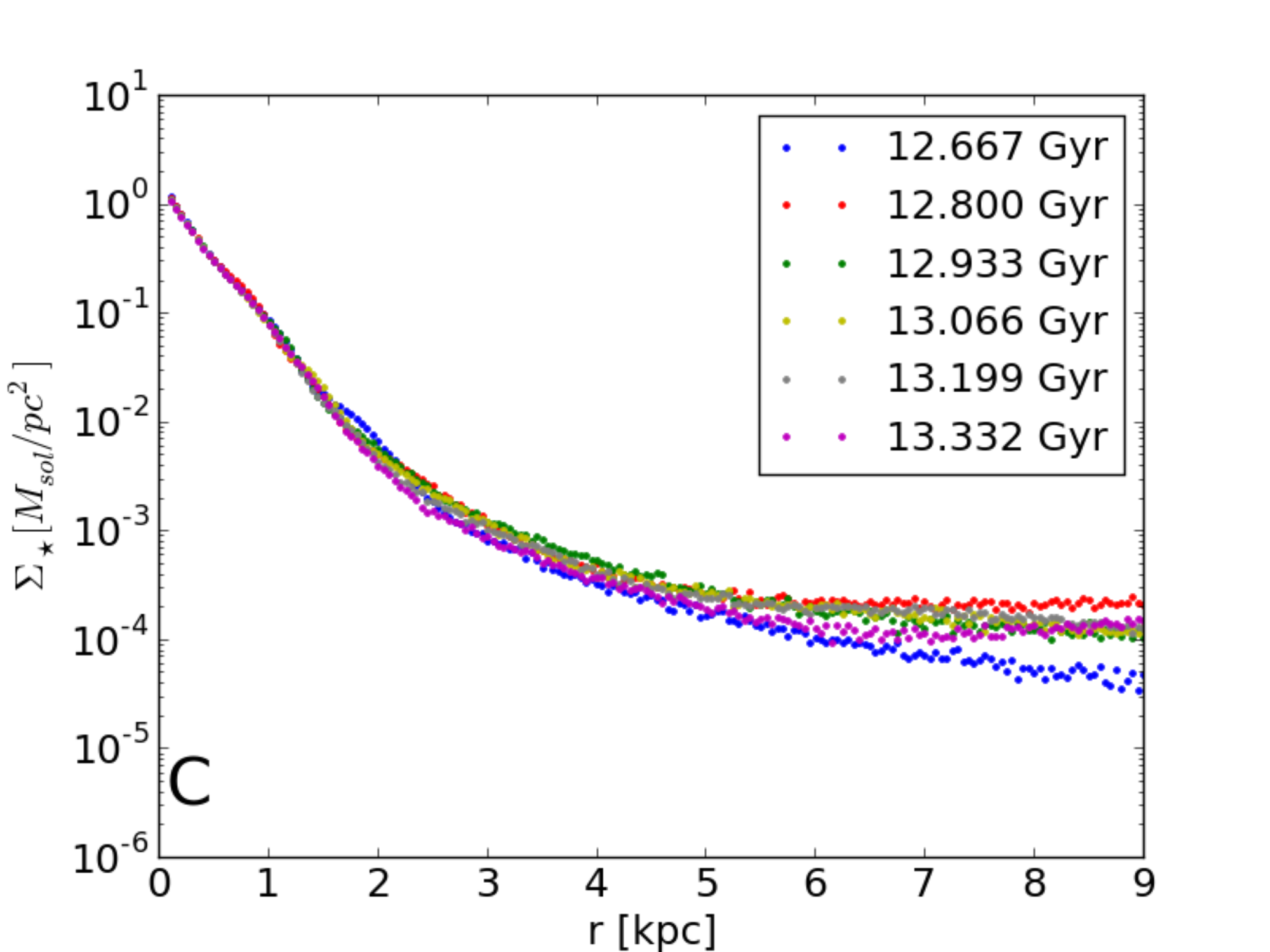}
\includegraphics[width = .4\textwidth]{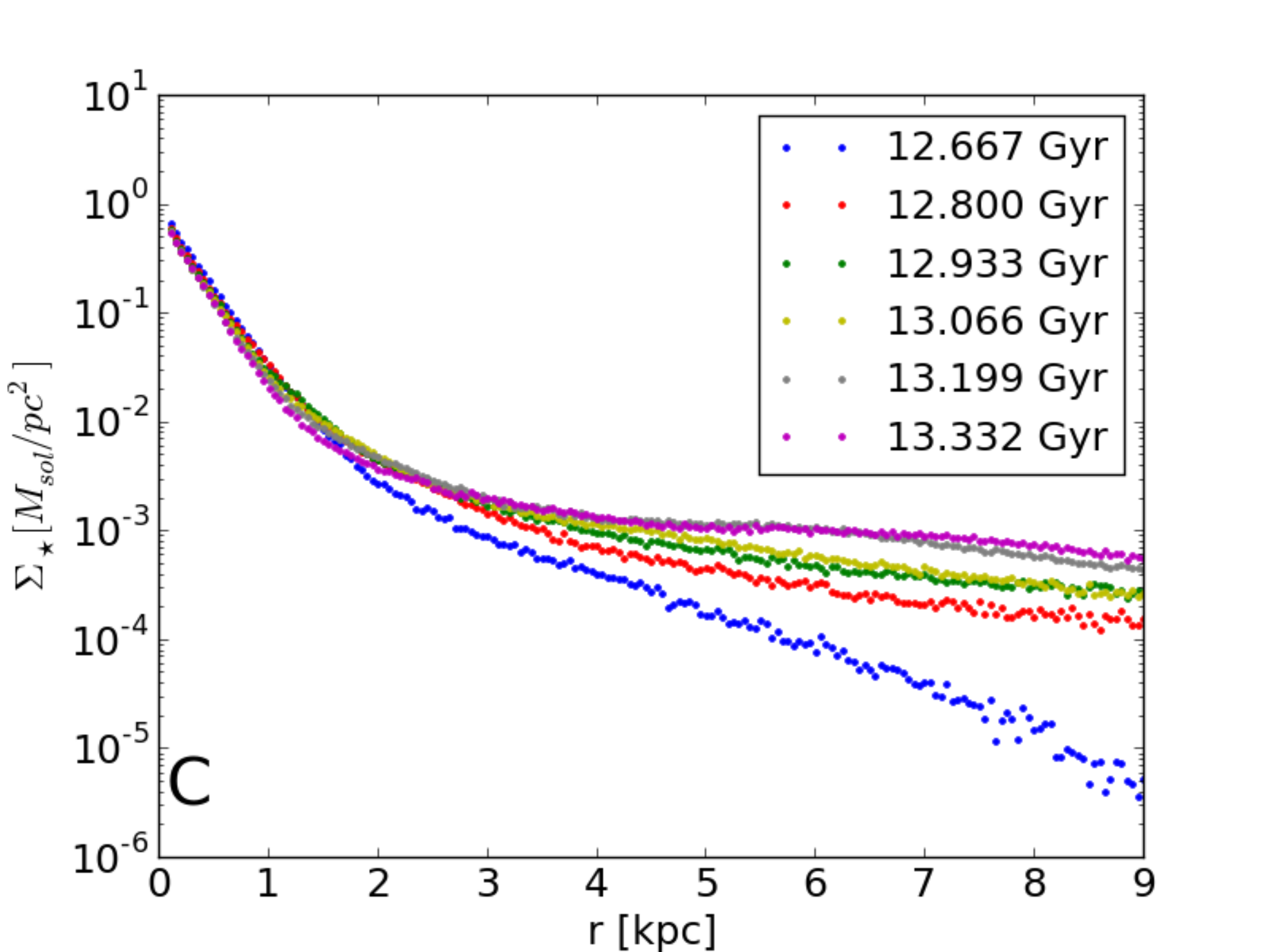}

\caption{\label{fig:figure21} The cylindrically averaged surface mass density profiles are presented.
 The top row illustrates the surface density of satellite B computed along all three principal axes.
 The bottom row displays the surface density of satellite C at different moments in it's orbital movement. 
 Left column corresponds to objects with an original cusp coefficient equal to 1.0. 
 Right column corresponds to objects with a cusp coefficient equal to 0.6. }
\end{figure}

Moreover in order for the velocity ratio of a tidally affected object to reach the $0.5$ threshold for a spheroidal 
it must lose more than 25$\%$ of its original stellar mass (figure \ref{fig:figure17}). The total fractional mass loss was 1/3 at a minimum 
(in the cuspy version of model satellite D). This occurs for both cuspy and core variants, and implies a prominent tidal stream has to form 
if the object is transformed into a dSph. Detecting tidal features or tails might be difficult though because, as our simulations show, 
tails and streams show up at a surface density that is between $10^3$ and $10^4$ times lower than the central
surface density of the object (\ref{fig:figure14}, \ref{fig:figure21}), which corresponds to 7-10 magnitudes lower relative to the central surface brightness. 
The lowest values apply to cuspy models in which stellar removal is less efficient (figure \ref{fig:figure14} and \ref{fig:figure21}). 
This confirms earlier results of \cite{2002MNRAS.336..119M}. The transition to a marginally bound tidal tail and fully unbound stream
following the orbital path is marked by the flattening of the profile at large radii, typically occurring at 3-6 kpc, or 3-6 times 
the half-light radius (figures \ref{fig:figure14} and \ref{fig:figure21}), confirming past work with a variety of simulation techniques and models 
(eg \cite{1999MNRAS.302..771J}; \cite{2002MNRAS.336..119M}). This means a very wide field of view
would be needed to unequivocally detect the tidal debris surrounding the dwarf in addition to very deep photometry. 
The actual shape of the stellar profile varies also with projection and over time (figure \ref{fig:figure14} ), but the basic facts just highlighted hold. 
We also notice that in some cases the change of slope of the surface density profile between the inner bound and outer unbound stellar component
is more evident in the cuspy variant, probably because in the cored models tidal mass loss occurs all the way to the very center, leading
to more continuous removal of stars as a function of radius.

The evolution of the observable structural parameters for the simulated objects shown in figures 
\ref{fig:figure18}, \ref{fig:figure19} and figure \ref{fig:figure20} also highlights
that cored models can evolve enough in stellar mass and radius to end up in the region populated by Ultra Faint Dwarfs (UFDs), 
confirming a previous claim of \cite{2012ApJ...751L..15L} obtained with idealized simulations.  
In general the objects evolve from the proximity of the locus populated by dwarf irregulars towards the locus of ultra-faint dwarf spheroidals 
by passing through the classical dwarf spheroidal locus. 
Except for the satellite A pair all galaxies can be found in the locus of classical dwarf spheroidals and the transition region between the 
faintest irregular dwarf galaxies and spheroidals. Rather than being necessarily reionization fossils, as it has been claimed in the literature
(\cite{2010AdAst2010E..33R}), some UFDs might be thus just more extreme outcomes of tidal stirring. 
This scenario has the attractive aspect of allowing to explain naturally why there
is a substantial continuum between the properties of classical dSphs and those of UFDs.
Such a population of UFDs would have cored profiles, while on the contrary a population of pristine
UFDs formed by gas collapse in mini-halos would likely maintain its original cuspy profile as feedback
from star formation would have a negligible effect (\cite{2012ApJ...759L..42P}).
 
Despite the low final stellar mass such a population would yield no tension with the proposed core formation mechanism via feedback (\cite{2012ApJ...759L..42P}) because it would be the remains of a population of much more massive dwarf galaxies that were severely affected by the tides of the host. Indeed
among our remnants the UFD-like object originates from one of the most massive infalling dwarfs,
whose initial luminosity would be comparable to Fornax, exactly the mass scale at which the effect of 
feedback should be important (\cite{2012MNRAS.422.1231G}; \cite{2014ApJ...792...99S}). Finding UFDs with cored
profiles is thus a potential test of our scenario. UFDs forming from an early infall population of
satellites would also be consistent with the notion that these galaxies typically lack an intermediate
or young population (\cite{2012ApJ...753L..21B}) as star formation would be truncated soon after infall as gas is rapidly removed
by tides and ram pressure when the cosmic ionizing background is still at its peak amplitude (\cite{2007Natur.445..738M}).

The smallest two objects replaced merge while orbiting the host galaxy at significantly larger pericenter
and apocenter distances in comparison to the other satellites. The effect of the tides
is minimal with insignificant stellar mass loss and small effects in the shape and velocity ratios for the merged objects with $\gamma$ = 1.0.
In the case of the $\gamma$ = 0.6 mergers the velocity ratio further decreases due to the gravitational interaction with the host.
If one considers the condition $v_{rot}$/$\sigma_{\star} < 0.5$ as the definition of a dwarf spheroidal, then the merging pair
with $\gamma$ = 0.6 clearly reaches this state at the end.
At the same time the merger remnant of galaxies with $\gamma$ = 1.0 fails
the velocity criterion, although the later appears shapewise to be closer to a spheroid than the former.
The formation of a significantly round object albeit with rotation still dynamically important
cannot be caused by tidal stirring only since in that case the transformation of the initial disky shape
is correlated with the redistribution of angular momentum due to tidally induced instabilities and tidal heating.

Hence finding a dwarf with photometric properties analogous to those of classical dSphs but with a significant
rotational velocity might be the signature of dwarf-dwarf mergers. Indeed low luminosity ellipticals, whose kinematics is reproduced
well by mergers of disks with intermediate masses, do retain significant rotation (\cite{2011MNRAS.416.1680C}).
One of such rare objects might indeed be the Tucana  dSph, which is on a very wide trajectory around the primary and exhibits appreciable rotation
(\cite{2009A&A...499..121F}).

Figure \ref{fig:figure14} also shows that the change of slope in the case of the merger remnant occurs much more gradually,
with a clear flattening not appearing before 6-7 kpc from the center.
This is not surprising as in in this case the morphological evolution is not accompanied by significant tidal mass removal.
The different shape of the profile relative to tidally stirred remnants is very interesting because it hints at a further
possible way to distinguish between a dSph formed by tidal stirring and one formed via dwarf galaxies mergers.
The presence of a swift change from one exponential to another should be the signature of a tidally affected stellar system.

\section{Conclusion and Caveats}

To the authors' knowledge the current work addresses for the first time  the importance of the dark matter density
 profile for the dynamical evolution and morphological transformation of disky satellites into spheroidals in
a cosmological context. Therefore the work bridges the gap between detailed simulations of dwarf-host interaction and fully cosmological
hydrodynamical simulations of MW-sized galaxies which do not have at the moment the resolution to reliably address this subject.
This is an important step forward since the assembly history of the host, the infall orbits of the satellites and
the structure of the potential well of the host, which all concur to determine the effect of tidal shocks, are all
self-consistently taken into account. As a by product, dwarf-dwarf interactions are also take into account, and
we witnessed a particular example of an interaction which gave rise to a merger.

An important overall conclusion of this work is that
tidal stirring is confirmed to be an effective mechanism for transforming disky field dwarfs into objects with 
properties resembling dwarf spheroidals and even
Ultra Faint Dwarfs (UFDs). Most importantly, it now appears clear that the morphological transformation induced
by tidal stirring is a natural and inevitable consequence of hierarchical structure formation.
The transformation is enhanced when the satellites have shallow dark matter density profiles,
as in our cored models, confirming previous results based on idealized simulations of individual dwarf-host
interaction (\cite{2013ApJ...764L..29K}). 
In the latter case, both tidal mass loss of the stellar component as well as tidal heating and internal redistribution are augmented
as the response of the galaxy is impulsive down to its central regions. As a consequence remnants can become as faint
and puny as UFDs, which does not happen among the set of cuspy models.

We also found that major mergers of satellite galaxies with shallow density profiles can effectively evolve into spheroidals on wide orbits.  Since the merger happens before the satellites are accreted by the host,
when they have lower velocities,it naturally leads
to a dwarf spheroidal orbiting far from the center of the host. Especially in the cuspy case the remnant retains
significant rotation despite achieving a spheroidal shape, which would not happen in tidally stirred remnants
This is a plausible scenario for the origin
of distant dSphs such as Tucana and Cetus, and confirms a previous claim based on idealized simulations (\cite{2011ApJ...726...98K}).

Another important conclusion of this work is the identification of two signatures that could separate dwarf spheroidals 
formed due to a merger on large orbits from the ones formed due to the strong tides generated by the host galaxy. 
The first signature is the presence of a tidal stellar stream with a total mass comparable to the associated satellite. 
The second signature can be found in the stellar density/surface brightness profile of the object. 
It should appear as a fast transition on lengths of a few hundreds parsecs from the steep central exponential to a significantly shallower one.   
The objects that are not the source of such a stream or do not have a fast departure from the central brightness exponential decay most probably 
did not become spheroidal due to the gravity of their current host. They were probably formed through the merger of comparable galaxies.

There are of course important caveats in the simulations, which primarily stem from the fact that they are carried out in the
dissipationless approximation, namely without treating hydrodynamics, star formation and feedback.
One caveat is the inexistence of the baryonic disk of the primary. This would increase significantly the central density
of the host, within 10-20 kpc, enhancing tidal mass loss and tidal stirring for satellites with orbits having pericenters
smaller than $\sim 10-20$ kpc \cite{2010ApJ...709.1138D}. Given the orbits of the satellites analysed in this paper, this effect
would play a role at least for some of them (eg model A and model B, which have the smallest pericenters).
The neglected effect should enhance the tidal transformation, but may lead to complete tidal disruption of objects
that would have otherwise ended up as UFDs, or produce UFDs out of cuspy progenitors (\cite{2010MNRAS.406.1290P}).
Another caveat is the lack of a a gaseous component in the
disky dwarf progenitors, but previous work has shown that this would be readily removed by ram pressure in the galactic diffuse corona
over 1-3 Gyr for the mass range considered here ($V_{max} < 45$ km/s), especially at  $z >1$ ,  when the cosmic photoionizing background
maintains the ISM warm and loosely bound, thus suppressing further star formation even for the gas that has not been removed yet
(\cite{2010HiA....15..193M}). 

Furthermore, our modelling approach relies on the assumption that the progenitors of dwarf spheroidal satellites started out their life as disky dwarf galaxies. Cosmological hydrodynamical simulations show that this is a well justified assumption for dwarf galaxies forming in halos in the range $10^9-10^{10} M_{\odot}$ (\cite{2010Natur.463..203G}; \cite{2014ApJ...792...99S}), while at lower masses field dwarfs can have a rather thick and turbulent disk, with a ratio $v_{rot}/\sigma_{*} \sim 1$ (Shen et al., in preparation),
which appears consistent with the kinematics of the faintest dIrrs in the Local Group (\cite{1998ARA&A..36..435M}; \cite{2012AJ....144....4M}).
A thicker, kinematically hotter disk could have a different response to tidal shocks. On the one hand the triggering of
non-axisymmetric instabilities, such as bars and spiral arms, would be weaker due to the reduced self-gravity, on the
other hand the disk binding energy will be lower, so that the direct effect of tidal heating could be stronger.
\cite{2011ApJ...726...98K} have considered thicker disks in the wide range of initial conditions they explored in their
tidal stirring simulations, finding small differences in the transformation as the initial disk thickness was varied
by a factor 3, with only a slight tendency of thicker disks to transition faster to a nearly isotropic configuration.
However, future cosmological hydrodynamical simulations with enough resolution to study
the morphological evolution of satellites with the same level of detail as done here will be required to
assess the impact of disk thickness and kinematics, of both gas and stars, on the efficiency of  the transformation.

Finally, here we have considered a relatively small sample of satellites as we focused on the earliest infalling objects with
a sizable mass, large enough to host luminous dwarfs and be sufficiently well resolved. The advantage of the small sample
is that we could study in detail the different dynamical evolution between cuspy and cored variants of the same object.
In addition, since these early infallers are exposed for the maximum amount of time possible to the tides of the primary we
can safely assume that the results provide an upper limit on how deeply the structure of the dwarfs can be modified
by tidal shocks in cored versus cuspy variants. As a result, we inferred that the population of satellites of the Milky Way
should be dominated by remnants of cored objects otherwise we would see a population of dwarfs with intermediate properties
between dwarf irregulars and dwarf spheroidals at $\sim 100-200$ kpc galactocentric distances, which is not seen neither
in the MW or the M31 halo.  In a forthcoming paper we will study a
much larger sample of satellites which includes objects that are accreted at low redshifts. The size of the sample that we will
present will be of the same order of the number of luminous satellites known so far to lie within the putative virial radius
of the Milky Way, thus allowing to make statistically meaningful statements on the evolution of the entire population of 
satellites.

\acknowledgments
The first author is grateful to Annalisa Pillepich, Davide Fiacconi, Michela Mapelli, Doug Potter, Stelios Kazantzidis and Rok Ro$\check{s}$kar for their help.
Funding for the work was provided by the Institutes for Theoretical Physics and Computational Science at the University of Zurich.

%%%%%%%%%%%%%%%%%
%%TABLES%%%%%%%%%
%%%%%%%%%%%%%%%%%

\begin{deluxetable}{l c c c}
\tablecaption{\label{table:table1} The virial quantities of the replaced halos and their distance from the center of the host.}
\tablehead{
    \colhead{Satellite} & \colhead{$M_{vir}$ [$10^9 M_{\odot}$]} & \colhead{$R_{vir} [kpc]$} & \colhead{ $d_{host} [kpc]$ }\\ 
          }
    \startdata
    A & 3.83 & 13.0 & 140\\ 
    B & 3.04 & 12.0 & 75 \\ 
    C & 1.51 & 9.5  & 82 \\ 
    D & 0.94 & 8.1  & 157\\ 
    E & 0.40 & 6.1  & 165\\ 
    F & 0.33 & 5.7  & 159
    \enddata
\end{deluxetable}

\clearpage

\begin{deluxetable}{l c c c c c c }

\tablecaption{\label{table:table2}Scale quantities and concentrations of the different halos and the properties of their associated stellar disks.}
\tablehead{
	\colhead{Satellite} & \colhead{A} & \colhead{B} & \colhead{C} & \colhead{D} & \colhead{E} & \colhead{F}
	      }
	      
	\startdata
	
    c & 6.0  & 6.1 & 6.3 & 6.4 & 6.8  & 6.8\\ 
    $r_s$ [kpc] & 2.18  & 2.00 & 1.53 & 1.27 & 0.91  & 0.85\\ 
    $\sigma_s^{\gamma = 1.0} [km s^{-1}]$ & 88.4  & 82.0 & 65.2 & 55.9 & 42.4  & 40.0\\ 
    $\sigma_s^{\gamma = 0.6} [km s^{-1}]$ & 84.7  & 78.6 & 62.4 & 53.4 & 40.4  & 38.0\\ 
    $M_{stars}$ [$10^6 M_{\odot}$]        & 4.50  & 3.57 & 1.75 & 1.08 & 0.44  & 0.37\\ 
    $R_{d} [pc]$ & 80  & 80 & 80 & 80 & 80  & 80  \\
	$h_{d} [pc]$ & 10  & 10 & 10 & 10 & 10  & 10  
    \enddata

\end{deluxetable}

\clearpage

\begin{deluxetable}{l c c c c}
\tablecaption{\label{table:table3}Masses and force resolutions of dark matter and stellar particles.}
\tablehead{
		\colhead{Satellite} & \colhead{$M_{sp}$ $[M_{sol}]$} & \colhead{$M_{dmp} [M_{sol}]$} & \colhead{$\epsilon_{sp} [pc]$} & \colhead{$\epsilon_{dmp} [pc]$}
          }
	\startdata
    A & 4.50  & 3830 & 9.6 & 93.8\\ 
    B & 3.57  & 3040 & 8.9 & 86.9\\ 
    C & 1.75  & 1510 & 7.1 &68.8\\ 
    D & 1.08  & 940 & 6.0 & 58.8\\ 
    E & 0.44  & 400 & 4.5 & 44.2\\ 
    F & 0.37  & 330 & 4.2 & 41.6
    \enddata
\end{deluxetable}

\clearpage

\begin{deluxetable}{ l c c c c}
\tablecaption{\label{table:table4}Bound stellar and total masses inside the 500 pc radius spheres centered on the density peaks of the satellite galaxies with steep (left) and shallow (right) density profiles at z = 2.87.}

\tablehead{ \colhead{Satellite} & \colhead{$M_{500} [10^7 M_{\odot}]$} & \colhead{$M_{500}^{\star} [10^5 M_{\odot}]$} & \colhead{ $M_{500} [10^7 M_{\odot}]$} & \colhead{$M_{500}^{\star} [10^5 M_{\odot}]$}
		 }
		\startdata
         A & 8.75 & 6.26 & 4.32 & 6.09 \\
         B & 7.98 & 5.91 & 3.99 & 5.82 \\
         C & 6.08 & 4.88 & 3.37 & 4.80 \\
         D & 4.96 & 4.06 & 2.91 & 4.09 \\
         E & 3.44 & 2.67 & 2.21 & 2.69 \\
	 	 F & 3.17 & 2.40 & 2.06 & 2.40 
        \enddata
\end{deluxetable}

\clearpage

\begin{deluxetable}{l c c c c c}
\tablecaption{\label{table:table5} Characteristic masses and mass ratios at last epoch for the galaxies with original $\gamma$ = 1.0 (upper rows) and $\gamma$ = 0.6 (lower rows)}
	 
	\tablehead{ \colhead{Satellite} & \colhead{A} & \colhead{B} & \colhead{C} & \colhead{D} & \colhead{EF} 
     }
     
     \startdata
		
	 $M_{500} [10^7 M_{\odot}]$             & 0.64 & 3.62 & 3.21 &  3.07 & 4.44\\
	 $M_{500}^{\star} [10^5 M_{\odot}]$     & 0.25 & 3.49 & 3.23 &  3.37 & 3.50\\
	 $M_{500}^{total}/M_{500}^{\star} $   & 256  & 104  & 99   &  91   & 127\\
	 $M_{bound}^{total}/M_{bound}^{\star}$& 238  & 102  & 112  &  128  & 370\\
	 			\\
	 $M_{500} [10^7 M_{\odot}]$             & 0.14 & 0.39 & 0.60 &  0.71 & 2.61\\
	 $M_{500}^{\star} [10^5 M_{\odot}]$     & 0.06 & 0.72 & 1.42 &  1.75 & 3.42\\
	 $M_{500}^{total}/M_{500}^{\star} $   & 234  & 54   & 42   &  41   & 76\\
	 $M_{bound}^{total}/M_{bound}^{\star}$& 234  & 39   & 42   &  47   & 312
	 
	 \enddata

\end{deluxetable}

\clearpage

\begin{deluxetable}{ lccccc}
\tablecaption{\label{table:table6} Characteristic velocities and velocity ratio at the last epoch for the galaxies with original $\gamma$ = 1.0 (upper rows) and $\gamma$ = 0.6 (lower rows)}
			\tablehead{ \colhead{Satellite} & \colhead{A} & \colhead{B} & \colhead{C} & 				\colhead{D} & \colhead{EF} 
                      }
	 	\startdata
		$\sigma_{\star} [km/s]$        & 6.20  & 11.04 & 9.88 & 10.01 & 11.46\\
		$v_{circ} [km/s]$              & 9.90  & 18.85 & 17.33& 16.99 & 20.96\\
		$\sigma_{\star} $ / $v_{circ}$ & 0.63  & 0.59  & 0.57 & 0.59  & 0.55\\
                      \\
		$\sigma_{\star} [km/s]$        & -  & 3.52 & 4.54 & 4.92 & 8.95\\
		$v_{circ} [km/s]$              & -  & 6.0  & 8.0  & 8.1  & 17.8\\
		$\sigma_{\star} $ / $v_{circ}$ & -  & 0.59 & 0.57 & 0.61 & 0.50
		\enddata

\end{deluxetable}

\clearpage

%%%%%%%%%%%%%%%%%%%%%%%%%%%%%
%%%%FIGURES%%%%%%%%%%%%%%%%%%
%%%%%%%%%%%%%%%%%%%%%%%%%%%%%

\clearpage

\clearpage

%%%%%%%%%%%%%%%%%%%%%%%%

\clearpage

\clearpage

\clearpage
%%%%%%%%%%%%%%%%%%%%%%%

\clearpage
%%%%%%%%%%%%%%%%%%%%%%%%%%%%

\clearpage
%%%%%%%%%%%%%%%%%%%%%%%%%%%%%

\clearpage

%%%%%%%%%%%%%%%%%%%%%%%%%%%%%%%%

%%%%%%%%%%%%%%%%%%%%%%%%%%%%%%%%

\clearpage

\clearpage
%%%%%%%%%%%%%%%%%%%%%%%%%%%%%%%%%%

\clearpage

\clearpage

%%%%%%%%%%%%%%%%%%%%%%%%%%%%%%%%%%%%%%%%%

\clearpage

\clearpage

\clearpage

%%%%%%%%%%%%%%%%%%%%%%%%%%%%%%%%%%%%%%%%%%%%%%

\clearpage

\clearpage

\clearpage

\clearpage

\clearpage

%%%%%%%%%%%%%%%%%%%%%%%%%%%%%%%%%%%%%%%%%%%%%%

\clearpage
%%%%%%%%%%%%%%%%%%%%%%%%%%%%%%%%%%:%%%%%%%%%%%%%

\def\apj{ApJ}
\def\apjl{ApJL}
\def\aj{AJ}
\def\mnras{MNRAS}
\def\aap{A\&A}
\def\nat{nature}
\def\araa{ARAA}
\def\pasa{PASA}
\bibliography{ads}

\begin{thebibliography}{}
\expandafter\ifx\csname natexlab\endcsname\relax\def\natexlab#1{#1}\fi

\bibitem[{{Arvo}(1991)}]{arvo1991}
{Arvo}, J. 1991, Graphics Gems II, 355

\bibitem[{{Behroozi} {et~al.}(2013){Behroozi}, {Wechsler}, \&
  {Conroy}}]{2013ApJ...770...57B}
{Behroozi}, P.~S., {Wechsler}, R.~H., \& {Conroy}, C. 2013, \apj, 770, 57

\bibitem[{{Bird} {et~al.}(2013){Bird}, {Kazantzidis}, {Weinberg}, {Guedes},
  {Callegari}, {Mayer}, \& {Madau}}]{2013ApJ...773...43B}
{Bird}, J.~C., {Kazantzidis}, S., {Weinberg}, D.~H., {et~al.} 2013, \apj, 773,
  43

\bibitem[{{Brooks} {et~al.}(2013){Brooks}, {Kuhlen}, {Zolotov}, \&
  {Hooper}}]{2013ApJ...765...22B}
{Brooks}, A.~M., {Kuhlen}, M., {Zolotov}, A., \& {Hooper}, D. 2013, \apj, 765,
  22

\bibitem[{{Brown} {et~al.}(2012){Brown}, {Tumlinson}, {Geha}, {Kirby},
  {VandenBerg}, {Mu{\~n}oz}, {Kalirai}, {Simon}, {Avila}, {Guhathakurta},
  {Renzini}, \& {Ferguson}}]{2012ApJ...753L..21B}
{Brown}, T.~M., {Tumlinson}, J., {Geha}, M., {et~al.} 2012, \apjl, 753, L21

\bibitem[{{Bryan} \& {Norman}(1998)}]{1998ApJ...495...80B}
{Bryan}, G.~L., \& {Norman}, M.~L. 1998, \apj, 495, 80

\bibitem[{{Cappellari} {et~al.}(2011){Cappellari}, {Emsellem}, {Krajnovi{\'c}},
  {McDermid}, {Serra}, {Alatalo}, {Blitz}, {Bois}, {Bournaud}, {Bureau},
  {Davies}, {Davis}, {de Zeeuw}, {Khochfar}, {Kuntschner}, {Lablanche},
  {Morganti}, {Naab}, {Oosterloo}, {Sarzi}, {Scott}, {Weijmans}, \&
  {Young}}]{2011MNRAS.416.1680C}
{Cappellari}, M., {Emsellem}, E., {Krajnovi{\'c}}, D., {et~al.} 2011, \mnras,
  416, 1680

\bibitem[{{Colpi} {et~al.}(1999){Colpi}, {Mayer}, \&
  {Governato}}]{1999ApJ...525..720C}
{Colpi}, M., {Mayer}, L., \& {Governato}, F. 1999, \apj, 525, 720

\bibitem[{{Di Cintio} {et~al.}(2013){Di Cintio}, {Knebe}, {Libeskind}, {Brook},
  {Yepes}, {Gottl{\"o}ber}, \& {Hoffman}}]{2013MNRAS.431.1220D}
{Di Cintio}, A., {Knebe}, A., {Libeskind}, N.~I., {et~al.} 2013, \mnras, 431,
  1220

\bibitem[{{Diemand} {et~al.}(2007){Diemand}, {Kuhlen}, \&
  {Madau}}]{2007ApJ...667..859D}
{Diemand}, J., {Kuhlen}, M., \& {Madau}, P. 2007, \apj, 667, 859

\bibitem[{{D'Onghia} {et~al.}(2010){D'Onghia}, {Springel}, {Hernquist}, \&
  {Keres}}]{2010ApJ...709.1138D}
{D'Onghia}, E., {Springel}, V., {Hernquist}, L., \& {Keres}, D. 2010, \apj,
  709, 1138

\bibitem[{{Dressler}(1980)}]{1980ApJ...236..351D}
{Dressler}, A. 1980, \apj, 236, 351

\bibitem[{{Dubinski}(1999)}]{1999ASPC..182..491D}
{Dubinski}, J. 1999, in Astronomical Society of the Pacific Conference Series,
  Vol. 182, Galaxy Dynamics - A Rutgers Symposium, ed. D.~R. {Merritt},
  M.~{Valluri}, \& J.~A. {Sellwood}, 491

\bibitem[{{Fraternali} {et~al.}(2009){Fraternali}, {Tolstoy}, {Irwin}, \&
  {Cole}}]{2009A&A...499..121F}
{Fraternali}, F., {Tolstoy}, E., {Irwin}, M.~J., \& {Cole}, A.~A. 2009, \aap,
  499, 121

\bibitem[{{Gill} {et~al.}(2004){Gill}, {Knebe}, {Gibson}, \&
  {Dopita}}]{2004MNRAS.351..410G}
{Gill}, S.~P.~D., {Knebe}, A., {Gibson}, B.~K., \& {Dopita}, M.~A. 2004,
  \mnras, 351, 410

\bibitem[{{Governato} {et~al.}(2010){Governato}, {Brook}, {Mayer}, {Brooks},
  {Rhee}, {Wadsley}, {Jonsson}, {Willman}, {Stinson}, {Quinn}, \&
  {Madau}}]{2010Natur.463..203G}
{Governato}, F., {Brook}, C., {Mayer}, L., {et~al.} 2010, \nat, 463, 203

\bibitem[{{Governato} {et~al.}(2012){Governato}, {Zolotov}, {Pontzen},
  {Christensen}, {Oh}, {Brooks}, {Quinn}, {Shen}, \&
  {Wadsley}}]{2012MNRAS.422.1231G}
{Governato}, F., {Zolotov}, A., {Pontzen}, A., {et~al.} 2012, \mnras, 422, 1231

\bibitem[{{Grebel} {et~al.}(1999){Grebel}, {Seitzer}, {Dolphin}, {Geisler},
  {Guhathakurta}, {Hodge}, {Karachentsev}, {Karachentseva}, {Sarajedini}, \&
  {Sharina}}]{1999AAS...195.0803G}
{Grebel}, E.~K., {Seitzer}, P., {Dolphin}, A.~E., {et~al.} 1999, in Bulletin of
  the American Astronomical Society, Vol.~31, American Astronomical Society
  Meeting Abstracts, 1380

\bibitem[{{Guedes} {et~al.}(2011){Guedes}, {Callegari}, {Madau}, \&
  {Mayer}}]{2011ApJ...742...76G}
{Guedes}, J., {Callegari}, S., {Madau}, P., \& {Mayer}, L. 2011, \apj, 742, 76

\bibitem[{{Guedes} {et~al.}(2013){Guedes}, {Mayer}, {Carollo}, \&
  {Madau}}]{2013ApJ...772...36G}
{Guedes}, J., {Mayer}, L., {Carollo}, M., \& {Madau}, P. 2013, \apj, 772, 36

\bibitem[{{Jetley} {et~al.}(2008){Jetley}, F., {Mendes}, {Kale}, \&
  {Quinn}}]{changa2008}
{Jetley}, P., F., G., {Mendes}, C., {Kale}, L., \& {Quinn}, T. 2008, IEEE
  International Parallel and Distributed Processing Symposium, 1

\bibitem[{{Jetley} {et~al.}(2010){Jetley}, F., {Mendes}, {Kale}, \&
  {Quinn}}]{changa2010}
---. 2010, IEEE International Conference for High Performance Computing,
  Networking, Storage and Analysis (SC), 1

\bibitem[{{Johnston} {et~al.}(1999){Johnston}, {Sigurdsson}, \&
  {Hernquist}}]{1999MNRAS.302..771J}
{Johnston}, K.~V., {Sigurdsson}, S., \& {Hernquist}, L. 1999, \mnras, 302, 771

\bibitem[{Kale \& Krishnan(1996)}]{CharmppPPWCPP96}
Kale, L.~V., \& Krishnan, S. 1996, in Parallel Programming using C++, ed. G.~V.
  Wilson \& P.~Lu (MIT Press), 175--213

\bibitem[{{Kaufmann} {et~al.}(2007){Kaufmann}, {Wheeler}, \&
  {Bullock}}]{2007MNRAS.382.1187K}
{Kaufmann}, T., {Wheeler}, C., \& {Bullock}, J.~S. 2007, \mnras, 382, 1187

\bibitem[{{Kazantzidis} {et~al.}(2011{\natexlab{a}}){Kazantzidis}, {{\L}okas},
  {Callegari}, {Mayer}, \& {Moustakas}}]{2011ApJ...726...98K}
{Kazantzidis}, S., {{\L}okas}, E.~L., {Callegari}, S., {Mayer}, L., \&
  {Moustakas}, L.~A. 2011{\natexlab{a}}, \apj, 726, 98

\bibitem[{{Kazantzidis} {et~al.}(2013){Kazantzidis}, {{\L}okas}, \&
  {Mayer}}]{2013ApJ...764L..29K}
{Kazantzidis}, S., {{\L}okas}, E.~L., \& {Mayer}, L. 2013, \apjl, 764, L29

\bibitem[{{Kazantzidis} {et~al.}(2011{\natexlab{b}}){Kazantzidis}, {{\L}okas},
  {Mayer}, {Knebe}, \& {Klimentowski}}]{2011ApJ...740L..24K}
{Kazantzidis}, S., {{\L}okas}, E.~L., {Mayer}, L., {Knebe}, A., \&
  {Klimentowski}, J. 2011{\natexlab{b}}, \apjl, 740, L24

\bibitem[{{Klimentowski} {et~al.}(2007){Klimentowski}, {{\L}okas},
  {Kazantzidis}, {Prada}, {Mayer}, \& {Mamon}}]{2007MNRAS.378..353K}
{Klimentowski}, J., {{\L}okas}, E.~L., {Kazantzidis}, S., {et~al.} 2007,
  \mnras, 378, 353

\bibitem[{{Klypin} {et~al.}(2011){Klypin}, {Trujillo-Gomez}, \&
  {Primack}}]{2011ApJ...740..102K}
{Klypin}, A.~A., {Trujillo-Gomez}, S., \& {Primack}, J. 2011, \apj, 740, 102

\bibitem[{{Knollmann} \& {Knebe}(2009)}]{2009ApJS..182..608K}
{Knollmann}, S.~R., \& {Knebe}, A. 2009, \apjs, 182, 608

\bibitem[{{Kuhlen} {et~al.}(2013){Kuhlen}, {Guedes}, {Pillepich}, {Madau}, \&
  {Mayer}}]{2013ApJ...765...10K}
{Kuhlen}, M., {Guedes}, J., {Pillepich}, A., {Madau}, P., \& {Mayer}, L. 2013,
  \apj, 765, 10

\bibitem[{{Kuijken} \& {Dubinski}(1995)}]{1995MNRAS.277.1341K}
{Kuijken}, K., \& {Dubinski}, J. 1995, \mnras, 277, 1341

\bibitem[{{Lisker} {et~al.}(2006){Lisker}, {Grebel}, \&
  {Binggeli}}]{2006AJ....132..497L}
{Lisker}, T., {Grebel}, E.~K., \& {Binggeli}, B. 2006, \aj, 132, 497

\bibitem[{{{\L}okas} {et~al.}(2012){{\L}okas}, {Kazantzidis}, \&
  {Mayer}}]{2012ApJ...751L..15L}
{{\L}okas}, E.~L., {Kazantzidis}, S., \& {Mayer}, L. 2012, \apjl, 751, L15

\bibitem[{{Madau} {et~al.}(2014){Madau}, {Weisz}, \&
  {Conroy}}]{2014ApJ...790L..17M}
{Madau}, P., {Weisz}, D.~R., \& {Conroy}, C. 2014, \apjl, 790, L17

\bibitem[{{Mastropietro} {et~al.}(2005){Mastropietro}, {Moore}, {Mayer},
  {Debattista}, {Piffaretti}, \& {Stadel}}]{2005MNRAS.364..607M}
{Mastropietro}, C., {Moore}, B., {Mayer}, L., {et~al.} 2005, \mnras, 364, 607

\bibitem[{{Mateo}(1998)}]{1998ARA&A..36..435M}
{Mateo}, M.~L. 1998, \araa, 36, 435

\bibitem[{{Mayer}(2010)}]{2010HiA....15..193M}
{Mayer}, L. 2010, Highlights of Astronomy, 15, 193

\bibitem[{{Mayer}(2011)}]{2011EAS....48..369M}
{Mayer}, L. 2011, in EAS Publications Series, Vol.~48, EAS Publications Series,
  ed. M.~{Koleva}, P.~{Prugniel}, \& I.~{Vauglin}, 369--381

\bibitem[{{Mayer}(2012)}]{2012ASPC..453..289M}
{Mayer}, L. 2012, in Astronomical Society of the Pacific Conference Series,
  Vol. 453, Advances in Computational Astrophysics: Methods, Tools, and
  Outcome, ed. R.~{Capuzzo-Dolcetta}, M.~{Limongi}, \& A.~{Tornamb{\`e}}, 289

\bibitem[{{Mayer} {et~al.}(2001{\natexlab{a}}){Mayer}, {Governato}, {Colpi},
  {Moore}, {Quinn}, {Wadsley}, {Stadel}, \& {Lake}}]{2001ApJ...559..754M}
{Mayer}, L., {Governato}, F., {Colpi}, M., {et~al.} 2001{\natexlab{a}}, \apj,
  559, 754

\bibitem[{{Mayer} {et~al.}(2001{\natexlab{b}}){Mayer}, {Governato}, {Colpi},
  {Moore}, {Quinn}, {Wadsley}, {Stadel}, \& {Lake}}]{2001ApJ...547L.123M}
---. 2001{\natexlab{b}}, \apjl, 547, L123

\bibitem[{{Mayer} {et~al.}(2007){Mayer}, {Kazantzidis}, {Mastropietro}, \&
  {Wadsley}}]{2007Natur.445..738M}
{Mayer}, L., {Kazantzidis}, S., {Mastropietro}, C., \& {Wadsley}, J. 2007,
  \nat, 445, 738

\bibitem[{{Mayer} {et~al.}(2006){Mayer}, {Mastropietro}, {Wadsley}, {Stadel},
  \& {Moore}}]{2006MNRAS.369.1021M}
{Mayer}, L., {Mastropietro}, C., {Wadsley}, J., {Stadel}, J., \& {Moore}, B.
  2006, \mnras, 369, 1021

\bibitem[{{Mayer} {et~al.}(2002){Mayer}, {Moore}, {Quinn}, {Governato}, \&
  {Stadel}}]{2002MNRAS.336..119M}
{Mayer}, L., {Moore}, B., {Quinn}, T., {Governato}, F., \& {Stadel}, J. 2002,
  \mnras, 336, 119

\bibitem[{{McConnachie}(2012)}]{2012AJ....144....4M}
{McConnachie}, A.~W. 2012, \aj, 144, 4

\bibitem[{{Menon} {et~al.}(2015){Menon}, {Wesolowski}, {Zheng}, {Jetley},
  {Kale}, {Quinn}, \& {Governato}}]{2015ComAC...2....1M}
{Menon}, H., {Wesolowski}, L., {Zheng}, G., {et~al.} 2015, Computational
  Astrophysics and Cosmology, 2, 1

\bibitem[{{Moore} {et~al.}(1996){Moore}, {Katz}, \&
  {Lake}}]{1996ApJ...457..455M}
{Moore}, B., {Katz}, N., \& {Lake}, G. 1996, \apj, 457, 455

\bibitem[{{O{\~n}orbe} {et~al.}(2015){O{\~n}orbe}, {Boylan-Kolchin}, {Bullock},
  {Hopkins}, {Ker{\v e}s}, {Faucher-Gigu{\`e}re}, {Quataert}, \&
  {Murray}}]{2015arXiv150202036O}
{O{\~n}orbe}, J., {Boylan-Kolchin}, M., {Bullock}, J.~S., {et~al.} 2015, ArXiv
  e-prints, arXiv:1502.02036

\bibitem[{{Pe{\~n}arrubia} {et~al.}(2010){Pe{\~n}arrubia}, {Benson}, {Walker},
  {Gilmore}, {McConnachie}, \& {Mayer}}]{2010MNRAS.406.1290P}
{Pe{\~n}arrubia}, J., {Benson}, A.~J., {Walker}, M.~G., {et~al.} 2010, \mnras,
  406, 1290

\bibitem[{{Pe{\~n}arrubia} {et~al.}(2012){Pe{\~n}arrubia}, {Pontzen}, {Walker},
  \& {Koposov}}]{2012ApJ...759L..42P}
{Pe{\~n}arrubia}, J., {Pontzen}, A., {Walker}, M.~G., \& {Koposov}, S.~E. 2012,
  \apjl, 759, L42

\bibitem[{{Pillepich} {et~al.}(2014){Pillepich}, {Kuhlen}, {Guedes}, \&
  {Madau}}]{2014ApJ...784..161P}
{Pillepich}, A., {Kuhlen}, M., {Guedes}, J., \& {Madau}, P. 2014, \apj, 784,
  161

\bibitem[{{Pontzen} \& {Governato}(2012)}]{2012MNRAS.421.3464P}
{Pontzen}, A., \& {Governato}, F. 2012, \mnras, 421, 3464

\bibitem[{{Rashkov} {et~al.}(2013){Rashkov}, {Pillepich}, {Deason}, {Madau},
  {Rockosi}, {Guedes}, \& {Mayer}}]{2013ApJ...773L..32R}
{Rashkov}, V., {Pillepich}, A., {Deason}, A.~J., {et~al.} 2013, \apjl, 773, L32

\bibitem[{{Ricotti}(2010)}]{2010AdAst2010E..33R}
{Ricotti}, M. 2010, Advances in Astronomy, 2010, 33

\bibitem[{{Shen} {et~al.}(2014){Shen}, {Madau}, {Conroy}, {Governato}, \&
  {Mayer}}]{2014ApJ...792...99S}
{Shen}, S., {Madau}, P., {Conroy}, C., {Governato}, F., \& {Mayer}, L. 2014,
  \apj, 792, 99

\bibitem[{{Skillman} {et~al.}(2014){Skillman}, {Hidalgo}, {Weisz}, {Monelli},
  {Gallart}, {Aparicio}, {Bernard}, {Boylan-Kolchin}, {Cassisi}, {Cole},
  {Dolphin}, {Ferguson}, {Mayer}, {Navarro}, {Stetson}, \&
  {Tolstoy}}]{2014ApJ...786...44S}
{Skillman}, E.~D., {Hidalgo}, S.~L., {Weisz}, D.~R., {et~al.} 2014, \apj, 786,
  44

\bibitem[{{Stadel}(2001)}]{2001PhDT........21S}
{Stadel}, J.~G. 2001, PhD thesis, UNIVERSITY OF WASHINGTON

\bibitem[{{Swaters} {et~al.}(2009){Swaters}, {Sancisi}, {van Albada}, \& {van
  der Hulst}}]{2009A&A...493..871S}
{Swaters}, R.~A., {Sancisi}, R., {van Albada}, T.~S., \& {van der Hulst}, J.~M.
  2009, \aap, 493, 871

\bibitem[{{Trujillo-Gomez} {et~al.}(2015){Trujillo-Gomez}, {Klypin},
  {Col{\'{\i}}n}, {Ceverino}, {Arraki}, \& {Primack}}]{2015MNRAS.446.1140T}
{Trujillo-Gomez}, S., {Klypin}, A., {Col{\'{\i}}n}, P., {et~al.} 2015, \mnras,
  446, 1140

\bibitem[{{Widrow} \& {Dubinski}(2005)}]{2005ApJ...631..838W}
{Widrow}, L.~M., \& {Dubinski}, J. 2005, \apj, 631, 838

\bibitem[{{Widrow} {et~al.}(2008){Widrow}, {Pym}, \&
  {Dubinski}}]{2008ApJ...679.1239W}
{Widrow}, L.~M., {Pym}, B., \& {Dubinski}, J. 2008, \apj, 679, 1239

\end{thebibliography}

\end{document}